\documentclass{aastex631}

\usepackage{rotating}
\usepackage{soul}

\let\splitbox\relax
\usepackage{adjustbox}

\def\ltsima{$\; \buildrel < \over \sim \;$}
\def\simlt{\lower.5ex\hbox{\ltsima}}
\def\gtsima{$\; \buildrel > \over \sim \;$}
\def\simgt{\lower.5ex\hbox{\gtsima}}
\def\kpc{{\rm\,kpc}}

\def\g{{\rm\,g}}

\def \r {r$^{1/4}$ }

\def\vs{\textit{vs}\/}

\shorttitle{On the Use of Field RR Lyrae as Galactic Probes - VIII}
\shortauthors{Bono et al.}

\begin{document}

\title{On the Use of Field RR Lyrae as Galactic Probes - VIII. Early Formation of the Galactic Spheroid\footnote{Based on observations collected at the European Southern Observatory (ESO) with the ERIS spectrograph available at the Very Large Telescope (VLT), Paranal Observatory, Chile (ESO programs: 0111.D-0233(A), 0112.D-2235(A), 0113.D-2226(A), PI: G. Bono), and with WINERED spectrograph available as a visitor instrument at the ESO New Technology Telescope (NTT), La Silla, Chile (ESO program: 098.D-0878(A), PI: G. Bono).
Based in part on observations made with the Southern African Large Telescope (SALT):  
Program IDs: 2017-2-SCI-041, 2018-1-SCI-018, 2018-2-SCI-025, 2019-1-SCI-013, 2021-2-SCI-028, 2022-2-DDT-001, PI: B. Chaboyer).
Based in part on data obtained in the Observatorios de Canarias del Instituto de Astrofisica de Canarias (IAC) with: the STELLA robotic telescope, an AIP facility jointly operated by AIP and IAC at the Teide Observatory in Tenerife, Spain; the Italian Telescopio Nazionale Galileo (TNG) operated by the Fundaci{\'o}n Galileo Galilei of the INAF, the Nordic Optical Telescope (NOT) owned in collaboration by the University of Turku and Aarhus University and operated jointly by Aarhus University, the University of Turku and the University of Oslo, representing Denmark, Finland and Norway, the University of Iceland and Stockholm University, and the Mercator telescope operated by the Flemish Community, all at the Observatorio del Roque de los Muchachos of the IAC, La Palma, Spain (Program IDs: 101-MULTIPLE-4/21B, 107-MULTIPLE-4/22A, 120-MULTIPLE-2/23B, 165-Stella12/20A, PI: M. Monelli; Program ID: 108-MULTIPLE-2/25B, PI: M. S{\'a}nchez-Benavente).}}

\correspondingauthor{G. Bono}
\email{bono@roma2.infn.it}

\author[0000-0002-4896-8841]{G. Bono}
\affiliation{Dipartimento di Fisica, Università di Roma Tor Vergata, via della Ricerca Scientifica 1, I-00133 Roma, Italy}
\affiliation{INAF - Osservatorio Astronomico di Roma, via Frascati 33, I-00078 Monte Porzio Catone, Italy}

\author{V. F. Braga}
\affiliation{INAF - Osservatorio Astronomico di Roma, via Frascati 33, I-00078 Monte Porzio Catone, Italy}

\author{M. Fabrizio}
\affiliation{ASI - Space Science Data Center, via del Politecnico snc, I-00133 Roma, Italy}
\affiliation{INAF - Osservatorio Astronomico di Roma, via Frascati 33, I-00078 Monte Porzio Catone, Italy}

\author{M. Tantalo}
\affiliation{INAF - Osservatorio Astronomico di Roma, via Frascati 33, I-00078 Monte Porzio Catone, Italy}

\author{K. Baeza-Villagra}
\affiliation{Dipartimento di Fisica, Università di Roma Tor Vergata, via della Ricerca Scientifica 1, I-00133 Roma, Italy}
\affiliation{INAF - Osservatorio Astronomico di Roma, via Frascati 33, I-00078 Monte Porzio Catone, Italy}

\author{J. Crestani}
\affiliation{Dipartimento di Fisica, Università di Roma Tor Vergata, via della Ricerca Scientifica 1, I-00133 Roma, Italy}

\author{V. D'Orazi}
\affiliation{Dipartimento di Fisica, Università di Roma Tor Vergata, via della Ricerca Scientifica 1, I-00133 Roma, Italy}
\affiliation{INAF - Osservatorio Astronomico di Padova, vicolo dell'Osservatorio 5, I-35122 Padova, Italy}

\author{M. Dall'Ora}
\affiliation{INAF - Osservatorio Astronomico di Capodimonte, Salita Moiariello 16, I-80131 Napoli, Italy}

\author{M. Di Criscienzo}
\affiliation{INAF - Osservatorio Astronomico di Roma, via Frascati 33, I-00078 Monte Porzio Catone, Italy}

\author{G. Fiorentino}
\affiliation{INAF - Osservatorio Astronomico di Roma, via Frascati 33, I-00078 Monte Porzio Catone, Italy}

\author{M. Gholami}
\affiliation{INAF - Osservatorio Astronomico di Capodimonte, Salita Moiariello 16, I-80131 Napoli, Italy}

\author{M. Marengo}
\affiliation{Florida State University, Department of Physics, 77 Chieftain Way, Tallahassee, FL 32306, USA}

\author{C. E. Mart{\'{\i}}nez-V{\'a}zquez}
\affiliation{NSF NOIRLab, 670 N. A'ohoku Place, Hilo, Hawai'i 96720, USA}

\author{M. Monelli}
\affiliation{INAF - Osservatorio Astronomico di Roma, via Frascati 33, I-00078 Monte Porzio Catone, Italy}
\affiliation{Instituto de Astrofísica de Canarias, Calle Via Lactea s/n, E-38205 La Laguna, Tenerife, Spain}
\affiliation{Departamento de Astrofísica, Universidad de La Laguna (ULL), E-38200, La Laguna, Tenerife, Spain}

\author{J. P. Mullen}
\affiliation{Department of Physics and Astronomy, Vanderbilt University, Nashville, TN 37240, USA} 

\author{A. Nunnari}
\affiliation{Dipartimento di Fisica, Università di Roma Tor Vergata, via della Ricerca Scientifica 1, I-00133 Roma, Italy}
\affiliation{INAF - Osservatorio Astronomico di Roma, via Frascati 33, I-00078 Monte Porzio Catone, Italy}

\author{V. D. Pipwala}
\affiliation{Dipartimento di Fisica, Università di Roma Tor Vergata, via della Ricerca Scientifica 1, I-00133 Roma, Italy}
\affiliation{INAF - Osservatorio Astronomico di Roma, via Frascati 33, I-00078 Monte Porzio Catone, Italy}

\author{Z. Prudil}
\affiliation{European Southern Observatory, Karl-Schwarzschild-Straße 2, D-85748 Garching, Germany}

\author{C. Sneden}
\affiliation{Department of Astronomy and McDonald Observatory, The University of Texas, Austin, TX 78712, USA}

\author{G. Altavilla}
\affiliation{INAF - Osservatorio Astronomico di Roma, via Frascati 33, I-00078 Monte Porzio Catone, Italy}
\affiliation{ASI - Space Science Data Center, via del Politecnico snc, I-00133 Roma, Italy}

\author{M. Bergemann}
\affiliation{Max-Planck-Institut f{\"u}r Astronomie, K{\"o}nigstuhl 17, D-69117 Heidelberg, Germany}

\author{G. B{\"o}cek Topcu}
\affiliation{Dipartimento di Fisica, Università di Roma Tor Vergata, via della Ricerca Scientifica 1, I-00133 Roma, Italy}
\affiliation{INAF - Osservatorio Astronomico di Roma, via Frascati 33, I-00078 Monte Porzio Catone, Italy}

\author{R. Buonanno}
\affiliation{INAF - Osservatorio Astronomico di Roma, via Frascati 33, I-00078 Monte Porzio Catone, Italy}

\author{A. Calamida}
\affiliation{Space Telescope Science Institute, 3700 San Martin Drive, Baltimore, MD 21218}

\author{E. Carretta}
\affiliation{INAF - Osservatorio di Astrofisica e Scienza dello Spazio di Bologna, Via P. Gobetti 93/3, I-40129 Bologna, Italy}

\author{G. Ceci}
\affiliation{Dipartimento di Fisica, Università di Roma Tor Vergata, via della Ricerca Scientifica 1, I-00133 Roma, Italy}
\affiliation{Dipartimento di Fisica, Sapienza Università di Roma, P.le A. Moro 5, I-00185 Roma, Italy}
\affiliation{INAF - Osservatorio Astronomico di Roma, via Frascati 33, I-00078 Monte Porzio Catone, Italy}

\author{B. Chaboyer}
\affiliation{Department of Physics and Astronomy, Dartmouth College, 6127 Wilder Laboratory, Hanover, NH 03755, USA}

\author{M. Correnti}
\affiliation{INAF - Osservatorio Astronomico di Roma, via Frascati 33, I-00078 Monte Porzio Catone, Italy}
\affiliation{ASI - Space Science Data Center, via del Politecnico snc, I-00133 Roma, Italy}

\author{R. da Silva}
\affiliation{INAF - Osservatorio Astronomico di Roma, via Frascati 33, I-00078 Monte Porzio Catone, Italy}
\affiliation{ASI - Space Science Data Center, via del Politecnico snc, I-00133 Roma, Italy}

\author{I. Ferraro}
\affiliation{INAF - Osservatorio Astronomico di Roma, via Frascati 33, I-00078 Monte Porzio Catone, Italy}

\author{F. A. G{\'o}mez}
\affiliation{Departamento de Astronom{\'{\i}}a, Universidad de La Serena, Av. Ra{\'u}l Bitr{\'a}n 1305, La Serena, Chile}

\author{G. Iannicola}
\affiliation{INAF - Osservatorio Astronomico di Roma, via Frascati 33, I-00078 Monte Porzio Catone, Italy}

\author{R.-P. Kudritzki}
\affiliation{Institute for Astronomy, University of Hawai'i at Manoa, Honolulu, HI 96822, USA}
\affiliation{LMU M{\"u}nchen, Universit{\"a}tssternwarte, Scheinerstr. 1, 81679 M{\"u}nchen, Germany}

\author{A. Kunder}
\affiliation{Saint Martin’s University, 5000 Abbey Way SE, Lacey, WA 98503, USA} 

\author{S. Kwak}
\affiliation{Leibniz Institut f\"{u}r Astrophysik Potsdam (AIP), An der Sternwarte 16, D-14482, Potsdam, Germany}

\author{M. Marconi}
\affiliation{INAF - Osservatorio Astronomico di Capodimonte, Salita Moiariello 16, I-80131 Napoli, Italy}

\author{S. Marinoni}
\affiliation{INAF - Osservatorio Astronomico di Roma, via Frascati 33, I-00078 Monte Porzio Catone, Italy}
\affiliation{ASI - Space Science Data Center, via del Politecnico snc, I-00133 Roma, Italy}

\author{N. Matsunaga}
\affiliation{Department of Astronomy, School of Science, The University of Tokyo, 7-3-1 Hongo, Bunkyo-ku, Tokyo 113-0033, Japan}

\author{F. Matteucci}
\affiliation{INAF - Osservatorio Astronomico di Trieste, via G.B. Tiepolo 11, I-34131 Trieste, Italy}
\affiliation{Dipartimento di Fisica, Sezione di Astronomia, Università di Trieste, Via G. B. Tiepolo 11, I-34143 Trieste, Italy}
\affiliation{INFN - Sezione di Trieste, via Valerio 2, I-34134 Trieste, Italy}

\author{A. Monachesi}
\affiliation{Departamento de Astronom{\'{\i}}a, Universidad de La Serena, Av. Ra{\'u}l Bitr{\'a}n 1305, La Serena, Chile} 

\author{I. Musella}
\affiliation{INAF - Osservatorio Astronomico di Capodimonte, Salita Moiariello 16, I-80131 Napoli, Italy}

\author{M. G. Navarro Ovando}
\affiliation{INAF - Osservatorio Astronomico di Roma, via Frascati 33, I-00078 Monte Porzio Catone, Italy}

\author{G. W. Preston}
\affiliation{Carnegie Observatories, 813 Santa Barbara Street, Pasadena, CA 91101-1292, USA}

\author{V. Ripepi}
\affiliation{INAF - Osservatorio Astronomico di Capodimonte, Salita Moiariello 16, I-80131 Napoli, Italy}

\author{M. Salaris}
\affiliation{Astrophysics Research Institute, Liverpool John Moores University, 146 Brownlow Hill, Liverpool L3 5RF, United Kingdom}
\affiliation{INAF - Osservatorio Astronomico d’Abruzzo, via M. Maggini, s/n, I-64100, Teramo, Italy}

\author{M. S{\'a}nchez-Benavente}
\affiliation{Instituto de Astrof{\'{\i}}sica de Canarias, Calle Via Lactea s/n, E-38205 La Laguna, Tenerife, Spain}
\affiliation{Departamento de Astrof{\'{\i}}sica, Universidad de La Laguna (ULL), E-38200, La Laguna, Tenerife, Spain}

\author{E. Spitoni}
\affiliation{INAF - Osservatorio Astronomico di Trieste, via G.B. Tiepolo 11, I-34131 Trieste, Italy}

\author{P. B. Stetson}
\affiliation{Herzberg Astronomy and Astrophysics, NRC, 5071 West Saanich Road, Victoria, British Columbia V9E 2E7, Canada}

\author{F. Thévenin}
\affiliation{Université de Nice Sophia-Antipolis, CNRS, Observatoire de la C{\^o}te
d’Azur, Laboratoire Lagrange, BP4229, F-06304 Nice, France }

\author{I. B. Thompson}
\affiliation{Carnegie Observatories, 813 Santa Barbara Street, Pasadena, CA 91101-1292, USA}

\author{P. B. Tissera}
\affiliation{Instituto de Astrof{\'{\i}}sica, Pontificia Universidad Cat{\'o}lica de Chile, Av. Vicu{\~n}a Mackenna 4860, Santiago, Chile}
\affiliation{Centro de AstroIngenier{\'{\i}}a, Pontificia Universidad Cat{\'o}lica de Chile, Av. Vicu{\~n}a Mackenna 4860, Santiago, Chile}

\author{T. Tsujimoto}
\affiliation{National Astronomical Observatory of Japan, Mitaka, Tokyo 181-8588, Japan}

\author{E. Valenti}
\affiliation{European Southern Observatory, Karl-Schwarzschild-Straße 2, D-85748 Garching, Germany}
\affiliation{Excellence Cluster ORIGINS, Boltzmann–Straße 2, D–85748 Garching bei München, Germany}

\author{A. K. Vivas}
\affiliation{Cerro Tololo Inter-American Observatory/NSF NOIRLab, Casilla 603, La Serena, Chile}

\author{A. R. Walker}
\affiliation{Cerro Tololo Inter-American Observatory/NSF NOIRLab, Casilla 603, La Serena, Chile}

\author{M. Zoccali}
\affiliation{Instituto de Astrof{\'{\i}}sica, Pontificia Universidad Cat{\'o}lica de Chile, Av. Vicu{\~n}a Mackenna 4860, Santiago, Chile}
\affiliation{Millennium Institute of Astrophysics, Av. Vicu{\~n}a Mackenna 4860, Santiago, Chile}

\author{A. Zocchi}
\affiliation{Department of Astrophysics, T{\"u}rkenschanzstraße 17 (Sternwarte), 1180
Wien, Austria}

\collaboration{57}{(AAS Journals Data Editors)}



\begin{abstract}

We introduce a new photometric catalog of RR Lyrae variables 
(RRLs, $\sim$300,000) mainly based on data available in public datasets.
We also present the largest and most homogeneous spectroscopic dataset of RRLs 
and Blue Horizontal Branch [BHB] stars ever collected. 
This includes radial velocity measurements ($\sim$16,000) 
and iron abundances ($\Delta$S method for 8,140~RRLs, plus $547$~from literature). 
Elemental abundances based on high-resolution spectra are provided for 487~RRLs and 64~BHB stars.
We identified candidate RRLs associated to the main Galactic components and their iron distribution 
function (IDF) becomes more metal-rich when moving from the Halo ([Fe/H]=-1.56) to the 
Thick (TCD; [Fe/H]=-1.47) and Thin (TND; [Fe/H]=-0.73) disk. Furthermore, Halo RRLs 
and RRLs in retrograde orbits are $\alpha$-enhanced ([$\alpha$/Fe]=0.27, $\sigma$=0.18), 
while TCD~RRLs are either $\alpha$-enhanced ([Fe/H]$\le$-1.0) or $\alpha$-poor ([Fe/H]$>$-1.0), 
and TND~RRLs are mainly $\alpha$-poor ([$\alpha$/Fe]=-0.01, $\sigma$=0.20). 
We also identified RRLs associated to the main stellar streams (Gaia-Sausage-Enceladus [GSE]; Sequoia, 
Helmi, Sagittarius) and we found that their IDFs are quite similar to Halo~RRLs. However, 
GSE RRLs lack the metal-poor/metal-rich tails and their $\alpha$-element distribution is quite compact. 
The iron radial gradient in Galactocentric distance for TND, TCD and Halo~RRLs is negative and it 
decreases from -0.026, to -0.010, and to -0.002~dex/kpc. The iron radial gradient based on dry Halo 
(Halo without substructures) RRLs is, within the errors, equal to the global Halo.   
We also found a strong similarity between iron and [$\alpha$/Fe] radial gradients of Milky Way RRLs 
and M31 globular clusters throughout the full range of galactocentric distances covered by the two samples.  
\end{abstract}



\defcitealias{crestani2021a}{C21a}
\defcitealias{crestani2021b}{C21b}
\defcitealias{fabrizio2021}{F21}
\defcitealias{gilligan2021}{G21}
\defcitealias{dorazi2024}{D24}
\defcitealias{dorazi2025}{D25}

\section{Introduction} \label{sec:intro}

In spite of the relevant improvements in the understanding of the early 
formation and evolution of Milky Way (MW), we still lack a “canonical” model, i.e. 
a comprehensive view of the physical mechanisms and time scales behind the 
formation of the old Galactic component (Halo). The two most popular theories 
concerning the early formation of the stellar Halo date back to more than 
40 years ago. The first was introduced by \citet{eggen1962} and suggests a dissipative 
process on a rapid time scale, while the second was suggested by \citet{searle78} and 
relies on a dissipation-less hierarchical mechanism and a slower formation.

A modern approach to cast this long-standing problem is to establish on a quantitative 
basis the fraction of stars that formed {\it in situ} from those that were accreted 
(or formed {\it ex situ}). In a dissipative scenario a significant fraction 
of the Halo stellar content was formed inside the Galaxy in the early epochs of its 
formation \citep{zolotov09, cooper15}. In a dissipation-less scenario a (significant) 
fraction of the Halo stellar content was formed {\it ex situ} by the accretion and tidal 
disruption of dwarf galaxies. The latter mechanism is the hierarchical assembly 
predicted by cold dark matter (CDM) cosmological models \citep{bullock05, font11} 
and it is an ongoing mechanism taking place on very long time scales. The key 
prediction of this merging scenario is that stellar halos should host a large 
number of stellar streams and of small satellites in the process to be accreted and 
covering a large range in Galactocentric distances \citep{helmi1999, gomez13}. 
Although, the CDM cosmology is nowadays considered the “concordance model”, 
the current empirical scenario for hierarchical galaxy formation is far from being 
settled. To investigate the hierarchical merging mechanism(s), their timescale and the mass 
assembly history of galaxies, one can analyze the metallicity distribution and kinematics 
of field stars. Below, we outline some of the most recent findings in this area, along 
with their respective strengths and limitations.

{\em Chemical-orbital properties}-- The current investigations are mainly based on spectroscopy 
of bright field red giant (RG) stars \citep{carollo07,carollo10,liu2018}, but their 
individual distances can hardly be more accurate than 10--20\%  \citep{schoenrich2011}.
\citet{carollo07} identified two Halo components, a more spherical and prograde "inner Halo" 
that is “metal-rich”  ([Fe/H]$\sim$-1.6) and with highly eccentric orbits, and an 
"outer Halo"  that is retrograde and more metal-poor ([Fe/H]$\sim$-2.2). Note that their 
targets are mainly located in the solar neighborhood (heliocentric distance d$\le$ 4~kpc), 
therefore they were able to investigate a large Halo volume only by taking into account for 
the maximum vertical extent of their orbits. This is the reason why the transition 
between inner and outer Halo ranges from 15 to 35 kpc \citep{kim2019, dietz2020}.
\citet{conroy2019b} found, using high-resolution (HR) spectra for more than 4200 field RGs, 
an unimodal Iron Distribution Function (IDF) peaking at -1.2 dex. This estimate is 
at least 0.3 dex more metal-rich than previous estimates based on low-resolution (LR) spectra 
and a broad range of stellar tracers. 
Recent investigations based on field RR Lyrae (RRL) and Blue (hot) Horizontal Branch (BHB) stars 
further support the dual nature of the Halo \citep{hattori11,kinman12a}.
They found that inner and outer Halo have different density profiles
\citep[flattened \vs\ spherical,][]{mc12}, kinematics and metallicity: 
the inner halo is more metal-rich and with a modest net prograde 
rotation, while the outer halo has a net retrograde rotation and the peak in the
metallicity distribution is 1/3 more metal-poor than the inner halo.
Although,  \citet{iorio2019} using RRLs from Gaia DR2 detected a transition 
from a spherical-to-triaxial Halo, moving from the inner to the outer regions.
Furthermore, \citet{nissen2010} investigated the 
{\it in-situ} and the {\it ex-situ} stellar components of the Galactic halo 
by using accurate elemental abundances and kinematics of nearby dwarf stars. 

{\em Chemical tagging}-- Chemical tagging plays a crucial role to trace in space 
and in time the chemical enrichment of the different Galactic components. 
Among the different Galactic components (Halo, Bulge, Thick/Thin disk [TCD, TND]) 
hosting truly old (t $>$ 10 Gyr) stellar populations, the Halo plays a 
crucial role in constraining its early formation. The CDM models predict 
that pristine dwarf galaxies -- clustering around mini dark matter halos -- 
are the main building blocks of large galaxies \citep{tolstoy09}.
This means that the Halo hosts the relics of accreted dwarf galaxies,
preserved as stellar streams with peculiar kinematics, chemical 
enrichment histories \citep{belokurov06},  and distinct pulsation properties \citep{luongo2024}.

Metallicity estimates for field RRL stars are somewhat heterogeneous,
relying primarily on the analysis of the light curves \citep{mateu2012,mullen2021,mullen2023},
or on pulsation properties \citep{sarajedini2024,sarajedini2025}, 
or on machine learning algorithms \citep{muraveva2025},
or on metallicity indicators such as hydrogen lines and the Ca II K line 
\citep[$\Delta$S method,][]{preston1959, Suntzeff91, Suntzeff94, layden94, layden1995}.
More recently, metallicities for field RRLs were derived either from the
Sloan Digital Sky Survey/Sloan Extension for Galactic Understanding and Exploration
\citep[SDSS/SEGUE, DR8;][]{lee2008_sspp,drake13a},
or from the Dark Energy Spectroscopic Instrument \citep[DESI,][]{medina2025a} LR spectra. 
However, their mean iron abundances
were estimated using several spectroscopic indicators and spectra
collected at different pulsation phases. Therefore, they are very inhomogeneous
and a detailed abundance analysis of MW RRLs is still lacking.

A new spin on the Halo IDF has been recently provided by 
\citet[][hereinafter~\citetalias{crestani2021a,crestani2021b}]{crestani2021a,crestani2021b}
and by \citet[][hereinafter~\citetalias{fabrizio2021}]{fabrizio2021} using a homogeneous 
sample of more than 8,000 field RRLs. Their RRL spectroscopic catalog includes both 
HR ($\sim$220), and SDSS/SEGUE and the 
Large Sky Area Multi-Object Fiber Spectroscopic Telescope (LAMOST) LR spectra ($\sim$8,000). 
They found that the IDF is unimodal and it peaks at [Fe/H]$\sim$-1.55. Although
field RRLs are less numerous compared with RG and MS stars, they trace 
a significant fraction of the MW spheroid (5$\le R_G \le$100 kpc) with high 
accuracy (3-5\% on individual distance, \citealt{Bragaetal2021,mullen2023}). 

{\em Metallicity gradient}-- The possible occurrence of a Halo metallicity 
gradient makes the MW assembly scenario more complex. 
Indeed, \citet{layden1995} found, using LR spectra for 302 field RRLs, 
that the Halo iron gradient is quite flat between 10 and 40 kpc with a 
typical value of [Fe/H]$\sim$-1.65. Moreover, he found evidence of a steady increase
in metallicity within the solar circle ($R_G\le$8 kpc). 
More recently,  \citet{xue2015} selected $\sim$1,800 K-giant stars with [Fe/H]$\le$-1.2
(SDSS/SEGUE) and found evidence of a shallow metallicity gradient showing a 
decrease in the mean metallicity of 0.1-0.2 dex when moving from $R_G\sim$10 to 
100 kpc. On the other hand, \citet{conroy2019b} based on HR optical spectra 
(H3 survey, \citealt{conroy2019a}) of $\sim$4,200 giants, and \citet{fernandezalvar2017}
based on HR near-infrared spectra of $\sim$3,900 RGs 
(Apache Point Observatory Galactic Evolution Experiment [APOGEE]), found no evidence 
of a gradient over a wide range of Galactocentric distances (10-80 kpc). 

The main aim of this investigation is to address the chemo--dynamical properties 
of field RRLs collected by our group
(\citealt{for11}; \citealt{chadid2017}; \citealt{sneden2017};
\citetalias{crestani2021a,crestani2021b,fabrizio2021};
\citealt[][hereinafter \citetalias{gilligan2021}]{gilligan2021};
\citealt[][hereinafter \citetalias{dorazi2024}]{dorazi2024};
\citealt[][hereinafter \citetalias{dorazi2025}]{dorazi2025})
for which we do have spectroscopic estimates of chemical abundance(s) and/or radial velocity (RV).
The current sample was complemented with data already available in the 
literature for BHB stars based on both LR and HR spectra.
We also take advantage of the astrometric properties (proper motion, position) 
provided by Gaia DR3 to perform a solid selection of RRLs belonging to the 
different Galactic components and to the most populous stellar streams. 
Moreover, we also perform a detailed analysis of the metallicity distribution
functions to constrain their early formation and evolution. 

In more detail the structure of the paper is the following. 
In \S~2 we introduce the different photometric datasets adopted in the 
Photometric Rome RRL catalog (PR3C), and discuss in more detail the spectroscopic datasets based 
on LR, medium-resolution (MR) and HR spectra we collected for the 
Spectroscopic Rome RRL Catalog (SR3C, see Appendix~\ref{sec:SR3C}). 
In this section, the RV and abundance measurements are also outlined. 
The criteria adopted to perform the kinematic selection of the RRLs belonging to the main Galactic components 
and to the most populous stellar streams are discussed in \S~3 together with their metallicity distributions. 
The comparison with the metallicity distributions available in the literature for field 
stars, globular clusters and dwarf galaxies are presented in \S~4. 
The radial metallicity gradients for the main Galactic components and the major stellar streams, 
together with linear and logarithmic fits and the comparison with similar Galactic estimates 
are introduced and discussed in \S~5. \S~6 is focused on the comparison between MW radial gradients 
based on RRLs with similar estimates available in the literature, while in \S~7 we discuss the 
comparison between MW and M31 radial gradients for the global halo, the dry halo and the major 
stellar streams. 
The summary of the results are included in \S~8, where we also briefly 
outline the near future developments of the current project.

\begin{table}[htbp]
\caption{Spectroscopic datasets included in SR3C. Listed from left to right are: the name of each dataset, the number of RRLs and HB stars with available RV measurements from either LR or HR spectra, the number of RRLs and HB stars with iron abundance measurements based on LR ($\Delta S$ method) or HR spectroscopy, and the number of RRLs and HB stars with [$\alpha$/Fe] abundance measurements derived from HR spectra. 
The numbers listed in this table do refer to the number of objects of each dataset that contributes to SR3C according to the priority list for RVs and chemical abundances discussed in \S~\ref{vrad} and in \S~\ref{iron_dis}, respectively.} 
\label{tbl:spectro_dfatasets}
\begin{center}
\begin{tabular}{lccccc}
\hline
\hline
Dataset & \multicolumn{2}{c}{RV} & \multicolumn{2}{c}{[Fe/H]} & [$\alpha$/Fe] \\
& LR & HR & LR & HR & HR \\
\hline
\multicolumn{6}{c}{---RRL---} \\
Crestani & \ldots & 576\footnote{\citet{bono2020,braga2021}; \citetalias{dorazi2024}} & 8117\footnote{\citetalias{crestani2021a,fabrizio2021}} & 159+78\footnote{\citetalias{crestani2021a}} & 201\footnote{\citetalias{crestani2021b}} \\
\citetalias{dorazi2024} & \ldots & \ldots & \ldots & 78     & 60     \\
\citetalias{dorazi2025} & \ldots & \ldots & \ldots & 3      & \ldots \\
\citet{sesar13b}        & \ldots & \ldots & 21     & \ldots & \ldots \\
\citet{liu2020}         & \ldots & \ldots & 303    & \ldots & \ldots \\
LAMOST                  & 5436   & \ldots & \ldots & \ldots & \ldots \\
SDSS                    & 1688   & \ldots & \ldots & \ldots & \ldots \\
Gaia                    & 5895   & \ldots & \ldots & \ldots & \ldots \\
GALAH                   & \ldots & 2307   & \ldots & 119    & 51     \\
\citet{zinn2020}        & 104    & \ldots & 195    & \ldots & \ldots \\
\citet{kinman12a}       & \ldots & \ldots & 51     & \ldots & \ldots \\
\citet{dambis2013,dambis14} & \ldots & \ldots & 9  & \ldots & \ldots \\
RAVE                    & \ldots & \ldots & 3      & \ldots & \ldots \\
\citet{Duffau2014}      & \ldots & 39     & \ldots & 36     & \ldots \\
\citet{hansen2011}      & \ldots & 2      & \ldots & \ldots & \ldots \\
\citet{nemec2013}       & \ldots & 11     & \ldots & \ldots & \ldots \\
DESI                    & 1227   & \ldots & \ldots & \ldots & \ldots \\
\citetalias{gilligan2021} & \ldots & \ldots & \ldots & 14   & \ldots \\
\citet{medina2021}      & \ldots & 15     & \ldots & \ldots & \ldots \\
\multicolumn{6}{c}{---HB---} \\
\citet{ForSneden2010}   & \ldots & 45     & \ldots & 46     & 39     \\
Kinman\footnote{\citet{kinman1994,kinman2000,kinman2007,kinman12a}} & 123 & 18 & \ldots & 9 & 7 \\ 
\hline
\end{tabular}
\end{center}
\end{table}

\section{Photometric and spectroscopic datasets} \label{dataset}

To provide firm constraints on the metallicity distribution across the 
Galactic spheroid we used different photometric and spectroscopic catalogs 
based either on proprietary data or available in the literature. Note that 
the key advantage of the current approach is that we use homogeneous estimates 
for individual distances and reddening together with an homogeneous 
metallicity scale for both LR (iron) and HR (iron, $\alpha$-elements) spectra.  

\subsection{Photometric datasets}\label{phot}

Long--term photometric surveys provided a 
wealth of new information concerning the identification and the characterization 
of variable stars in the Galaxy and in nearby dwarf galaxies.

In the last ten years, we have assembled a large photometric catalog of field, 
cluster and dwarf galaxy
RRLs \citep{fiorentino15a,fiorentino15b,Bragaetal2016,Bragaetal2021,fabrizio2019,fabrizio2021},
called the PR3C catalog. It includes more than 300,000 sources.
The astrometric reference frame and the foundation of PR3C are based on  
Gaia~DR3 \citep{gaia_dr3} and its comprehensive RRL catalog 
\citep{clementini2023}, but it includes RRLs from almost all the known photometric 
surveys. They can be listed according to the wavelength range (see list below). 
Note that PR3C will be made available to the community in a forthcoming
investigation (Braga et al. 2025, in preparation) together with a more
detailed discussion concerning the cross-match procedure adopted to build-up
the catalog \citep{fabrizio2019}.

{\em Optical}-- 
The main optical catalogs we took into account are: 
Catalina Sky Survey \citep[CSS,][]{drake2009,drake2017},
All Sky Automated Survey \citep[ASAS,][]{pojmanski1997,pojmanski2002,pojmanski2014},
ASAS for SuperNovae \citep[ASAS-SN,][]{jayasinghe2019}  
and the General Catalogue of Variable Stars \citep[GCVS,][]{samus2017}. 
In this context, large extragalactic surveys 
have also played a crucial role, since they released detailed catalog of nearby 
variable stars as ancillary results: 
SDSS \citep{ivezic2000} and the Panoramic Survey Telescope and Rapid Response System
\citep[Pan-STARRS1,][]{Abbas14} survey.
The same outcome applies to field variables provided by large photometric surveys 
interested in the identification either of nearby moving objects
(Lincoln Near-Earth Asteroid Research [LINEAR], \citealt{stokes2000,Sesar2011b}; 
Lowell Observatory Near-Earth-Object Search [LONEOS], \citealt{miceli08}),
or in transient sources 
(Robotic Optical Transient Search Experiment-I [ROTSE-I], \citealt{akerlof2000,kinemuchi2006}; 
LINEAR+CSS+PTF, \citealp{sesar13b};
Zwicky Transient Facility [ZTF], \citealt{ztf2019a,ztf2019b}), 
or in the identification of stellar streams in the 
Galactic halo (La Silla QUEST RRL Star Survey [QUEST], \citealt{vivaszinn06,zinn14}).
Moreover, we also took into account RRLs from the
Optical Gravitational Lensing Experiment IV (OGLE~IV) for the Bulge, 
the TND \citep{soszynski2019c} and for the Magellanic Clouds \citep{soszynski2019b}. 
As a whole the apparent optical magnitude ranges from 
G$\sim$7.62 to G$\sim$21.19 mag. The reader interested in more details 
is referred to \citet{fabrizio2019}, \citet{braga2021} and to \citetalias{fabrizio2021}. 

\begin{figure*}[]
\includegraphics[width=14cm]{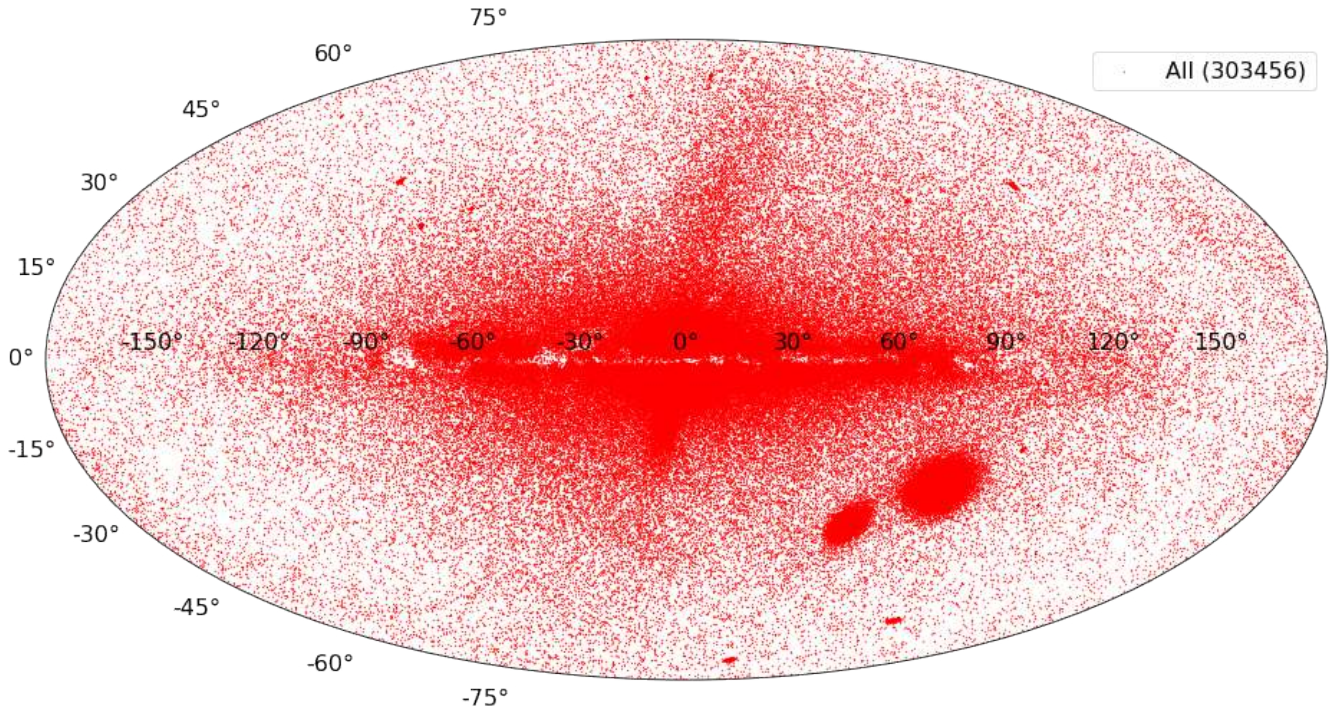}{}
\caption{Aitoff projection in Galactic coordinates of the RRLs in the PR3C. Isolated 
over--densities are associated either to the Magellanic Clouds, or to nearby stellar 
systems or to stellar streams.}
\label{fig:aitoff_all}
\end{figure*}

\begin{figure*}[]
\includegraphics[width=10cm]{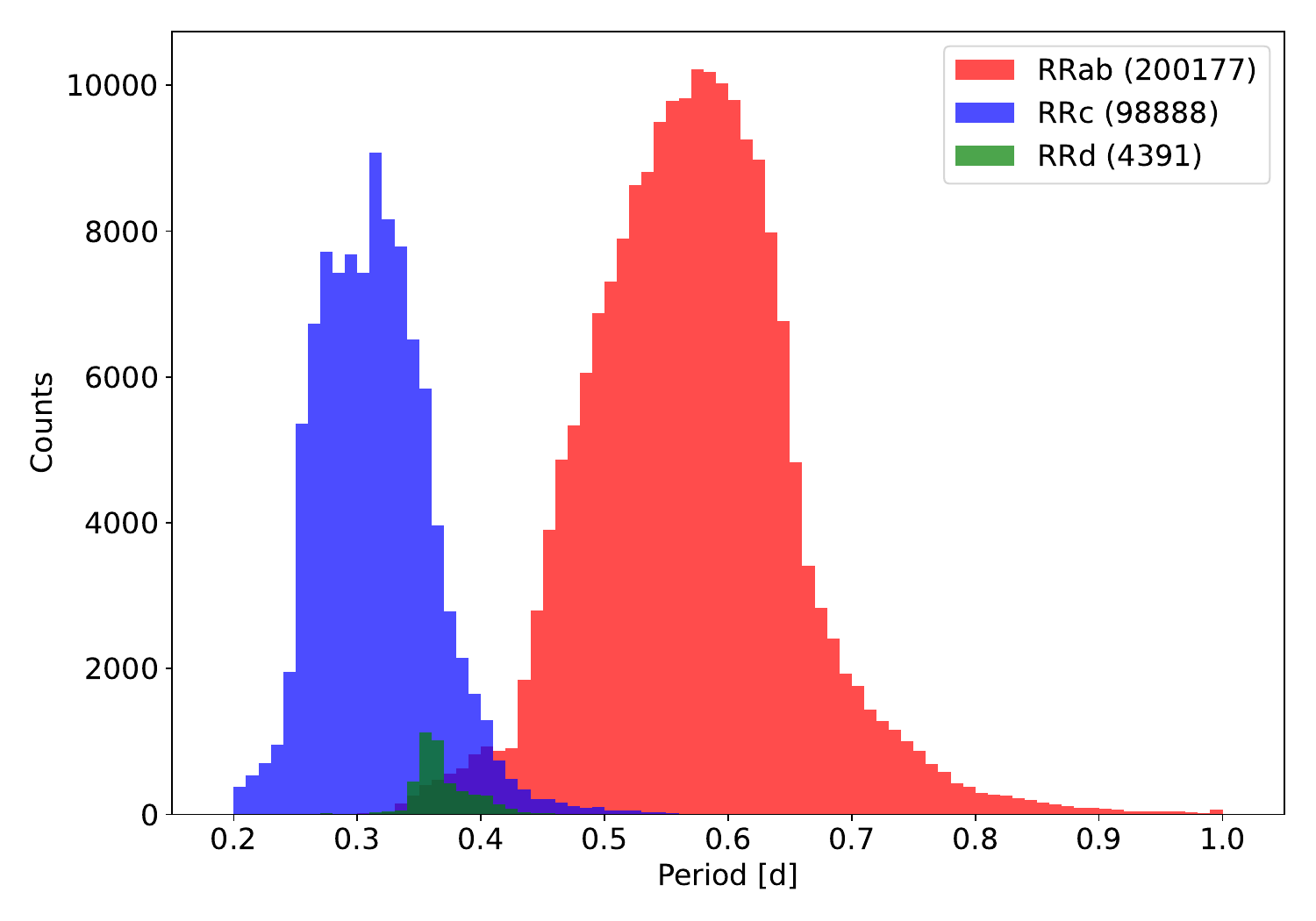}{}
\caption{Period distribution of the RRLs in the PR3C. 
Fundamental (RRab), first overtone (RRc), and mixed mode (RRd) RRLs are 
plotted with different colors (see labels).}
\label{fig:perdistrib_all}
\end{figure*}

{\em Near-Infrared}--
Similar datasets have also been collected in the NIR bands by the Two Micron All Sky Survey
\citep[2MASS,][]{skrutskie2006,layden2019}, the NIR surveys of the Galactic plane 
(UKIRT Deep Sky Survey [UKIDSS], \citealt{warren2007,lucas2008}; UKIRT Hemisphere Survey [UHS], \citealt{dye2018}),
and the VISTA Hemisphere Survey \citep[VHS,][]{mcmahon13,mcmahon2021}. 
The apparent NIR magnitudes of the current dataset range from 
K$\sim$4.35 to K$\sim$19.31 mag. 

{\em Mid-Infrared}-- We also collected datasets in the MIR: 
Wide-field Infrared Survey Explorer \citep[WISE,][]{wright2010}, 
NEO-WISE \citep{mainzer2011}, 
AllWISE \citep{cutri2013}, 
and SPITZER \citep{werner2004}.
The apparent MIR magnitudes of the current dataset range from 
W1$\sim$5.94 to W1$\sim$19 mag.

{\em Ultraviolet}-- 
Finally, we also took into account datasets in the UV bands from 
the Galaxy Evolution Explorer \citep[GALEX,][]{Welsh2005,kinman14}.
The apparent NUV magnitudes of the current dataset range from 
NUV$\sim$12.5 to NUV$\sim$23.2 mag.

\begin{figure*}[]
\includegraphics[height=18cm]{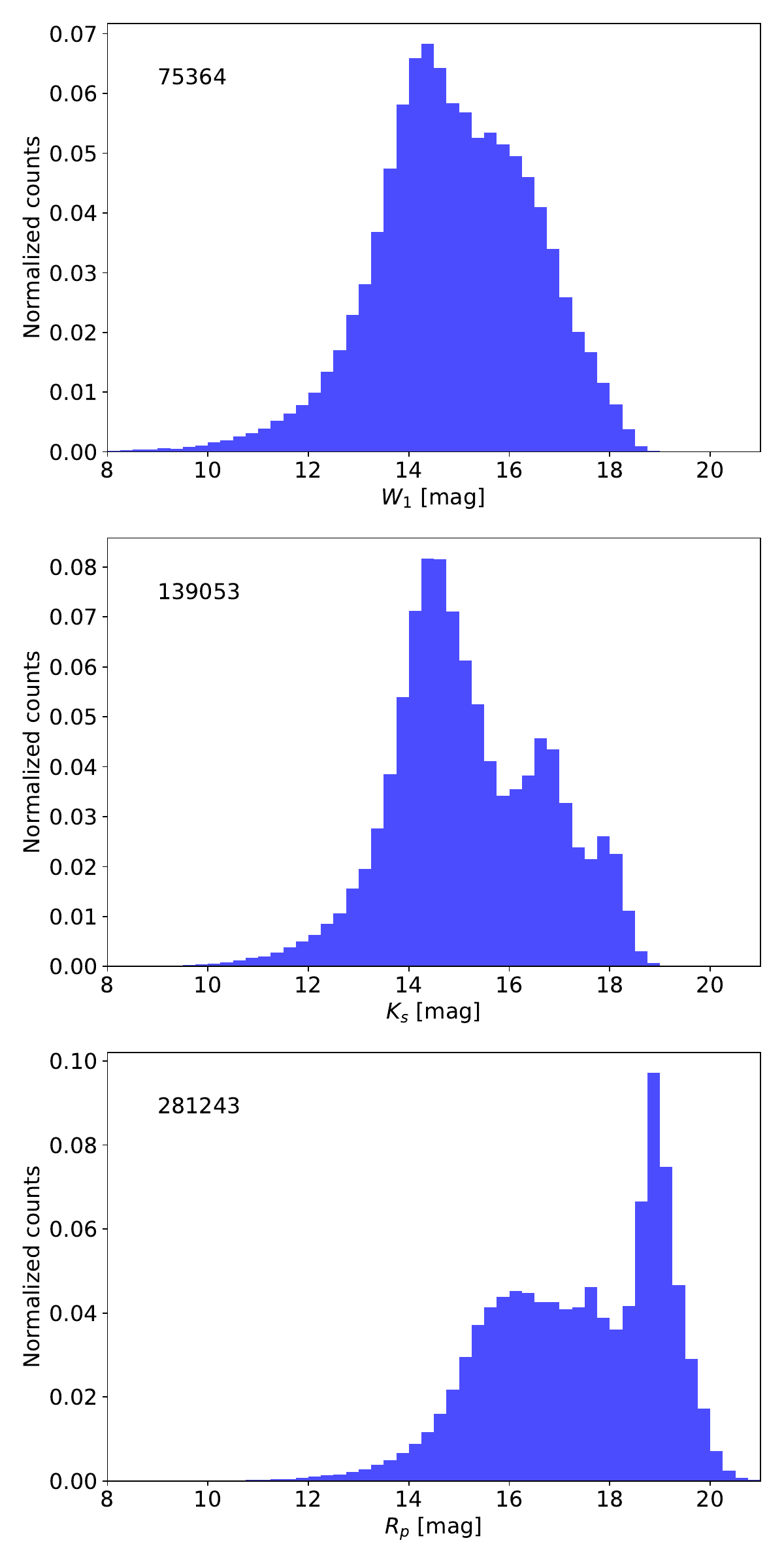}{}
\caption{Selected magnitude distributions for RRLs in the PR3C. 
From top to bottom the histograms display the mid-infrared $W1$-band, the 
near-infrared  $K$-band, and the Gaia optical $R_P$-band magnitudes.}
\label{fig:magdistrib_all}
\end{figure*}

All in all, the PR3C catalog includes more than 302,000 field and cluster RRLs and
Fig.~\ref{fig:aitoff_all} shows the distribution in Galactic coordinates 
of the entire photometric catalog. The over--densities are associated either to nearby 
dwarf galaxies (Magellanic Clouds, dwarf spheroidals) or to Galactic Globular Clusters (GGCs) 
or to stellar streams. To characterize the pulsation properties of the PR3C catalog, 
Fig.~\ref{fig:perdistrib_all} shows the period distribution for RRLs pulsating in the 
fundamental mode (FU, RRab; red), in the first overtone (FO, RRc; blue), and simultaneously 
in the first two radial modes (mixed mode [MM], RRd; green). Typically, the mode identification 
is based on the same approach suggested by \citet{clementini19}, but we double checked the 
location of individual RRLs in the Bailey diagram and some outliers were fixed according to 
visual inspection of their light curves. 
Fig.~\ref{fig:magdistrib_all} displays from top to bottom the magnitude distribution of 
RRLs in the PR3C catalog in the MIR ($W_1$, top), in the NIR ($K_s$, middle), and in the 
optical ($R_p$, bottom). Although RRLs are, on average, half magnitude brighter in NIR/MIR 
bands where they are also significantly less affected by reddening,
the optical bands allow us to reach the outskirts of the Galactic halo
(R$_G\approx$100 kpc, $R_P\sim$20 mag).

\begin{figure*}[]
\includegraphics[width=12cm]{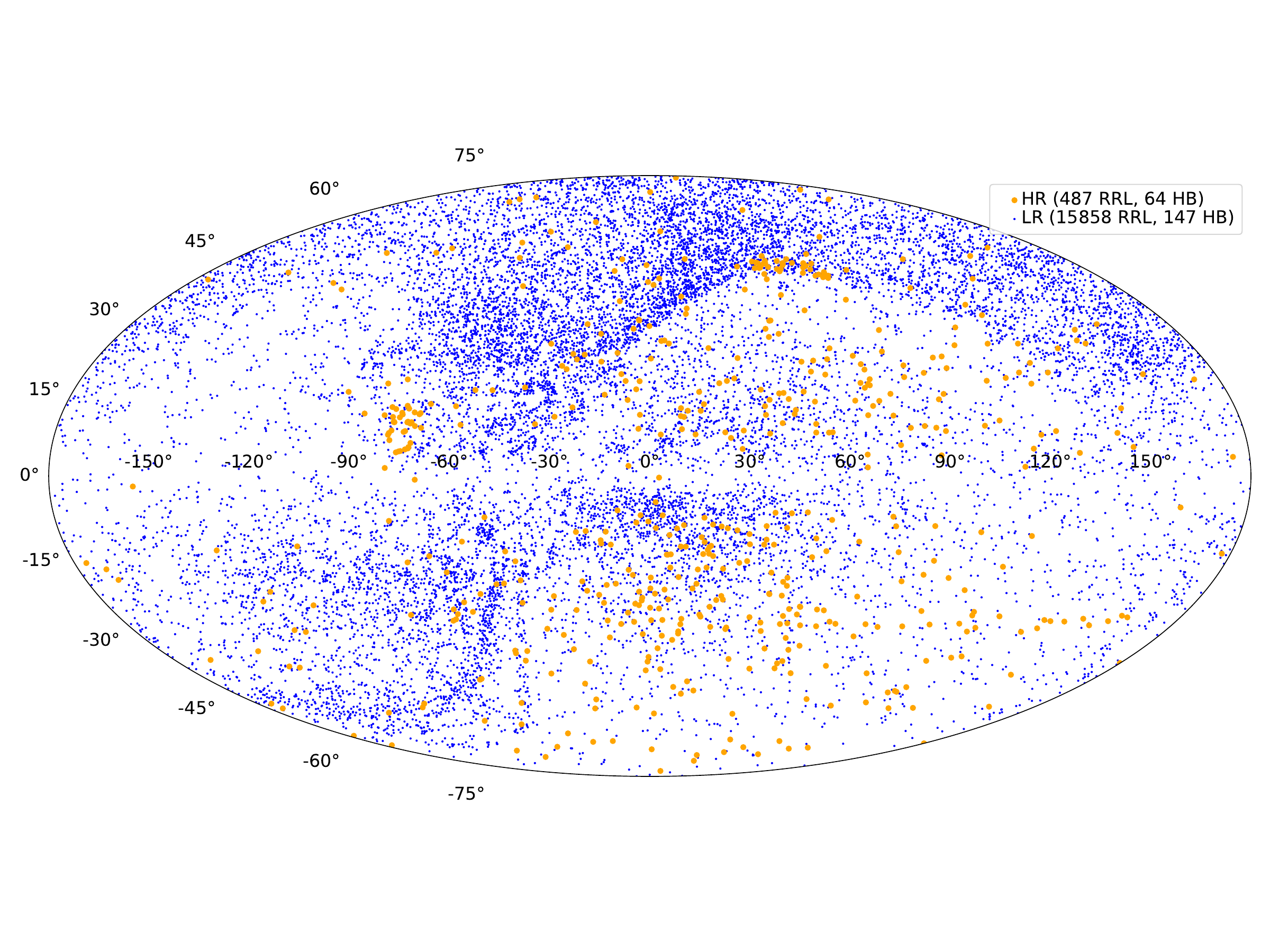}{}
\caption{Aitoff projection in Galactic coordinates of the RRLs in the SR3C. 
Yellow and blue dots mark RRLs and HB stars for which are available 
either HR or LR spectra, respectively. The number of 
RRLs and HBs with HR and LR spectra is labeled.}
\label{fig:aitoffSR3C}
\end{figure*}

\subsection{Spectroscopic datasets}\label{spec}

Our group secured a large sample of HR, MR and LR spectra for field and cluster RRLs. 
Moreover, RRLs in PR3C were cross-matched with spectroscopic catalogs available in the 
literature. The subsample of field RRLs for which we do have spectroscopic information
is the SR3C catalog.

Fig.~\ref{fig:aitoffSR3C} shows the distribution in Galactic coordinates of the RRL 
in SR3C for which we do have a spectroscopic measurement of either the iron abundance 
or the RV. As a whole, we are dealing with more than 16,000 field RRLs 
for which are available either LR/MR/HR spectra.

\begin{figure*}[]
\includegraphics[height=18cm]{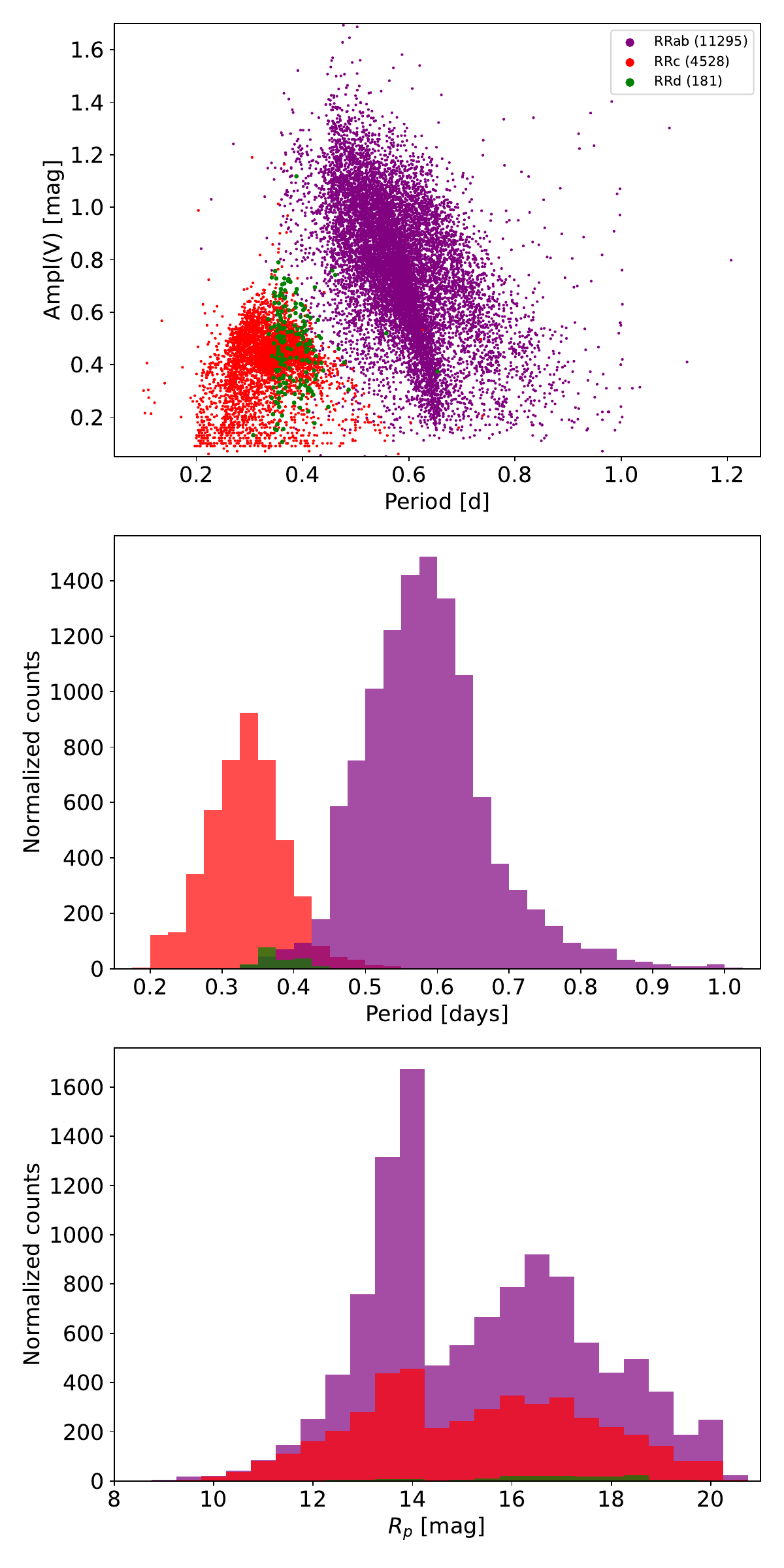}{}
\caption{
Top: V-band luminosity amplitude \vs\ period (Bailey diagram) for RRLs in SR3C. 
RRLs that are pulsating in the fundamental mode (RRab) are highlighted in purple, 
those pulsating in the first overtone (RRc) in red and the RRLs that are simultaneously 
pulsating in the first two radial modes (RRd, mixed mode) in dark green. The numbers 
close to the acronyms show the sample size. 
Middle: Same as the top, but for the period distribution. 
Bottom: Same as the top, but for the mean $R_P$-band magnitude distribution.}
\label{fig:bailey}
\end{figure*}

To characterize the spectroscopic catalog, Fig.~\ref{fig:bailey} shows from top to bottom 
the Bailey diagram, the period distribution and the $R_P$-band magnitude distribution 
of the RRLs in SR3C. Data plotted in this figure display that RRLs in SR3C cover the 
typical range in period and in amplitudes for fundamental (RRab, purple), first overtone 
(RRc, red) and mixed mode (RRd, dark green) variables.

This sample was complemented with a sample of field BHB stars for which 
are also available either LR or HR spectra (see Table~\ref{tbl:spectro_dfatasets}). 
Blue HB stars are, together with RRLs,
solid tracers of old stellar populations and they can be safely identified in the field, 
since their colors are typically bluer than F-, G-type and M-type field stars.  
In the literature, there are several investigations concerning red HB stars, but the
selection criteria (photometry, kinematics) adopted for their identification are not 
very solid yet. We only adopted a few red HB stars for which it is available a 
spectroscopic characterization (\citetalias[][and references therein]{crestani2021b}).

\subsubsection{Individual radial velocities}\label{vrad}

In this investigation we are dealing with RV measurements 
coming from a wide range of HR, MR and LR spectra. 

{\em--RV curve--} The RV measurements based on HR spectra were performed using 
synthetic spectra with a spectral resolution and wavelength range similar to 
the observed ones. The reader interested in a more detailed discussion is referred to 
\citetalias{crestani2021a}, \citetalias{crestani2021b} and \citetalias{fabrizio2021}.
A handful of field RRLs have a full coverage of the pulsation cycle 
\citep{bono2020} based on HR spectra and their $\gamma$-velocity was 
estimated  with an analytical fit of the RV curve based on weak 
metallic lines. The RV error associated with these measurements 
is the standard deviation of the analytical fit. 
Note that this sample also includes two relevant literature contributions.  

{\em Gaia} --  Gaia performed accurate estimates of the $\gamma$ 
velocity for 1096 RRLs, by performing an analytical fit of the RV curves. 
The reader interested in more details is referred to \citet{clementini2023}. 

{\em Baade--Wesselink} -- For roughly three dozen of field Baade--Wesselink (BW) 
RRLs accurate optical-NIR light curves and also very well--sampled 
RV curves, based on weak metallic lines \citep[][and references therein]{bono94c,cacciari03a}, are available. 

{\em HR spectra} -- This sample mainly includes the spectra we collected to measure 
individual elemental abundances. They are mainly a single spectrum per object, with a small 
fraction with two/three spectra. Thanks to the high spectral resolution, the error
on the RV for the RRLs with single spectra is the standard deviation among the different 
metallic lines. The error on the RV for RRLs with a few HR spectra is the standard 
deviation of the mean. Note that this sample also includes the BHB stars observed 
by \citet{ForSneden2010} and by \citet[][and references therein]{kinman12a}.

{\em MR spectra} -- This sample mainly includes 5895 RRLs that have been observed by Gaia, but for which the RV values are the mean of the measurements and their error is the standard deviation of the measurements. 

This sample also includes a few (3) RRLs observed by the RAdial Velocity Experiment survey
\citep[RAVE,][]{steinmetz2006,kunder2014,kunder2017,steinmetz2020a}.
They are only based on MR spectra and they are quite limited (see Table~\ref{tbl:spectro_dfatasets}).  
They were downloaded from their website\footnote{\url{https://www.rave-survey.org/}}. The RVs and the associated errors were measured following the same approach for the HR spectra. 

{\em LR spectra} --
This is the largest sample of RVs and they mainly come from public spectroscopic surveys and the $\gamma$ velocity is either the mean or the median of the measurements. The criteria adopted for defining the priority list are the following: 

{\em a) Number of measurements} -- The highest priority among the 
LR spectra is given to the datasets for which 
multiple individual measurements are available. For these objects the mean 
RV was estimated as the mean of the individual measurements and the 
associated error is the standard deviation of the measurements. 
In case the multiple measurements have been co-added either to 
a single reference spectrum or measured on co-added intra-night 
spectra, they have lower priorities. 

{\em b) Wavelength coverage} -- Spectra covering a broader range 
in wavelength have higher priorities. 

{\em c) Sample size} -- Higher priority is given to datasets 
including larger samples of field RRLs. 

Note that the LR sample also includes a sizable sample of
spectra collected with du Pont coming from the RRL Carnegie dataset. 
They are HR spectra, but their individual SNR is modest and they do 
not allow us  to perform accurate elemental abundances. Therefore, 
their original spectral resolution was decreased by roughly a factor 
of ten and used as LR spectra (\citetalias{crestani2021a, crestani2021b})
to estimate both RVs and iron abundances ($\Delta$S method).

A large sample (see Table~\ref{tbl:spectro_dfatasets}) of HR spectra collected
by the GALactic Archaeology with HERMES (GALAH) survey
was also used to estimate RVs, as their individual SNR is 
modest/poor. These spectra were not used to estimate the iron abundances with 
the $\Delta$S method, because they do not cover the Ca~H\&K lines. 

{\em Literature} -- This sample is a mixed bag, since it includes 
RV measurements available in the literature based 
on spectra covering a broad range of spectral resolutions and 
number of measurements. 

According to these selection criteria the priority list is the following:
a) RV curve fit from our HR spectra, 
b) RV curve fit from Gaia measurements,
c) \citet{zinn2020}, 
d) averaged RVs from DESI, 
e) averaged RVs from Gaia,  
f) averaged RVs from our HR spectra,
g) averaged RVs from LAMOST MR spectra,
h) averaged RVs from LAMOST LR spectra, 
i) averaged RVs from SDSS spectra, 
j) \citet{Duffau2014}, 
k) \citetalias{dorazi2024}, 
l) averaged RVs based on Gaia, LAMOST and SDSS (when less than 2 data points are available from the different datasets),  
m) \citet{medina2023}, 
n) RV from stacked GALAH spectra, 
o) \citet{hansen2011}, 
p) \citet{nemec2013}, 
q) averaged $H_{\alpha}$, $H_{\beta}$, $H_{\gamma}$ RVs from our HR spectra, when metallic lines are not available. 

\begin{figure*}[]
\includegraphics[height=18cm]{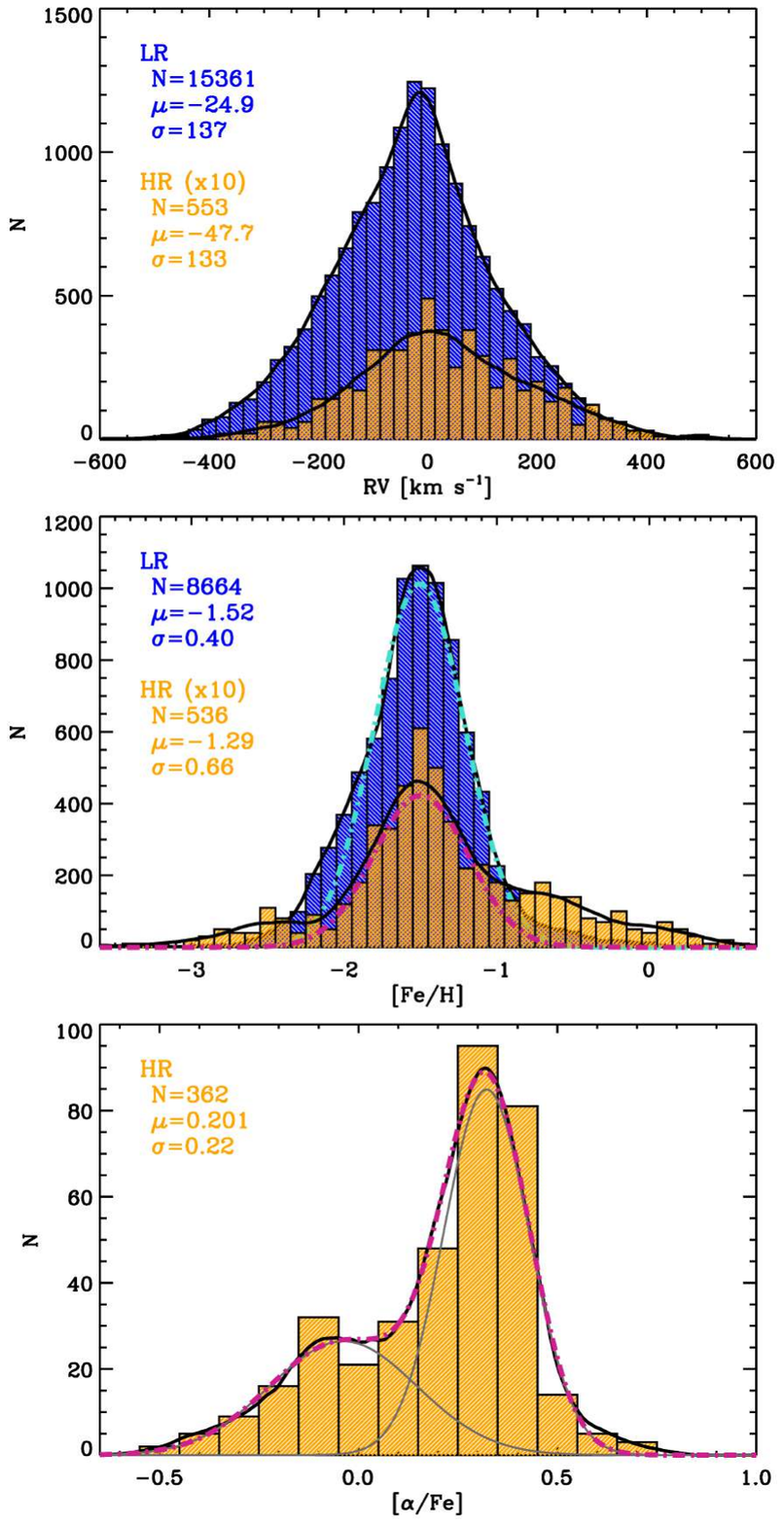}{}
\caption{Top: mean RVs for spectroscopic RRLs based either on 
LR (blue) or on HR (orange) spectra. The RV distribution of the latter sample was 
amplified by a factor of ten to improve the visibility. The black solid lines display 
the smoothed distributions based on a Gaussian kernel. The number of RRLs in the two 
different samples are labeled together with the mean and the standard deviation 
of the two different samples.
Middle: Same as the top, but for the iron distribution based either on LR or on 
HR spectra. The black solid lines display the smoothed distributions, while the 
dashed--dotted lines show the fit with a Gaussian function.
Bottom: Same as the top, but for the [$\alpha$/Fe] distribution only based on HR 
spectra. The black solid lines shows the smoothed distribution, the dashed--dotted 
line the fit with two Gaussian functions and the thin gray lines the individual 
Gaussian functions.}
\label{fig:global_spectro}
\end{figure*}

The top panel of Fig.~\ref{fig:global_spectro} shows the RV distribution 
of both HR (dark yellow) and LR (blue) sample. As expected the radial 
velocity distribution of field RRLs is quite symmetric around zero with two well extended 
tails approaching $\pm$400 km/sec. The black solid line shows the smoothed distribution 
estimated by using a Gaussian kernel with unitary weight and $\sigma$ equal to the 
Poisson error of the individual bins. Although the LR sample is more than one order of 
magnitude larger than the HR sample, the standard deviations of the two RV distributions 
have quite similar values (see labeled values). This indicates that the RV distributions 
are dominated by RRLs located across the Solar circle.

\subsubsection{Individual iron abundances}\label{iron_dis}

The selection criteria and the datasets adopted for estimating
individual iron abundances are very similar to those adopted 
for the RVs. The iron abundances of the HR sample were estimated using 
individual FeI and FeII lines. The reader interested in a more 
detailed discussion is referred to \citetalias{crestani2021a},
\citetalias{crestani2021b}, \citetalias{gilligan2021}, \citetalias{dorazi2024} and to \citetalias{dorazi2025}.
The priority list for the metallicity estimates is the following:
a) \citetalias{dorazi2024, dorazi2025},
b) \citetalias{crestani2021b}, 
c) \citetalias{gilligan2021},
d) \citet{Duffau2014}, 
e)  GALAH DR3, 
f) \citetalias{crestani2021a}, 
g) \citet{zinn2020}, 
h) \citet{liu2020}, 
i) \citet{dambis2013}, 
j) \citet{sesar13b}, 
k) RAVE DR6,
l) \citet{kinman12a}. 

In passing we note that the metallicity scale adopted in the 
quoted investigations is the same adopted by \citet{magurno2018,magurno2019}
in dealing with cluster RRL variables.
The reader interested in a more detailed discussion concerning 
field and cluster RRLs is referred to the quoted papers.

The bulk of the current iron abundances is based on a new 
version of the $\Delta$S method provided by \citetalias{crestani2021a,crestani2021b}
and originally suggested by \citet{preston1959}. This spectroscopic diagnostic 
was applied to all the LR spectra we collected for the RV measurements. 
The individual spectra with a SNR smaller than 15 were neglected. This is 
the reason why the number of RRLs with iron abundances is smaller than the 
RRLs with RV measurements.
The error associated to the iron abundances of RRLs with multiple spectra 
is the standard deviation. The error associated to RRLs for which we do 
have a single spectrum was estimated as the mean error of the spectroscopic 
dataset. The reader interested in a more detailed discussion is referred to 
\citetalias{fabrizio2021}.
 
Note that SR3C includes 181 mixed mode variables (RRd) and they were 
treated as first overtone variables because the dominant mode is typically 
the first overtone \citepalias{fabrizio2021}. The application of the $\Delta$S 
method to a field mixed-mode variable (ASAS J183952-3200.9) for which we 
collected a dozen of HR spectra along the pulsation cycle 
\citepalias[][see their section 7.2]{crestani2021a} supports this assumption.

The spectroscopic iron abundances available in the literature were 
estimated using a broad variety of $\Delta$S methods or similar 
diagnostics based either on LR or on MR spectra. The different sets 
of iron abundances were included in the current catalog only if 
we had a sizable sample of objects in common, that we defined 
as local calibrators. We performed a linear analytical fit of the difference 
in iron abundance between our estimates and those available in 
the literature, and these relations were used to move the iron abundances 
of the entire sample into our metallicity scale (see Table~1 in \citetalias{fabrizio2021}). 

The current spectroscopic dataset, compared to \citetalias{fabrizio2021}, 
includes the following new datasets: \citet{kinman12a}, \citetalias{gilligan2021},
\citet{medina2023}, GALAH~DR3 \citep{galah_dr3}, \citetalias{dorazi2024} and \citetalias{dorazi2025}.
Note that we could not rescale the \citet{medina2023} sample 
because we found no match with any of our RRLs. Also, \citetalias{gilligan2021}, \citetalias{dorazi2024} and \citetalias{dorazi2025} 
are already in our metallicity scale. The linear analytical relations used to convert the 
metallicities of \citet{kinman12a} and GALAH~DR3 into our metallicity scale are 
the following linear relations~\ref{eqn:eq_galahconv} and \ref{eqn:eq_kinmanconv}:

\begin{equation}
\label{eqn:eq_galahconv}
[Fe/H]_{our} = -0.497\pm0.047 + 0.770\pm0.049 \cdot [Fe/H]_{GALAH}\, (0.172)
\end{equation}
\begin{equation}
\label{eqn:eq_kinmanconv}
[Fe/H]_{our} = -0.37\pm0.22 + 0.75\pm0.14 \cdot [Fe/H]_{K12}\, (0.30)
\end{equation}

where the symbols have their usual meaning and the number in parentheses 
are the standard deviations. 
As a whole, the HR sample compared with \citetalias{fabrizio2021} was 
increased by more than a factor of two (550 \vs\ 246), while the LR sample 
was minimally changed (8800 \vs\ 8769). 

Note that in the current analysis we are not taking  account for 
iron estimates of field RRL stars based either on Fourier 
parameters of optical \citep{dekany2018,nemec2011,garofalo13,pietrukowicz2015,torrealba2015}
and NIR/MIR  \citep{mullen2021,mullen2023,zoccali2024} light curves, 
or on the inversion of the PLZ relations \citep{martinezvazquez16a,Bragaetal2016}, 
or on the $K_{15}$ parameter based on luminosity amplitudes \citep{bono07}, or 
on the {\tt REDIME} algorithm based on optical and NIR mean magnitudes 
\citep{bono2019}. They will be addressed in a forthcoming paper. 

The middle panel of Fig.~\ref{fig:global_spectro} displays the IDF 
based either on LR- (blue) or on HR-spectra (dark yellow). 
The black solid lines show the smoothed distributions estimated by using a 
Gaussian kernel. We associated to each RRL in the current sample a Gaussian 
function with unitary weight and $\sigma$ equal to the error of the individual 
estimates and the black lines are the cumulative distribution of the individual 
Gaussian functions. 
The two IDFs agree within the errors; we fit the two smoothed distributions 
with a Gaussian function and we found that the peak and the $\sigma$ are: 
-1.51$\pm$0.01, $\sigma$=0.30$\pm$0.01 (LR) and -1.50$\pm$0.03, 
$\sigma$=0.31$\pm$0.04 (HR), respectively. The two IDFs display a symmetrical 
distribution marginally skewed in the metal-poor regime and a well--defined peak.   
The difference in the metal-rich and in the metal-poor tails between the two 
distributions is mainly due to the difference in sample size. In passing, we note 
that the current findings fully support earlier results based on smaller spectroscopic 
samples from \citetalias{crestani2021a}, \citetalias{crestani2021b} and \citetalias{fabrizio2021}.

\subsubsection{Individual $\alpha$-element abundances}\label{alpha_dis}
The bottom panel of Fig.~\ref{fig:global_spectro} shows the [$\alpha$/Fe] abundance 
ratio for field RRL and BHB stars based on HR spectra. The mean $\alpha$-element 
abundance includes three different elements (Ca, Mg, Ti), since these are the three 
elements measured by \citetalias{crestani2021b}.
The black line shows the smoothed distribution computed 
using a Gaussian kernel, the histogram shows quite clearly that the [$\alpha$/Fe] 
distribution is bi-modal. We fit the distribution
with two Gaussian functions and we found that the main peak is located at 
[$\alpha$/Fe]$\sim$0.32$\pm$0.01, $\sigma$=0.11$\pm$0.01; while the secondary
peak is located at [$\alpha$/Fe]$\sim$-0.04$\pm$0.01, $\sigma$=0.19$\pm$0.01. 
According to previous results obtained by \citet{for11}, \citet{chadid2017}, 
\citet{sneden2017}, \citetalias{crestani2021a}, \citetalias{crestani2021b}, 
\citetalias{dorazi2024} and \citetalias{dorazi2025} we can safely associate the 
main peak with metal-poor RRLs, showing a well--defined $\alpha$-enhancement, 
while the secondary peak is mainly associated with metal-rich RRLs that are 
either $\alpha$-poor or even $\alpha$-depleted.

\section{Kinematic selection}\label{chapt_kin}

The RRL kinematics, and in particular their orbital integration, was 
investigated by using the MW-Potential2014 \citep{bovy2015} implemented in the 
code {\tt galpy}. Moreover, we
adopted a solar distance from the Galaxy center of 8.122~kpc \citep{gravitycollaboration2018}, an
height of the Sun above the Galactic plane of $z_\sun=20.8$~pc
\citep{bennett2019}, the solar peculiar motion by \citet{schoenrich2011},
and the orbital velocity at the solar distance suggested by
\citet{drimmelandpoggio2018} that is $(U_\sun,V_\sun,W_\sun)=(12.9,245.6,7.78)$~km~s$^{-1}$. 
Among the input parameters adopted to estimate individual orbits, Galactocentric
distances and RVs have been discussed in Appendix~\ref{sec:rrldistance} and in 
\S~\ref{vrad}, while the sky coordinates and the proper motions come from Gaia~DR3.

Special attention was paid to estimate the uncertainties associated with
the orbital parameters. To constrain on a quantitative basis the errors
associated to the orbital properties, we performed a bootstrap analysis
in which the initial conditions for the six input parameters 
(position, proper motion, RV, distance) were extracted one thousand times, 
with a Monte-Carlo, by using Gaussian distributions with means equal to the 
adopted values and $\sigma$ equal to their intrinsic errors. Moreover, to properly
trace possible orbital variations on individual RRL, we integrated the initial
conditions forward for 10~Gyr with steps of 1~Myr. Note that these
simulations were performed by taking into account the correlation
between the components of the proper motion.
Moreover and even more importantly,
the kinematic properties of the individual RRLs were estimated as the
median of the 1,000 realizations and their associated errors are the
16th and the 84th percentile of their distributions.

We found that the error budget in the kinematic properties of the
current RRL sample is dominated by uncertainties in proper motion. This
evidence was somehow expected, since our sample covers a wide range in
Galactocentric distances. To overcome thorny problems in the analysis of
the kinematic properties we decide to be conservative and to include in
the ``solid sample" the RRLs up to 98th percentile in the proper motion
errors distribution. This means a cut at $\epsilon PM_{RA}=0.57$~mas/yr
and $\epsilon PM_{DEC}=0.50$~mas/yr. The ``solid sample" includes 15,550~RRLs 
and the subsequent chemo-dynamical analysis is only based on these objects.  

\subsection{Identification of the main Galactic components}\label{sec_Galcomp}

To provide a robust selection of the main Galactic components, Halo, TND and 
TCD, we adopted the same selection criteria, based on relative probability 
ratios, suggested by \citet{bensby2014}. It is worth noticing that stars that share 
a common origin according to the adopted selection criteria will be discussed in \S~\ref{sec_stream}. 

\begin{figure*}[] 
\centering
\includegraphics[width=0.9\textwidth]{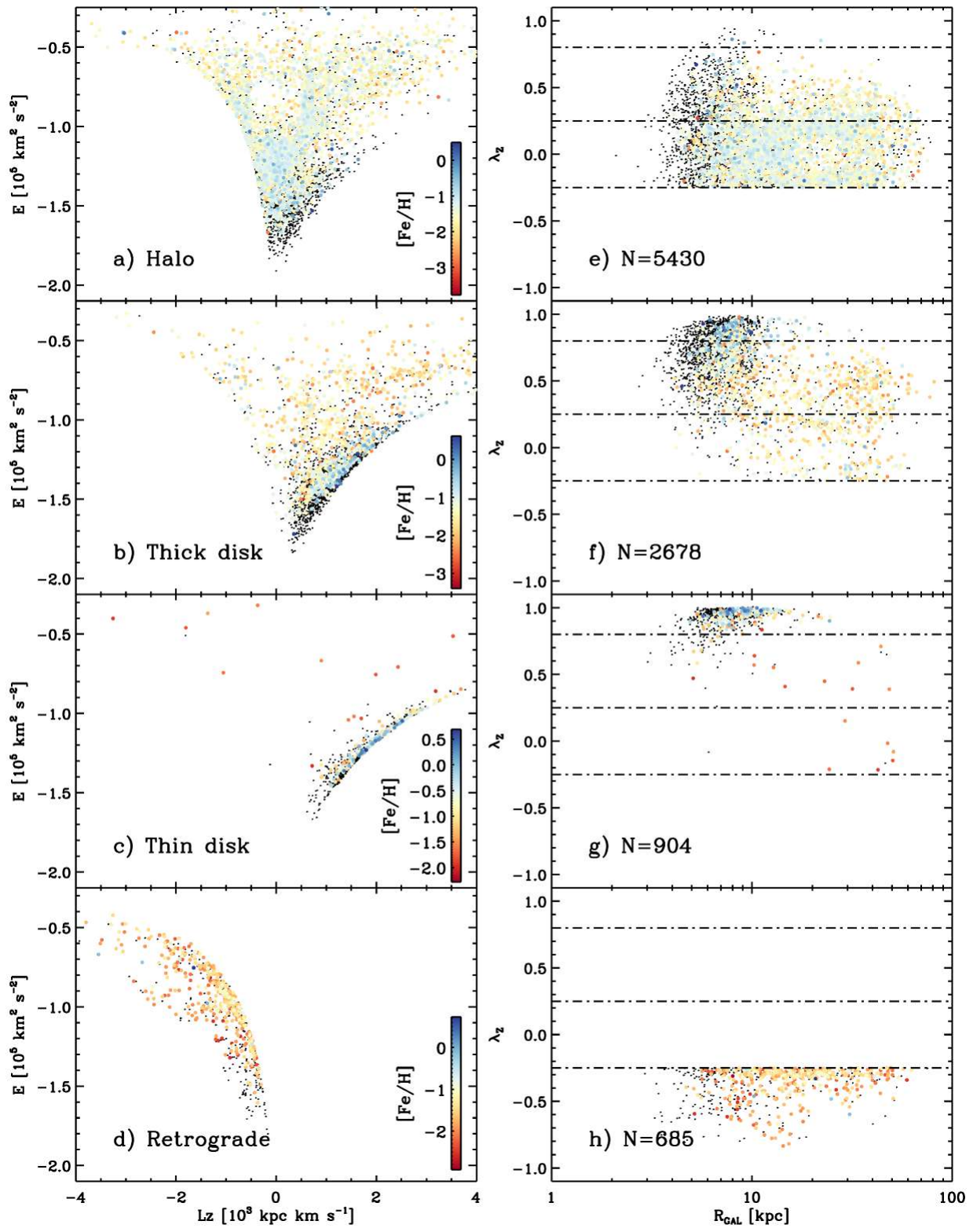}
\caption{Left -- Panels a), b), and c) show the Lindblad diagram for the three 
Galactic components (Halo, TCD and TND) selected according to the 
probabilistic criteria suggested by \citet{bensby2014}. Panel d) shows the RRLs 
in retrograde orbits. The symbols are color-coded according to the iron abundance 
(see color bar in the inset).  
Right -- Same as the left, but the circularity of the orbits ($\lambda_Z$)  as a 
function of the Galactocentric distance. The number of RRLs in the different 
Galactic components are labeled. See text for more details.
}
\label{fig:bensby}
\end{figure*}

The left panels of Fig.~\ref{fig:bensby} display from top to bottom the RRLs belonging  
to the Halo, the TCD, the TND and in retrograde orbits.
The symbols are color coded according to the iron abundance and
they show quite clearly the systematic drift 
from the more metal-poor RRLs of the Halo to the more metal-rich component of the TND.
The RRLs associated with the TCD trace quite well the transition between 
the more metal-poor and the more metal-rich regime. There is solid evidence that the 
TCD sample includes a sizable sample of RRLs with solar iron abundance.  

In our previous investigation (\citetalias{dorazi2024}), we adopted the circularity of the orbits, 
defined as the angular momentum J$_z$ along the vertical z-axis normalized
by the angular momentum of a circular orbit with the same binding
energy E ($\lambda_z=J_z/J_{max}$(E)), as a function of the Galactocentric distance 
to identify the different Galactic components.  
Following \citet{zhu2018} and \citet{santucci2023} the stellar orbits were divided 
into four different stellar components: a cold component with near
circular orbits and strong rotation,  a warm component with weak rotation, 
a hot component with radial orbits and random motions, and a counter-rotating 
component with a strong-to-weak rotation in between the warm and the cold 
component.

To validate the current selection criteria, the right panels of 
Fig.~\ref{fig:bensby} show from top to bottom the same RRLs, but in the circularity 
of the orbits \vs\ the Galactocentric distance ($\lambda_z$ \vs\ R$_G$).  
Data plotted in this plane display that there is a fair agreement between the two 
different approaches. Indeed, more than 
73\% of Halo RRLs have hot orbits (-0.25$< \lambda_z \le$0.25),  and 
R$_G$ ranging from a few kpc to almost 100 kpc;  
59\% of TCD RRLs have warm orbits (0.25$\le$ $\lambda_z$ $\le$0.8), 
and R$_G$ on average smaller than 40 kpc; 
and 92\% of TND RRLs have cold orbits (0.80$\le$ $\lambda_z$) and 
R$_G$ at most of 10/20 kpc. According to this selection criteria, RRLs in 
retrograde orbits have a circularity smaller than $\lambda_z$ $\le$-0.25. 
Data plotted in the bottom panels of Fig.~\ref{fig:bensby} show that RRL in 
retrograde orbits are roughly the 7\% of the entire sample, moreover, 
they attain R$_G$ distances very similar to Halo RRLs and similar 
metal-poor iron abundances.

\begin{figure*}[] 
\centering
\includegraphics[width=0.9\textwidth]{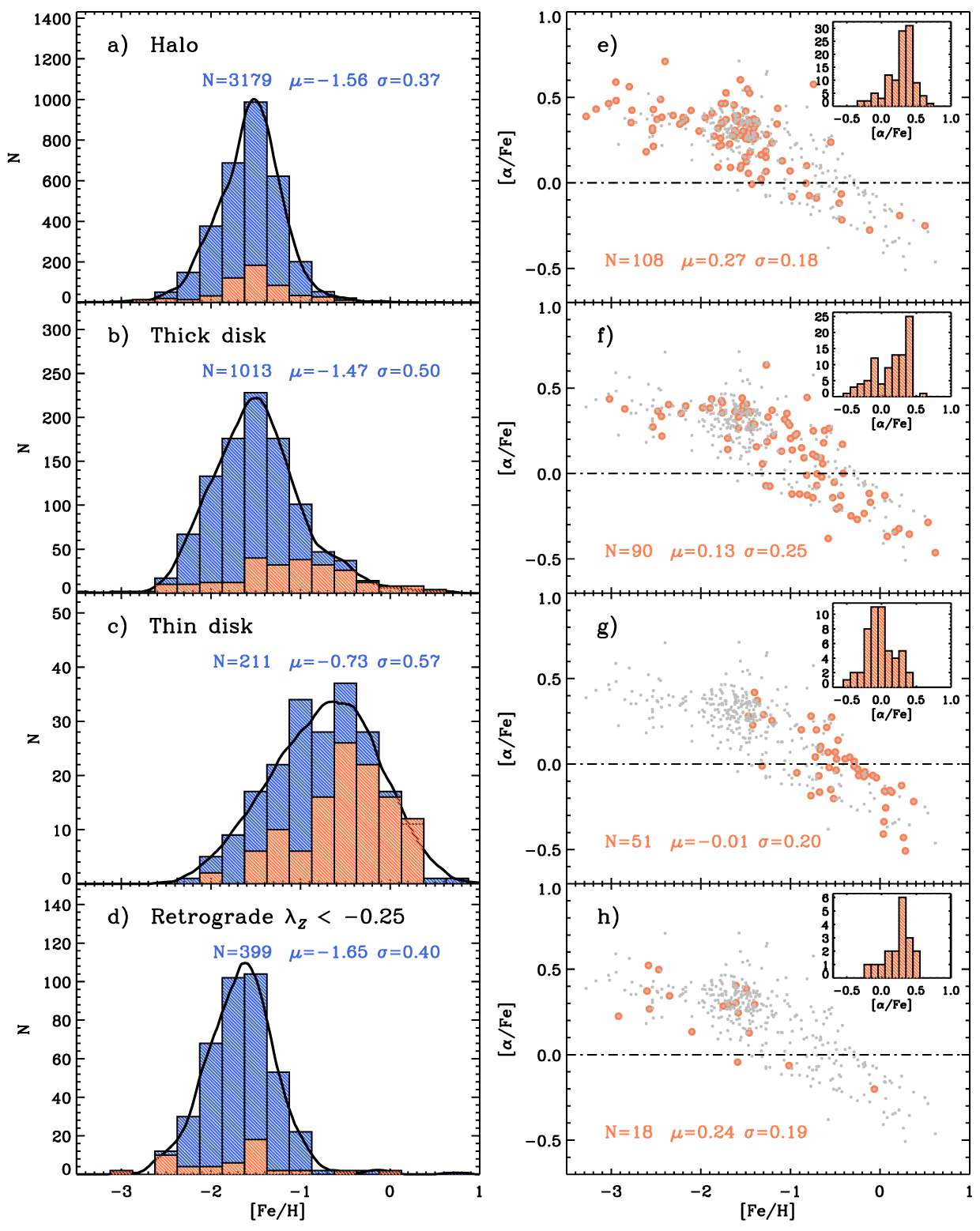}
\caption{Left -- Iron distribution function, based on LR (blue) and on HR (orange) spectra, 
for the four Galactic components (see labeled values) 
identified in Fig.~\ref{fig:bensby}. The black lines show the running average over the 
entire sub-sample. The number of RRLs, the mean and the standard deviations are also labeled. 
Right -- Same as the left, but for individual [$\alpha$/Fe] abundance ratios as a function of 
the iron abundance. The gray dots display the entire sample, while the orange circles individual 
abundances in the four Galactic components. The inset shows the [$\alpha$/Fe] distribution function. 
The mean and the standard deviations are also labeled. 
}
\label{fig:circ_histo}
\end{figure*}

\begin{figure*}[] 
\centering
\includegraphics[width=0.9\textwidth]{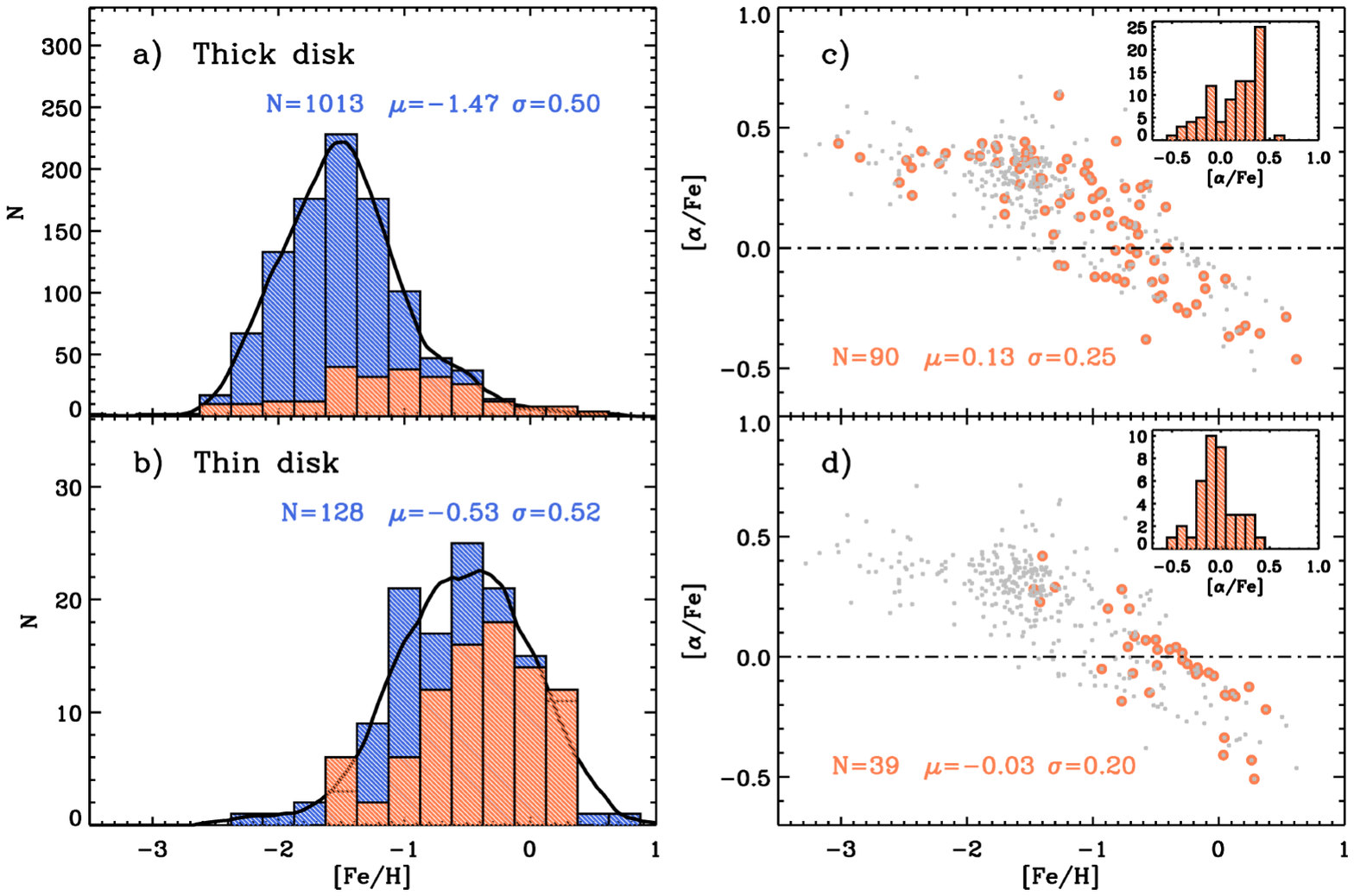}
\caption{Left -- Iron distribution function for TND and TCD  RRLs 
(see labeled values) identified following the criteria provided by  
\citet{bensby2014} joined with the criteria suggested by \citet{fernandezalvar2024} 
[$J_\phi/J_{tot}>0.95$, $|(J_Z-J_R)/J_{tot}|<0.05$ and $|Z_{MAX}|<1.5$~kpc]. 
The black lines show the running average over the entire sub-sample. The number 
of RRLs, the mean and the standard deviations are also labeled. 
Right -- Same as the left, but for the [$\alpha$/Fe] distribution function.
}
\label{fig:circ_histo_fernandez}
\end{figure*}

To investigate on a more quantitative basis the metallicity distribution functions, 
the left panels of Fig.~\ref{fig:circ_histo} display from top to bottom the IDFs of the 
different Galactic components. 
It should be noted that the peak of the IDF steadily shifts towards more metal-rich chemical 
compositions when moving from the Halo to the TND. Moreover, the IDF becomes 
broader and broader; indeed, the standard deviation of the distributions increases 
by 50\% when moving from the Halo to the TND (0.37 \vs\ 0.57). 
It is worth mentioning that TCD RRLs also show a broad IDF with a 
well--defined metal-rich tail, while the IDF of RRLs in retrograde orbits is 
very similar to the Halo IDF. 

The $\alpha$-element
distributions plotted in the right panels of Fig.~\ref{fig:circ_histo} 
are only based on HR spectra and they fully support the variations in the IDFs. Indeed, 
Halo RRLs are $\alpha$-enhanced since their mean $\alpha$-element abundance ratio is 
[$\alpha$/Fe]=0.27 ($\sigma$=0.18), and they only include a few outliers that are 
$\alpha$-poor. The TCD RRLs include RRLs that are either $\alpha$-enhanced 
for [Fe/H]$\le$-1.0 or $\alpha$-poor/$\alpha$-depleted in the more iron-rich regime. 
The reader interested in a more quantitative discussion concerning the impact that 
radial migrations have on the decrease in the [$\alpha$/Fe] abundance ratio of the 
TCD is referred to \citet{aumer2017a} and \citet{aumer2017b}.
Interestingly enough, TND RRLs are mainly $\alpha$-poor/$\alpha$-depleted, and 
indeed, their mean $\alpha$-element abundance ratio is [$\alpha$/Fe]=0 ($\sigma$=0.20).
However, there is a small group ($\sim$10 out of 50) that are either moderately 
$\alpha$-enhanced ([Fe/H]=$\sim$-1) or $\alpha$-enhanced ([Fe/H]$\sim$-1.5). 
The sample of $\alpha$ measurements for RRLs in retrograde orbits is limited (18), 
but they show, within the errors, an $\alpha$-element distribution similar to Halo RRLs. 
 
To provide a more stringent selection for Thin disk RRLs, we decided to use the selection criteria 
recently suggested by \citet{fernandezalvar2024}. They adopted two different 
selections for TND stars. They should be located in the right cusp 
of the action diamond (i.e. those with circular orbits on the Galactic plane) 
with L$_z$/L$_{tot}>$0.95 and $|(J_z - J_r)/J_{tot}|<$0.05. 
Moreover, they only selected stars with small distances from the Galactic 
plane $|Z_{max}|<1.5$~kpc. The bottom left panel of Fig.~\ref{fig:circ_histo_fernandez} 
displays TND stars once we applied the quoted selection criteria. 
The number of RRLs with TND kinematics decreased, as expected, from 211 to 128, 
while the mean iron abundance moves from [Fe/H]=-0.73 to [Fe/H]=-0.53, but the 
standard deviations are quite similar (0.57 vs 0.52).  Concerning the 
$\alpha$-element distribution (bottom right panel), the number of RRLs that are 
either moderately $\alpha$-enhanced ([Fe/H]=$\sim$-1) or $\alpha$-enhanced 
([Fe/H]$\sim$-1.5) decreased, but the relative fraction is roughly the same 
(7 out of 39). It is clear that TND stars, that are either moderately or 
$\alpha$-enhanced, are peculiar from the chemical point of view. They might be 
a limitation of the adopted kinematic selection criteria. It goes without saying 
that they are in a high-priority watch list and they are going to be further 
investigated for a more detailed analysis of their elemental abundances.
The impact of the new selection on the metallicity distribution functions of 
TCD stars is, within the errors, negligible (see top panels of 
Fig.~\ref{fig:circ_histo_fernandez}). 

The main result of the current analysis concerning the chemical enrichment of the 
different Galactic components is a steady variation from the iron-poor regime of 
the Halo to the metal-rich regime of the TND. This continuous  variation  
is further supported by the variation in the  $\alpha$-element abundance ratio, 
indeed it moves from $\alpha$-enhanced in the Halo to $\alpha$-poor in the TND,
with the TCD hosting both $\alpha$-enhanced and $\alpha$-poor/$\alpha$-depleted RRLs.
 
In this context it is worth discussing the variation in the 
metallicity distribution function when moving from Halo to TND RRLs. 
The occurrence of a metal-rich tail in the metallicity distribution function 
of field RRLs dates back to \citet{baade58c} in his seminal paper on Bulge/Disk 
RRLs. The very first solid spectroscopic identification of metal-rich RRLs was 
provided by \citet{preston1959,preston1961}, and later supported by 
\citet{butler1976} and by \citet{saha1984} using metallicity estimates 
based on the $\Delta$S method and by \citet{lub1977} on the basis of the 
Walraven photometry. Subsequent spectroscopic investigations for Bulge 
\citep{walker1991a} and field \citep{layden94,layden1995} RRLs
fully supported these preliminary evidence. Measurements of RRLs with either 
solar or super-solar iron abundance were provided by \citet{chadid2017,sneden2017,magurno2018,magurno2019}
and more recently by \citetalias[][]{crestani2021a,crestani2021b}
and by \citetalias{dorazi2024}. These investigations and the current findings indicate that 
the enhancement in $\alpha$-element abundances steadily decreases when moving from 
metal-intermediate to metal-rich RRLs. In passing we also note that the decrease in 
$\alpha$-enhancement does not imply a steady decrease in age. This circumstantial 
evidence is supported not only by Galactic chemical evolution models 
\citep[][see their figure~8]{micali2013,spitoni2019}, but also on the detailed 
chemo-dynamical analysis of TND and TCD stars recently provided by 
\citet{borbolato2025}.

The literature concerning the evolutionary channel producing metal-rich RRLs is wide
and dates back to \citet{taam1976}. On the basis of pioneering evolutionary calculations
of horizontal branch models at solar iron abundance and plain physical arguments
concerning the efficiency of the mass loss along the red giant branch (RGB), they found that
the probability to form metal-rich RRLs is significantly lower when compared with
typical metal-poor and metal-intermediate RRLs. Subsequent evolutionary and pulsation
prescriptions based on a wide range of chemical compositions ranging from the very
metal-poor to super-solar iron abundances \citep{bono97a,bono1997f,bono97b} suggested
that metal-rich RRLs can be formed according to canonical assumptions concerning old
progenitors and the efficiency of mass loss along the RGB. To take account
for the recent discovery of a large fraction of metal-rich RRLs it was also suggested
that they are significantly younger than canonical RRLs and they are the aftermath of
binary evolution \citep{bobrick2024}. Spectroscopic investigations based on HR spectra
suggest that three metal-poor/metal-intermediate RRLs out of $\approx$480 that have
been investigated so far display evidence of an enrichment in neutron capture elements
\citep{dorazi2025b}. This evidence indicates that they evolved in an interacting binary
system and the fraction is similar to non-variable stars. However, the identification
of binary RRLs via radial velocity variations is quite difficult due to their intrinsic
radial velocity displacements. The approach suggested by \citet{kervella2019,kervella2019a}
appears very solid and very promising, since it is based on the signature of the presence of
a companion on the RRL proper motion. In their seminal investigations they identified
seven spatially resolved, probably bound, systems out of the 789 RRLs that have been
investigated. This fraction should be considered as an upper limit, since only a minor
fraction is in interacting binaries. The role played by the iron content in metal-rich
RRLs deserves more detailed spectroscopic investigations.

\begin{figure*}[] 
\centering
\includegraphics[width=0.9\textwidth]{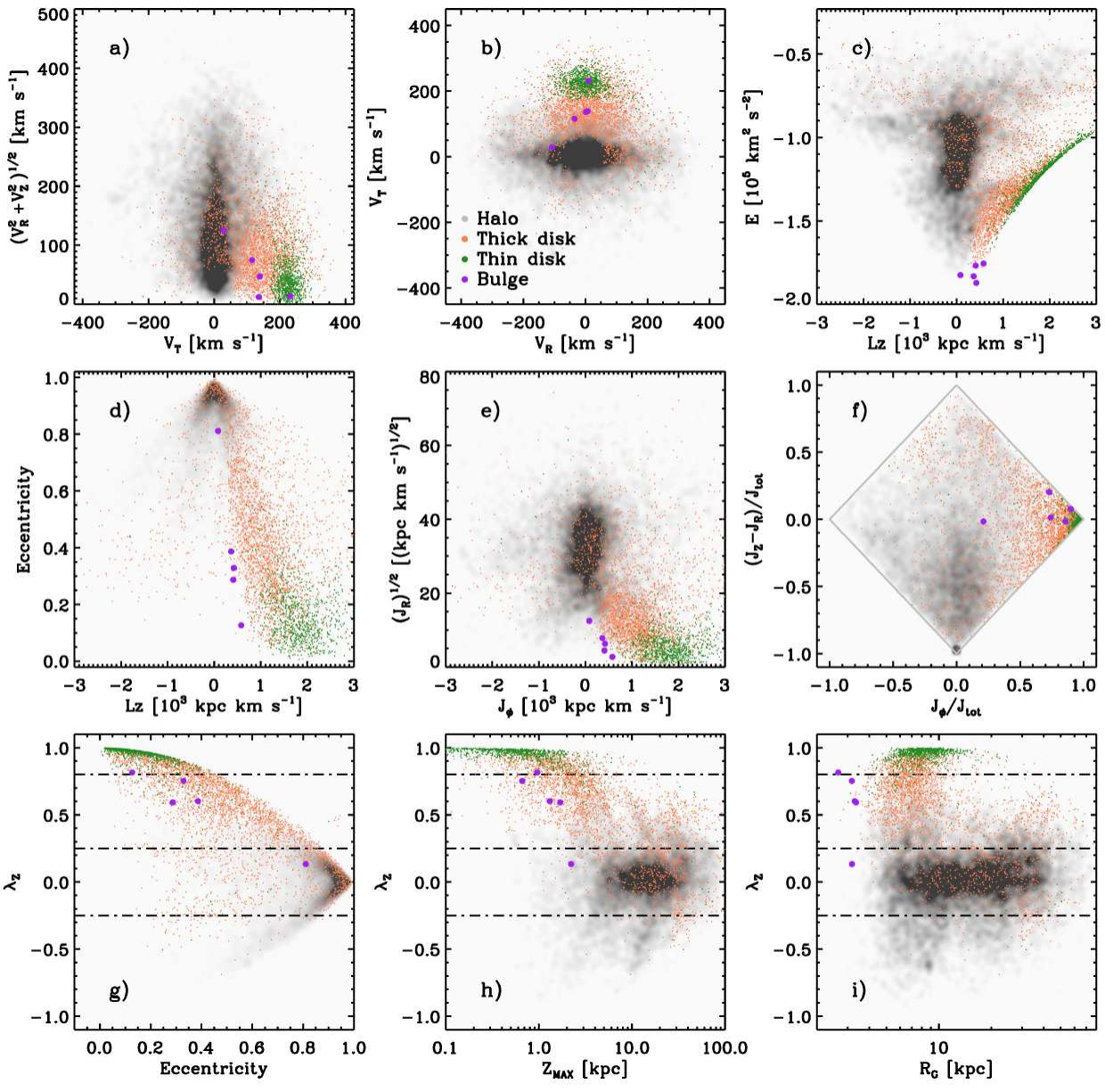}
\caption{Kinematic planes for RRLs in the main Galactic components. 
Panel a)--[Toomre diagram]: RRL transversal velocity as a function of the sum in 
quadrature of radial and vertical velocity components ($\sqrt{V^2_R+V^2_Z}$). 
TND and TCD RRLs are marked with green and orange dots, while Bulge RRLs 
are marked with magenta dots. The shaded gray area shows the distribution of Halo 
RRLs and RRLs in retrograde orbits. RRLs in the different Galactic components 
were selected according to \citet{bensby2014}, to circularity (retrograde), 
and to apocentric radius (Bulge). See text for more details.
Panel b): RRL transversal velocity as a function of the RV ($V_T$ \vs\ $V_R$).  
Panel c)--[Lindblad diagram]: RRL orbital energy as a function of the vertical 
component of the angular momentum ($E$ \vs\ L$_Z$). 
Panel d):  Eccentricity \vs\ $L_Z$.
Panel e): the square root of the radial action ($\sqrt{J_R}$) \vs\ $J_\phi=L_Z$. 
Panel f)--[Action diamond diagram]: $(J_Z-J_R)/J_{tot}$ \vs\ $J_\phi/J_{tot}$, 
where $J_{tot}=|J_\phi|+J_R+J_Z$).
Panel g): RRL circularity as a function of eccentricity. 
Panel h): RRL circularity as a function of the maximum height above the Galactic 
plane (logarithmic axis). 
Panel i): RRL circularity as a function of the Galactocentric distance 
(logarithmic axis).
}
\label{fig:gal_dynamic}
\end{figure*}

To properly identify RRLs with common origin we adopted the same dynamical 
planes suggested by \citet{lane2022}.
The first two diagrams are based on velocity information, in particular, 
panel a)  of  Fig.~\ref{fig:gal_dynamic} shows the Toomre diagram: 
the RRL transversal velocity as a function of the perpendicular velocity, i.e. 
the sum in quadrature of radial and vertical velocity components 
($\sqrt{V^2_R+V^2_Z}$) in Galactocentric cylindrical coordinates; 
panel b) shows the RRL in the V$_R$ \vs\ V$_T$ plane.  
Three panels are based on the vertical component of the angular momentum 
(L$_Z$), namely 
panel c) shows the Lindblad diagram, i.e. the RRL orbital energy--E--\vs\ L$_Z$, 
panel d) the RRL eccentricity \vs\ L$_Z$,
and panel e) the square root of the radial action ($\sqrt{J_R}$) 
versus $J_\phi=L_Z$.
Moreover, panel f) shows the "action diamond", i.e. 
$(J_Z-J_R)/J_{tot}$ \vs\ $J_\phi/J_{tot}$,  where $J_{tot}=|J_\phi|+J_R+J_Z$.
Together with these classical kinematic planes we also took advantage 
of three additional planes based on the circularity of the orbits. 
Panel g) shows the RRL circularity--$\lambda_Z$--as a function of the 
eccentricity, panel h) shows the circularity--$\lambda_Z$--as a function of Z$_{max}$, 
and panel i) displays the  circularity--$\lambda_Z$--as a function of the Galactocentric 
distance. We added these three planes because RRL cover a very broad range in 
Galactocentric distances, indeed they have been identified from the very center 
of the Galaxy \citep{minniti2016} to the outskirts of the Galactic Halo 
\citep{sesar2012a,medina2025a}. Moreover, they are ubiquitous across the Galactic spheroid 
and the three adopted parameters significantly change when moving from TND and TCD 
to Halo RRLs. 

To validate the criteria adopted to select RRLs in the main Galactic components, 
they have been plotted in the above kinematic planes. In particular, the shaded
gray area shows the distribution of Halo RRLs and RRLs with retrograde orbits. 
TND and TCD RRLs are marked with orange and green dots. 
Candidate TND and TCD RRLs can be easily identified in the Toomre diagram (panel a),  
since due to their relatively cold vertical velocity they cluster in the 
bottom right corner with $V_T$ $\simeq$250\,km~s$^{-1}$ and $V_T$ $\simeq$150\,km~s$^{-1}$, 
respectively. 
The same outcome applies for the candidate Halo RRLs and to RRLs in retrograde 
orbits, since they cover a very wide range in perpendicular velocities and they 
also show either vanishing or negative transversal velocities. 
The separation between TND/TCD and Halo RRLs is also quite 
clear in the V$_R$ \vs\ V$_T$ plane, since the three different components 
display a well--defined ranking in V$_T$ and a steady increase in RV.

In this context it is worth addressing on a more quantitative basis the 
RRLs with hot orbits. Indeed, according to the adopted selection criteria 
RRLs with hot orbits might belong either to the Halo or to the Bulge. 
To quantify the fraction of RRLs  that are 
confined to the Bulge, we identified the RRLs that along their orbits have 
an apocentric radius smaller than r$_{apo}$ $\le$3.5~kpc. Note that this search was 
performed on the one thousand realizations we computed to investigate their 
kinematic properties. We found that only five RRLs in our sample are candidate 
bulge RRLs, since their apocentric radius in the median value of one thousand 
orbits is smaller than 3.5~kpc. They are plotted as magenta diamonds in the 
dynamical planes of  Fig.~\ref{fig:gal_dynamic}. In particular, they can be easily 
identified in the Lindblad diagram (panel c), since they are characterized by 
very low orbital energies and vanishing vertical angular momentum and in the 
$\sqrt{J_R}$) \vs\ $J_\phi$ plane (panel e), since they also attain very low values 
in the square root of the radial action.

TND RRLs cluster (green dots), as expected, in the right corner of the action diamond
(panel f), since their action budget is mainly dominated by L$_Z$. Moreover, they are 
also distributed on a very tiny and compact slice in the Lindblad diagram with 
(panel c) very high L$_Z$ values and intermediate orbital energies. At the same time, 
TND RRLs display large circularities, small eccentricities (panel g) and small 
Z$_{max}$ (panel h) while in the eccentricity \vs\ $L_Z$ plane (panel d), and in 
the $\sqrt{J_R}$) \vs\ $J_\phi$ plane (panel e) they cluster in the bottom right corner.  

Candidate TCD RRLs (orange dots) share similar kinematic features, and they can also be 
easily identified, since they display a smooth transition between TND and 
Halo RRLs in the different kinematic planes. 

It is worth mentioning that the distribution of Halo RRLs in the 
circularity \vs\ Z$_{max}$ plane (panel h) is far from being homogeneous. 
They show four distinct stellar arcs that 
move from circularities typical of TCD ($\lambda_Z>$0.25) stars into 
circularities typical of RRLs with retrograde orbits ($\lambda_Z<$-0.25). 
Moving from the outskirts to smaller Z$_{max}$ distances they are located at 
Z$_{max}\sim$30--40~kpc, 10--20~kpc, 7--9~kpc, and 4--5~kpc. This appears as 
a new evidence suggesting that RRL radial distribution might be affected either 
by early merging events (see next section) or by peculiar radial motions. 
Note that the lack of these features in the circularity \vs\ Galactocentric distance 
plane indicates that they are mainly associated to variations 
perpendicular to the Galactic plane. 

To investigate on a more quantitative basis the possible 
difference in kinematic properties introduced by the adopted MW potential we also 
performed an independent selection of the four Galactic components by using the  
MW potential of \citet{mcmillan2017} as implemented in AGAMA \citep{vasiliev2019}. 
Data plotted in Fig.~\ref{fig:circ_histo_McMillan} (Appendix~\ref{sec:gal_pote}) 
show that the mean values and the standard deviations of both iron and $\alpha$-element 
distribution functions for the four Galactic components are very similar. The 
similarity also applies to the different kinematic planes, indeed 
Fig.~\ref{fig:gal_dynamic_McMillan}  (Appendix~\ref{sec:gal_pote}) shows  
that the distribution of the RRLs associated to the four Galactic components are, 
within the errors, very similar (Fig.\ref{fig:gal_dynamic}). Note that there is, 
as expected, a difference in the energy adopted in the Lindblad plane (panel c) 
due to the mass differences between the MWPotential2014 and the McMillan17 potential.

\subsection{Identification of stellar streams}\label{sec_stream}

The advent of homogeneous kinematical measurements for sizable sample of field 
stars together with accurate and homogeneous abundance measurements from 
ground-based spectroscopic surveys gave the impetus to a wide literature 
concerning the chemo-dynamical criteria to identify stars with a common origin
\citep[][and references therein]{helmi2020,antoja2020,feuillet2021,massari2025}.
In the following, we decided to use selection criteria that have already been 
suggested in the literature and the reason is twofold. 

{\em i)}-- We are planning to use the integrals of motion to identify 
cluster of RRLs in the different dynamical planes in order to have an objective 
and quantitative insight into the different identifications. However, 
a detailed clustering analysis is beyond the aim of the current investigation. 

{\em ii)}-- This is the very first time that the chemo-dynamics of a large sample 
of field RRLs is investigated on a quantitative basis and we are interested 
in the comparison with previous MW investigations based on different stellar tracers. 

For the identification of the main stellar streams, such as
Gaia-Sausage-Enceladus (GSE) and Sequoia, we adopted the same selection 
criteria adopted by \citet{feuillet2021}. A similar approach was also undertaken 
by \citet{bonifacio2024} in their analysis of high speed stars. Moreover, we also followed the
identification of the Sagittarius (SGR) stream recently provided by \citet{antoja2020}, using 
overdensities in proper motion distributions provided by Gaia DR2. Note that for 
this stream we did not apply kinematic selections, we only performed a cross-correlation 
between the SR3C catalog and their catalog of SGR candidates. Concerning the identification 
of RRLs in the Helmi stream we adopted the kinematic criteria recently suggested by 
\citet{koppelman2019a} and \citet{Horta2023}.

\begin{figure*}[] 
\centering
\includegraphics[width=0.9\textwidth]{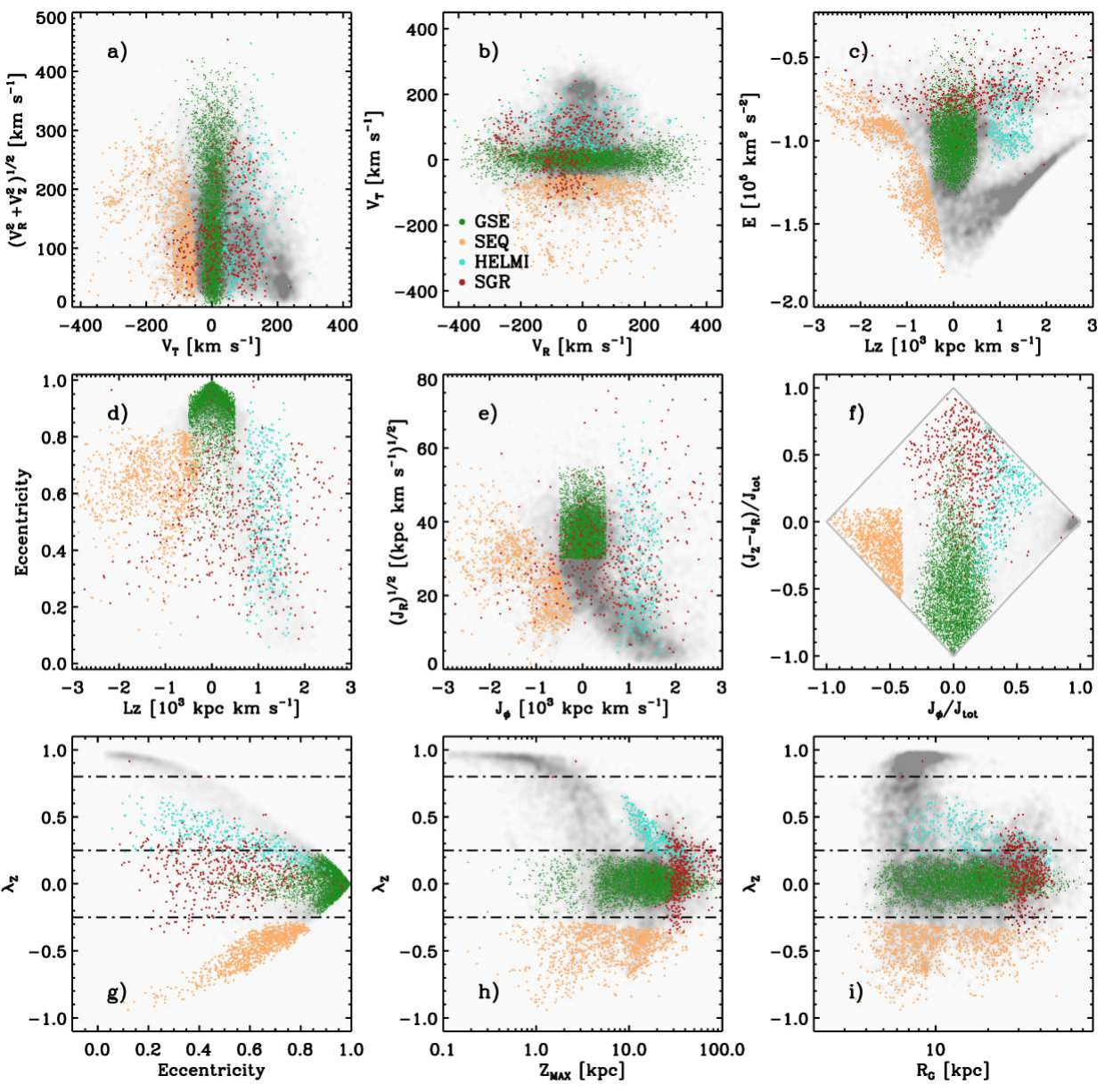}
\caption{Kinematic planes for RRLs in the four selected 
stellar streams. The green dots mark candidate GSE RRLs, orange
dots display candidate SEQ1 and SEQ2 RRLS according to \citet{feuillet2021}. 
The turquoise dots mark Helmi RRLs, according to \citet{koppelman2019a} and \citet{Horta2023}.
The red dots show candidate SGR RRLs according to \citet{antoja2020}.
The gray shaded area shows the same RRLs belonging to the main Galactic 
components plotted in Fig.~\ref{fig:gal_dynamic}
}
\label{fig:stream_dynamic}
\end{figure*}

Fig.~\ref{fig:stream_dynamic} shows the identification of the different stellar 
streams in the same kinematic planes adopted to identify the Galactic 
components. RRLs that according to the kinematic criteria belong to GSE have 
been marked with green dots, while RRLs associated to the different Galactic 
components (see Fig.~\ref{fig:gal_dynamic}) are shown as a shaded gray area.
Candidate GSE RRLs can be easily identified, since they display a well 
defined over-density not only in the Toomre (panel a) and in the 
Lindblad (panel c) diagram, but also in the action diamond (panel f). 
The compact distribution that GSE RRLs show in panel h) further supports 
the evidence that they originated from a single merger event and their 
clustering in the bottom corner of the action diamond is probably 
tracing the actions of the progenitor.
Candidate GSE RRLs also show a compact distribution in the square root 
of the radial action \vs\ the vertical angular momentum (panel e) 
as originally suggested by \citet{feuillet2021}. 
Indeed, the GSE peaks at radial action values that are systematically larger than 
Halo RRLs. Moreover, they also differ when compared with RRLs with retrograde 
orbits and with TND/TCD RRLs. Similar peaks for GSE RRLs  can be 
easily identified in the eccentricity \vs\ L$_Z$ plane (panel d) and in 
the circularity \vs\ eccentricity plane (panel g).  

The identification of the Sequoia stream is more complex. Candidate 
Seq RRLs, marked with orange dots, cluster in the left corner of the action diamond, since this is 
the plane originally adopted by \citet{feuillet2021}.  However,  the 
Sequoia sample clearly splits into two different groups in the Lindblad diagram 
(at energy E$\simeq$-1.1 10$^5$~km$^2$~s$^{-2}$), 
in the eccentricity \vs\ L$_Z$ plane, and in the  $\sqrt{J_R}$ \vs\ $L_Z$ plane.
It is worth mentioning that Sequoia RRLs display a well--defined minimum 
for $R_G\sim$12~kpc in the circularity \vs\ Galactocentric plane (panel i). 
We adopted this criterion to split the Sequoia candidate into 
Seq.~~1 (R$_G\le12$ kpc) and Seq.~~2 (R$_G >12$ kpc) RRLs. In passing we note that according to 
the separation in the Lindblad diagram there are solid reasons to believe 
that Seq.~1 RRLs might be associated with Thamnos RRLs (\citetalias{dorazi2024}). 

Candidate SGR RRLs (red dots) do not show clear over-densities in the kinematic planes. 
There are only two exceptions. They are relatively concentrated in the 
circularity \vs\ Z$_{max}$ plane and in the circularity \vs\ R$_G$ plane.  
Indeed, they cover a narrow range in Galactocentric distances (R$_G$=30--40~kpc) 
and in maximum height above the Galactic plane (Z$_{max}$=20--30~kpc). 
Moreover, there is also evidence for clustering in the top corner of the action 
diamond, but in the other kinematic planes they overlap with canonical outer Halo 
RRLs. 
The compact distribution in Galactocentric distance and in Z$_{max}$ further 
supports the evidence that SGR candidates in the current catalog mainly trace 
the SGR stream, with a minimal, if any, contribution from the SGR core.

Candidate Helmi RRLs (turquoise dots) display mixed behavior in the different
kinematic planes. They show a smooth distribution in the Toomre diagram 
(panel a) and in the transversal velocity plane (panel b), but they show a 
well--defined concentration in the Lindblad plane across  
E$\sim$-1.0~10$^5$~km$^2$~s$^{-2}$, L$_Z\sim$1.0~10$^3$~kpc~km~s$^{-1}$ 
and a high-energy tail extending from E$\sim$-0.9~10$^5$~km$^2$~s$^{-2}$.
The distribution is relatively compact in the action-diamond plane (panel f) and 
in the circularity \vs\ Z$_{max}$ plane (panel h)
and in the circularity \vs\ R$_G$ plane (panel i). Data plotted in the 
$\lambda_Z$ \vs\ Z$_{max}$ plane indicate that candidate Helmi RRLs  
are mainly located in the TCD, since Z$_{max}$ ranges from a few kpc to $\sim$20 kpc 
and $\lambda_Z$ between 0.7 and 0.2. Further improvements in the kinematics selection 
of this stellar stream are highly recommended, since it is distributed between the 
TCD and the Halo.

\begin{figure*}[] 
\centering
\includegraphics[width=0.9\textwidth]{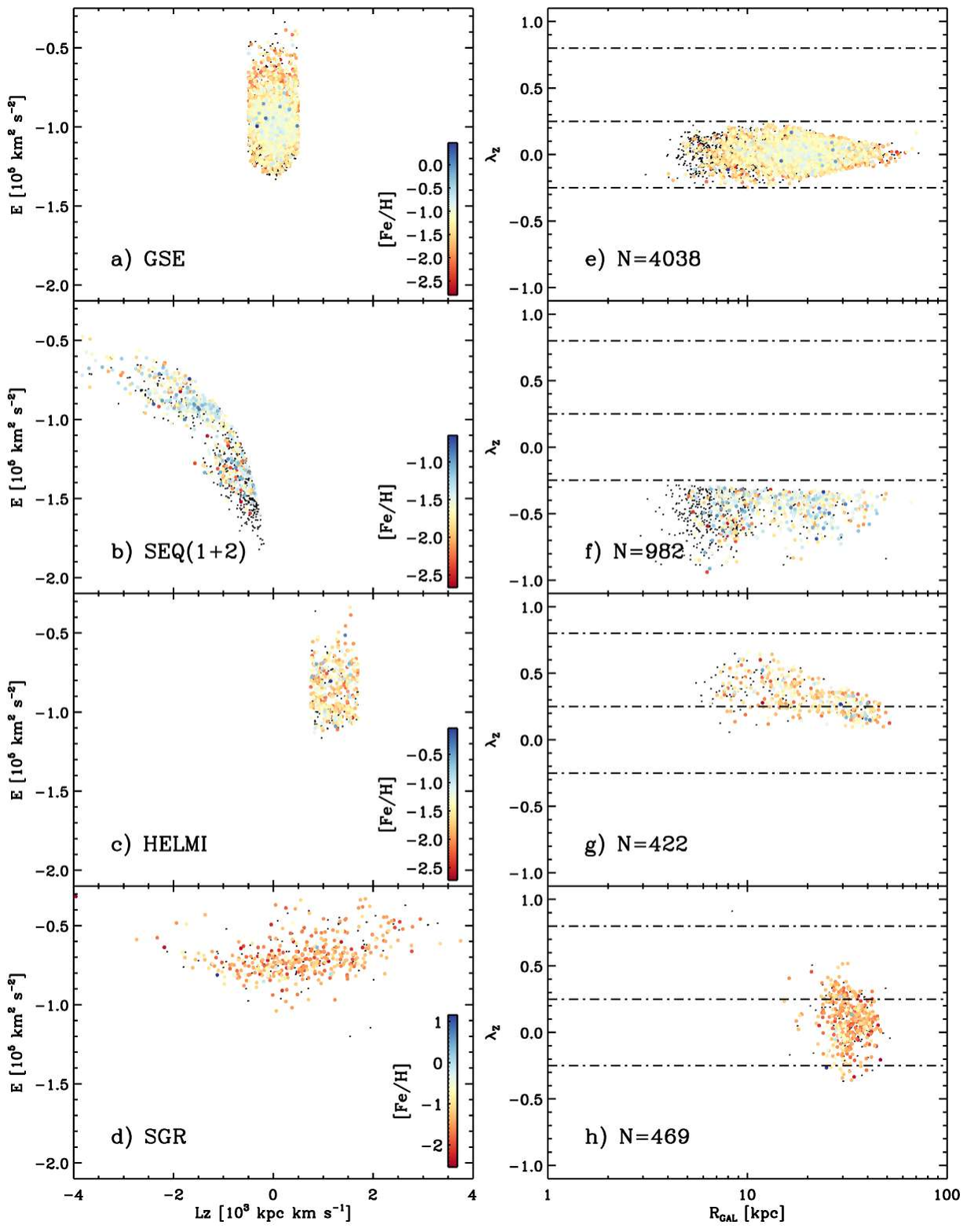}
\caption{Left -- Lindblad diagram for the four streams identified in 
Fig.~\ref{fig:stream_dynamic}. The symbols are color-coded according to the iron 
abundance (see color bar in the inset).  
Right -- Same as the left, but the circularity of the orbits ($\lambda_Z$)  is 
plotted as a function of the Galactocentric distance. The number of RRLs in 
the different streams are labeled. See text for more details.
}
\label{fig:streams}
\end{figure*}

\begin{figure*}[] 
\centering
\includegraphics[width=0.9\textwidth]{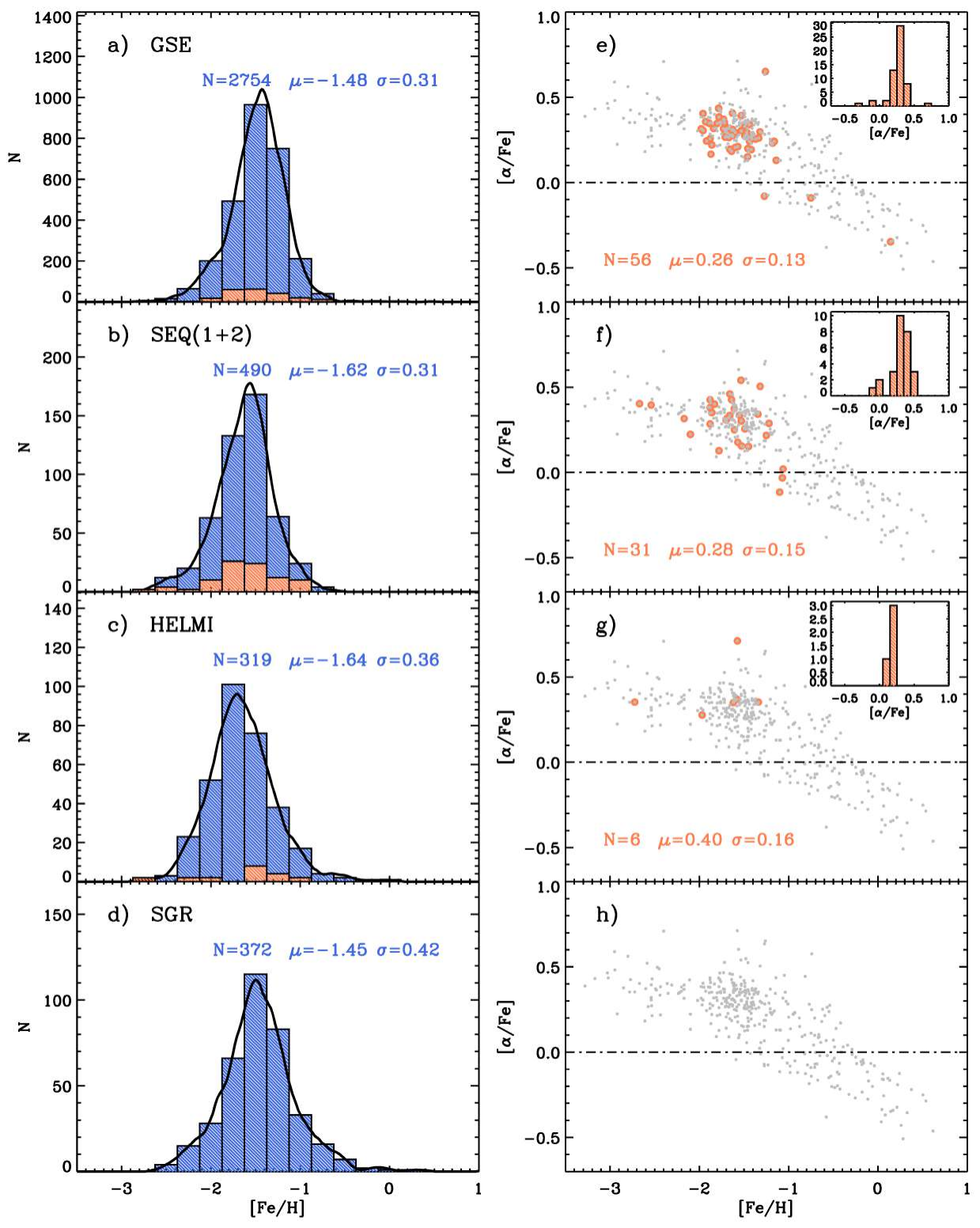}
\caption{Left -- Iron distribution function for the four stellar streams (see labeled names) 
identified in Fig.~\ref{fig:stream_dynamic}. The black lines shows the running average over the 
entire sub-sample. The number of RRLs, the mean and the standard deviations are also labeled. 
Right -- Same as the left, but for the [$\alpha$/Fe] abundance ratios as a function of the iron 
abundance. The gray dots display the entire Galactic sample, while the orange circles individual 
abundances in the four stellar streams. The inset shows the [$\alpha$/Fe] distribution function.}
\label{fig:stream_idf}
\end{figure*}

The circularity \vs\ Z$_{max}$ plane is a very interesting diagnostic to investigate 
the spatial distribution of stellar streams. Indeed, Seq.~1 and Seq.~2 RRLs have 
retrograde orbits and display in this plane a well-defined separation also in Z$_{max}$. In particular, 
the Seq.~2 RRLs attain maximum heights above the Galactic plane  that are larger than 
9--10~kpc, while Seq.~1 RRls are located closer to the Galactic plane. It is worth mentioning, 
that in this plane Sequoia, and SGR RRLs appear to be associated with the stellar arcs 
discussed in the identification of the main Galactic component. SGR RRLs are associated with 
the stellar arc located at larger Galactocentric distances (Z$_{max}\sim$30--40~kpc) 
while Helmi and Seq.~2 RRL are associated with the stellar arc located at Z$_{max}\sim$10--20~kpc, 
and Seq.~1 to the stellar arcs located at 7--9~kpc, and 4--5~kpc.
The main difference is that candidate Sequoia RRLs are in retrograde motion, while 
Helmi RRLs have prograde motion. A detailed discussion of these kinematic features 
is beyond the aim of the current investigation.

To investigate the global properties of the selected streams, Fig.~\ref{fig:streams} 
shows the candidate RRLs in the Lindblad diagram (left panels) and in the 
circularity \vs\ Galactocentric distance (right panels). In analogy with the analysis 
we performed to characterize the different Galactic components (see Fig.~\ref{fig:bensby}) 
the symbols were color-coded according to the iron abundance (color bar in the inset). 
Data plotted in panels a) and e) display that GSE RRLs are quite homogeneous  
in iron abundance, indeed they are metal-intermediate and show a homogeneous chemical 
enrichment not only in the range of energies covered in the Lindblad diagram, but also 
in the wide range of Galactocentric distances they are covering (from $\sim$5 to $\sim$50 kpc). 

Sequoia RRLs display a more complex distribution, since the Seq.~1 candidates 
located at modest Galactocentric distances display a broader iron distribution 
when compared with Seq.~2 candidates. This trend becomes even more clear for candidate Helmi RRLs, 
indeed RRLs either with TND kinematics or approaching TND kinematics are systematically 
more metal-rich than RRLs with TCD kinematics located at larger Galactocentric distances.   
The candidate SGR RRLs display a homogeneous spatial distribution with a well-defined mix 
between metal-poor and metal-intermediate RRLs. 

To further constrain the accuracy of the kinematic selections for the stellar 
streams, we decide to follow the same approach we adopted for the Galactic 
components. The left panels of Fig.~\ref{fig:stream_idf} show from top to 
bottom the IDFs for the selected stellar streams. GSE RRLs show an IDF that is, 
within the errors, quite similar to Halo RRLs. The similarity also applies to the 
standard deviation (0.31 \vs\ 0.37). Note that this comparison is quite robust, 
since we are comparing two samples with a similar number of RRLs (2754 \vs\ 3179)
and with a similar range in Galactocentric distances. However, GSE RRLs lack 
the metal-poor/metal-rich tails and their $\alpha$-element 
distribution (top right panel of Fig.~\ref{fig:stream_idf}) is quite compact.   

The Seq.~1 and Seq.~2 RRLs are also quite similar to Halo RRLs, indeed their IDFs 
peak at [Fe/H]=-1.68 and -1.60. However, Seq.~2 RRLs have a more symmetrical IDF, 
and in particular, a smaller standard deviation (0.28 \vs\ 0.38). Seq.~2 RRLs also 
lack the metal-poor/metal-rich tails, and show a compact distribution in 
$\alpha$-elements. The chemical evidence for Seq.~1 RRLs is less clear; however, 
the evidence that Seq.~1 and Seq.~2 RRLs have retrograde orbits and cover 
limited ranges in Galactocentric distances further supports the solid identification 
of this stellar stream.

The Helmi RRLs display an IDF function very similar to Halo RRLs, indeed 
their IDF peaks at [Fe/H]=-1.64 (versus -1.56) and their standard deviation 
is $\sim$0.36 (versus 0.37).

The SGR RRLs also have very similar IDF and standard deviation (0.42 \vs\ 0.37) 
to the Halo RRLs, with marginal evidence for a metal-poor/metal--rich tail. Moreover, 
they are located at large Galactocentric distances and we still lack measurements of 
their $\alpha$-element abundances. 

Finally, to further constrain possible systematics in the kinematic 
properties of candidate RRLs associated to the four stellar streams we also performed an 
independent orbital integration by using the MW potential of \citet{mcmillan2017}. 
Data plotted in Fig.~\ref{fig:stream_dynamic_McMillan} (Appendix~\ref{sec:gal_pote})
shows that four stellar streams display in the different kinematic planes distributions 
that are, within the errors, quite similar to the distributions based on the MW potential 
of \citet{bovy2015} (see Fig.~\ref{fig:stream_dynamic}).

\section{Comparison with different stellar tracers}\label{comp_chem}


To validate the current metallicity scale we compare the chemo-dynamical properties 
of old stellar tracer like RRLs with similar properties for Galactic Globular 
Clusters (GGCs) that are popular tracers of old stellar populations. The kinematic 
properties of GGCs were estimated by using proper motions and distances provided by
\citet{baumgardt2019} together with the same Galactic potential adopted 
for field RRLs. The association with the different Galactic components was 
performed by using the same criteria adopted for field RRLs (see \S~\ref{sec_Galcomp}). 
Concerning the association of GGCs to the different stellar streams, we took advantage 
of the recent detailed chemo-dynamical investigation provided by \citet{massari2025}. 

Several investigations in the literature have recently provided identification 
of globulars likely sharing common origins by using either different 
chemical and dynamical criteria or different input parameters (distances, 
kinematics, metallicity distribution functions) or different theoretical 
frameworks. Typically, they reach different conclusions concerning the 
co-natal globulars associated to the different stellar streams. Moreover and even 
more importantly, there is mounting theoretical evidence that accreted globular 
clusters do not show dynamical coherence, since they do not cluster in typical 
kinematic spaces \citep{pagnini2023}.

\begin{figure*}[] 
\centering
\includegraphics[width=0.9\textwidth]{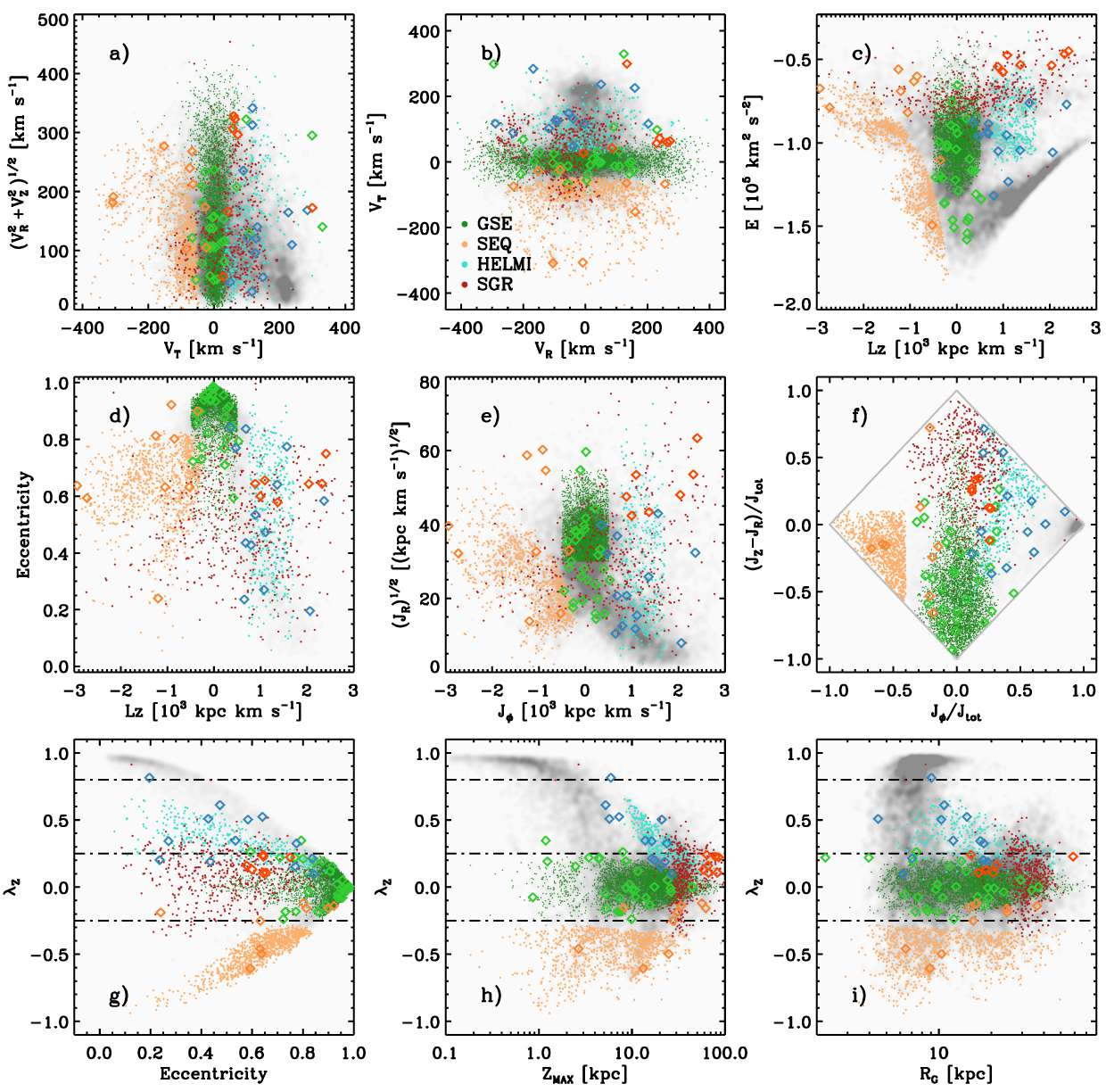}
\caption{Comparison in the kinematic planes between field RRLs in the selected 
stellar streams and Galactic Globular Clusters that have been associated to 
the same stellar streams \citep{massari2025}. The symbols and the colors for field 
RRLs are the same as in Fig.~\ref{fig:stream_dynamic}. The GCs are marked with 
green (GSE), orange (Sequoia), cyan (Helmi) and red (SGR) diamonds.  
}
\label{fig:stream_ggc_dynamic}
\end{figure*}

In the following, we perform a preliminary comparison between field RRLs
and GGCs in the canonical kinematic planes and in the radial gradients for a consistency
check between old stellar tracers. 

For this purpose, we put together a catalog of GGCs by collecting data from the literature, more 
specifically 165 from \citet{massari2025} and 19 from \citet{harris96,harris10}.
The metallicities and $[\alpha/Fe]$ values were adopted from \citet[][plus Carretta private communication, 27 GCs]{carretta09}
and from \citet[][plus Zoccali private communication, four Bulge GCs]{zoccali2004}.

Fig.~\ref{fig:stream_ggc_dynamic} shows the comparison in the same kinematic planes 
between the four stellar streams and the GGCs that according to kinematic selections 
have been associated to the same stellar streams. Data plotted in this figure 
bring forward a few empirical circumstances: 

a) {\it GSE} -- Field RRLs and GGCs associated to GSE show a very similar distribution 
not only in the classical six dynamical planes, but also in the planes based on the 
circularity \vs\ eccentricity, Z$_{max}$ and Galactocentric distance (panels g,h,i). 
There is evidence of two outliers (Djorgovski~1, NGC~6333 [M9]) in the Toomre diagram (panel a) and 
in the radial \vs\ vertical velocity (panel b), since they attain transversal 
velocities that are systematically larger than the bulk of the stream. A few outliers are 
also present in the action diamond plane (panel f), indeed a few GGCs associated to GSE 
attain $J_\phi/J_{tot}$ values that are either systematically smaller 
(NGC~288, NGC~6205 [M13], NGC~7099 [M30]) or larger (Djorgovski~1) than the bulk of the stream. 
Moreover, Ter~10, Djorgovski~1 and NGC~6333 appear to be outliers in the eccentricity \vs\ 
Galactocentric distance,  since their distances are systematically smaller than the bulk of the GSE stream. 

b) {\it Sequoia} -- There are nine GGCs that have been associated  to the Sequoia stellar stream
and their distribution in the kinematic planes marginally follows the RRLs associated to the 
same stream. Several outliers are present not only in the Toomre diagram (panel c), in the Lindblad 
diagram (panel c), but also in the action plane (panel f) in which they attain $J_\phi/J_{tot}$ values 
that are either systematically larger  than the bulk of the stream (AM4, IC~4499, Pal~13, NGC~2298, 
NGC~5466, NGC~7006 [C42]). Data plotted in the circularity versus Galactocentric distance plane  
(panel i) are suggesting that three GGCs NGC~5139, NGC~6101 and NGC~3201 might be associated with Seq.~1, 
while the data plotted in the circularity \vs\ Z$_{max}$ (panel h) are suggesting that NGC~3201 and 
NGC~6101 are associated to Seq.~2 and NGC~5139 to Seq.~1. 
Moreover, at least five (AM4, Pal~13, NGC~2298, NGC~5466, NGC~7006) out of the nine GGCs associated 
to this stream are located, according to the current classification, at the transition between 
prograde and retrograde orbits. In order to constrain on a quantitative basis this issue we 
estimated the kinematic properties of both field RRLs and GGCs by using the Galactic potential 
suggested by \citet{mcmillan2017} and we found that the quoted clusters are located once again 
at the transition between prograde and retrograde orbits.

c) {\it Helmi} -- The comparison in the kinematic planes indicates that field RRLs and 
GGCs associated to the Helmi stream share similar distributions. There are a few outliers that 
show up not only in the Toomre (panel a) and in the Lindblad (panel c) diagram (NGC~4590 [M68], E~3),
but also in the eccentricity \vs\ L$_Z$ and in the action diamond 
plane (NGC 6426, NGC~7078 [M15]). The comparison is hampered by the fact that RRLs 
associated to the Helmi stream cover a very wide range in Galactocentric distances.  

d) {\it Sagittarius} -- There are seven GGCs that have been associated to the SGR stream. 
The globular Pal~12 seems to be an outlier in the Toomre diagram (panel a) and 
in the radial \vs\ vertical velocity (panel b), but it follows the bulk of the SGR stream
in the other dynamical planes. 
Note that the association between NGC~6715 (M54) and SGR is supported not only by the Lindblad 
diagram and by the action-diamond plane, but also by the plane showing the circularity of the 
orbit as a function of Z$_{max}$. They typically display a compact spatial distribution 
(bottom panels). However, the distribution in the circularity \vs\ R$_G$ (panel i) is 
suggesting that five GGCs (Pal~12, NGC~6715, Ter~7, Ter~8, Arp~2) are associated to 
the core of the SGR stellar stream (R$_G\le$25~kpc) while the other two 
(Laevens~3, NGC~2419), located at larger Galactocentric distances, to the stream of this  
disrupting galaxy.

\begin{figure*}[] 
\centering
\includegraphics[width=0.95\textwidth]{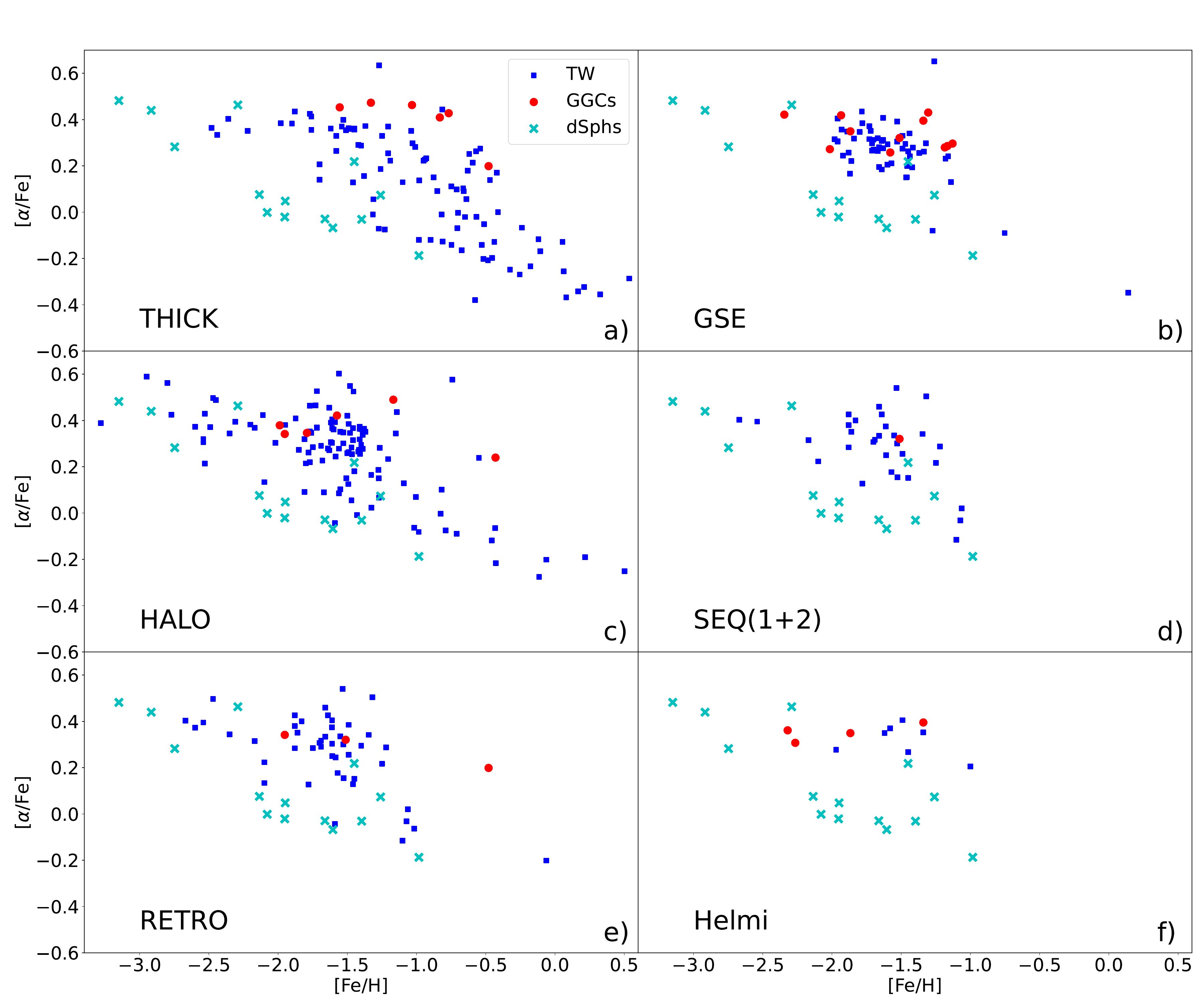}
\caption{Comparison in the chemical plane ([$\alpha$/Fe] \vs\ [Fe/H]) among field 
RRLs (blue squares) in different Galactic components (left panels) and in three  
stellar streams (right panels) with GGCs (red circles) 
and nearby dwarf galaxies (light blue crosses).  RRLs with retrograde orbits 
are plotted in the bottom left panel as orange squares.
}
\label{fig:stream_ggc_chemistry}
\end{figure*}

The same comparison was also performed using the chemical planes [$\alpha$/Fe] \vs\ [Fe/H]. 
Panel a) of Fig.\ref{fig:stream_ggc_chemistry} shows that candidate TCD RRLs cover a broad  
range in metallicity, but they are systematically more $\alpha$-enhanced than nearby dwarf 
galaxies. Interestingly enough, GGCs that have been associated to the TCD 
\citep[red circles,][]{callingham2022} only cover the metal-intermediate regime 
and they are, at fixed iron abundance, more $\alpha$-enhanced than TCD RRLs.  

The comparison with Halo RRLs (panel c) is more homogeneous, indeed Halo RRLs 
and RRLs with retrograde orbits (panel e) show as expected a similar trend in  
[$\alpha$/Fe] \vs\ [Fe/H] and agree quite well with GGCs. Moreover, in the 
metal-poor tail ([Fe/H]$\le$-2) they also agree with nearby dwarf galaxies, while 
in the metal-intermediate regime field RRLs and GGCs are systematically more 
$\alpha$-enhanced (see for more details \citealt{fabrizio15}). 

The comparison of RRLs associated to the three stellar streams and nearby stellar 
systems (right panels of Fig.\ref{fig:stream_ggc_chemistry}) shows quite clearly 
that RRLs associated with GSE and Sequoia attain, at fixed iron abundance, 
$\alpha$-element abundances similar to the GGCs associated to these streams. Thus 
supporting the current associations. The  dwarf galaxies are once again 
systematically more $\alpha$-poor in the metal--poor/metal--intermediate regime. 
The comparison with nearby stellar systems is partially hampered either by the lack 
or by the limited number of dwarf galaxies and GGCs more metal-rich than 
[Fe/H]$\ge$-1.

\begin{figure*}[] 
\centering
\includegraphics[width=0.9\textwidth]{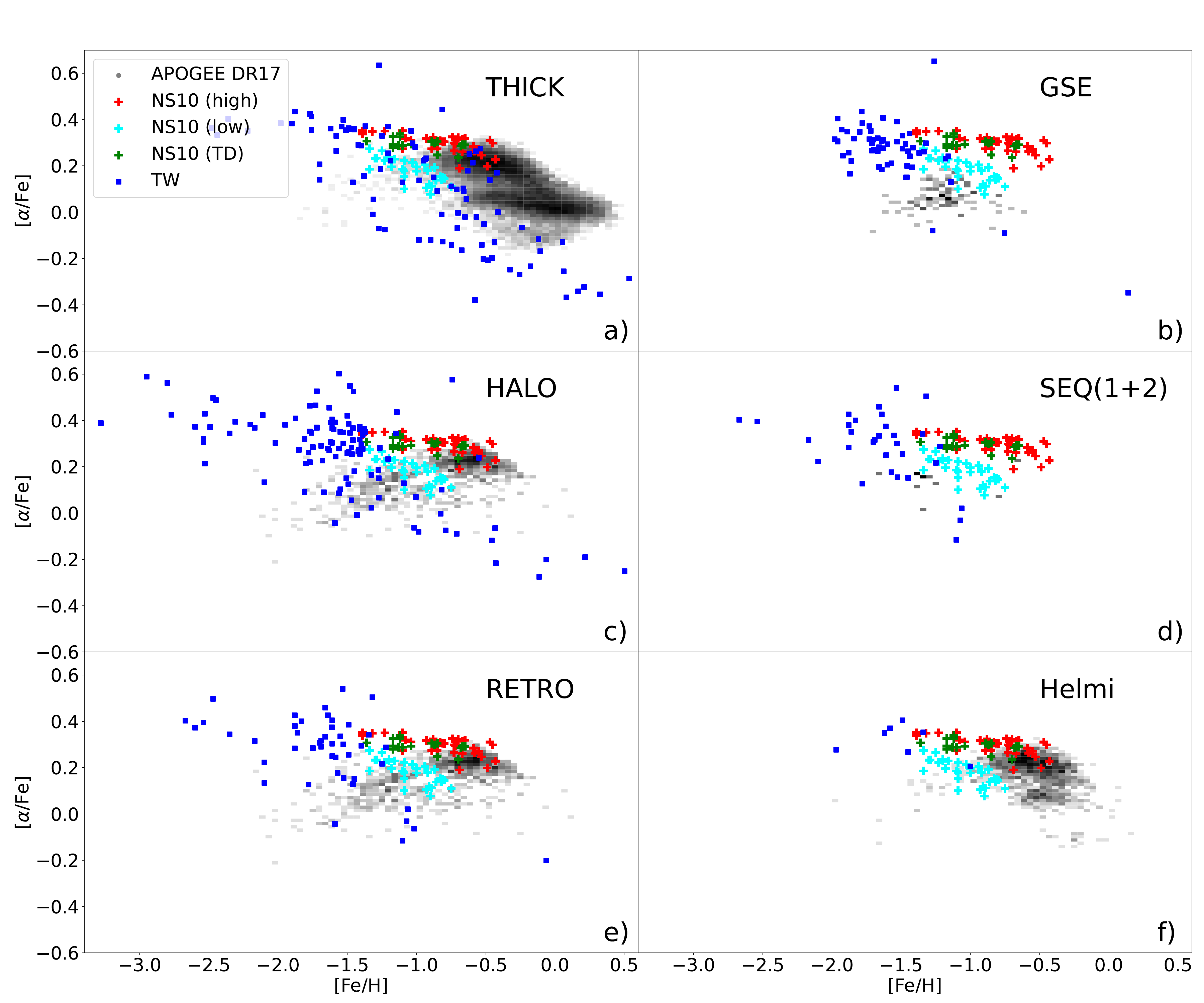}
\caption{Same as Fig.~\ref{fig:stream_ggc_chemistry}, but the comparison is between field 
RRLs in the current catalog (blue squares) and field stars observed by APOGEE \citep[grey shaded area,][]{apogee_dr14} and by \citet{nissen2010}.
The latter sample is plotted with different symbols: red crosses show stars with an high-$\alpha$ abundance, light blue crosses stars with low-$\alpha$ abundance, and green crosses are candidate
TCD stars.
}
\label{fig:stream_field_chemistry}
\end{figure*}

To overcome this limitation, panel a) of Fig.~\ref{fig:stream_field_chemistry} shows a similar
comparison as in Fig.~\ref{fig:stream_ggc_chemistry}, but with field stars observed
either by APOGEE \citep{apogee_dr14} or by \citet[][NS10]{nissen2010} in the solar
neighborhood. The former sample is plotted as a gray shaded area, while the latter is plotted with different symbols: red crosses show stars with a high-$\alpha$ abundance, 
light blue crosses stars with low-$\alpha$ abundance, and green crosses candidate
TCD stars.  

The TCD RRLs (panel a) display a trend quite similar to the low-$\alpha$ subsample collected by 
NS10. The main difference is that TCD RRLs move into the metal-rich regime and display, 
at fixed iron content, a larger spread in  $\alpha$ abundance.  The similarity is also 
quite good with APOGEE. Indeed, they attain similar $\alpha$-values in the 
metal-intermediate and in the metal-rich regime. Field RRLs display a plume of 
field RRLs across solar iron abundance that is systematically more $\alpha$-poor 
when compared with APOGEE RRLs. 
  
Panel c) and Panel e) of the same figure show that Halo and retrograde RRLs are, 
as expected, systematically more metal-poor when compared with APOGEE and 
SN subsamples. Moreover, in the metal-intermediate 
regime (-1.5$\le$[Fe/H]$\le$-1.0) Halo RRLs display a continuous transition
between the $\alpha$-rich  and the $\alpha$-poor regime.
Data plotted in the left panels indicate that field RRLs show a steady decrease in 
$\alpha$ abundances when moving from the metal-poor to the metal-rich regime, 
while field stars display a well--defined dichotomy in $\alpha$ abundance.
Note that this evidence applies to TCD, Halo and Retro RRLs. 

The right panels show  that RRLs in the three stellar streams minimally overlap 
with literature samples. Indeed, GSE RRLs (panel b) are mainly located in the 
metal-intermediate regime and they are systematically more $\alpha$-enhanced 
than both APOGEE and SN samples. The same outcome applies to Sequoia RRLs 
(panel d). The Helmi RRLs only marginally overlap with field stars in the 
metal-intermediate regime, but their trend is similar to  Halo+Retro RRLs 
(panel f).

\section{Radial metallicity gradient}\label{radial_grad}

\begin{figure*}[] 
\includegraphics[height=0.80\textheight, width=0.95\textwidth]{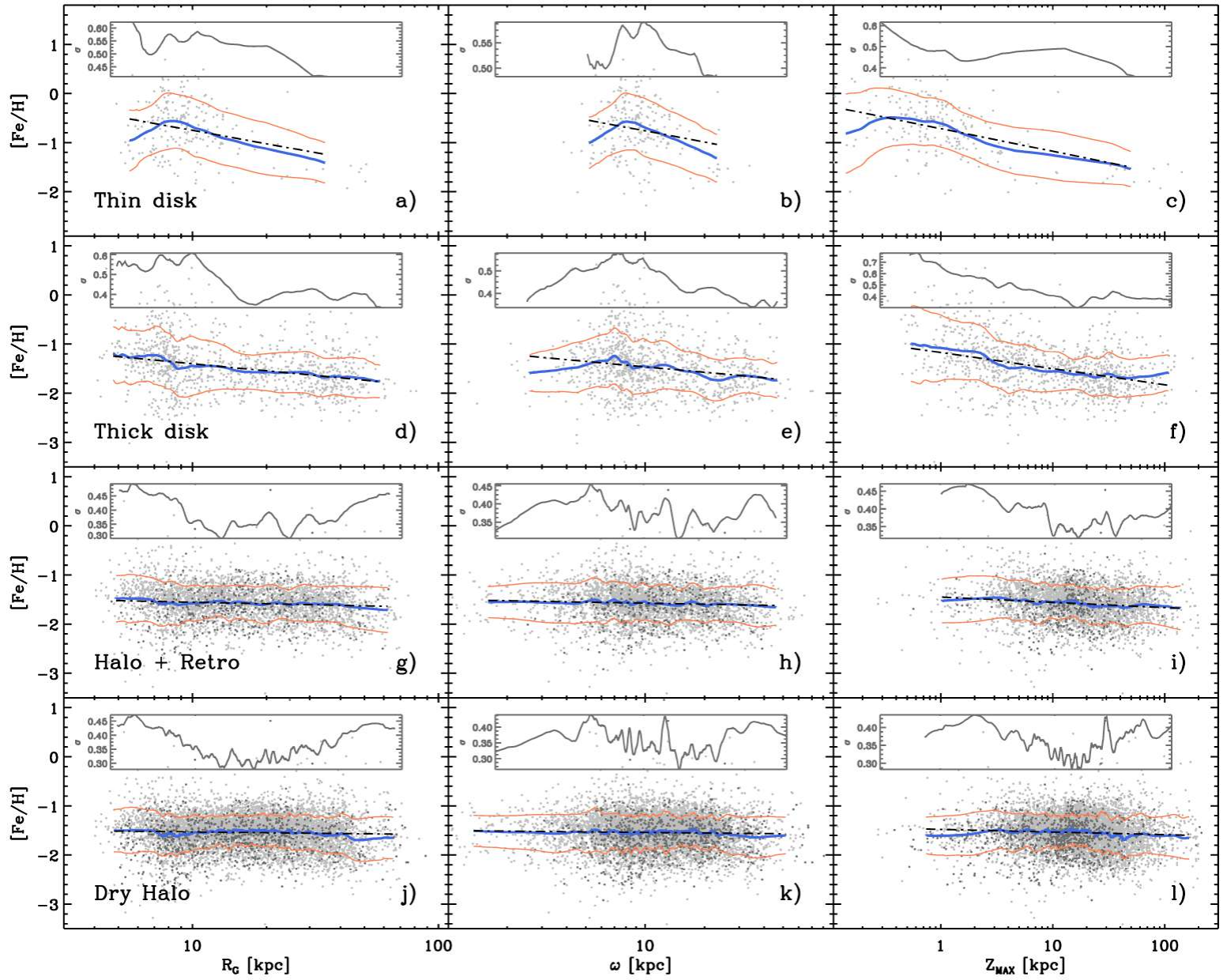}
\caption{Panels a,b,c -- Iron abundance for TND RRLs as a function of Galactocentric 
distance (panel a), distance projected onto the Galactic plane (panel b), and 
maximum height above the Galactic plane (panel c).  The X-axes are in logarithmic  
scale. The blue lines display the running average, while the orange lines the 
standard deviation of the running average. The dashed-dotted lines show 
the linear fit over the different sub-samples. The insets display the standard 
deviation of the spectroscopic sample as a function of distance.
Panels d,e,f -- Same as the top, but for TCD RRLs. 
Panels g,h,i -- Same as the top, but for Halo+Retro RRLs. 
Panels j,k,l -- Same as the top, but for dry Halo+Retro RRLs. See text for more details.
}
\label{fig:iron_radial_gradient}
\end{figure*}

\subsection{RRLs in Galactic components}\label{radial_grad_MW}

Key information concerning the mass assembly history of the different Galactic components 
are implanted in radial metallicity gradients. Fig.~\ref{fig:iron_radial_gradient} shows 
from top to bottom the iron radial gradient for the four different Galactic components 
as a function of Galactocentric distance (panels a, d, g), distance projected onto the 
Galactic plane (panels b, e, h), and as a function of the maximum height above the 
Galactic plane (panels c, f, i). Data plotted in this figure bring forward several 
interesting features that are discussed in more detail in the following.

{\em i)}-- TCD -- 
The TCD (middle panels [d,e,f] of Fig.~\ref{fig:iron_radial_gradient}) 
shows well--defined radial gradients as a function of R$_G$, $\omega$, and  Z$_{max}$. 
It is worth mentioning that TCD RRLs display a significant increase in the dispersion 
of iron abundances for Galactocentric distances smaller than the solar circle (R$_G\sim$8--10~kpc). 
A similar increase in the spread of the iron abundance is also present in $\omega$ 
and in Z$_{max}$, but at slightly different radial distances. The standard deviations 
in iron abundance plotted in the insets show that the increase in these regions is a 
factor of two larger when compared with larger distances. 

The TCD RRLs were ranked as a function of the Galactocentric distance ($R_G$) 
to avoid spurious fluctuations in the mean iron abundance and we estimated the 
running average by using the first 80 RRLs in the list. 
The mean distance and the mean iron abundance ($<[Fe/H]>$) inside the box were 
estimated as the mean over the individual distances and iron abundances of the 
same 80 RRLs. 
Subsequently, we estimated the same quantities by moving outward the box by one 
object in the ranked list until we took account for the last 80 RRLs in the TCD 
sample with the largest distance. The same approach and the same number of objects 
in the running box, was used to estimate the averages as a function of 
$\omega$, and  Z$_{max}$ and they are plotted as a blue lines, while their 
standard deviations as orange lines in the middle panels of 
Fig.~\ref{fig:iron_radial_gradient}. We  performed a number of simulations by 
changing the number of RRLs included in the running box and the number of RRLs 
adopted to move outwards the box and we found that both the current running 
average and standard deviations are quite solid.

In order to provide a more quantitative analysis of the radial gradients, we  
performed a linear fit over the three different distances and they are plotted 
as dashed-dotted lines in the middle panels. The iron gradient is as expected 
negative and the slope is steeper in R$_G$ and in $\omega$ than in Z$_{max}$ 
(see Table~\ref{tbl:radial_iron_gradient}). 
Interestingly enough, the iron gradient is changing from metal-intermediate 
([Fe/H]$\sim$-1) in the innermost TCD regions to metal-poor ([Fe/H]$\sim$-2) in the 
outermost TCD regions. 

The transition across the solar circle is even more clear in 
Galactocentric distance and in cylindrical coordinates, and indeed the RRL iron 
gradient becomes steeper for R$_G$ and $\omega\le$10 kpc (panels d,e). 
Note that the lack of objects with $R_G$ 
smaller than $\sim$5 kpc indicates that our sample includes a large number 
of RRLs that are located in the innermost Galactic regions, but above/below 
the Galactic Bulge ($R_G$[Bulge]$\lesssim$ 3.5 kpc, \citealt{zoccali2024}).

{\em ii)}-- TND -- 
There is evidence that RRLs associated to the TND (top panels [a,b,c]  
of Fig.~\ref{fig:iron_radial_gradient}) also show well--defined radial gradients 
as a function of R$_G$, $\omega$, and in particular as a function of Z$_{max}$. 
The sample in the outer disk is limited, but the decrease toward more metal-poor 
abundances is quite clear. This circumstantial evidence supports the working 
hypothesis that the inside-out scenario for the chemical enrichment of TND was 
already in place in the early phase of MW formation. The dashed-dotted lines 
display the linear fit across TND RRLs. The change in the slope across the Solar 
circle was neglected, since it is mainly caused by the significant increase in 
metallicity dispersion. Note that the iron gradients associated to TND RRLs 
typically cover a narrower range in distances, but they are a factor of 
two/three steeper when compared with the iron gradients of TCD RRLs. 

{\em iii)}-- Halo$+$Retro-- Global Halo RRLs display a very mild radial gradient when 
moving from the innermost to the outermost regions (panels [g,h,i] of
Fig.~\ref{fig:iron_radial_gradient}). The spread in iron abundance plotted in the 
insets is large not only across the Solar circle, but also for R$_G$ $\sim$40~kpc. 
The linear fits (dashed-dotted lines) to the iron gradients give slopes that are, 
on average, more than one order of magnitude shallower than TND RRLs, and at 
least a factor of five shallower than TCD RRLs. 
The iron abundances of the RRLs with retrograde orbits (black dots) display 
similar radial trends. 

The Halo$+$Retro  RRLs show a steady increase in iron content across 
the solar circle. The mean iron abundance increases by more than 0.3 dex 
when moving from $R_G\sim$10 \kpc $\ $to $R_G\sim$7 \kpc. This evidence
supports previous findings by \citet{layden94}, \citet{Suntzeff94} and 
more recently by \citet{kinman12a}. The change across the solar circle 
is also supported by the gradient projected onto the Galactic 
plane (panel h).

{\em iv)}-- Dry Halo-- The above discussion concerning the iron radial gradients is based on the 
global Halo sample, i.e. the RRLs that according to our selection criteria 
have an halo kinematics (see \S~\ref{sec_Galcomp}). However, this sample also includes RRLs 
that have been associated to four different stellar streams (see \S~\ref{sec_stream}). 
We call Dry Halo the sample of pure Halo RRLs i.e. those that have not been 
associated to any of the quoted stellar streams. We performed once again the linear 
fits to the iron radial gradients and, interestingly enough, we found that iron 
gradients are similar when compared with the Global Halo (see Table~\ref{tbl:radial_iron_gradient}).
This suggests that the inclusion of RRLs belonging to stellar streams does not change the slope. 
The standard deviations of the linear relations of the Dry Halo are slightly 
smaller (0.36 \vs\ 0.38).

\begin{figure*}[] 
\centering
\includegraphics[width=0.9\textwidth]{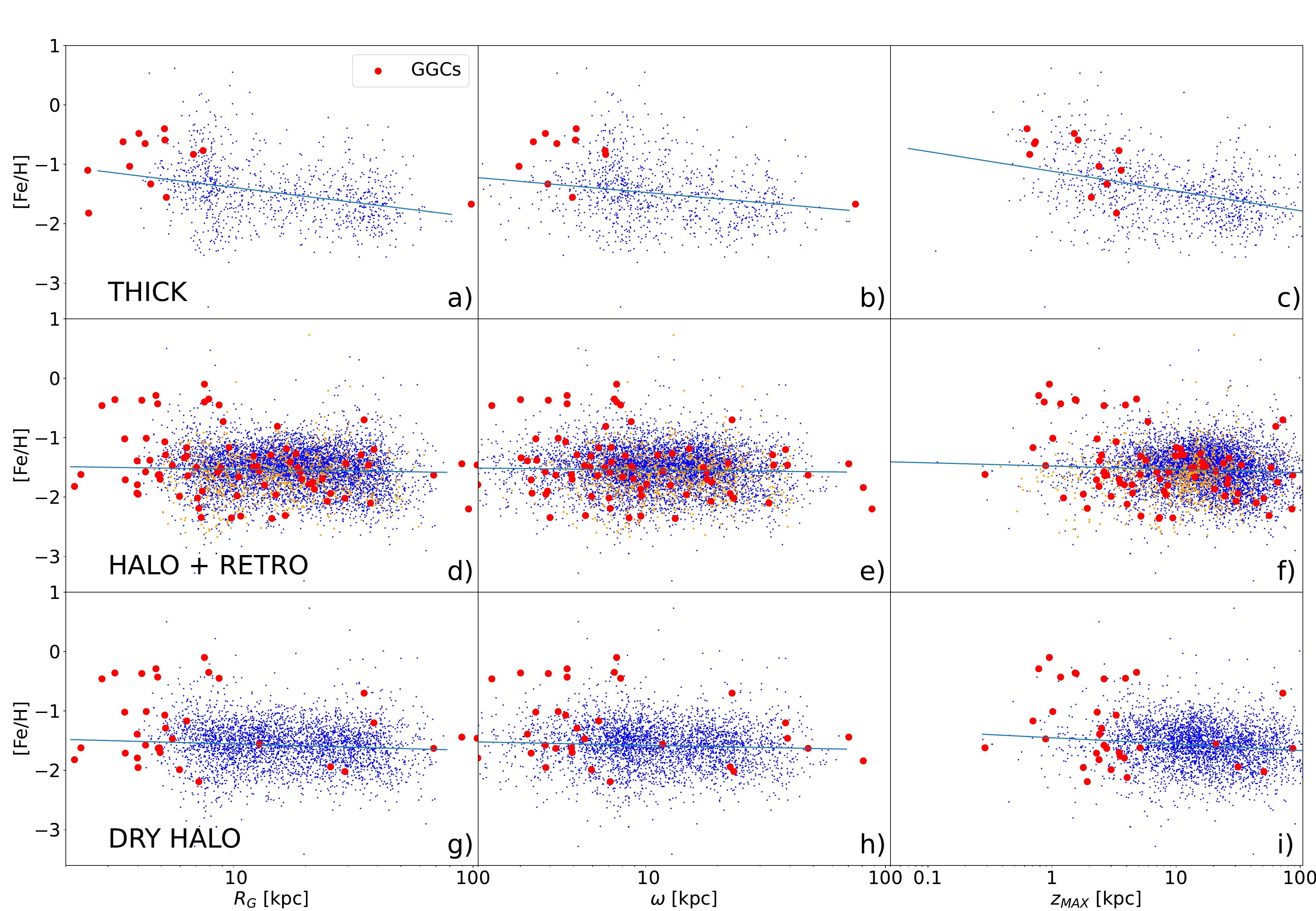}
\caption{Comparison between the iron radial gradients of field RRLs and GGCs.  From top 
to bottom the different panels display the Galactic components identified and discussed in 
\S~\ref{sec_Galcomp}. The red circles mark GGCs associated, according to
the current kinematic criteria, to the different Galactic components.
}
\label{fig:iron_gradient_comparison_gc}
\end{figure*}

Furthermore, we performed, as a consistency check, the comparison between the 
iron radial gradient of field RRLs (see, e.g., Fig.~\ref{fig:iron_radial_gradient}) 
with GGCs associated with the Galactic components. The data plotted in 
Fig.~\ref{fig:iron_gradient_comparison_gc} display a global agreement 
in R$_G$, $\omega$ and in particular in Z$_{max}$. 
There is only a limited sample of GGCs associated to the TCD, but the radial trends are 
quite similar for the different Galactic components and for the Dry Halo. 

The variation of the radial metallicity gradients as a function of 
stellar age has been investigated in a number of empirical and theoretical investigations 
\citep{roskar2008}. They are mainly focused on the Galactic TND and they 
typically indicate an overall flattening and an increase in spread of the radial 
gradients with time due to stellar radial heating and stellar migration 
\citep{DaflonCunha2004, Ratcliffe2023}. However, the possible presence of a break 
at intermediate-ages \citep{anders2017,willett2023} is still debated. 

In the current investigation we are only using old stellar tracers, and the 
comparison with theoretical predictions is beyond the aim of the current 
investigation. However, it is worth mentioning that \citet{tissera14} investigated 
the metallicity profiles of six simulated MW--like galaxies from the Aquarius Project 
and their predicted slopes are consistent with the current halo estimates. 
In particular, they found that flat slopes are often found in haloes made up 
by a mixture of satellites with different masses, i.e. a more balance 
contribution of small and massive satellites.

In closing this section, it is worth mentioning that the main reason why we adopted 
a logarithmic fit is based on plain physical arguments. Radial abundance gradients 
mainly trace the density 
gradient across the Galactic spheroid. The density typically follows a power law 
distribution and a linear fit in a log-log plane is supporting this global trend. 
Moreover, the residuals display an increase in dispersion at small and large 
distances. We performed a number of numerical simulations using several 
analytical functions, and we found that the use of a logarithmic distance 
improves the fit at small and large distances. The current evidence 
indicates that in these regions the density profile might be more complex than 
predicted by a simple power law. However, for the sake of comparison 
with similar estimates available in the literature, we also performed a fit 
of the radial gradient by using linear distances (see 
Fig.~\ref{fig:iron_radial_gradient_linear} in the Appendix~\ref{sec:fit_rad_grad}).
The global standard 
deviations of logarithmic and linear fits are, as expected, quite similar, 
indeed the number of RRLs located at small and large distances are a modest 
sub--sample. The coefficients and the standard deviations of the logarithmic and
linear fits are listed in Table~\ref{tbl:radial_iron_gradient}.

\subsection{RRLs in Stellar streams}\label{radial_grad_stream}

\begin{figure*}[] 
\centering
\includegraphics[width=0.9\textwidth]{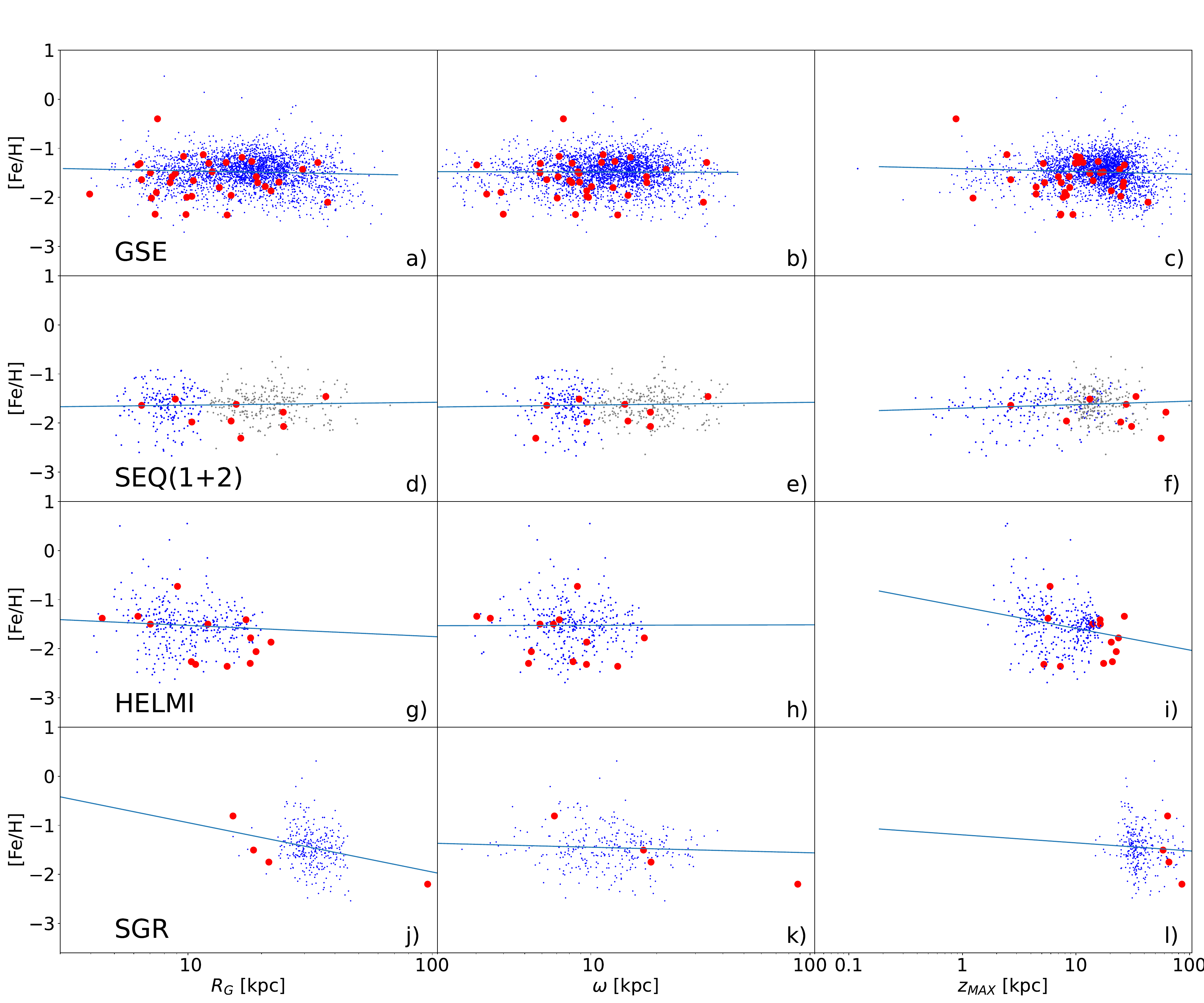}
\caption{Same as Fig.~\ref{fig:iron_gradient_comparison_gc}, but for the four different 
stellar streams discussed in \S~\ref{sec_stream}.  
}
\label{fig:iron_gradient_streams}
\end{figure*}

In this section we are dealing with the radial iron gradient in the stellar streams 
discussed in \S~\ref{sec_stream}. Data plotted in the top panels of Fig.~\ref{fig:iron_gradient_streams}
show that GSE RRLs do not display any radial gradient and its spread is quite 
constant over the entire range in distance covered by the sample. However, the distribution of
RRLs as a function of Z$_{max}$ (panel c) suggests two well--defined gaps. The former 
one located between  Z$_{max}$ $\sim$1 and 8 kpc and the latter between 
Z$_{max}$ $\sim$10 and 12 kpc. There are reason to believe that the occurrence of
these gaps are mainly a consequence of the accuracy of individual RRL distances,
and they suggest that the current identification of GSE might include different 
substreams. 
The comparison between GGCs associated to GSE and candidate GSE RRLs is quite good, since 
they overlap with each other. Interestingly enough the globulars associated with 
GSE display a spread in iron abundance similar to GSE RRLs.

On the other hand, Seq.~1 RRLs (panels d,e,f of Fig.~\ref{fig:iron_gradient_streams}) 
display a very compact spatial distribution, while Seq.~2 RRLs a broader radial distribution, 
thus further supporting the separation suggested on the basis of the kinematic 
planes. It is worth mentioning that there is mild evidence either for a 
well-defined minimum or a gap for R$_G$ $\sim$12~kpc and $\omega$ $\sim$10~kpc, 
but a solid identification awaits for larger samples. 
The comparison between Sequoia RRLs and GGCs indicates that they cover a very
similar spatial distribution and iron abundances, and they possibly have a positive radial gradient ($\beta=0.001$).

The candidate Helmi RRLs and GGCs (panels g,h,i) display, once again, very similar spatial 
distributions and iron abundances. The main difference is that GGCs associated 
to the Helmi stream cover a range in Z$_{max}$ larger than field 
RRLs. They show a mild evidence of a positive radial gradient ($\beta=0.001$) and 
the difference might be due to incompleteness in the current RRL sample.

Finally, Sgr RRLs (panels j,k,l) display, as expected, a very compact spatial 
distributions in R$_G$ and in Z$_{max}$, while in $\omega$ they cover a wide range. 
This further supports the high level of collimation of this stellar streams.

The coefficients and the standard deviations of the performed logarithmic and
linear fits for the four substructures are listed in Table~\ref{tbl:radial_iron_gradient_streams}.

\begin{table*}[htbp]
\scriptsize
\caption{Radial iron gradients for TND, TCD and Halo RRLs by using either
logarithmic ([Fe/H]=$\alpha$+$\beta\times\log(dis)$) or linear 
([Fe/H]=$\alpha$+$\beta\times dis$) distances (kpc), where the distance is  
either the Galactocentric distance (R$_G$), or the distance projected onto the Galactic plane 
($\omega$), or Z$_{max}$ or the height above the Galactic plane (Z) in absolute value.}  
\label{tbl:radial_iron_gradient}
\begin{center}
\begin{tabular}{lllllllll}
\hline
\hline
Dis.          & \multicolumn{2}{c}{Thin Disk}               & \multicolumn{2}{c}{Thick Disk}             & \multicolumn{2}{c}{Global Halo} & \multicolumn{2}{c}{Dry Halo} \\
\hline
                                                  \multicolumn{9}{c}{Logarithmic} \\
R$_G$          & 0.158$\pm$0.206 & -0.907$\pm$0.206(0.55)\footnote{The numbers in parentheses are the standard deviations (dex).} &-0.922$\pm$0.060 &-0.479$\pm$0.050(0.49) & -1.450$\pm$0.029 & -0.104$\pm$0.024(0.38) & -1.474$\pm$0.022 & -0.057$\pm$0.017(0.36)\\
$\omega$       &-0.006$\pm$0.245 & -0.754$\pm$0.250(0.56) &-1.096$\pm$0.060 &-0.369$\pm$0.056(0.50) & -1.503$\pm$0.024 & -0.071$\pm$0.022(0.38) & -1.501$\pm$0.018 & -0.041$\pm$0.016(0.36)\\
Z$_{max}$      &-0.715$\pm$0.035 & -0.461$\pm$0.062(0.51) &-1.172$\pm$0.029 &-0.328$\pm$0.026(0.47) & -1.451$\pm$0.019 & -0.106$\pm$0.015(0.37) & -1.478$\pm$0.014 & -0.055$\pm$0.011(0.36)\\
$\vert Z\vert$ &-0.787$\pm$0.038 & -0.335$\pm$0.058(0.54) &-1.272$\pm$0.023 &-0.286$\pm$0.025(0.48) & -1.540$\pm$0.015 & -0.039$\pm$0.014(0.38) & -1.539$\pm$0.010 & -0.005$\pm$0.009(0.36)\\ 
                                                      \multicolumn{9}{c}{Linear} \\
                                                      
R$_G$          &-0.456$\pm$0.066 & -0.026$\pm$0.005(0.54) &-1.294$\pm$0.026 & -0.010$\pm$0.001(0.49) &-1.531$\pm$0.012 & -0.002$\pm$0.001(0.38) & -1.505$\pm$0.008 & -0.002$\pm$0.001(0.36)\\
$\omega$       &-0.392$\pm$0.094 & -0.034$\pm$0.009(0.56) &-1.328$\pm$0.027 & -0.011$\pm$0.002(0.50) &-1.545$\pm$0.012 & -0.002$\pm$0.001(0.38) & -1.518$\pm$0.008 & -0.002$\pm$0.001(0.36)\\
Z$_{max}$      &-0.674$\pm$0.039 & -0.013$\pm$0.002(0.54) &-1.381$\pm$0.020 & -0.005$\pm$0.001(0.49) &-1.543$\pm$0.009 & -0.001$\pm$0.001(0.37) & -1.516$\pm$0.006 & -0.001$\pm$0.001(0.35)\\
$\vert Z\vert$ &-0.658$\pm$0.039 & -0.031$\pm$0.005(0.53) &-1.345$\pm$0.021 & -0.012$\pm$0.001(0.49) &-1.544$\pm$0.010 & -0.002$\pm$0.001(0.38) & -1.520$\pm$0.007 & -0.002$\pm$0.001(0.36)\\
\hline 
\end{tabular}
\end{center}
\end{table*}

\begin{table*}[htbp]
\scriptsize
\caption{Radial iron gradients for substructures by using either 
logarithmic ([Fe/H]=$\alpha$+$\beta \times \log(dis)$) or linear 
([Fe/H]=$\alpha$+$\beta \times dis$) distances (kpc), where the distance is  
either the Galactocentric distance (R$_G$), or the distance projected onto the Galactic plane 
($\omega$), or Z$_{max}$ or the height above the Galactic plane (Z) in absolute value.}  
\label{tbl:radial_iron_gradient_streams}
\begin{center}
\begin{tabular}{llllrrllr}
\hline
\hline
Dis.          & \multicolumn{2}{c}{GSE}      & \multicolumn{2}{c}{Sequoia}     & \multicolumn{2}{c}{SGR}    & \multicolumn{2}{c}{Helmi} \\
\hline
                                                  \multicolumn{9}{c}{Logarithmic} \\
R$_G$          &-1.374$\pm$0.038 & -0.087$\pm$0.030(0.31)\footnote{The numbers in parentheses are the standard deviations (dex).} &-1.705$\pm$0.071 & 0.066$\pm$0.060(0.32) & 0.055$\pm$0.408 & -1.002$\pm$0.270(0.42)  &-1.775$\pm$0.124 & 0.097$\pm$0.093(0.37)\\
$\omega$       &-1.476$\pm$0.028 & -0.006$\pm$0.026(0.31) &-1.700$\pm$0.070 & 0.064$\pm$0.062(0.32) & -1.334$\pm$0.120 & -0.112$\pm$0.110(0.42) &-1.662$\pm$0.105 & 0.013$\pm$0.092(0.37)\\
Z$_{max}$      &-1.418$\pm$0.023 & -0.054$\pm$0.019(0.31) &-1.697$\pm$0.037 & 0.072$\pm$0.036(0.32) & -1.199$\pm$0.168 & -0.161$\pm$0.105(0.42) &-1.625$\pm$0.134 &-0.016$\pm$0.097(0.37)\\
$\vert Z\vert$ &-1.474$\pm$0.014 & -0.009$\pm$0.014(0.31) &-1.669$\pm$0.026 & 0.060$\pm$0.032(0.32) & -0.556$\pm$0.287 & -0.619$\pm$0.197(0.42) &-1.712$\pm$0.059 & 0.059$\pm$0.050(0.37)\\ 
                                                      \multicolumn{9}{c}{Linear} \\
R$_G$          &-1.413$\pm$0.014 & -0.004$\pm$0.001(0.31) &-1.651$\pm$0.029 & 0.001$\pm$0.001(0.32) & -0.981$\pm$0.123 & -0.015$\pm$0.004(0.42) &-1.680$\pm$0.048 & 0.001$\pm$0.002(0.37)\\
$\omega$       &-1.457$\pm$0.014 & -0.002$\pm$0.001(0.31) &-1.657$\pm$0.029 & 0.002$\pm$0.002(0.32) & -1.413$\pm$0.053 & -0.003$\pm$0.004(0.42) &-1.637$\pm$0.043 &-0.001$\pm$0.003(0.37)\\
Z$_{max}$      &-1.419$\pm$0.011 & -0.003$\pm$0.001(0.31) &-1.639$\pm$0.022 & 0.001$\pm$0.001(0.32) & -1.453$\pm$0.022 & -0.000$\pm$0.001(0.42) &-1.645$\pm$0.041 &-0.000$\pm$0.001(0.37)\\
$\vert Z\vert$ &-1.443$\pm$0.010 & -0.003$\pm$0.001(0.31) &-1.635$\pm$0.023 & 0.001$\pm$0.002(0.32) & -1.133$\pm$0.096 & -0.011$\pm$0.003(0.42) &-1.692$\pm$0.041 & 0.003$\pm$0.002(0.37)\\
\hline 
\end{tabular}
\end{center}
\end{table*}

\normalsize


\section{Comparison with MW metallicity gradients}\label{sec_comp_MW}

The occurrence of a Halo metallicity gradient has been a controversial issue 
for several decades. Some pioneering works using GCs as tracers
of the Stellar Halo have claimed both the existence of
a metallicity gradient \citep{harris79} as well
as the lack of it \citep[see e.g.][and references
therein]{searle78,armandroff92,alfaro93}.
The possible occurrence of an iron gradient was also 
investigated by \citet{butler1982}, \citet{saha1984} and \citet{saha1985}, by 
using samples of field RRLs covering R$_G$=10-40 kpc. In these 
pioneering investigations they did not find evidence 
of a gradient with the Galactocentric distance and the distance 
from the Galactic plane, but a very large spread in 
iron abundance ranging from [Fe/H]=-1.0 to [Fe/H]=-2.2. 
In particular \citet{saha1984} and \citet{saha1985} brought forward, for the 
first time, the possible occurrence of moderately metal-rich 
([Fe/H]$\sim$-0.5) RRLs in the Halo. 
Indeed, the estimate and the comparison of different radial gradients 
are typically affected by three main drawbacks: the spatial distribution, 
the accuracy of individual distances and the homogeneity of the adopted 
stellar tracer. 

\citet{carollo07} and \citet{carollo10} used orbital properties 
of local Halo stars from SDSS data to measure the metallicity
and inferred from this sample the metallicity of the outer halo. They 
claimed that the MW halo has a strong negative metallicity gradient, with 
the median metallicity changing from $\sim$-1.6 in the Solar
neighborhood to -2.2 for Galactocentric distances larger than 
15 kpc. However, these results suffer from important biases. Their magnitude 
limited sample includes only luminous low metallicity stars at
large distances, imposing an artificial metallicity gradient
\citep{schoenrich2011}. This emphasizes the need for more representative
samples of distant halo stars, a requirement that only recently has 
been met. 
The duality of the Galactic halo was also supported by \citet{dejong2010} by using 
metallicity distribution function based on photometric indices. They found evidence 
of a radial metallicity gradient in the MW stellar halo with a mean metallicity of 
[Fe/H]$\sim$-1.6 for R$\lesssim$15~kpc and a mean metallicity of [Fe/H]$\sim$-2.2 at 
larger distances. \citet{sesar2011a} used near-turnoff main sequence
stars out to  35 kpc from CFHT and SDSS photometry together with 
photometric metallicities based on the $u$-band \citep{ivezic2008a} 
and they did not find a metallicity gradient. Subsequently, \citet{xue2015} 
used a sample of SEGUE K-giants halo stars covering a range in Galactocentric 
distances between 10 to 50 kpc together with LR spectra collected by 
SDSS and they found a weak metallicity gradient between the more metal-poor 
and the more metal-rich subsamples.  

\begin{figure*}[] 
\centering
\includegraphics[height=0.95\textheight]{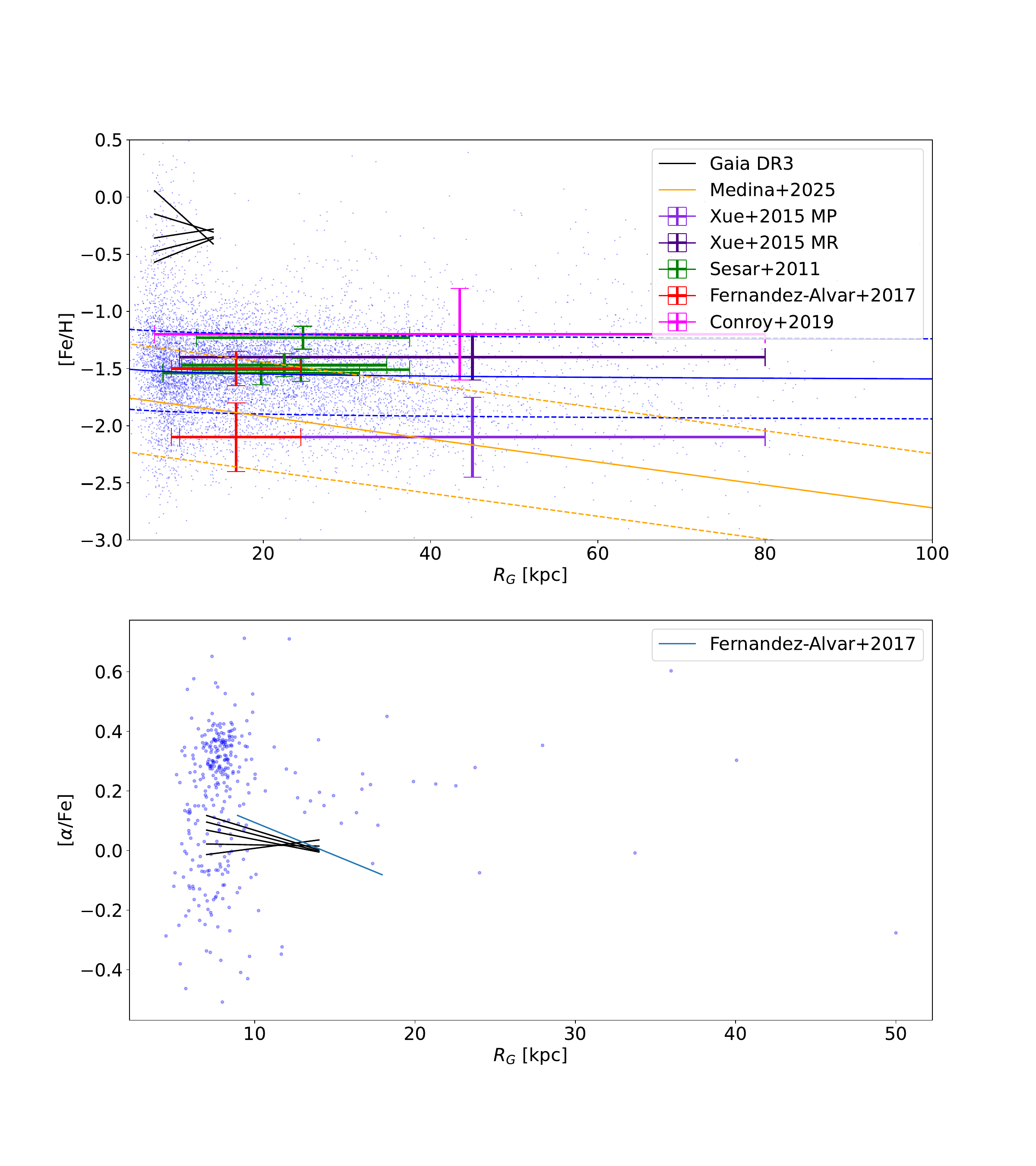}
\caption{
Top -- Comparison between radial iron gradients available in the literature and current 
Halo+Retro RRLs. The horizontal bars display the range in Galactocentric distances covered 
by the different estimates. The vertical bars show either the standard deviations 
or the mean error on iron abundance. The blue solid line shows our iron radial gradient based 
on linear distances, while the dashed lines display its standard deviation. The solid black 
lines display the iron abundance gradients at different distances (Z) above the Galactic 
plane provided by \citet{gaiacollaboration23}.
Bottom -- Same as the top, but for the [$\alpha$/Fe] radial gradient.}
\label{fig:iron_gradient_MW_lit}
\end{figure*}

Data plotted in the top panel of Fig.~\ref{fig:iron_gradient_MW_lit}) show that 
there is a global agreement between the current RRL sample and the estimates 
provided by both \citet{sesar2011a} and \citet{xue2015}. A more quantitative
comparison is hampered by the lack of accurate individual distances of their
targets. 

The Halo gradient has also been investigated by the Gaia 
collaboration \citep{gaiacollaboration23} using nearby stars with accurate 
estimates of kinematic properties, distances and iron abundances based on 
MR spectra. 
They found well--defined metallicity gradients that changed from a positive 
to a negative slope when moving from small to large above the Galactic plane.
The solid black lines plotted in the top panel of Fig.~\ref{fig:iron_gradient_MW_lit}
show that these gradients mainly trace the increase in the spread in iron abundance 
across the solar circle.  

\citet{fernandezalvar2017} performed a detailed analysis of chemical trends 
in the Galactic halo using field RGs and abundances based on HR $H$-band 
spectra provided by APOGEE. They found a dichotomic distribution in iron abundance  peaking 
at [M/H]$\sim$-1.5 and $\sim$-2.1. However, they did not find evidence of a radial gradient.
Furthermore,  \citet{conroy2019b} used HR optical spectra (H3 survey) for a very large 
sample of field RG stars ($\sim$4,200) to constrain the metallicity distribution 
function of the Halo. They found that the MW stellar halo is relatively metal-rich ([Fe/H]=-1.2) 
and at the same time they did not find evidence of a metallicity gradient for Galactocentric 
distances ranging from $\sim$6 up to $\sim$80 kpc. Data plotted in the top panel of 
Fig.~\ref{fig:iron_gradient_MW_lit}, shows that their estimates covers the upper envelope 
of the iron gradient based on RRLs. However, they found evidence of a complex metallicity 
distribution with well--defined metal-rich and metal-poor components located at Galactocentric 
distances smaller than 10 kpc and larger than 30 kpc. The former preliminary evidence is fully 
supported by the current RRL sample. Moreover they found, in fair agreement with the current 
analysis, that the very metal-poor ([Fe/H]$\le$-2) Halo component is a minority fraction. 

Even more recently, \citet{medina2025a,medina2025b} using a large and homogeneous spectroscopic dataset based on LR spectra collected by DESI found that the inner Halo (R$_G\lesssim$50~kpc) is mainly dominated by RRLs with [Fe/H]=-1.5 associated with the GSE merger event. Moreover, they also found that, once the GSE RRLs are removed, the dry halo displays a well--defined radial gradient that appears to be slightly steeper than the current one. 

The bottom panel of Fig.~\ref{fig:iron_gradient_MW_lit} shows the [$\alpha$/Fe] radial gradient. 
The range in Galactocentric distances covered by this sample is significantly smaller when compared with 
the iron radial gradient, since the measurement of $\alpha$-element abundances requires HR 
spectra. Data plotted in this panel display that [$\alpha$/Fe] abundances cover almost 0.8 dex across 
the solar circle, since they range from very $\alpha$-enhanced ([$\alpha$/Fe]$\sim$0.4) to very 
$\alpha$-depleted ([$\alpha$/Fe]$\sim$-0.2/-0.3). The [$\alpha$/Fe] abundances attain an almost 
constant value for Galactocentric distances ranging from $\sim$10 to $\sim$25 kpc, thus suggesting 
a steady decrease in the $\alpha$-enhancement when moving toward larger radial distances. This negative 
trend is soundly supported by the radial gradient in $\alpha$-elements provided by 
\citet{fernandezalvar2017} by using APOGEE spectra. The main difference is that Dry Halo RRLs indicate, 
in the same range of Galactocentric distances, a shallower slope. The current sample is too limited 
to have more quantitative constraints on $\alpha$-element abundances in the outer Halo. 

\section{Comparison with metallicity gradients in M31}\label{sec_comp_MW_M31}

M31 is the external galaxy with the most detailed abundance analysis of its different 
components. Dating back to more than half century ago the metallicity distribution of 
both the Halo and the disks in M31 has been investigated by using photometric indices 
and spectroscopy for a wide range of stellar tracers. 
The opportunity to use wide field imagers and multi-object spectrographs at the 4-8m 
class telescopes and space facilities gave new impetus to these investigations. 
In the following we perform a detailed comparison between radial metallicity gradients 
in MW and M31 Galactic components and in their stellar streams.

\begin{figure*}[] 
\centering
\includegraphics[width=0.9\textwidth]{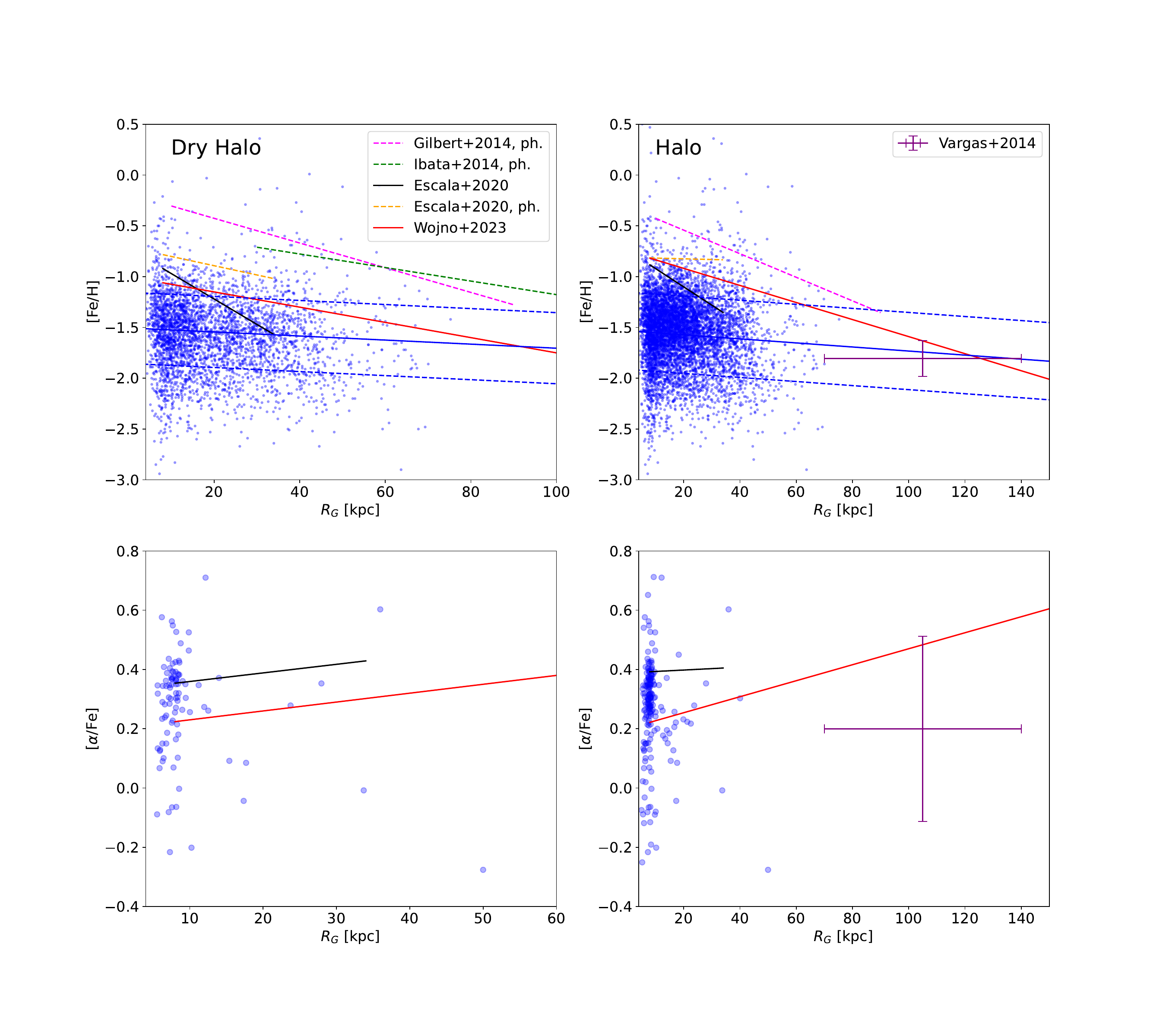}
\caption{Top -- Iron abundance as a function of the galactocentric distance
(linear scale). The blue dots display the distribution of Milky Way
RRLs associated to the Dry Halo (left) and to the global Halo (right). 
The blue solid lines show the linear gradient for the same samples and 
the blue dotted lines their standard deviations (see \S~\ref{radial_grad_MW}). 
Iron gradients for M31 based either on photometric indexes (dashed) or on 
spectroscopic (solid) measurements are plotted with different colors (see labels).   
Bottom -- Same as the top but for the [$\alpha$/Fe] gradients. See text for more details.}
\label{fig:iron_gradient_M31_lit}
\end{figure*}

The top left panel of Fig.~\ref{fig:iron_gradient_M31_lit} shows the comparison 
between the iron radial gradient of dry Halo$+$Retro RRLs  with similar estimates 
for the global M31 halo available in the literature. Data plotted in this panel 
show that M31 iron radial gradients based on photometric indices (dashed lines) 
appear to be systematically more metal-rich than suggested by RRLs. The difference 
is mainly caused by a difference in the zero-point of the gradient. However, 
M31 iron gradients based on spectroscopic measurements (solid lines) agree quite 
well with the current global sample of RRLs. In particular, the iron 
gradient provided by \citet{Wojno2023} covers the upper envelope of the distribution, 
but the slope is quite similar within $1\sigma$ (-0.008 \vs\ -0.002 dex/kpc),
while the spectroscopic iron gradient provided by \citet{escala2020b} is even closer to
metallicity distribution of the inner Halo traced by RRLs. 

The top right panel of Fig.~\ref{fig:iron_gradient_M31_lit} shows the same comparison, 
but for the global Halo+Retro RRLs. Data plotted in this panel further support the shift in 
the zero-point of iron radial gradients based on photometric indexes (dashed line). 
The agreement with M31 spectroscopic iron gradient provided by \citet{Wojno2023} 
is even better since they overlap, within the errors, on a wide range of galactocentric 
distances. The same agreement applies to the iron abundances measured by \citet{vargas14}
for four field RGs in the M31 halo by using LR Keck/DEIMOS spectra. 

The bottom panels of the same figure display the comparison between M31 and MW halos 
for the [$\alpha$/Fe] radial gradients. Data plotted in these panels show that there 
is evidence of a difference in the zero-point of 0.2 dex with the radial gradient 
provided by \citet{escala2020a}. There is also a reasonable agreement between the 
[$\alpha$/Fe] radial gradient provided by \citet{Wojno2023} and current [$\alpha$/Fe] 
abundances. The M31 radial gradient measured by \citet{Wojno2023} is positive and 
suggests a steady increase in $\alpha$ enhancement in the outer halo. On the other hand, 
the mean [$\alpha$/Fe] abundance for the four RGs provided by \citet{vargas14} indicates, 
within the errors, a modest $\alpha$-enhancement in the outer halo.
For the sake of clearness, in Table~\ref{tbl:M31_radial_iron_gradient} we have summarized
the metallicity radial gradients for M31 and its substructures, from both
spectroscopic and photometric investigations, available in the literature.

This global agreement between M31 and MW halos comes as a surprise, since current 
empirical and theoretical investigations indicate that the M31 is significantly 
more massive than the MW. Moreover, there is evidence that the assembly history 
of M31 has been mainly driven by several events of major mergings, while for the 
MW it has been suggested a single event of major merging plus a number of minor 
merging \citep{fiorentino2017b}. Finally, it is worth mentioning that spectroscopic 
abundances for M31 halo are based on field RGs. These stars cover a broad range in stellar ages, 
but the agreement in the radial iron gradients further supports the evidence that 
M31 and MW halos are dominated by old stellar populations that experienced similar 
chemical enrichment. 

\begin{figure*}[] 
\centering
\includegraphics[width=0.9\textwidth]{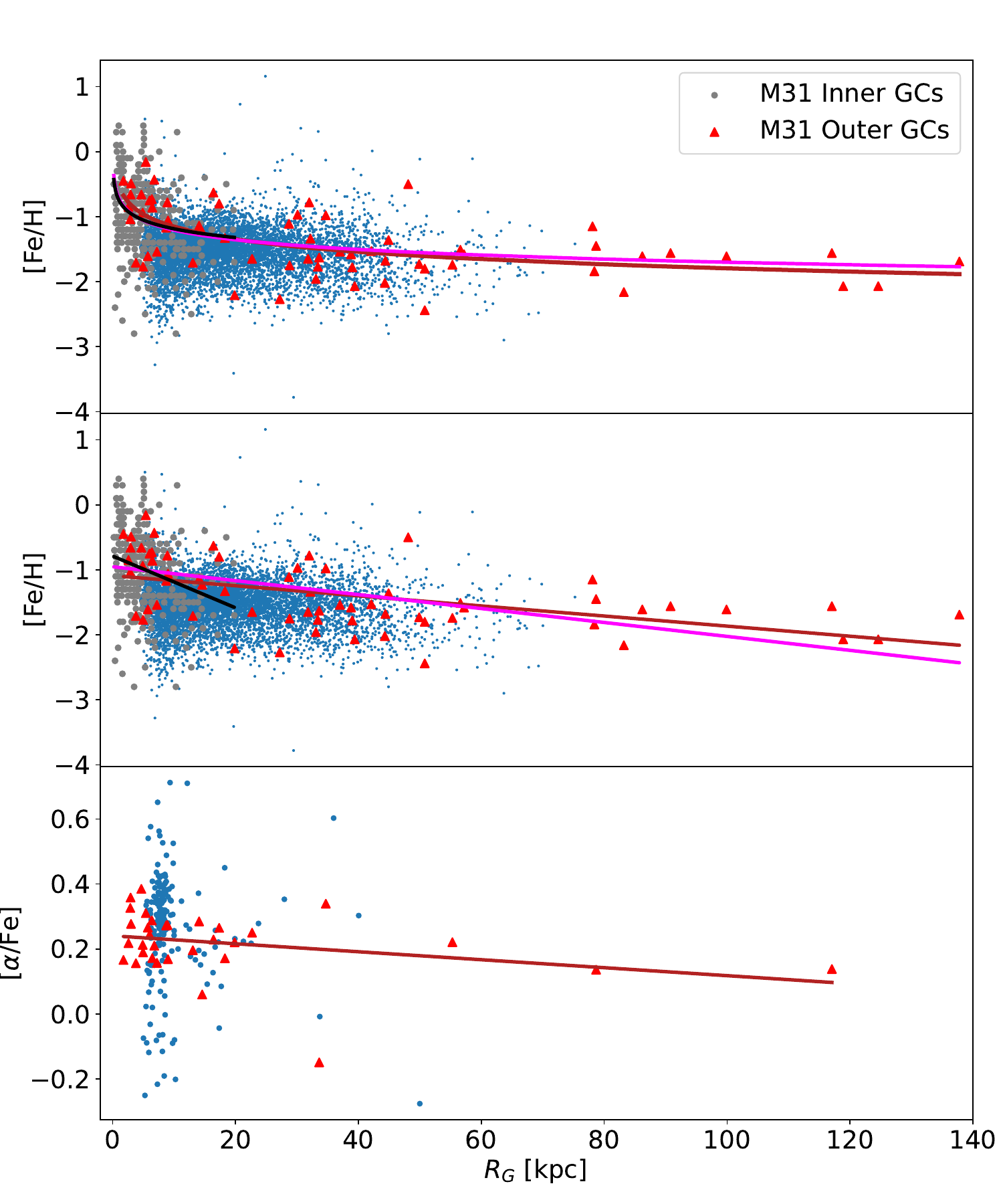}
\caption{Top -- Iron abundance \vs\ galactocentric distances. 
Blues dots mark Milky Way global Halo$+$Retro RRLs, while gray circles 
and red triangles inner and outer halo M31 GCs. (see Table~\ref{tbl:M31_GCs_abundances}). 
The black and the magenta lines display the fit of logarithmic gradients for inner and 
outer Halo M31 GCs, while the brown line the fit for the global sample. 
Middle -- Same as the top, but for the linear fits.
Bottom -- Same as the top, but for the [$\alpha$/Fe] linear gradients. 
See text for more details.
}
\label{fig:fealpha_gcM31}
\end{figure*}


To further constrain the similarity between the iron radial gradients of MW and 
M31 halos we took advantage of the sample of M31 GCs, for which both
accurate spectroscopic abundances of iron and $\alpha$-elements 
\citep{caldwell2011,colucci2014,sakari2015,sakari2021,sakari2022},
based on HR integrated-light spectra, and homogeneous 
individual projected distances \citep{colucci2014,caldwell2016,mackey2019a,sakari2021} are available. 
The quoted parameters of the entire sample are listed in Table~\ref{tbl:M31_GCs_abundances}. 

To make a more quantitative comparison with Halo RRLs, we performed 
an analytical fit by using both linear and logarithmic distances on M31 GCs.
The coefficients of the fits are listed in Table~\ref{tbl:m31_gcs_gradients}.  
The top panel of Fig.~\ref{fig:fealpha_gcM31} shows a remarkable agreement 
between the iron gradient traced by M31 GCs and Halo RRLs. Moreover and even more 
importantly, the good agreement applies not only to the innermost regions in which 
the two samples display a similar spread in iron abundance but also over the entire 
range of galactocentric distances covered by the two samples. The agreement is fully 
supported by the [$\alpha$/Fe] radial gradient plotted in the bottom panel of the 
same figure in which the two samples attain, within the errors, very similar values.

\begin{figure*}[] 
\centering
\includegraphics[width=0.9\textwidth]{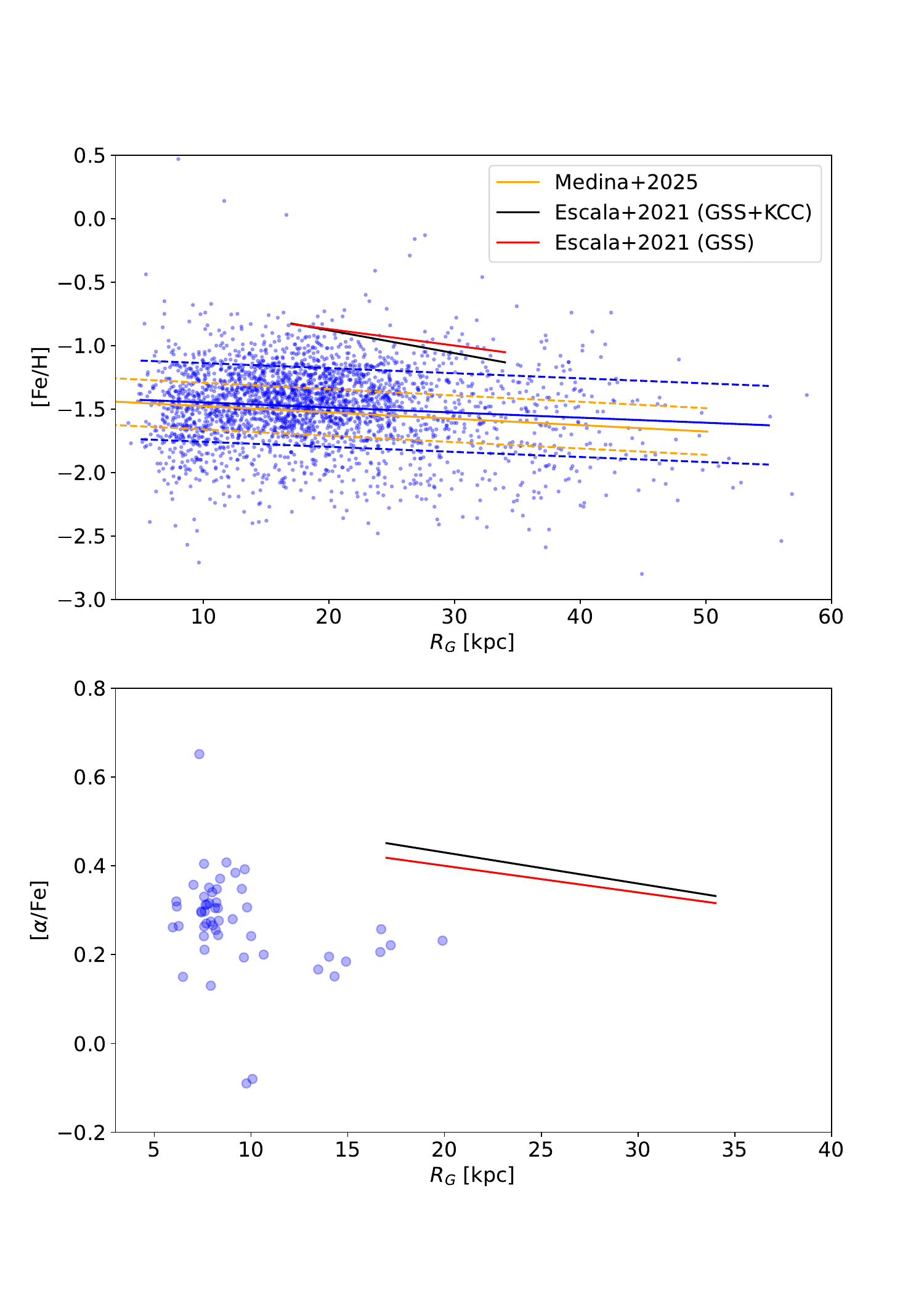}
\caption{Top -- Iron \vs\ galactocentric distance (linear scale). 
Blues dots display candidate GSE RRLs, while the blue solid line shows the linear fit of the 
iron radial gradient and the dashed lines its standard deviation. 
The red solid line shows the linear iron radial gradient for the GSS
in M31, while the black line the same gradient but for stars in GSS plus the Kinematic 
Cold Component (KCC) provided by \citet{escala2021}. The yellow lines display the radial 
gradient (solid) for GSE RRLs provided by \citet{medina2025a} together with its standard 
deviation (dashed lines). 
Bottom -- Same as the top, but for the  [$\alpha$/Fe] gradient.
}
\label{fig:iron_gradient_GSS_GSE}
\end{figure*}


The identification and characterization of substructures in M31 has been lively 
discussed in several empirical \citep[][and references therein]{Brown06, Brown09} 
and theoretical \citep[][and references therein]{bullock05,tissera14,monachesi2019} 
papers. Here we took advantage of the long term project SPLASH to compare on a more 
quantitative basis the abundance gradients in the two largest stellar streams 
identified in M31 (Giant Stellar Stream, GSS) and in the MW (GSE). 

Data plotted in the top panel of Fig.~\ref{fig:iron_gradient_GSS_GSE} show the comparison 
between candidate GSE RRLs, selected according to the criteria discussed in 
\S~\ref{sec_stream}, and the iron radial gradient in GSS (red line) and in GSS plus 
the Kinematic Cold Component (black line) provided by \citet{escala2021}
by using a large set of spectroscopic measurements.
There is evidence of a difference in the zero-point of 
the order of 0.5 dex, but the slopes in the same range of Galactocentric distances 
are quite similar (-0.016 \vs\ -0.004 dex/kpc). This global agreement also applies to the 
[$\alpha$/Fe] radial gradient, indeed, if we account for the difference in the zero-point 
of the iron gradient, the radial gradient of the $\alpha$-elements are quite similar. 
In passing we also note that the [$\alpha$/Fe] radial gradients provided by \citet{escala2021}
indicate a decrease in the $\alpha$-enhancement when moving toward the outer halo.

\begin{table*}[htbp]
\scriptsize
\caption{Metallicity gradients and metallicity estimates for M31 and its substructures.
They have been estimated through a linear analytical fit 
[Fe/H]=$\alpha$+$\beta\times dis$,  where the $dis$ (kpc) is  the  projected distance.}  
\label{tbl:M31_radial_iron_gradient}
\begin{center}
\begin{tabular}{lrllllcc}
\hline
\hline
Gal. Comp.     &   range\footnote{Range in projected distance (kpc)}        &  $\alpha$      &  $\beta$     & $\alpha$     &    $\beta$        &  tracer        & Ref.\footnote{References:
 1) \citet{gilbert2019}; 2) \citet{escala2020a}; 3) \citet{escala2020b}; 4) \citet{gilbert2014}; 5) \citet{escala2021}; 6) \citet{escala2022}; 7) \citet{Wojno2023}; 8) \citet{sakari2022}; 9) \citet{gilbert2009}; 10) \citet{fardal2012}; 11) \citet{gilbert2007}; 12) \citet{gregersen2015}. 
} \\
\hline
                                       &          & \multicolumn{2}{c}{[Fe/H]}  &   \multicolumn{2}{c}{[$\alpha$/Fe]} &                &        \\ 
\hline
                                      \multicolumn{8}{c}{Spectroscopy: Halo}    \\ 
Dry Halo\footnote{Halo neglecting substructures.}& 17  & -0.94$\pm$0.43 & \ldots  & \ldots  & \ldots  &  RGs  &     1    \\ 
Dry Halo                               & 12   & -1.30$\pm$0.11    & \ldots & 0.45$\pm$0.13   &\ldots  &  RGs  &     2     \\
Dry Halo                               & 17   & -1.04$\pm$0.09    & \ldots & 0.53$\pm$0.07   &\ldots  &  RGs  &     3    \\
Dry Halo                               & 18   & -1.00$\pm$0.06    & \ldots & 0.40$\pm$0.04   &\ldots  &  RGs  &     3    \\
Dry Halo                               & 22   & -0.66$\pm$0.18    & \ldots & 0.49$\pm$0.06   &\ldots  &  RGs  &     2     \\
Dry Halo                               & 26   & -1.00$\pm$0.19    & \ldots & 0.55$\pm$0.13   &\ldots  &  RGs  &     2     \\
Dry Halo                               & 33   & -1.48$\pm$0.13    & \ldots & 0.41$\pm$0.07   &\ldots  &  RGs  &     3    \\
Dry Halo                               & 8-34 & -0.72$\pm$0.03    & -0.025$\pm$0.002 & \ldots   &0.0029$\pm$0.0027\footnote{Not statistically significant} &  RGs  &     3     \\
Dry Halo                               &  8-177 & \textit{-1.00$\pm$0.05}\footnote{The value highlighted in italic were calculated by using the slope, the [Fe/H] and/or the [$\alpha$/Fe] at other distances from the reference.} & -0.0075$\pm$0.0012 & \textit{0.2}  & 0.0030$\pm$0.0009 &  RGs  & 7     \\
Halo                                   & 9    & -0.93$\pm$0.09    & \ldots & 0.32$\pm$0.08   &\ldots  &  RGs  &     3     \\
Halo                                   & 23   & -1.54$\pm$0.14    & \ldots & 0.43$\pm$0.12   &\ldots  &  RGs  &     2     \\
Halo                                   & 23   & -1.64$\pm$0.11    & \ldots & 0.39$\pm$0.10   &\ldots  &  RGs  &     3    \\
Halo                                   & 31   & -1.35$\pm$0.10    & \ldots & 0.40$\pm$0.10   &\ldots  &  RGs  &     3    \\
Halo                                   & 8-34 & -0.74$\pm$0.03\footnote{Intercept at r$_{proj}$=0 kpc}   & -0.018$\pm$0.001 & \ldots & 0.00048$\pm$0.00261   &  RGs  &     3     \\
Halo                                   & 8-177 & \textit{-0.75$\pm$0.05} & -0.0084$\pm$0.0008 & \textit{0.2} & 0.0027$\pm$0.0005 & RGs & 7     \\
                                             \multicolumn{8}{c}{Spectroscopy: Discs}      \\
Outer disk                             & 26   & -0.82$\pm$0.09    & \ldots              & 0.60$\pm$0.10   &\ldots  &  RGs  &     2     \\
Disk (phot)                            & 4-20 & \ldots            & -0.020$\pm$0.004&    &\ldots  &  RGs  &     12     \\
                                        \multicolumn{8}{c}{Spectroscopy: Substructures}  \\
GSS\footnote{Giant Stellar Stream.}    & 17       & -0.75$\pm$0.37 &                  &  \footnote{$\alpha$-enhanced for [Fe/H]$\le$-0.9, decreasing $\alpha$-enhancement with increasing iron.}&                &  RGs  &     1   \\ 
GSS                                    & 17   & -0.87$\pm$0.10    & \ldots & 0.44$\pm$0.05   &\ldots  &  RGs  &     3    \\
GSS                                    & 22   & -1.02$\pm$0.15    & \ldots & 0.38$\pm$0.19   &\ldots  &  RGs  &     2,3    \\
GSS                                    & 33   & -1.11$\pm$0.13    & \ldots & 0.34$\pm$0.09   &\ldots  &  RGs  &     3    \\
KCC\footnote{Kinematically Cold Component.}&17& -0.61$\pm$0.40   & \ldots  & \ldots &    \ldots       &  RGs  &     1    \\
KCC                                    & 17   & -0.79$\pm$0.07    & \ldots & 0.54$\pm$0.06   &\ldots  &  RGs  &     3    \\
KCC                                    & 22   & -0.71$\pm$0.11    & \ldots & 0.35$\pm$0.09   &\ldots  &  RGs  &     2,3    \\
GSS+KCC                                & 17-33& -0.52$\pm$0.08   & -0.018$\pm$0.003 & 0.57$\pm$0.17 & -0.007$\pm$0.007  &  RGs  &  5  \\
GSS                                    & 17-33& -0.61$\pm$0.09   & -0.016$\pm$0.004 & 0.52$\pm$0.14 & -0.006$\pm$0.007  &  RGs  &  5  \\
GSS+KCC                                & 17-58& -0.96$\pm$0.06   & \ldots           & 0.40$\pm$0.05 & \ldots            &  RGs  &  5   \\
GSS                                    & 17-58& -1.03$\pm$0.07   & \ldots           & 0.39$\pm$0.06 & \ldots            &  RGs  &  5   \\
SES\footnote{Southern-East Shelf}      & 12   & -1.30$\pm$0.13    & \ldots & 0.53$\pm$0.10   &\ldots  &  RGs  &     2,3   \\
SES                                    & 18   & -0.71$\pm$0.07    & \ldots & 0.41$\pm$0.05   &\ldots  &  RGs  &     3    \\
SES                                    & 17-58& -0.89$\pm$0.08  & \ldots            & 0.45$\pm$0.05 & \ldots            &  RGs  &  5   \\
Substructures                          & 8-90 & \ldots & -0.0086$\pm$0.0023 & \ldots  & 0.0011$\pm$0.0021 &  RGs  & 7     \\
                                           \multicolumn{8}{c}{Photometry: Halo}  \\ 

Dry Halo                               & 8-34 & -0.81$\pm$0.03    & -0.00070$\pm$0.0016 & \ldots & \ldots &  RGs  &     3     \\
Dry Halo                               &10-90 & \textit{-0.75$\pm$0.03} & -0.0105$\pm$0.0013  & \ldots & \ldots &  RGs  &     4     \\
Halo                                   & 8-34 & -0.71$\pm$0.04    & -0.0091$\pm$0.0019  & \ldots & \ldots &  RGs  &     3     \\
Halo                                   &10-90 & \textit{-0.31} & -0.0110$\pm$0.0007 & \ldots & \ldots &  RGs  &     4     \\

                                        \multicolumn{8}{c}{Photometry: Substructures}  \\
NES\footnote{Northern-East Shelf}      & 15-29 & -0.53$\pm$0.02  & \ldots & \ldots   & \ldots &  RGs  &     6    \\                 
GSS                                    & 17-69 & -0.56$\pm$0.02  & \ldots & \ldots   &\ldots  &  RGs  &     9,6     \\
WS\footnote{West Shelf}                & 10-30 & -0.67$\pm$0.03  & \ldots & \ldots   &\ldots  &  RGs  &     10,6     \\
SES                                    & 9-30  & -0.50$\pm$0.02  & \ldots & \ldots   &\ldots  &  RGs  &   11,6     \\
\hline 
\end{tabular}
\end{center}
\end{table*}

\begin{table*}[htbp]
\caption{Iron, $\alpha$-elements abundances and projected distances of the M31 GCs\footnote{Table~\ref{tbl:M31_GCs_abundances} is published in its entirety in machine-readable format. A portion is shown here for guidance regarding its form and content.}.} \label{tbl:M31_GCs_abundances}
\begin{center}
\begin{tabular}{lcccccc}
\hline
\hline
Name     &      R.A.       &   DEC.      &   R$_{pro}$   &   [Fe/H]   &   [$\alpha$/Fe]\footnote{$\alpha$-element abundances were derived by averaging (in logarithmic scale), the Ca, Mg and TiII abundances found in the literature.}   &  Ref.\footnote{References: 1) \citet{colucci2014}; 2) \citet{caldwell2011}; 3) \citet{caldwell2016}; 4) \citet{sakari2021}; 5) \citet{sakari2022}; 6) \citet{mackey2019a}; 7) \citet{sakari2015}.} \\
         &  \multicolumn{2}{c}{(J2000)}  &       (kpc)         &            &                   &  \\
\hline
                                      \multicolumn{7}{c}{Inner halo GCs}    \\ 
B006-G058   & 00:40:26.48  & 41:27:26.7  & 6.39  & -0.73$\pm$0.22  &  0.29$\pm$0.03   &   1    \\ 
B012-G064   & 00:40:32.46  & 41:21:44.2  & 5.74  & -1.61$\pm$0.21  &  0.27$\pm$0.17   &   1    \\
B029-G090   & 00:41:17.82  & 41:00:23.0  & 6.78  & -0.43$\pm$0.11  &  0.21$\pm$0.13   &   1    \\
B034-G096   & 00:41:28.12  & 40:53:49.6  & 6.02  & -0.75$\pm$0.14  &  0.24$\pm$0.06   &   1    \\
B045-G108   & 00:41:43.11  & 41:34:20.3  & 4.90  & -0.94$\pm$0.22  &  0.21$\pm$0.11   &   1    \\
                                      \multicolumn{7}{c}{Outer halo GCs}    \\ 
G001	    & 00:32:46.50  & 39:34:40.6  & 34.70 & -0.98$\pm$0.05  &   0.34$\pm$0.05  &   4    \\
G002	    & 00:33:33.80  & 39:31:18.5  & 33.62 & -1.63$\pm$0.16  &  -0.15$\pm$0.19  &   1    \\
B514-MCGC4  & 00:31:09.85  & 37:54:00.4  & 55.30 & -1.74$\pm$0.17  &   0.22$\pm$0.23  &   1    \\
MCGC5-H10   & 00:35:59.76  & 35:41:03.9  & 78.68 & -1.45$\pm$0.10  &   0.14$\pm$0.13  &   1    \\
MGC1	    & 00:50:42.50  & 32:54:59.6  & 117.00& -1.56$\pm$0.09  &   0.14$\pm$0.04  &   1    \\
\hline 
\end{tabular}
\end{center}
\end{table*}

\begin{table*}[htbp]
\scriptsize
\caption{Radial iron and $\alpha$-element gradients in $R_G$ for the GCs in M31
by using either a logarithmic ([Fe/H]=$\alpha$+$\beta \times \log(dis)$)
or a linear ([Fe/H]=$\alpha$+$\beta \times dis$) fit. 
They have been estimated by using iron and $\alpha$-elements abundances 
for M31 GCs listed in Table~\ref{tbl:M31_GCs_abundances}.}  \label{tbl:m31_gcs_gradients}
\begin{center}
\begin{tabular}{ccccccc}
\hline
\hline
Gradient          & \multicolumn{2}{c}{Inner GCs}      & \multicolumn{2}{c}{Outer GCs}     & \multicolumn{2}{c}{Global}  \\
\hline
                                                  \multicolumn{7}{c}{Logarithmic} \\
$[Fe/H]$ & -0.742$\pm$  0.059 &  -0.446$\pm$  0.084 (0.57)\footnote{The numbers in parentheses are the standard deviations (dex).} &-0.517$\pm$  0.154 &  -0.639$\pm$  0.108 (0.42) & -0.713$\pm$  0.050 &  -0.494$\pm$  0.057 (0.54) \\
                                                      \multicolumn{7}{c}{Linear} \\
$[Fe/H]$ & -0.788$\pm$  0.053 &  -0.040$\pm$  0.008 (0.57) & -1.090$\pm$  0.086 &  -0.008$\pm$  0.002 (0.46) &  -0.952$\pm$  0.034 &  -0.011$\pm$  0.002  (0.56) \\
$[\alpha/Fe]$   & \ldots &  \ldots &  0.241$\pm$  0.021 &  -0.001$\pm$  0.001 (0.10) &  \ldots &  \ldots \\
\hline 
\end{tabular}
\end{center}
\end{table*}

\section{Summary and conclusions}\label{sec_summa}

We introduce the largest and most homogeneous photometric catalog of RRLs
(PR3C). PR3C includes more than 300,000 RRLs, among them roughly two thirds 
pulsate in the fundamental mode (RRab), one third in the first overtone (RRc), 
and $\approx$10\% as mixed mode RRL (RRd). PR3C includes both field RRLs 
and RRLs associated to stellar systems (globular clusters, dwarf galaxies). 
The catalog is mainly based on data available in public datasets. 
PR3C provides astrometric parameters, pulsation parameters 
(period, pulsation mode, luminosity amplitudes), mean apparent magnitudes, 
heliocentric and Galactocentric distances and reddening together with 
their errors.

We also present the largest and most homogeneous spectroscopic dataset of old 
stellar tracers (RRLs, BHB) ever collected (SR3C). SR3C includes more than 
16,000 field RRLs and for all of them we provide a RV measurement. 
SR3C is mainly based on LR spectra collected by public 
spectroscopic surveys. For 8,140 of them that cover the Ca~II~H\&K lines, we 
provide an estimate of the iron abundance by using the $\Delta$S  method 
developed by \citetalias{crestani2021a} and \citetalias{fabrizio2021}.
This sample was complemented with 547 iron abundance estimates available 
in the literature.

For a small fraction of field stars (487 RRLs, 64 BHB), 
HR spectra collected with 1m to 8m class telescopes are also available. 
Our group obtained and analyzed a significant fraction of them (342). 
For these objects, SR3C includes individual elemental abundances for iron 
and $\alpha$-elements. They were also complemented with similar estimates 
available in the literature. Measurements from the literature were only included 
if the individual samples overlapped with our dataset and could have been moved 
to the same metallicity scale.  SR3C also includes astrometric and kinematic 
parameters (position, proper motion) based on Gaia~DR3. 

The main results of the current investigation are the following. 


The probabilistic criteria suggested by \citet{bensby2014} were adopted to select 
the RRLs associated either to the main Galactic components --Halo, TND 
and TCD-- or in retrograde orbits. To validate the 
quoted selections, we also adopted the more stringent criteria suggested by 
\citet{fernandezalvar2024} for the identification of TND stars. 

By using two different diagnostics --the Lindblad diagram and the circularity 
of the orbits as a function of the Galactocentric distance-- we found that 
field RRLs display a systematic drift from the more metal-poor Halo to the 
more metal-rich component of the TND. The RRLs associated with 
the TCD trace quite well the transition between the more metal-poor 
and the more metal-rich regimes, since several of them have solar iron 
abundance. 

The peak in the IDF becomes systematically more metal-rich when moving from 
the Halo ([Fe/H]=-1.56) to the TCD ([Fe/H]=-1.47) and to the TND 
([Fe/H]=-0.73). Moreover, the IDF becomes broader and the standard deviation 
increases by about 50\%  between the Halo and the TND (0.37 \vs\ 0.57).
RRLs in retrograde orbits are similar to Halo RRLs, since they show similar 
Galactocentric distances, similar peaks in the IDF, and similar standard 
deviations.  

The separation among the different Galactic components is also supported 
by [$\alpha$/Fe] abundances. Halo RRLs are $\alpha$-enhanced, since their abundance 
ratio is [$\alpha$/Fe]=0.27 ($\sigma$=0.18). The TCD RRLs are either 
$\alpha$-enhanced ([Fe/H]$\le$-1.0) or $\alpha$-poor/$\alpha$-depleted in the more 
iron-rich regime, while TND RRLs are mainly $\alpha$-poor/$\alpha$-depleted 
with [$\alpha$/Fe]=-0.01 ($\sigma$=0.20). RRLs in retrograde orbits appear to be 
quite similar to Halo RRLs, since they show similar $\alpha$-element distributions.

To validate the selection criteria adopted to identify the different Galactic 
components we adopted the dynamical planes suggested by \citet{lane2022} together 
with three additional kinematic planes showing the circularity of the orbits 
($\lambda_Z$) as a function of eccentricity, the maximum height above the Galactic 
plane (Z$_{max}$) and the Galactocentric distance.   
Candidate TND and TCD RRLs can be easily identified in the Toomre diagram, since 
they cluster around V$_T \approx$ 250 km~s$^{-1}$ and V$_T \approx$ 150 km~s$^{-1}$, 
respectively. The same outcome applies for 
candidate Halo RRLs and  RRLs in retrograde orbits, since they attain either 
vanishing or negative transversal velocities. The same outcome applies to 
the V$_R$ \vs\ V$_T$ plane and to the Lindblad diagram. Special attention was 
paid to characterize RRLs with hot orbits 
(Halo \vs\ Bulge). To identify Bulge RRLs we selected RRLs that, along their 
orbits, have an apocentric radius smaller than r$_{ap}\le$3.5~kpc. Bulge RRLs
can easily be identified in the Lindblad diagram, since they are characterized
by very low orbital energies and vanishing vertical angular momentum, and in the
$\sqrt{J_R}$ \vs\ L$_Z$ plane due to their very low values of the square root of the 
radial action. We found, as expected, that only five RRLs in our sample are 
candidate Bulge RRLs. Candidate TND RRLs can also be easily identified as they 
cluster in the right corner of the action diamond, since their action budget is 
dominated by L$_Z$. Moreover, they cover a very tiny and compact slice in the Lindblad 
diagram, together with large circularities, small eccentricities, and small Z$_{max}$ 
values. Candidate TCD RRLs display a smooth transition between TND 
and Halo RRLs in the different kinematic planes.

Interestingly enough, we also found evidence that the distribution of Halo RRLs in the 
circularity of the orbits \vs\ Z$_{max}$ plane is far from being homogeneous. They display 
four distinct stellar arcs that move from TCD RRLs ($\lambda_Z >$0.25) to RRLs with 
retrograde orbits ($\lambda_Z<$ -0.25). Moving from larger to smaller Z$_{max}$ distances 
they are located at Z$_{max}\sim$30–40 kpc, 10–20 kpc, 7–9 kpc, and 4–5 kpc. This is a new 
evidence driven by the accuracy on the individual RRL distances suggesting that RRL radial 
distribution might be affected either by early merging events or by peculiar radial motions. 
Moreover, the lack of these features in the circularity \vs\ Galactocentric distance plane indicates 
that these stellar arcs appear to have radial distributions perpendicular to the Galactic plane.

We also performed a chemo-dynamical investigation of the substructures in the MW.
To identify candidate GSE and Sequoia RRLs we adopted the 
same selection criteria suggested by \citet{feuillet2021}, while for RRLs in the 
SGR stream we followed the selections from \citet{antoja2020} 
and for RRLs in the Helmi stream the kinematic criteria recently 
provided by \citet{koppelman2019a} and \citet{Horta2023}. 

Candidate GSE RRLs can be easily identified, since they cluster not only in the Toomre
and in the Lindblad diagram, but also in the action diamond plane. Moreover, the 
compact distribution that GSE RRLs show in the circularity \vs\ Z$_{max}$ plane further 
supports the evidence that they originated from a single major merging event. 
Furthermore GSE RRLs peak, in the square root of the radial action \vs\ the vertical 
angular momentum,  at radial action values that are systematically larger than Halo RRLs.  
The chemical characterization, the IDF and the standard deviations of GSE RRLs are, 
within the errors, quite similar to Halo RRLs. However, GSE RRLs lack the metal-poor/metal-rich 
tails, and their $\alpha$-element distribution is quite compact. They are 
metal-intermediate and show a homogeneous chemical enrichment not only in the range of energies 
covered in the Lindblad diagram but also in the wide range of Galactocentric distances 
they are covering (5--50~kpc).

Candidate Seq RRLs cluster in the left corner of the action diamond. However, the Sequoia 
sample clearly splits into two different sub-groups in the Lindblad diagram, in the 
eccentricity \vs\ L$_Z$ plane, and in the $\sqrt{J_R}$ \vs\ L$_Z$  plane.  Sequoia RRLs 
display a well--defined minimum for R$_G\sim$12 kpc in the circularity \vs\ 
Galactocentric. This evidence was adopted to split the Sequoia candidate into 
Seq.~1 and Seq.~2 RRLs. The IDFs of Seq.~1 and Seq.~2 RRLs peak at [Fe/H]=-1.68 and -1.60 
and they are also similar to Halo RRLs, but their standard deviation is smaller (0.31 \vs\ 0.37). 

Candidate Helmi RRLs show a smooth distribution in the Toomre diagram and in the action diamond, 
but they show a well--defined concentration in the Lindblad plane across  
E$\sim$-1.0 $\cdot$ 10$^5$~km$^2$~s$^{-2}$, L$_Z\sim$1.0 $\cdot$ 10$^3$~kpc~km~s$^{-1}$ 
and a high-energy tail extending from E$\sim$-0.9 $\cdot$ 10$^5$~km$^2$~s$^{-2}$.
Data plotted in the $\lambda_Z$ \vs\ Z$_{max}$ plane indicate that candidate 
Helmi RRLs  are mainly located in the TCD, since Z$_{max}$ ranges from 
a few kpc to $\sim$20 kpc and $\lambda_Z$ between 0.7 and 0.2.
The IDF of Helmi RRLs peaks at [Fe/H]=-1.64, and it is once again similar to 
Halo RRLs (-1.56), with similar standard deviations (0.36 \vs\ 0.37).

Candidate SGR RRLs typically have a sparse distribution in the classical 
kinematic planes, but they are relatively concentrated in the circularity \vs\ Z$_{max}$ plane 
and in the circularity \vs\ R$_G$ plane. In particular, they cover a narrow range in 
Galactocentric distances (R$_G\sim$30–40 kpc) and in maximum height above the Galactic 
plane (Z$_{max}\sim$20-30 kpc). This finding further supports the evidence that SGR candidates 
in the current catalog are mainly associated to the SGR stream, with a minimal contribution from the SGR core.
The SGR RRLs display an IDF similar to Halo RRLs, with a marginal evidence for a metal-poor/metal–rich tail. They are located at large Galactocentric distances and we still lack measurements of their $\alpha$-element abundances.

The circularity \vs\ Z$_{max}$ plane is a very interesting diagnostic to investigate 
the spatial distribution of stellar streams. Indeed, Seq.~1 and Seq.~2 RRLs have 
retrograde orbits and display in this plane a well-defined separation also in Z$_{max}$.
In particular, the Seq.~2 RRLs attain maximum heights above the Galactic plane that are 
larger than 9--10~kpc, while Seq.~1 RRLs are located closer to the Galactic plane.
Moreover, we found that in this plane candidate Sequoia, Helmi and SGR RRLs appear to be associated with the stellar arcs 
previously identified in the different Galactic components. SGR RRLs are associated with 
the stellar arc located at larger Galactocentric distances (Z$_{max}\sim$30--40~kpc),
Helmi and Seq.~2 RRLs are associated with the stellar arc located at Z$_{max}\sim$10--20~kpc, 
while Seq.~1 RRLs to the stellar arcs located at 7--9~kpc, and 4--5~kpc. The main difference 
is that candidate Sequoia RRLs are in retrograde motion, while Helmi RRLs are in prograde motion. 


Furthermore, we performed a detailed comparison between the chemo-dynamical properties 
of old stellar tracers like RRLs with similar properties for GGCs.
The GGC kinematics was estimated by using proper motions and 
distances provided by \citet{baumgardt2019}, together with the same Galactic potential 
adopted for field RRLs. The association with the different Galactic components was 
performed by using the same criteria adopted for field RRLs, while for the association of 
individual GGCs to the different stellar streams we took advantage of the chemo-dynamical 
investigation recently provided by \citet{massari2025}. We found a global agreement 
among the different stellar streams and the associated GGCs. However, there is mounting 
evidence of several outliers, i.e. GGCs that in different kinematic planes do not cluster 
with the bulk of the candidate RRLs. This preliminary finding supports predictions
based on numerical simulations by \citet{pagnini2023} suggesting the possible lack of 
dynamical coherence. 


During the last few years, several theoretical and empirical investigations 
have been focused on the occurrence of radial metallicity gradients. The 
radial metallicity gradient is a solid diagnostic to constrain the mass 
assembly and the chemical enrichment histories of the different Galactic 
components. 
We found that TCD RRLs display a well--defined radial iron gradient.  
The iron gradient is as expected negative and the slope in R$_G$ 
and in $\omega$ (-0.010) is steeper than in Z$_{max}$.
The iron abundance changes from 
metal-intermediate ([Fe/H]$\sim$ -1) in the innermost TCD regions to metal-poor 
([Fe/H]$\sim$ -2) in the outermost TCD regions. There is evidence of a steepening 
in the iron gradient for R$_G$ and $\omega\le$10~kpc, and this transition occurs 
across the solar circle. 

TND RRLs also display a well--defined radial iron gradient, in particular, 
as a function of Z$_{max}$. The change in the slope across the Solar circle 
was neglected, since it is mainly driven by the increased dispersion in 
metallicity. This finding indicates that the inside-out scenario 
for the chemical enrichment of the TND was already in place in the early phase 
of the MW formation. Moreover, the iron gradients associated to TND RRLs typically 
cover a narrower range in distances, but they are steeper when compared with the 
iron gradients of TCD RRLs (-0.026 \vs\ -0.010).

Global Halo RRLs present a very mild radial gradient and a large spread in iron 
abundance over a significant fraction of Galactocentric distances. The slopes of 
the iron gradients are, on average, more than one order of magnitude shallower than 
TND RRLs (-0.002 \vs\ -0.026 dex/kpc), and a factor of five shallower than TCD RRLs 
(-0.002 \vs\ -0.010 dex/kpc). The iron abundances 
of the RRLs with retrograde orbits display similar radial trends.
The mean iron abundance increases by more than 0.3 dex between R$_G\sim$10~kpc  
and R$_G\sim$7~kpc. This evidence supports previous findings by \citet{Suntzeff94},  
\citet{layden1995}, and more recently by \citet{kinman12a}.

To constrain on a more quantitative basis the possible impact of the stellar streams
on the radial iron gradient, we performed once again the linear fits by using Dry Halo 
RRLs, i.e. the field RRLs that do not belong to the four selected stellar streams. 
Interestingly enough, we found that the iron gradient is, within the errors,
equal to the Global Halo. This suggests that 
the inclusion of RRLs belonging to stellar streams has a minimal impact on the slope, but
the standard deviations of the linear relations of the Dry Halo are slightly smaller 
(0.36 \vs\ 0.38). Furthermore, the comparison between the iron radial gradients of the 
main Galactic components and GGCs associated to the same components display a global 
agreement. Thus suggesting that they are co-natal and they have had a common chemical 
enrichment history. 

The possible occurrence of iron radial gradients was also investigated among the 
selected stellar streams. We found that candidate GSE RRLs display a very 
mild negative radial gradient ($\beta$=-0.004) over the entire range of Galactocentric 
distances covered by the current sample. Preliminary evidence indicates the occurrence 
of two gaps in radial distance, suggesting the possible presence of different 
sub-structures among GSE RRLs.

Sequoia RRLs display a well--defined separation in Seq.~1 and in Seq.~2 RRLs. 
The iron abundance is quite constant with mild evidence of a possible 
positive radial gradient ($\beta$=0.001). The Helmi RRLs cover a modest range 
in radial distances and show tentative evidence of a positive radial gradient 
($\beta$=0.001). 

As a whole, we found mild evidence of iron radial gradients among the selected stellar 
streams, but the analysis is partially hampered by the range in Galactocentric 
distances covered by the current sub-samples and by the limited number of stars 
with spectroscopic abundances.


The comparison between current metallicity gradients based on RRLs and similar estimates 
available in the literature is far from being an easy effort. The reason is twofold. 

i) -- Literature estimates are a mixed bag, since we move from the presence of a radial 
gradient to the lack of a gradient according to the adopted stellar tracer, the range in Galactocentric 
distances covered by the different samples, and the kinematic selection of the different 
Galactic components. 
More recently, estimates based on large spectroscopic samples of field 
stars brought forward evidence of duality in the halo metallicity distribution. The inner 
halo being at least half dex more metal-rich than the outer halo 
\citep{carollo07,carollo10,dejong2010,xue2015,sesar2011a}. 

ii) -- The homogeneity and the accuracy of individual iron abundances 
and distances affect the estimate of the radial gradients. 

The comparison between the current and literature estimates shows, within the errors, 
a global agreement over the range in Galactocentric distances covered by the different 
estimates. The agreement with recent estimates based on APOGEE measurements 
\citep{fernandezalvar2017} is quite good not only for the iron radial gradients, but 
also for the [$\alpha$/Fe] radial gradients. However, this sample only covers
a modest range in Galactocentric distances. We also found a reasonable agreement 
with the radial gradient for the MW dry halo recently derived by \citet{medina2025a,medina2025b}
using a large and homogeneous sample of LR spectra collected by DESI. 
However, their slope is steeper than the current one (-0.01 \vs\ -0.002 dex/kpc). 

The unique opportunity to deal with a large and homogeneous sample of old 
stellar tracers open the path for a detailed comparison between the MW 
radial gradients and similar estimates for M31. M31 has been the cross-road 
of several investigations based either on photometric indices or on spectroscopic 
measurements. M31 is also the only nearby galaxy for which these measurements 
are available not only for the main galactic components, but also for its 
substructures. We performed a detailed comparison and we found several 
interesting results. 

i) -- The iron radial gradients for M31 based on photometric indices 
\citep{ibata2014, gilbert2014, escala2020b} have slopes that, 
within $1\sigma$, agree with the current estimates based on MW field RRLs (-0.011~vs~-0.002 dex/kpc). 
However, the zero-points are systematically more metal-rich. The iron 
radial gradients based on spectroscopic measurements \citep{escala2020b,Wojno2023}
agree quite well, within $1\sigma$, both in slope and zero-points, with similar 
estimates based on RRLs (-0.008 \vs\ -0.002 dex/kpc). This outcome applies not only to the 
global halo, but also to the dry halo (-0.007 \vs\ -0.002 dex/kpc).

ii) -- The comparison with the [$\alpha$/Fe] gradient 
\citep{escala2020a, Wojno2023} discloses a global agreement between M31 
and the MW. However, the range in Galactocentric distances covered by dry 
halo RRLs with $\alpha$ abundances is too limited to assess  the possible 
presence of a gradient, but the global halo show a good similarity 
over the entire range of Galactocentric distances (see Fig.~\ref{fig:iron_gradient_M31_lit}). 

iii) -- We made a similar comparison between the iron gradient 
detected by \citet{escala2021} for GSS in M31 and that of
GSE in the MW and found a similar slope within 
$1\sigma$ (-0.016 \vs\ -0.004) and a well--defined difference in the zero-point. 
GSS appears systematically more metal-rich and more $\alpha$--poor at any 
galactocentric distance. Interestingly enough, the slope for GSE RRLs 
recently derived by \citet{medina2025a} agrees quite well with the current 
estimate of the GSE radial gradient (-0.005 \vs\ -0.004).  

iv) -- Furthermore, we performed a detailed comparison between iron and 
$\alpha$ radial gradients  of M31 GCs and field RRLs. The abundances for M31 GCs 
are based on integrated-light spectroscopy, and it came as a surprise the very 
good agreement over the entire range in galactocentric distances covered by the 
two samples. This outcome applies to both inner and outer halo GCs and to both 
the iron and $\alpha$ radial gradients. The slopes of the iron radial gradient
of the M31 GCs and field MW RRLs are in global agreement, indeed the slope for 
outer halo GCs is $\beta$=-0.008  and for the global sample  $\beta$=-0.011. 
The similarity is further reinforced, but there is evidence that MW RRLs minimally 
include objects associated with the Galactic bulge and poorly cover the outermost
Halo regions.
Moreover, the $\alpha$-element radial gradients in M31 and in the MW indicate that the 
outer halo shows a smaller dispersion and more $\alpha$-poor abundances when 
compared with inner halo objects. There is no reason why old stellar tracers 
in M31 and in the MW should show similar radial trends. The current findings 
suggest similar early enrichment histories. 


One the most attractive feature of HB (RRL, BHB)  stars is that they are
old (t$>$ 10 Gyr) stellar tracers. This means that their properties
can provide firm constraints concerning the early formation and
evolution of the Galactic spheroid.

Detailed predictions based on HR N-body
simulations plus semi-analytical models \citep{deLuciaHelmi08}
indicate that the halo should show a dichotomic metallicity
distribution with no clear evidence of a metallicity gradient and
a well--defined dominance of metal-rich objects.
However, they also mentioned that the metallicity distribution
function is strongly correlated with the total mass of the
satellites that have accreted.
 
More detailed simulations taking account of the nucleosynthesis
yields and the transition between pop III and pop II also suggest a
broad metallicity distribution approaching solar metal-abundances
(see Fig.~3 and Fig.~6 in \citealt{salvadori07}).
More recently, N-body simulations by \citet{zolotov09}, \citet{zolotov10}
and by \citet[][Fig. 5, 6]{font11} show evidence of a metallicity gradient
in the Galactic halo.

The comparison with the simulations provided by the latter group
is more complex since they do not separate bulge and halo
components. We performed the same comparison, but we only took into
account Galactocentric distances larger than 3.5~kpc to limit
the contribution of the bulge. We found that the predicted
gradient is systematically more metal-rich over the entire
distance range (see their Fig.~5). The difference might
be tightly connected with the fraction of SN type Ia they
adopted in their simulations (see their Fig.~6).

There is also a clear difference between the inner and the outer halo. 
The above simulations suggest that the inner halo formed in situ and their 
stellar populations are typically old (t$>$13 ~Gyr). On the other
hand, the ages of the outer halo are systematically younger
(6-8 Gyr), since these are the typical ages of the accreted
satellites. However, the current empirical evidence indicates
the presence of large sample of field RRLs in the outer halo.

The current findings are in global agreement with the chemical evolution models 
of the Galactic halo provided by \citet{brusadin13}. Indeed, they found that 
the two--infall model, developed by \citet{chiappini1997} and \citet{romano2010}, plus 
the outflow predicts a broad metallicity distribution function for the Halo. 
A more detailed comparison between theory and observations is planned 
in a forthcoming investigation. 

In passing we also note that the results concerning the negative iron radial 
gradients and the modest variation in [$\alpha$/Fe] abundance ratio 
agree quite well with similar findings for late type galaxies found by 
\citet[][and references therein]{Parikh2021}, that used a large sample of
stacked spectra from the SDSS-IV/MaNGA survey of galaxies covering a 
broad range in stellar mass. 

The analysis of stellar streams and Galactic components depends on the
criteria adopted to identify the different substructures. They are all 
affected by limitations; the main objective of this investigation was to provide 
a global interpretation of the current empirical evidence. There is no doubt 
that the improved accuracy of astrometric parameters, RVs 
and distances coming with the Gaia DR4 will allow us to perform a more detailed 
tomography of the Galactic spheroid. 

Seventy-five years after the early evidence for a revision of the cosmic distance 
scale brought forward by \citet[][pg.~100]{baade1963}, based on the non-detection of cluster 
type variables in the field of M31 in blue and red photographic plates collected with the 
Palomar telescope, Galactic RRLs and M31 appear to be once again crucial to improve 
our knowledge of evolutionary and pulsation properties of old stellar populations and 
galaxy formation.

\section*{Acknowledgements}

We thank, with an unusual pleasure, the anonymous referee for his/her positive words 
and warm appreciation of the entire project, and in particular, of the current content of
the paper. Moreover, we are also very grateful to the anonymous referee for his/her 
constructive and pertinent suggestions that improved its content and readability.
We also thank Jo Bovy for many useful discussions and suggestions 
concerning physical and numerical assumptions adopted in his code and 
Luca Ciotti for many useful suggestions concerning the 
content of an early version of this manuscript.
Several of us acknowledge support from Project PRIN MUR 2022 (code 2022ARWP9C) 
‘Early Formation and Evolution of Bulge and HalO (EFEBHO)’ (PI: M. Marconi), 
funded by the European Union—Next Generation EU, and from the Large grant 
INAF 2023 MOVIE (PI: M. Marconi).
G.F., A.N. and M.T. acknowledge support from NextGenerationEU funds within the National Recovery and Resilience Plan (PNRR), Mission 4 - Education and Research, Component 2 - From Research to Business (M4C2), Investment Line 3.1 - Strengthening and creation of Research Infrastructures, Project IR0000034 – “STILES - Strengthening the Italian Leadership in ELT and SKA”, CUP C33C22000640006.
M.F. acknowledges financial support from the ASI-INAF agreement no. 2022-14-HH.0.
M.Mo. acknowledges support from the Spanish Ministry of Science, Innovation and Universities (MICIU) through the Spanish State Research Agency under the grants "RR Lyrae stars, a lighthouse to distant galaxies and early galaxy evolution" and the European Regional Development Fund (ERDF) with reference PID2021-127042OB-I00 and from the Severo Ochoa Programe 2020-2023 (CEX2019-000920-S). 
M.Mo and G.F. acknowledge the INAF projects "Participation in LSST – Large Synoptic Survey Telescope" (LSST inkind contribution ITA-INA-S22, PI: G. Fiorentino), OB.FU. 1.05.03.06 and "MINI-GRANTS (2023) DI RSN2" (PI: G. Fiorentino), OB.FU. 1.05.23.04.02.
P.B.T. acknowledges partial funding by Fondecyt-ANID 1240465/2024 and ANID
Basal Project FB210003.
F.A.G. acknowledges support from the ANID BASAL project FB210003, from the ANID FONDECYT Regular grants 1251493 and from the HORIZON-MSCA-2021-SE-01 Research and Innovation Programme under the Marie Sklodowska-Curie grant agreement number 101086388.
A.M. acknowledges support from the ANID BASAL project FB210003, from the ANID FONDECYT Regular grant 1251882 and funding from the HORIZON-MSCA-2021-SE-01 Research and Innovation Programme under the Marie Sklodowska-Curie grant agreement number 101086388.

G.B. and G.F. dedicate this paper to Anita whose arrival has revolutionized our lives. 

This research has been supported by the Munich Institute for Astro-, Particle and BioPhysics (MIAPbP), which is funded by the Deutsche Forschungsgemeinschaft (DFG, German Research Foundation) under Germany’s Excellence Strategy— EXC-2094—390783311.

This research has made use of both the BaSTI-IAC database (http://basti-iac.oa-abruzzo.inaf.it/index.html) and the CMD 3.7 web interface (http://stev.oapd.inaf.it/cgi-bin/cmd). This research has also made use of the GaiaPortal catalogs access tool, ASI - Space Science Data Center, Rome, Italy (https://gaiaportal.ssdc.asi.it).

Some of the observations reported in this paper were obtained with the Southern African Large Telescope (SALT).

Guoshoujing Telescope (the Large Sky Area Multi-Object Fiber Spectroscopic Telescope: LAMOST) is a National Major Scientific Project built by the Chinese Academy of Sciences. Funding for the project has been provided by the National Development and Reform Commission. LAMOST is operated and managed by the National Astronomical Observatories, Chinese Academy of Sciences.

Funding for the Sloan Digital Sky Survey V has been provided by the Alfred P. Sloan Foundation, the Heising-Simons Foundation, the National Science Foundation, and the Participating Institutions. SDSS acknowledges support and resources from the Center for High-Performance Computing at the University of Utah. SDSS telescopes are located at Apache Point Observatory, funded by the Astrophysical Research Consortium and operated by New Mexico State University, and at Las Campanas Observatory, operated by the Carnegie Institution for Science. The SDSS web site is \url{www.sdss.org}.
SDSS is managed by the Astrophysical Research Consortium for the Participating Institutions of the SDSS Collaboration, including the Carnegie Institution for Science, Chilean National Time Allocation Committee (CNTAC) ratified researchers, Caltech, the Gotham Participation Group, Harvard University, Heidelberg University, The Flatiron Institute, The Johns Hopkins University, L'Ecole polytechnique f\'{e}d\'{e}rale de Lausanne (EPFL), Leibniz-Institut f\"{u}r Astrophysik Potsdam (AIP), Max-Planck-Institut f\"{u}r Astronomie (MPIA Heidelberg), Max-Planck-Institut f\"{u}r Extraterrestrische Physik (MPE), Nanjing University, National Astronomical Observatories of China (NAOC), New Mexico State University, The Ohio State University, Pennsylvania State University, Smithsonian Astrophysical Observatory, Space Telescope Science Institute (STScI), the Stellar Astrophysics Participation Group, Universidad Nacional Aut\'{o}noma de M\'{e}xico, University of Arizona, University of Colorado Boulder, University of Illinois at Urbana-Champaign, University of Toronto, University of Utah, University of Virginia, Yale University, and Yunnan University.

This work has made use of data from the European Space Agency (ESA) mission Gaia \url{https://www.cosmos.esa.int/gaia}, processed by the Gaia Data Processing and Analysis Consortium (DPAC, \url{https://www.cosmos.esa.int/web/gaia/dpac/consortium}). Funding for the DPAC has been provided by national institutions, in particular the institutions participating in the Gaia Multilateral Agreement.

This work made use of the Third Data Release of the GALAH Survey (Buder et al. 2021). The GALAH Survey is based on data acquired through the Australian Astronomical Observatory, under programs: A/2013B/13 (The GALAH pilot survey); A/2014A/25, A/2015A/19, A2017A/18 (The GALAH survey phase 1); A2018A/18 (Open clusters with HERMES); A2019A/1 (Hierarchical star formation in Ori OB1); A2019A/15 (The GALAH survey phase 2); A/2015B/19, A/2016A/22, A/2016B/10, A/2017B/16, A/2018B/15 (The HERMES-TESS program); and A/2015A/3, A/2015B/1, A/2015B/19, A/2016A/22, A/2016B/12, A/2017A/14 (The HERMES K2-follow-up program). We acknowledge the traditional owners of the land on which the AAT stands, the Gamilaraay people, and pay our respects to elders past and present. This paper includes data that has been provided by AAO Data Central (datacentral.org.au).

Funding for RAVE has been provided by: the Leibniz-Institut f{\"u}r Astrophysik Potsdam (AIP);
the Australian Astronomical Observatory; the Australian National University; the Australian
Research Council; the French National Research Agency
(Programme National Cosmology et Galaxies (PNCG) of
CNRS/INSU with INP and IN2P3, co-funded by CEA and
CNES); the German Research Foundation (SPP 1177 and SFB
881: Project-ID 138713538); the European Research Council
(ERC-StG 240271 Galactica); the Istituto Nazionale di
Astrofisica at Padova; The Johns Hopkins University; the
National Science Foundation of the USA (AST-0908326);
the W. M. Keck foundation; the Macquarie University; the
Netherlands Research School for Astronomy; the Natural
Sciences and Engineering Research Council of Canada; the
Slovenian Research Agency (research core funding no. P1-
0188); the Swiss National Science Foundation; the Science \&
Technology Facilities Council of the UK; Opticon; Strasbourg
Observatory; and the Universities of Basel, Groningen,
Heidelberg and Sydney.

This research used data obtained with the Dark Energy Spectroscopic Instrument (DESI). DESI construction and operations is managed by the Lawrence Berkeley National Laboratory. This research is supported by the U.S. Department of Energy, Office of Science, Office of High-Energy Physics, under Contract No. DE–AC02–05CH11231, and by the National Energy Research Scientific Computing Center, a DOE Office of Science User Facility under the same contract. Additional support for DESI is provided by the U.S. National Science Foundation, Division of Astronomical Sciences under Contract No. AST-0950945 to the NSF’s National Optical-Infrared Astronomy Research Laboratory; the Science and Technology Facilities Council of the United Kingdom; the Gordon and Betty Moore Foundation; the Heising-Simons Foundation; the French Alternative Energies and Atomic Energy Commission (CEA); the National Council of Science and Technology of Mexico (CONACYT); the Ministry of Science and Innovation of Spain; and by the DESI Member Institutions (\url{www.desi.lbl.gov/collaborating-institutions}). The DESI collaboration is honored to be permitted to conduct astronomical research on Iolkam Du'ag (Kitt Peak), a mountain with particular significance to the Tohono O'odham Nation. Any opinions, findings, and conclusions or recommendations expressed in this material are those of the author(s) and do not necessarily reflect the views of the U.S. National Science Foundation, the U.S. Department of Energy, or any of the listed funding agencies.

This publication makes use of data products from the Wide-field Infrared Survey Explorer (WISE), which is a joint project of the University of California, Los Angeles, and the Jet Propulsion Laboratory/California Institute of Technology, and NEOWISE, which is a project of the Jet Propulsion Laboratory/California Institute of Technology. WISE and NEOWISE are funded by the National Aeronautics and Space Administration.

This publication makes use of data products from the Two Micron All Sky Survey (2MASS), which is a joint project of the University of Massachusetts and the Infrared Processing and Analysis Center/California Institute of Technology, funded by the National Aeronautics and Space Administration and the National Science Foundation.

\appendix
%
\section{Individual RRL distances}\label{sec:rrldistance}

Precise and homogeneous individual RRL distances are mandatory
to constrain the radial gradients and the possible occurrence
of local over-densities. A detailed description of the distance estimate will be
provided in the next paper on the photometric catalog (Braga et al., in preparation), 
but here we discuss a few issues that are crucial for this work.
One fundamental subsample to be defined is that of the RRLs with trustable Gaia parallaxes.
These are all the RRLs that comply with the following selection criterion:

\noindent 
\texttt{gaiadr3.gaia\_source.ruwe < 1.4 AND \\
gaiadr3.gaia\_source.ipd\_frac\_multi\_peak $\leq$ 30 AND \\
gaiadr3.gaia\_source.parallax\_corrected > 0 AND \\
gaiadr3.gaia\_source.parallax\_over\_error\_corrected $\geq$ 10 }.

\texttt{Parallax\_corrected} and \texttt{parallax\_over\_error\_corrected} were obtained by adopting the zero points calculated with the Python package by \citet{lindegren2021}.

We label this subsample, that contains 6666 RRLs, as $good\_\pi$, meaning ``good parallax''. 

The estimate of the individual distances moves along two different paths.

{\em i)}--RRLs for which no iron measurement/estimate is available. 
This is the largest sample, and includes roughly 90\% of RRLs 
in PR3C. We label this sample $no\_iron$. 

{\em ii)}--RRLs for which an estimate of the iron abundance, based 
either on a spectroscopic measurement (LR, HR spectra) or on a photometric 
estimate (Fourier parameters), is available. We label this sample $has\_iron$. SR3C is a subsample of $has\_iron$.

Concerning the $no\_iron$ sample, we performed a test using predicted 
metallicity--independent Period-Luminosity (PL)
and metallicity--dependent Period-Luminosity-Metallicity (PLZ) relations
for fundamental (FU) and first overtone (FO) RRLs \citep{marconi15}.
To use the PLZ relations we adopted a mean metallicity of 
[Fe/H]= --1.56 that is the mean metallicity of Halo RRLs 
(see \S~\ref{sec_Galcomp}). Note that TND/TCD RRLs and RRLs in retrograde orbits display, 
within the errors, similar peaks in their metallicity distribution. 

\begin{figure*}[]
\centering
\includegraphics[width=0.8\textwidth]{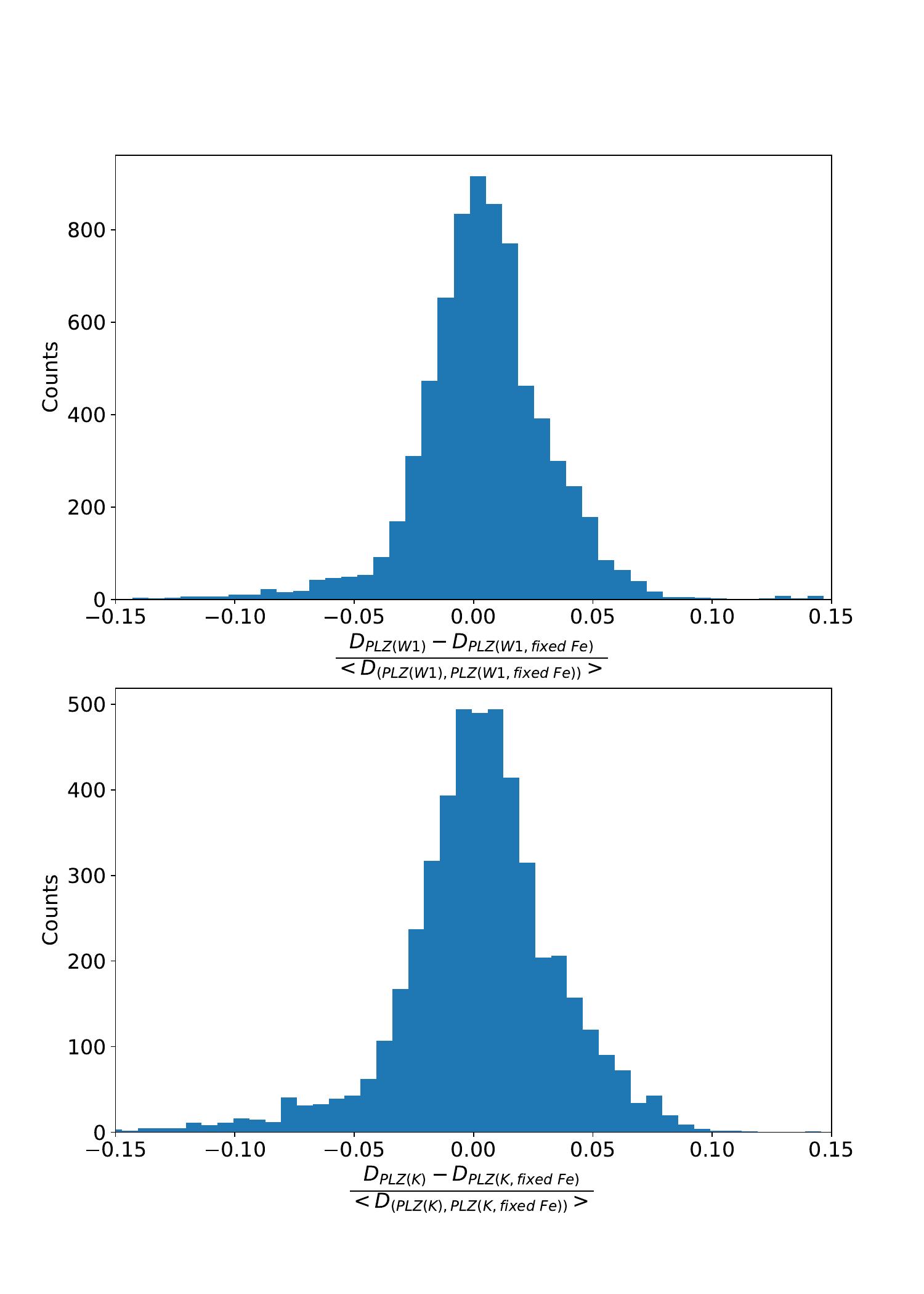}
\caption{
Top -- Fractional difference between distances based on the PLZ relation in the $W1$-band 
by either taking into account for the iron abundance or at a fixed [Fe/H] of --1.51 \citep{fabrizio2021}. 
Bottom -- Same as the top, but for the $K$-band.}
\label{fig:deltadistance_pl_plz}
\end{figure*}

Fig.~\ref{fig:deltadistance_pl_plz} shows the fractional difference for individual distances 
of FU and FO RRLs based on the PLZ and PL relations mentioned above ($W1$ in the top panel and $K_s$ in the bottom panel). The test was performed on the $no\_iron$ sample. In both cases, the displacement from 0 is smaller than the typical relative uncertainty of the distance estimates and is due to a small difference between the slope of the PL and PLZ relations. The mean and the standard deviations of the distributions are 0.0264$\pm$0.0064 ($W1$) and -0.00676$\pm$0.00010 ($K_s$). This means that  for the distance measurements of RRLs without an iron estimate we can safely adopt a PLZ by fixing [Fe/H] at --1.56.

The individual RRL distances for the $no\_iron$ sample were estimated using a mean iron abundance and the empirical PLZ relations provided either by \citet[][MIR]{mullen2023},
\citet[][$K$]{bhardwaj2022}, \citet[][$J$,$H$]{marconi15},
\citet[][$i$,$z$]{caceres2008}, \citet[][$I$]{neeley2019} or by 
\citet[][$G$,$G_{RP}$]{prudil2024}. 

The use of PLZ relations has several indisputable advantages when 
compared with the classical RRL absolute visual magnitude--metallicity ($M_V$--[Fe/H]) relation. 
{\em i)} They are linear over the entire period range.
{\em ii)} They are minimally affected by evolutionary effects (off-ZAHB evolution).
{\em iii)} They are less sensitive to metallicity uncertainties, since the 
coefficient of the metallicity term in optical/NIR/MIR PLZ relations is at least 
a factor of two smaller compared to the $M_V$--[Fe/H] relation. 
{\em iv)} The sensitivity to uncertainties in the reddening corrections 
is at least one order of magnitude smaller than in the optical bands.   

The individual distances of the RRLs in the $has\_iron$ sample were estimated by using the individual 
abundances and their intrinsic errors with the same approach adopted for the $no\_iron$ sample. 
This homogeneity is the most
important reason for which, in the end, we adopted PLZs for both groups.  

For each RRL, we adopted a well-defined priority list for the estimate 
of individual distances. The periods mainly come from the Gaia dataset; when 
this was not available, we adopted the period from the original catalog (OGLE, ASAS-SN...) and 
for a small sample we provided our own estimates. Concerning the apparent 
magnitudes, the highest priority was given to MIR and NIR magnitudes. 
The reasons are the following: 
{\em i)} The pulsation amplitudes in the NIR/MIR bands are at least a factor 
of four/five smaller ($\Delta$K=0.2--0.4 mag) than in the optical 
($\Delta$B=1.0--1.8) bands. This means that we can use either the mean 
of the measurements or the single epoch random measurements as a solid proxy 
of the mean magnitude based on the fit of the light curve \citep{Innoetal2013,braga2019}. 
{\em ii)} They are about one order of magnitude less affected by 
uncertainties in reddening corrections. 
{\em iii)} They are homogeneous, since 2MASS and WISE cover the entire sky 
and they are rooted on an accurate standard photometric systems.
{\em iv)} NIR/MIR PL/PLZ relations are steeper than in the optical regime. 
As a whole, 42\% of the RRLs in PR3C have individual distances based on 
NIR/MIR PLZ relations. 

The individual distances of the RRLs with no NIR/MIR measurements
are based either on the $i/z$-band PLZ relations or 
in the $G/G_{RP}$-band PLZ relation or by Gaia parallaxes of the $good\_\pi$ sample. 
The fraction of RRLs in PR3C with individual 
distances based on optical PLZ relations is 52\% .

Individual distances also require homogeneous reddening corrections. 
This is why we adopted the reddening map provided by \citet{schlafly11}.
Although this map is not optimal for the Bulge and TND, it is the best compromise
between sky coverage (full), homogeneity and reliability for Halo stars, which
represent our largest sample \citep[][, see their Fig.~5]{gonzalez2012}.
The reddening corrections were estimated in the different photometric bands 
by using the empirical reddening law provided by \citet{cardelli89}.

Concerning the SR3C, we ended up with a sample of 7708 RRL stars with 
individual distances estimated using either the NIR ($K$-band, 4\%; $H$-band, 1.5\%; 
$J$-band, 1.5\%) or the MIR PLZ relations ($W1$ and $W2$, 93\%; the $W1$ and 
$W2$ bands are centered at 3.6 and 4.5 $\mu$m, respectively). 

\begin{figure*}[]
\centering
\includegraphics[width=0.7\textwidth]{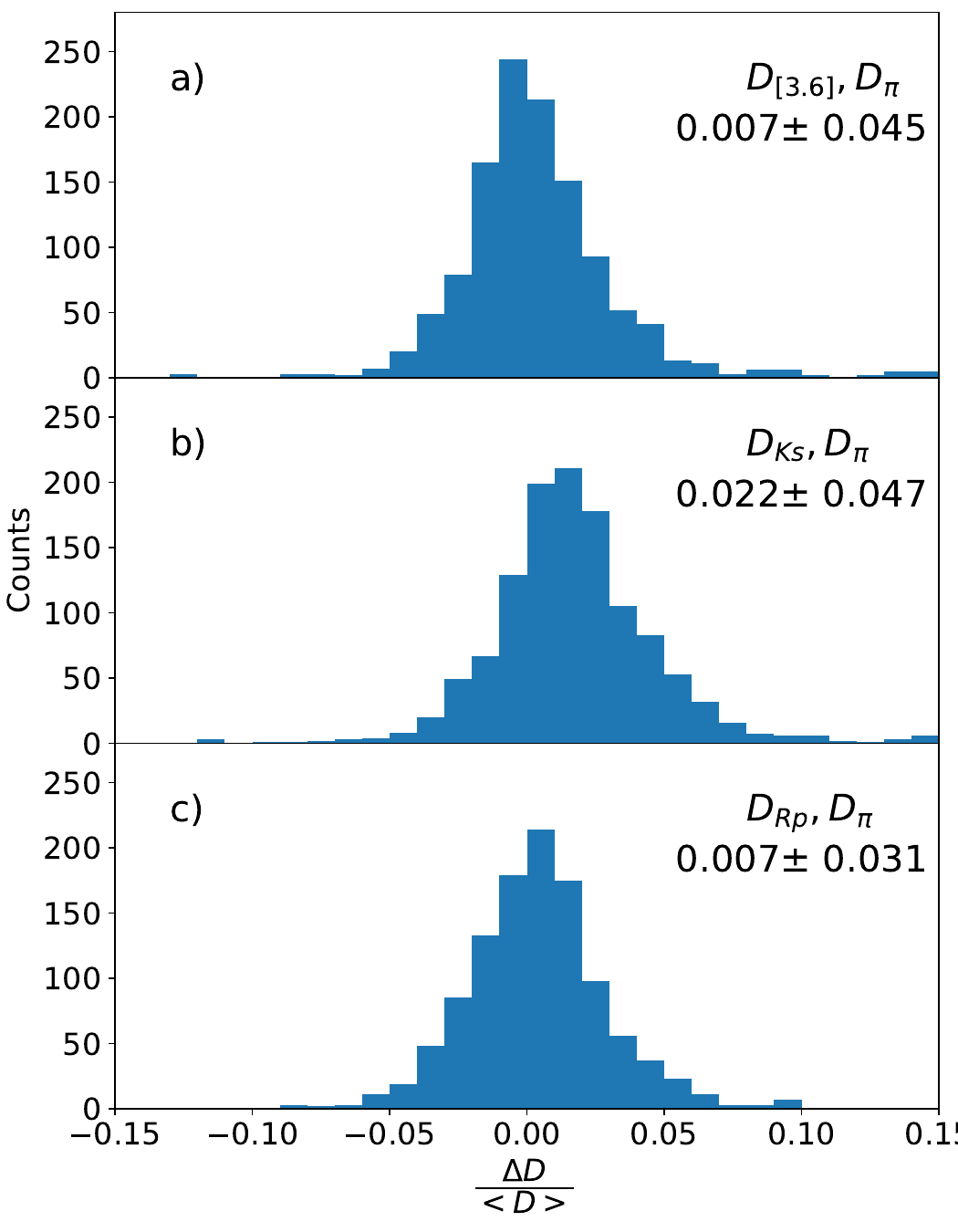}
\caption{Panel a) -- Distribution of the relative difference in distance derived from the 
PLZ([3.6]) relation and from the inverse Gaia~DR3 parallax, only for the intersection of 
the $good\_\pi$ and SR3C. The mean difference is labelled. 
Panel b) -- Same as Panel a), but for the PLZ($K$) relation.  
Panel c) -- Same as Panel a), but for PLZ($R_p$) relation.
}
\label{fig:deltadistance_w1}
\end{figure*}

The top panel of Fig.~\ref{fig:deltadistance_pl_plz} displays the fractional difference 
between the distances of RRLs of the $has\_iron$ sample obtained with the PLZ($W1$) by 
either fixing [Fe/H] at --1.56 (\citetalias{fabrizio2021}) or using the spectroscopic 
metallicity from our catalog.
Fig.~\ref{fig:deltadistance_w1} displays the same fractional difference 
in Galactocentric distance between Gaia parallaxes of the $good\_\pi$ sample 
and distances based on 
PLZ relations (from top to bottom, using the $W1$, $Ks$ and $Rp$ PLZs). 
The fractional differences are, as expected, on average smaller than the standard 
deviations (see labeled values). This plot serves as a validation for our set 
of distances, since we are comparing the best geometric estimates to our sample.

\section{Comparison between different Galactic potentials}\label{sec:gal_pote}

\begin{figure*}[] 
\centering
\includegraphics[width=0.9\textwidth]{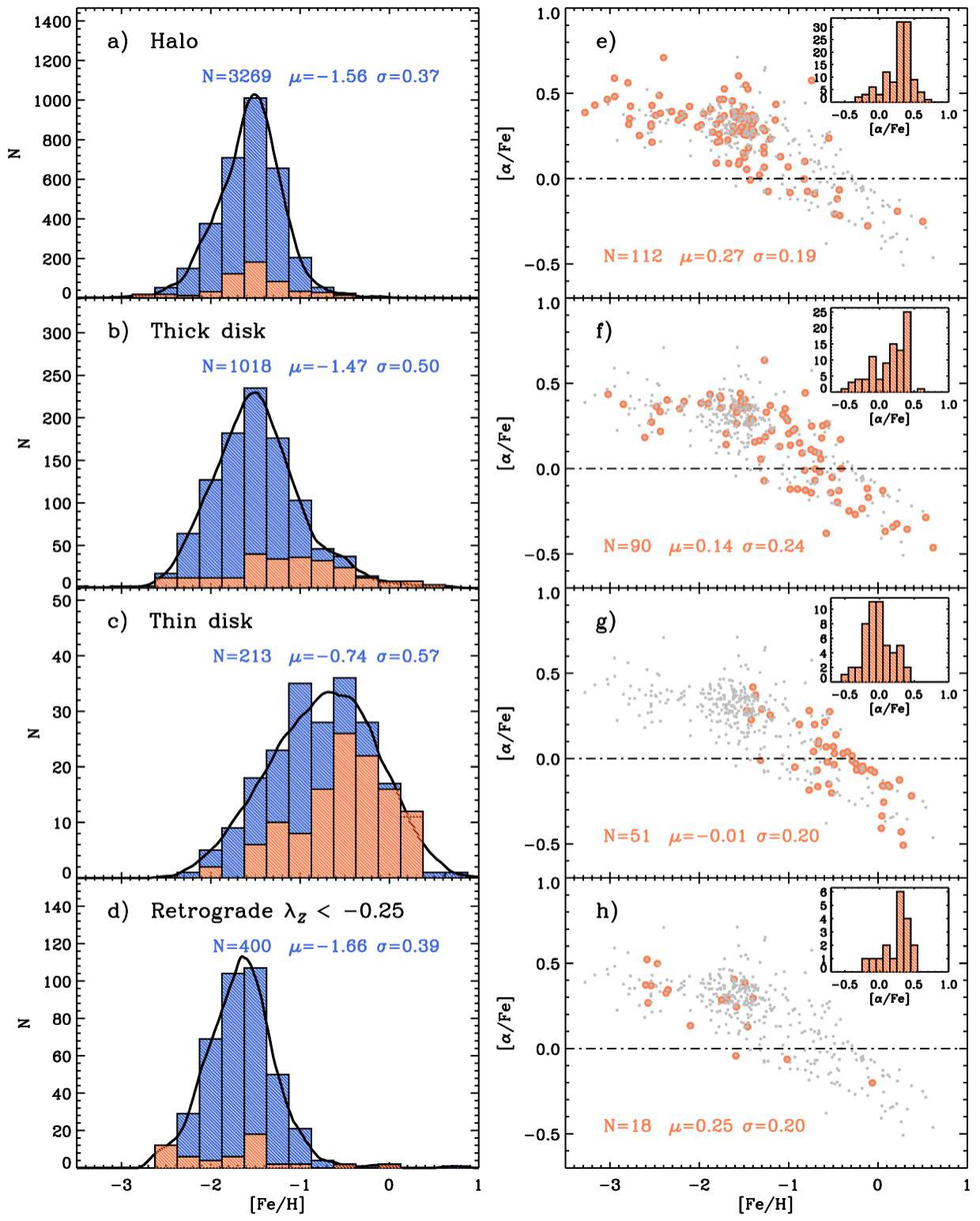}
\caption{Same as Fig.\ref{fig:circ_histo}, but the orbits were integrated by 
using the MW potential of \citet{mcmillan2017}.
}
\label{fig:circ_histo_McMillan}
\end{figure*}

\begin{figure*}[] 
\centering
\includegraphics[width=0.9\textwidth]{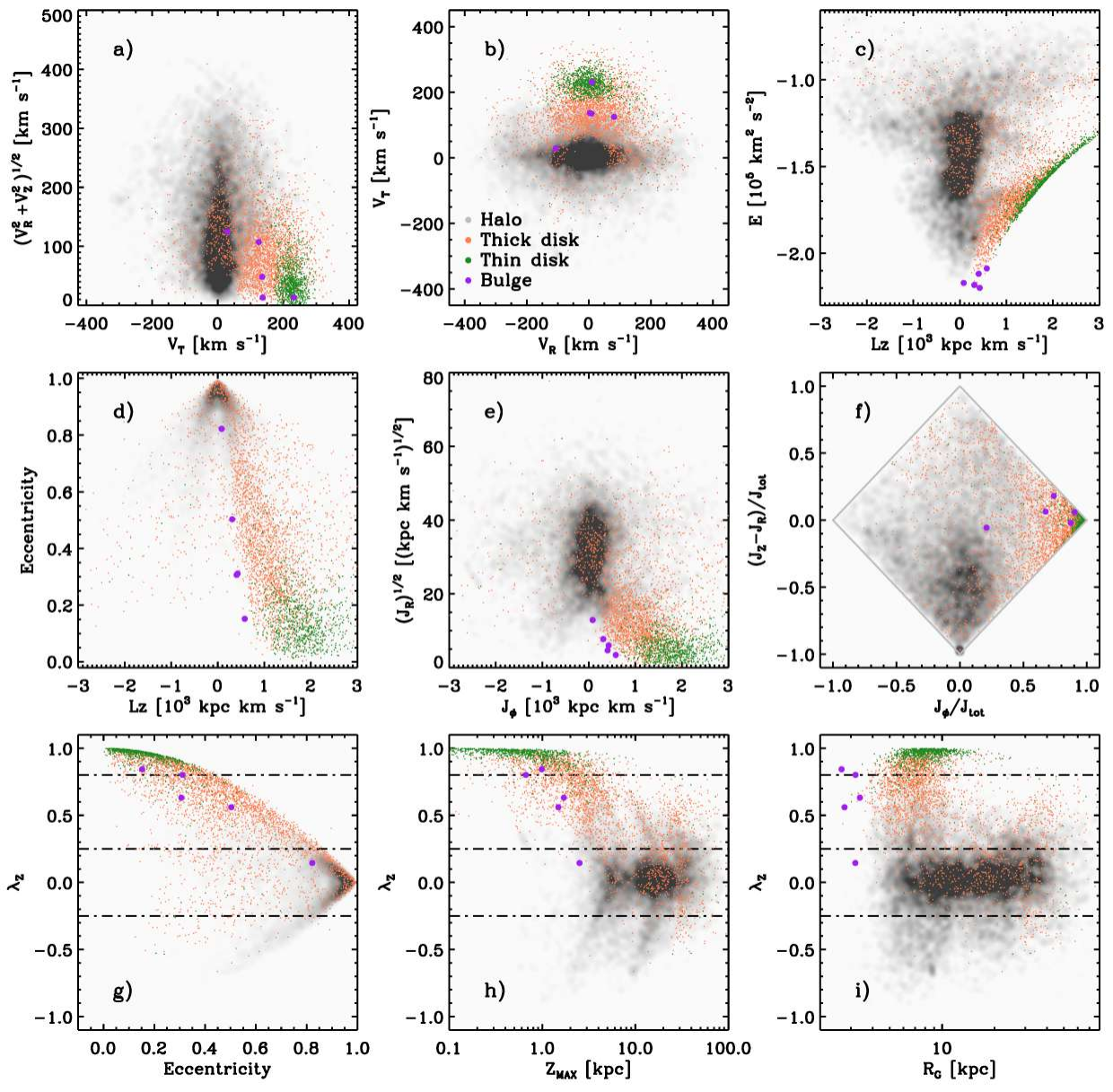}
\caption{Same as Fig.\ref{fig:gal_dynamic}, but the orbits were integrated by 
using the MW potential of \citet{mcmillan2017}.
}
\label{fig:gal_dynamic_McMillan}
\end{figure*}

\begin{figure*}[] 
\centering
\includegraphics[width=0.9\textwidth]{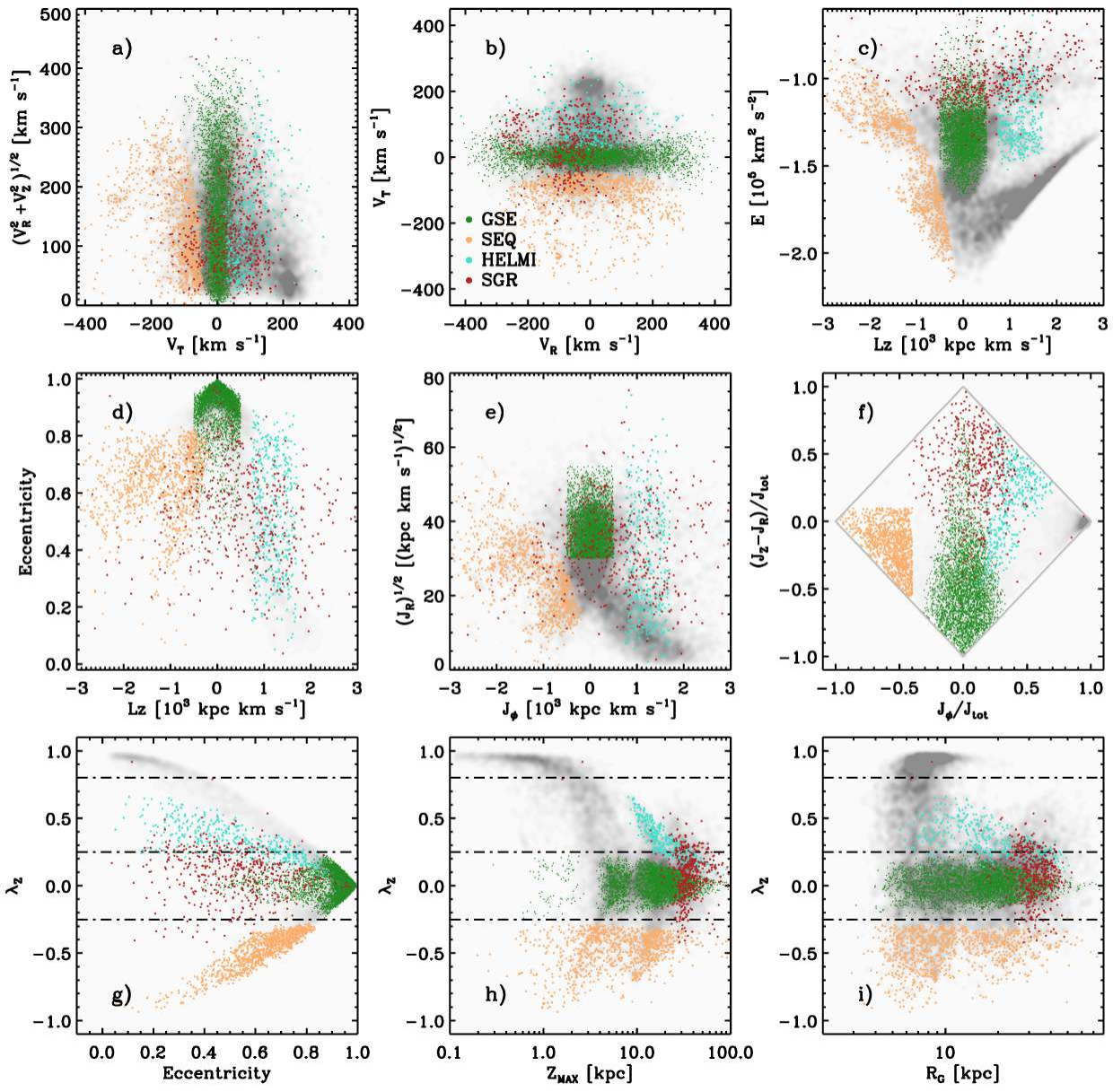}
\caption{Same as Fig.\ref{fig:stream_dynamic}, but the orbits were integrated by 
using the MW potential of \citet{mcmillan2017}.
}
\label{fig:stream_dynamic_McMillan}
\end{figure*}

In order to constrain the impact of the adopted MW potential on the RRLs associated to 
the main Galactic components and to the stellar streams, we performed an independent
orbital integration by using the MW potential of \citet{mcmillan2017} 
as implemented in AGAMA \citep{vasiliev2019}. Interestingly enough, we found that 
both iron and $\alpha$-element distribution functions based on the MW potential 
of \citet{bovy2015} (see Fig.~\ref{fig:circ_histo})  and of \citet{mcmillan2017} 
(see Fig.~\ref{fig:circ_histo_McMillan}) are quite similar.  The same outcome 
applies to the distribution in the different kinematic planes of the candidate 
RRLs associated to the main Galactic components (Fig.~\ref{fig:gal_dynamic} vs 
Fig.~\ref{fig:gal_dynamic_McMillan}) and to the selected stellar streams 
(Fig.~\ref{fig:stream_dynamic} vs Fig.~\ref{fig:stream_dynamic_McMillan}).


\section{Analytical fits of the iron radial gradients}\label{sec:fit_rad_grad}

\begin{figure*}[] 
\includegraphics[width=0.95\textwidth]{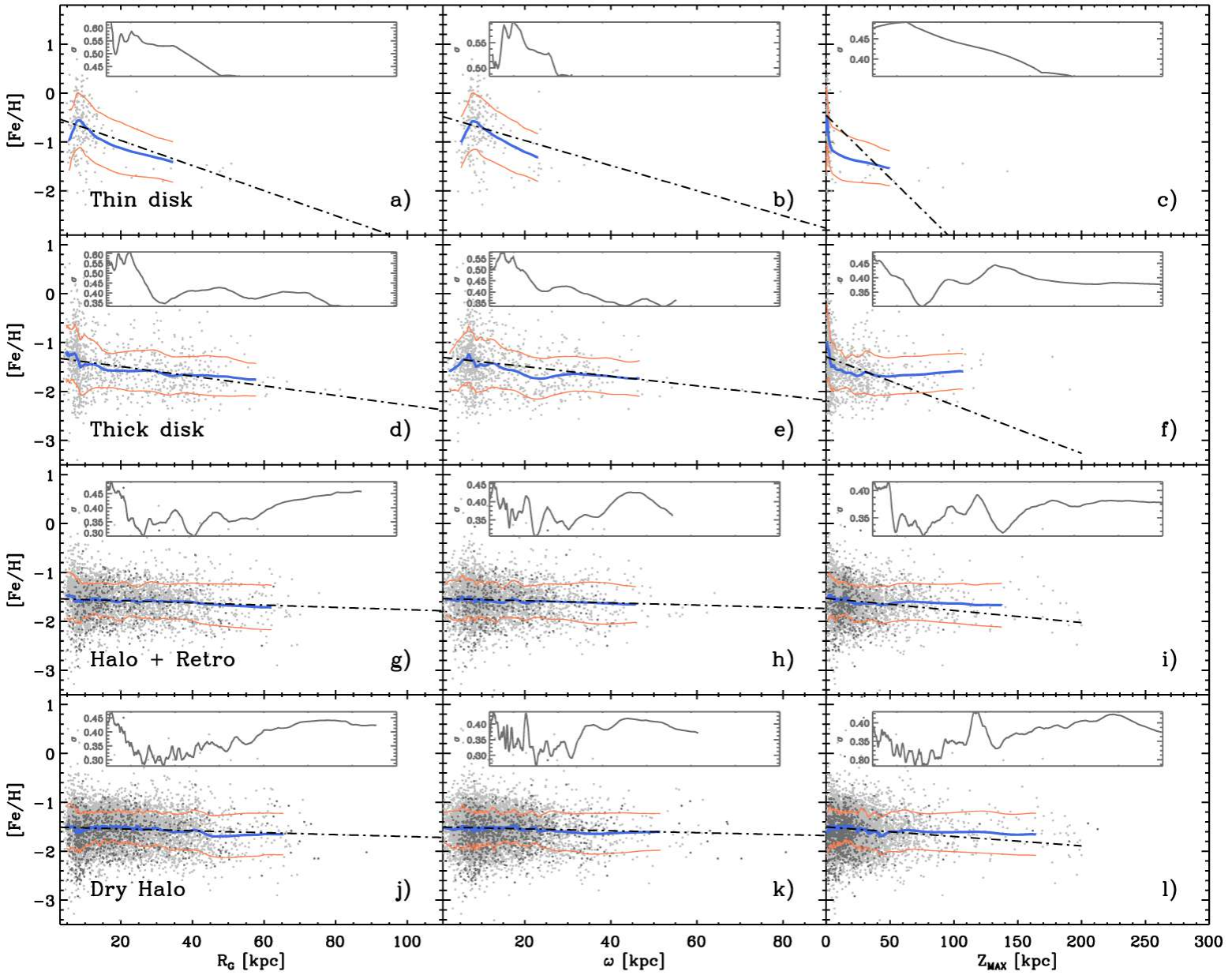}
\caption{Panels a,b,c -- Iron abundance for TND RRLs as a function of Galactocentric 
distance (panel a), distance projected onto the Galactic plane (panel b), and 
maximum height above the Galactic plane (panel c). The X-axes are in linear  
scale. The blue lines display the running average, while the orange lines the 
standard deviation of the running average. The dashed-dotted lines show 
the linear fit over the different sub-samples. The insets display the standard 
deviation of the spectroscopic sample as a function of distance.
Panels d,e,f -- Same as the top, but for TCD RRLs. 
Panels g,h,i -- Same as the top, but for Halo+Retro RRLs. 
Panels j,k,l -- Same as the top, but for dry Halo+Retro RRLs. See text for more details.
}
\label{fig:iron_radial_gradient_linear}
\end{figure*}

The analytical fits of the iron radial gradient in the different Galactic components 
were performed by using both linear and logarithmic Galactocentric distances 
(see \S~\ref{radial_grad_MW}). The coefficients of the analytical fits are listed 
in Table~\ref{tbl:radial_iron_gradient}, and the fits based on linear distances are shown in Fig.~\ref{fig:iron_radial_gradient_linear}.

\section{Spectroscopic Rome RR Lyrae Catalog (SR3C)}\label{sec:SR3C}

\subsection{Spectroscopic Rome RR Lyrae Catalog (SR3C$\_$ab)}\label{sec:SR3C_ab}
Tables~\ref{tbl:RRL_HR_iron_abundances} and~\ref{tbl:RRL_alfa_abundances} include individual 
elemental abundances of the RRLs in the SR3C based on HR spectra, and are published in their 
entirety in machine-readable format.

\begin{table*}[htbp]
\caption{Iron abundances from HR spectroscopy for the RRLs in the 
SR3C\footnote{Table~\ref{tbl:RRL_HR_iron_abundances} is published in its entirety in a 
machine-readable format. A portion is shown here for guidance regarding its form and content.}.}  
\label{tbl:RRL_HR_iron_abundances}
\begin{center}
\begin{tabular}{lccccc}
\hline
\hline
Gaia~DR3 ID   &  $\alpha$(ICRS)  &  $\delta$(ICRS)  &   [Fe/H]$_{lit}$\footnote{Iron abundances from the literature, when available.}   &   [Fe/H]$_{ours}$\footnote{Iron abundances either estimated for this work or converted into our metallicity scale.}   &  Ref.\footnote{References:
 1) \citetalias{crestani2021a}; 2) \citetalias{crestani2021b}; 3) \citet{fernley1996}; 4) \citet{nemec2013}; 5) \citet{pancino15}; 6) \citet{sneden2017}; 7) \citet{andrievsky2018};
 8) \citet{Duffau2014}; 9) \citet{galah_dr3}; 10) \citetalias{dorazi2024}; 11) \citetalias{gilligan2021}.} \\
\hline
6565527904791301504  & 324.51483  & -44.68671  & \ldots  	 & -0.47$\pm$0.04  & 1,2    \\
6380659528686603008  & 352.38918  & -72.54447  & \ldots  	 & -1.86$\pm$0.01  & 1,2  	\\
6321161342439508480  & 229.59119  &  -8.46186  & \ldots  	 & -0.54$\pm$0.01  & 1,2  	\\
4818348922610995200  &  73.31013  & -37.82105  & \ldots  	 & -0.93$\pm$0.03  & 1,2  	\\
3464204523694392192  & 177.48464  & -35.64746  & \ldots  	 & -0.77$\pm$0.11  & 1,2  	\\
\hline 
\end{tabular}
\end{center}
\end{table*}

\begin{table*}[htbp]
\caption{$\alpha$-element abundances from HR spectroscopy for the RRLs in the 
SR3C\footnote{Table~\ref{tbl:RRL_alfa_abundances} is published in its entirety 
in a machine-readable format. A portion is shown here for guidance regarding its form and content.}.}
\label{tbl:RRL_alfa_abundances}
\begin{center}
\begin{tabular}{lccccc}
\hline
\hline
Gaia~DR3 ID   &  $\alpha$(ICRS)  &  $\delta$(ICRS)  &   [$\alpha$/Fe]$_{lit}$\footnote{$\alpha$-element abundances from the literature, when available.}   &   [$\alpha$/Fe]$_{ours}$\footnote{$\alpha$-element abundances either estimated for this work or converted into our metallicity scale.}   &  Ref.\footnote{References:
 1) \citetalias{crestani2021b}; 2) \citetalias{dorazi2024}; 3) \citet{galah_dr3}.} \\
\hline
77849374617106176    &  30.13959  & 14.19830   & \ldots 	 & 0.13 $\pm$0.02  & 1,2  \\ 
80556926295542528    &  34.01548  & 17.53298   & \ldots 	 & 0.44 $\pm$0.12  & 1,2  \\
349612816093349120   &  28.78455  & 43.76568   & \ldots 	 & 0.07 $\pm$0.06  & 1,2  \\
584371601026374272   & 133.72292  & 6.43682    & \ldots 	 & 0.22 $\pm$0.09  & 1,2  \\
630421935431871232   & 151.93102  & 23.99171   & \ldots 	 & 0.26 $\pm$0.11  & 1,2  \\
\hline 
\end{tabular}
\end{center}
\end{table*}

\subsection{Spectroscopic Rome RR Lyrae Catalog (SR3C$\_$fe)}\label{sec:SR3C_fe}
Table~\ref{tbl:RRL_LR_iron_abundances} lists iron abundances of the RRLs in the SR3C based on the $\Delta$S method (MR or LR spectroscopy), and is published in its entirety in machine-readable format.

\begin{table*}[htbp]
\caption{Iron abundances from MR/LR spectroscopy for the RRLs in the SR3C\footnote{Table~\ref{tbl:RRL_LR_iron_abundances} is published in its entirety in a machine-readable format. A portion is shown here for guidance regarding its form and content.}.}  
\label{tbl:RRL_LR_iron_abundances}
\begin{center}
\begin{tabular}{lccccc}
\hline
\hline
Gaia~DR3 ID   &  $\alpha$(ICRS)  &  $\delta$(ICRS)  &   [Fe/H]$_{lit}$\footnote{Iron abundances from the literature, when available.}   &   [Fe/H]$_{ours}$\footnote{Iron abundances either estimated for this work or converted into our metallicity scale.}   &  Ref.\footnote{References:
 1) \citetalias{crestani2021a}; 2) \citet{zinn2020}; 3) \citet{liu2020}; 4) \citet{dambis2013}; 5) \citet{sesar13b}; 6) \citet{steinmetz2020b}; 7) \citet{medina2023}; 8) \citet{kinman12a}.} \\
\hline
780156482623646080   & 162.57884  & 42.56894   & \ldots 	 & -1.05$\pm$0.17  & 1  \\
777128427600786176   & 166.91592  & 40.56588   & \ldots 	 & -1.12$\pm$0.19  & 1  \\
776846196710082560   & 163.3049   & 41.31707   & \ldots 	 & -1.83$\pm$0.19  & 1  \\
765539437605028096   & 169.09549  & 41.23367   & \ldots 	 & -2.16$\pm$0.14  & 1  \\
717795623366814464   & 132.10925  & 36.33541   & \ldots 	 & -1.72$\pm$0.19  & 1  \\
\hline 
\end{tabular}
\end{center}
\end{table*}

\subsection{Spectroscopic Rome RR Lyrae Catalog (SR3C$\_$rv)}\label{sec:SR3C_rv}
Table~\ref{tbl:RRL_RV} includes the orbital and kinematic parameters, calculated using the MWPotential2014, and the RV measurements for the RRLs in the SR3C, and is published in its entirety in machine-readable format.

\begin{sidewaystable*} \tiny
\caption{Orbital and kinematic parameters, calculated by using the MWPotential2014, and center of mass RV ($\gamma$), from HR, MR and LR spectroscopy, for the RRLs in the SR3C, and their uncertainties.} 
\label{tbl:RRL_RV}
\begin{center}
\begin{adjustbox}{max width=\textheight}
\begin{tabular}{lcccccccccccccccc}
\hline
\hline
Gaia DR3 ID   &  $\alpha$(J2000)  &  $\delta$(J2000)  &   R$_{Gal}$\footnote{Galctocentric distance.}    &   Z$_{max}$\footnote{Maximum height above the Galactic plane.}    &
   $e$\footnote{Eccentricity of the orbit.}    &   V$_Z$\footnote{Vertical velocity component.}    &
      V$_T$\footnote{Transversal velocity component.}    &   V$_R$\footnote{Radial velocity component.}    &
         J$_R$\footnote{Radial orbital action.}    &   J$_{\Phi}$\footnote{Azimuthal orbital action.}     &
	    J$_Z$\footnote{Vertical orbital action.}    &   L$_Z$\footnote{Vertical angular momentum component.}     &
	        E\footnote{Orbital energy.}&   $\lambda_Z$\footnote{Circularity of the orbit.}&   $\gamma$\footnote{Center of mass RV.}     &
		   Ref.\footnote{References (referring to $\gamma$-velocity only):
 1) \citet{bono2020}; 2) \citet{gaia_dr3}; 3) \citet{zinn2020}; 4) Z. Prudil (private communication); 5) \citet{Duffau2014}; 6) \citet{medina2023}; 7) \citetalias{dorazi2024};
 8) \citet{nemec2013}; 9) \citet{hansen2011}; 10) \citet{galah_dr3}; 11) J. Crestani (private communication); TW) This work.} \\
\hline
6771307454464848768  & 293.08657  & -23.85379  & 7476.28999 $\pm$10.52805  & 0.45472$_{0.42397}^{0.48649}$   & 0.78209$_{0.77391}^{0.78966}$  & -25.80982$_{-27.27631}^{-24.24645}$	& 52.30631$_{50.32165}^{54.43635}$	 & 15.60223$_{13.30678}^{18.06603}$	  & 456.03313$_{449.80509}^{461.90186}$ & 390.86946$_{375.57891}^{407.31625}$	    & 6.45489$_{5.78387}^{7.13120}$	  & 390.86946$_{375.57891}^{407.31625}$       & -150673.22640$_{-150832.624496}^{-150490.30277}$ & 0.34957$_{0.33683}^{0.36312}$    & -60.44 $\pm$2.04  & 1  \\
2720896455287475584  & 333.85694  & 6.82262    & 8000.02440 $\pm$8.35446   & 0.50384$_{0.47806}^{0.53211}$   & 0.29926$_{0.29501}^{0.30354}$  & 21.41938$_{19.33163}^{23.50233}$	& 180.69917$_{179.18736}^{182.22681}$    & 59.89313$_{59.58380}^{60.21600}$	  & 97.15102$_{94.78423}^{99.54160}$    & 1444.65238$_{1432.86493}^{1456.54736}$    & 7.10791$_{6.48204}^{7.81510}$       & 1444.65238$_{1432.86493}^{1456.54736}$    & -130363.21989$_{-130605.382171}^{-130113.47739}$ & 0.90490$_{0.90134}^{0.90834}$    & -71.38 $\pm$2.46  & 1  \\
2857456211775108480  & 5.92951    & 29.40092   & 8315.02890 $\pm$8.11582   & 0.40298$_{0.38189}^{0.42601}$   & 0.12897$_{0.12664}^{0.13130}$  & -17.15670$_{-18.44854}^{-15.92234}$	& 220.93228$_{220.18918}^{221.68654}$    & -37.45245$_{-37.69041}^{-37.20558}$    & 20.84389$_{20.16101}^{21.53783}$     & 1836.28753$_{1829.35207}^{1843.32659}$    & 4.63224$_{4.21910}^{5.09731}$	  & 1836.28753$_{1829.35207}^{1843.32659}$    & -121414.00417$_{-121617.37576}^{-121205.73969}$  & 0.97888$_{0.97880}^{0.97893}$    & -20.1  $\pm$1.24  & 1  \\
5947570591534602240  & 269.79456  & -49.43350  & 7316.30910 $\pm$12.67856  & 0.70793$_{0.69542}^{0.72501}$   & 0.69600$_{0.68985}^{0.70232}$  & 37.01246$_{36.48057}^{37.54327}$	& 109.55267$_{107.84332}^{111.23729}$    & -178.85091$_{-180.51917}^{-177.21023}$ & 519.41094$_{510.19189}^{528.84601}$ & 801.32543$_{787.57781}^{814.90901}$ 	    & 9.26527$_{8.99720}^{9.53640}$	  & 801.32543$_{787.57781}^{814.90901}$       & -131058.15554$_{-131104.20290}^{-131003.82814}$  & 0.50813$_{0.49899}^{0.51722}$    & 183.29 $\pm$1.90  & 1  \\
6883653108749373568  & 315.37039  & -15.22995  & 7429.18591 $\pm$16.52036  & 10.82856$_{9.84896}^{11.92122}$ & 0.50039$_{0.46677}^{0.53291}$  & -186.12153$_{-190.79617}^{-181.41741}$  & -284.66154$_{-291.09982}^{-278.18329}$ & -58.20650$_{-61.39260}^{-55.01286}$	  & 525.11028$_{428.04163}^{636.45699}$ & -2107.48638$_{-2152.20299}^{-2062.39184}$ & 386.36100$_{371.60424}^{401.06324}$ & -2107.48638$_{-2152.20299}^{-2062.39184}$ & -92737.24997$_{-95552.73995}^{-89869.13967}$     & -0.63894$_{-0.66354}^{-0.61353}$ & -105.15$\pm$1.34  & 1  \\
\hline 
\end{tabular}
\end{adjustbox}
\end{center}
\end{sidewaystable*}

\bibliography{ms}{} 

@ARTICLE{fernandezalvar2024,
       author = {{Fern{\'a}ndez-Alvar}, Emma and {Kordopatis}, Georges and {Hill}, Vanessa and {Battaglia}, Giuseppina and {Gallart}, Carme and {Gonz{\'a}lez Rivera de la Vernhe}, Isaure and {Thomas}, Guillaume and {Sestito}, Federico and {Ardern-Arentsen}, Anke and {Martin}, Nicolas and {Viswanathan}, Akshara and {Starkenburg}, Else},
        title = "{The metal-poor edge of the Milky Way's ``thin disc''}",
      journal = {\aap},
         year = 2024,
        month = may,
       volume = {685},
          eid = {A151},
        pages = {A151},
          doi = {10.1051/0004-6361/202348918},
archivePrefix = {arXiv},
       eprint = {2402.02943},
 primaryClass = {astro-ph.GA},
       adsurl = {https://ui.adsabs.harvard.edu/abs/2024A&A...685A.151F}
}

@ARTICLE{luongo2024,
       author = {{Luongo}, E. and {Ripepi}, V. and {Marconi}, M. and {Prudil}, Z. and {Rejkuba}, M. and {Clementini}, G. and {Longo}, G.},
        title = "{An 'alien' called the Oosterhoff dichotomy?}",
      journal = {\aap},
         year = 2024,
        month = oct,
       volume = {690},
          eid = {L17},
        pages = {L17},
          doi = {10.1051/0004-6361/202451596},
archivePrefix = {arXiv},
       eprint = {2409.04259},
 primaryClass = {astro-ph.GA},
       adsurl = {https://ui.adsabs.harvard.edu/abs/2024A&A...690L..17L}
}

@ARTICLE{harris10,
       author = {{Harris}, William E.},
        title = "{A New Catalog of Globular Clusters in the Milky Way}",
      journal = {arXiv e-prints},
         year = 2010,
        month = dec,
          eid = {arXiv:1012.3224},
        pages = {arXiv:1012.3224},
          doi = {10.48550/arXiv.1012.3224},
archivePrefix = {arXiv},
       eprint = {1012.3224},
 primaryClass = {astro-ph.GA},
       adsurl = {https://ui.adsabs.harvard.edu/abs/2010arXiv1012.3224H}
}

@ARTICLE{massari2025,
       author = {{Massari}, Davide},
        title = "{Origin of the System of Globular Clusters in the Milky Way{\textemdash}Gaia eDR3 Edition}",
      journal = {Research Notes of the American Astronomical Society},
         year = 2025,
        month = mar,
       volume = {9},
       number = {3},
          eid = {64},
        pages = {64},
          doi = {10.3847/2515-5172/adc375},
archivePrefix = {arXiv},
       eprint = {2503.14657},
 primaryClass = {astro-ph.GA},
       adsurl = {https://ui.adsabs.harvard.edu/abs/2025RNAAS...9...64M}
}

@ARTICLE{zoccali2024,
       author = {{Zoccali}, M. and {Quezada}, C. and {Contreras Ramos}, R. and {Valenti}, E. and {Valenzuela-Navarro}, A. and {Olivares Carvajal}, J. and {Rojas Arriagada}, A. and {Minniti}, J.~H. and {Gran}, F. and {De Leo}, M.},
        title = "{VVV catalog of ab-type RR Lyrae in the inner Galactic bulge}",
      journal = {\aap},
         year = 2024,
        month = sep,
       volume = {689},
          eid = {A240},
        pages = {A240},
          doi = {10.1051/0004-6361/202450126},
archivePrefix = {arXiv},
       eprint = {2407.11226},
 primaryClass = {astro-ph.GA},
       adsurl = {https://ui.adsabs.harvard.edu/abs/2024A&A...689A.240Z}
}

@ARTICLE{bennett2019,
       author = {{Bennett}, Morgan and {Bovy}, Jo},
        title = "{Vertical waves in the solar neighbourhood in Gaia DR2}",
      journal = {\mnras},
         year = 2019,
        month = jan,
       volume = {482},
       number = {1},
        pages = {1417-1425},
          doi = {10.1093/mnras/sty2813},
archivePrefix = {arXiv},
       eprint = {1809.03507},
 primaryClass = {astro-ph.GA},
       adsurl = {https://ui.adsabs.harvard.edu/abs/2019MNRAS.482.1417B}
}

@ARTICLE{drimmelandpoggio2018,
       author = {{Drimmel}, Ronald and {Poggio}, Eloisa},
        title = "{On the Solar Velocity}",
      journal = {Research Notes of the American Astronomical Society},
         year = 2018,
        month = nov,
       volume = {2},
       number = {4},
          eid = {210},
        pages = {210},
          doi = {10.3847/2515-5172/aaef8b},
       adsurl = {https://ui.adsabs.harvard.edu/abs/2018RNAAS...2..210D}
}

@ARTICLE{schoenrich2011,
       author = {{Sch{\"o}nrich}, Ralph and {Asplund}, Martin and {Casagrande}, Luca},
        title = "{On the alleged duality of the Galactic halo}",
      journal = {\mnras},
         year = 2011,
        month = aug,
       volume = {415},
       number = {4},
        pages = {3807-3823},
          doi = {10.1111/j.1365-2966.2011.19003.x},
archivePrefix = {arXiv},
       eprint = {1012.0842},
 primaryClass = {astro-ph.GA},
       adsurl = {https://ui.adsabs.harvard.edu/abs/2011MNRAS.415.3807S}
}

@ARTICLE{gaiacollaboration23,
       author = {{Gaia Collaboration} and {Recio-Blanco}, A. and {Kordopatis}, G. and {de Laverny}, P. and {Palicio}, P.~A. and {Spagna}, A. and {Spina}, L. and {Katz}, D. and {Re Fiorentin}, P. and {Poggio}, E. and {McMillan}, P.~J. and {Vallenari}, A. and {Lattanzi}, M.~G. and {Seabroke}, G.~M. and {Casamiquela}, L. and {Bragaglia}, A. and {Antoja}, T. and {Bailer-Jones}, C.~A.~L. and {Schultheis}, M. and {Andrae}, R. and {Fouesneau}, M. and {Cropper}, M. and {Cantat-Gaudin}, T. and {Bijaoui}, A. and {Heiter}, U. and {Brown}, A.~G.~A. and {Prusti}, T. and {de Bruijne}, J.~H.~J. and {Arenou}, F. and {Babusiaux}, C. and {Biermann}, M. and {Creevey}, O.~L. and {Ducourant}, C. and {Evans}, D.~W. and {Eyer}, L. and {Guerra}, R. and {Hutton}, A. and {Jordi}, C. and {Klioner}, S.~A. and {Lammers}, U.~L. and {Lindegren}, L. and {Luri}, X. and {Mignard}, F. and {Panem}, C. and {Pourbaix}, D. and {Randich}, S. and {Sartoretti}, P. and {Soubiran}, C. and {Tanga}, P. and {Walton}, N.~A. and {Bastian}, U. and {Drimmel}, R. and {Jansen}, F. and {van Leeuwen}, F. and {Bakker}, J. and {Cacciari}, C. and {Casta{\~n}eda}, J. and {De Angeli}, F. and {Fabricius}, C. and {Fr{\'e}mat}, Y. and {Galluccio}, L. and {Guerrier}, A. and {Masana}, E. and {Messineo}, R. and {Mowlavi}, N. and {Nicolas}, C. and {Nienartowicz}, K. and {Pailler}, F. and {Panuzzo}, P. and {Riclet}, F. and {Roux}, W. and {Sordo}, R. and {Th{\'e}venin}, F. and {Gracia-Abril}, G. and {Portell}, J. and {Teyssier}, D. and {Altmann}, M. and {Audard}, M. and {Bellas-Velidis}, I. and {Benson}, K. and {Berthier}, J. and {Blomme}, R. and {Burgess}, P.~W. and {Busonero}, D. and {Busso}, G. and {C{\'a}novas}, H. and {Carry}, B. and {Cellino}, A. and {Cheek}, N. and {Clementini}, G. and {Damerdji}, Y. and {Davidson}, M. and {de Teodoro}, P. and {Nu{\~n}ez Campos}, M. and {Delchambre}, L. and {Dell'Oro}, A. and {Esquej}, P. and {Fern{\'a}ndez-Hern{\'a}ndez}, J. and {Fraile}, E. and {Garabato}, D. and {Garc{\'\i}a-Lario}, P. and {Gosset}, E. and {Haigron}, R. and {Halbwachs}, J. -L. and {Hambly}, N.~C. and {Harrison}, D.~L. and {Hern{\'a}ndez}, J. and {Hestroffer}, D. and {Hodgkin}, S.~T. and {Holl}, B. and {Jan{\ss}en}, K. and {Jevardat de Fombelle}, G. and {Jordan}, S. and {Krone-Martins}, A. and {Lanzafame}, A.~C. and {L{\"o}ffler}, W. and {Marchal}, O. and {Marrese}, P.~M. and {Moitinho}, A. and {Muinonen}, K. and {Osborne}, P. and {Pancino}, E. and {Pauwels}, T. and {Reyl{\'e}}, C. and {Riello}, M. and {Rimoldini}, L. and {Roegiers}, T. and {Rybizki}, J. and {Sarro}, L.~M. and {Siopis}, C. and {Smith}, M. and {Sozzetti}, A. and {Utrilla}, E. and {van Leeuwen}, M. and {Abbas}, U. and {{\'A}brah{\'a}m}, P. and {Abreu Aramburu}, A. and {Aerts}, C. and {Aguado}, J.~J. and {Ajaj}, M. and {Aldea-Montero}, F. and {Altavilla}, G. and {{\'A}lvarez}, M.~A. and {Alves}, J. and {Anders}, F. and {Anderson}, R.~I. and {Anglada Varela}, E. and {Baines}, D. and {Baker}, S.~G. and {Balaguer-N{\'u}{\~n}ez}, L. and {Balbinot}, E. and {Balog}, Z. and {Barache}, C. and {Barbato}, D. and {Barros}, M. and {Barstow}, M.~A. and {Bartolom{\'e}}, S. and {Bassilana}, J. -L. and {Bauchet}, N. and {Becciani}, U. and {Bellazzini}, M. and {Berihuete}, A. and {Bernet}, M. and {Bertone}, S. and {Bianchi}, L. and {Binnenfeld}, A. and {Blanco-Cuaresma}, S. and {Boch}, T. and {Bombrun}, A. and {Bossini}, D. and {Bouquillon}, S. and {Bramante}, L. and {Breedt}, E. and {Bressan}, A. and {Brouillet}, N. and {Brugaletta}, E. and {Bucciarelli}, B. and {Burlacu}, A. and {Butkevich}, A.~G. and {Buzzi}, R. and {Caffau}, E. and {Cancelliere}, R. and {Carballo}, R. and {Carlucci}, T. and {Carnerero}, M.~I. and {Carrasco}, J.~M. and {Castellani}, M. and {Castro-Ginard}, A. and {Chaoul}, L. and {Charlot}, P. and {Chemin}, L. and {Chiaramida}, V. and {Chiavassa}, A. and {Chornay}, N. and {Comoretto}, G. and {Contursi}, G. and {Cooper}, W.~J. and {Cornez}, T. and {Cowell}, S. and {Crifo}, F.},
        title = "{Gaia Data Release 3. Chemical cartography of the Milky Way}",
      journal = {\aap},
         year = 2023,
        month = jun,
       volume = {674},
          eid = {A38},
        pages = {A38},
          doi = {10.1051/0004-6361/202243511},
archivePrefix = {arXiv},
       eprint = {2206.05534},
 primaryClass = {astro-ph.GA},
       adsurl = {https://ui.adsabs.harvard.edu/abs/2023A&A...674A..38G}
}

@ARTICLE{gravitycollaboration2018,
       author = {{GRAVITY Collaboration} and {Abuter}, R. and {Amorim}, A. and {Anugu}, N. and {Baub{\"o}ck}, M. and {Benisty}, M. and {Berger}, J.~P. and {Blind}, N. and {Bonnet}, H. and {Brandner}, W. and {Buron}, A. and {Collin}, C. and {Chapron}, F. and {Cl{\'e}net}, Y. and {Coud{\'e} Du Foresto}, V. and {de Zeeuw}, P.~T. and {Deen}, C. and {Delplancke-Str{\"o}bele}, F. and {Dembet}, R. and {Dexter}, J. and {Duvert}, G. and {Eckart}, A. and {Eisenhauer}, F. and {Finger}, G. and {F{\"o}rster Schreiber}, N.~M. and {F{\'e}dou}, P. and {Garcia}, P. and {Garcia Lopez}, R. and {Gao}, F. and {Gendron}, E. and {Genzel}, R. and {Gillessen}, S. and {Gordo}, P. and {Habibi}, M. and {Haubois}, X. and {Haug}, M. and {Hau{\ss}mann}, F. and {Henning}, Th. and {Hippler}, S. and {Horrobin}, M. and {Hubert}, Z. and {Hubin}, N. and {Jimenez Rosales}, A. and {Jochum}, L. and {Jocou}, K. and {Kaufer}, A. and {Kellner}, S. and {Kendrew}, S. and {Kervella}, P. and {Kok}, Y. and {Kulas}, M. and {Lacour}, S. and {Lapeyr{\`e}re}, V. and {Lazareff}, B. and {Le Bouquin}, J. -B. and {L{\'e}na}, P. and {Lippa}, M. and {Lenzen}, R. and {M{\'e}rand}, A. and {M{\"u}ler}, E. and {Neumann}, U. and {Ott}, T. and {Palanca}, L. and {Paumard}, T. and {Pasquini}, L. and {Perraut}, K. and {Perrin}, G. and {Pfuhl}, O. and {Plewa}, P.~M. and {Rabien}, S. and {Ram{\'\i}rez}, A. and {Ramos}, J. and {Rau}, C. and {Rodr{\'\i}guez-Coira}, G. and {Rohloff}, R. -R. and {Rousset}, G. and {Sanchez-Bermudez}, J. and {Scheithauer}, S. and {Sch{\"o}ller}, M. and {Schuler}, N. and {Spyromilio}, J. and {Straub}, O. and {Straubmeier}, C. and {Sturm}, E. and {Tacconi}, L.~J. and {Tristram}, K.~R.~W. and {Vincent}, F. and {von Fellenberg}, S. and {Wank}, I. and {Waisberg}, I. and {Widmann}, F. and {Wieprecht}, E. and {Wiest}, M. and {Wiezorrek}, E. and {Woillez}, J. and {Yazici}, S. and {Ziegler}, D. and {Zins}, G.},
        title = "{Detection of the gravitational redshift in the orbit of the star S2 near the Galactic centre massive black hole}",
      journal = {\aap},
         year = 2018,
        month = jul,
       volume = {615},
          eid = {L15},
        pages = {L15},
          doi = {10.1051/0004-6361/201833718},
archivePrefix = {arXiv},
       eprint = {1807.09409},
 primaryClass = {astro-ph.GA},
       adsurl = {https://ui.adsabs.harvard.edu/abs/2018A&A...615L..15G}
}

@ARTICLE{bovy2015,
       author = {{Bovy}, Jo},
        title = "{galpy: A python Library for Galactic Dynamics}",
      journal = {\apjs},
         year = 2015,
        month = feb,
       volume = {216},
       number = {2},
          eid = {29},
        pages = {29},
          doi = {10.1088/0067-0049/216/2/29},
archivePrefix = {arXiv},
       eprint = {1412.3451},
 primaryClass = {astro-ph.GA},
       adsurl = {https://ui.adsabs.harvard.edu/abs/2015ApJS..216...29B}
}

@ARTICLE{antoja2020,
       author = {{Antoja}, T. and {Ramos}, P. and {Mateu}, C. and {Helmi}, A. and {Anders}, F. and {Jordi}, C. and {Carballo-Bello}, J.~A.},
        title = "{An all-sky proper-motion map of the Sagittarius stream using Gaia DR2}",
      journal = {\aap},
         year = 2020,
        month = mar,
       volume = {635},
          eid = {L3},
        pages = {L3},
          doi = {10.1051/0004-6361/201937145},
archivePrefix = {arXiv},
       eprint = {2001.10012},
 primaryClass = {astro-ph.GA},
       adsurl = {https://ui.adsabs.harvard.edu/abs/2020A&A...635L...3A}
}

@ARTICLE{feuillet2021,
       author = {{Feuillet}, Diane K. and {Sahlholdt}, Christian L. and {Feltzing}, Sofia and {Casagrande}, Luca},
        title = "{Selecting accreted populations: metallicity, elemental abundances, and ages of the Gaia-Sausage-Enceladus and Sequoia populations}",
      journal = {MNRAS},
         year = 2021,
        month = nov,
       volume = {508},
       number = {1},
        pages = {1489-1508},
          doi = {10.1093/mnras/stab2614},
archivePrefix = {arXiv},
       eprint = {2105.12141},
 primaryClass = {astro-ph.GA},
       adsurl = {https://ui.adsabs.harvard.edu/abs/2021MNRAS.508.1489F}
}

@ARTICLE{dorazi2025,
       author = {{D'Orazi}, V. and {Braga}, V. and {Bono}, G. and {Fabrizio}, M. and {Fiorentino}, G. and {Storm}, N. and {Pietrinferni}, A. and {Sneden}, C. and {S{\'a}nchez-Benavente}, M. and {Monelli}, M. and {Sestito}, F. and {J{\"o}nsson}, H. and {Buder}, S. and {Bobrick}, A. and {Iorio}, G. and {Matsunaga}, N. and {Marconi}, M. and {Marengo}, M. and {Mart{\'\i}nez-V{\'a}zquez}, C.~E. and {Mullen}, J. and {Takayama}, M. and {Testa}, V. and {Cusano}, F. and {Crestani}, J.},
        title = "{The elderly among the oldest: new evidence for extremely metal-poor RR Lyrae stars}",
      journal = {\aap},
         year = 2025,
        month = feb,
       volume = {694},
          eid = {A158},
        pages = {A158},
          doi = {10.1051/0004-6361/202453202},
archivePrefix = {arXiv},
       eprint = {2501.05807},
 primaryClass = {astro-ph.SR},
       adsurl = {https://ui.adsabs.harvard.edu/abs/2025A&A...694A.158D}
}

@ARTICLE{dorazi2025b,
       author = {{D'Orazi}, Valentina and {Iorio}, Giuliano and {Cseh}, Borb{\'a}la and {Sneden}, Chris and {Abdollahi}, Hedieh and {Moln{\'a}r}, L{\'a}szl{\'o} and {Bobrick}, Alexey and {Bono}, Giuseppe and {Braga}, Vittorio F. and {Karakas}, Amanda and {Lugaro}, Maria and {Campbell}, Simon W. and {Fabrizio}, Michele and {Fiorentino}, Giuliana and {Roederer}, Ian U. and {Storm}, Nicholas and {Tantalo}, Maria and {Crestani}, Juliana},
        title = "{Rare Find: Discovery and chemo-dynamical properties of two s-process enhanced RR Lyrae stars}",
      journal = {arXiv e-prints},
         year = 2025,
        month = oct,
          eid = {arXiv:2510.15723},
        pages = {arXiv:2510.15723},
          doi = {10.48550/arXiv.2510.15723},
archivePrefix = {arXiv},
       eprint = {2510.15723},
 primaryClass = {astro-ph.SR},
       adsurl = {https://ui.adsabs.harvard.edu/abs/2025arXiv251015723D}
}

@ARTICLE{escala2020a,
       author = {{Escala}, Ivanna and {Gilbert}, Karoline M. and {Kirby}, Evan N. and {Wojno}, Jennifer and {Cunningham}, Emily C. and {Guhathakurta}, Puragra},
        title = "{Elemental Abundances in M31: A Comparative Analysis of Alpha and Iron Element Abundances in the the Outer Disk, Giant Stellar Stream, and Inner Halo of M31}",
      journal = {\apj},
         year = 2020,
        month = feb,
       volume = {889},
       number = {2},
          eid = {177},
        pages = {177},
          doi = {10.3847/1538-4357/ab6659},
archivePrefix = {arXiv},
       eprint = {1909.00006},
 primaryClass = {astro-ph.GA},
       adsurl = {https://ui.adsabs.harvard.edu/abs/2020ApJ...889..177E}
}

@ARTICLE{escala2020b,
       author = {{Escala}, Ivanna and {Kirby}, Evan N. and {Gilbert}, Karoline M. and {Wojno}, Jennifer and {Cunningham}, Emily C. and {Guhathakurta}, Puragra},
        title = "{Elemental Abundances in M31: Properties of the Inner Stellar Halo}",
      journal = {\apj},
         year = 2020,
        month = oct,
       volume = {902},
       number = {1},
          eid = {51},
        pages = {51},
          doi = {10.3847/1538-4357/abb474},
archivePrefix = {arXiv},
       eprint = {2009.00529},
 primaryClass = {astro-ph.GA},
       adsurl = {https://ui.adsabs.harvard.edu/abs/2020ApJ...902...51E}
}

@ARTICLE{Wojno2023,
       author = {{Wojno}, J. Leigh and {Gilbert}, Karoline M. and {Kirby}, Evan N. and {Escala}, Ivanna and {Guhathakurta}, Puragra and {Beaton}, Rachael L. and {Kalirai}, Jason and {Chiba}, Masashi and {Majewski}, Steven R.},
        title = "{Elemental Abundances in M31: Individual and Coadded Spectroscopic [Fe/H] and [{\ensuremath{\alpha}}/Fe] throughout the M31 Halo with SPLASH}",
      journal = {\apj},
         year = 2023,
        month = jul,
       volume = {951},
       number = {1},
          eid = {12},
        pages = {12},
          doi = {10.3847/1538-4357/acd5d3},
archivePrefix = {arXiv},
       eprint = {2211.15288},
 primaryClass = {astro-ph.GA},
       adsurl = {https://ui.adsabs.harvard.edu/abs/2023ApJ...951...12W}
}

@ARTICLE{escala2022,
       author = {{Escala}, Ivanna and {Gilbert}, Karoline M. and {Fardal}, Mark and {Guhathakurta}, Puragra and {Sanderson}, Robyn E. and {Kalirai}, Jason S. and {Mobasher}, Bahram},
        title = "{Kinematics and Metallicity of Red Giant Branch Stars in the Northeast Shelf of M31}",
      journal = {\aj},
         year = 2022,
        month = jul,
       volume = {164},
       number = {1},
          eid = {20},
        pages = {20},
          doi = {10.3847/1538-3881/ac7146},
archivePrefix = {arXiv},
       eprint = {2203.16675},
 primaryClass = {astro-ph.GA},
       adsurl = {https://ui.adsabs.harvard.edu/abs/2022AJ....164...20E}
}

@ARTICLE{escala2021,
       author = {{Escala}, Ivanna and {Gilbert}, Karoline M. and {Wojno}, Jennifer and {Kirby}, Evan N. and {Guhathakurta}, Puragra},
        title = "{Elemental Abundances in M31: Gradients in the Giant Stellar Stream}",
      journal = {\aj},
         year = 2021,
        month = aug,
       volume = {162},
       number = {2},
          eid = {45},
        pages = {45},
          doi = {10.3847/1538-3881/abfec4},
archivePrefix = {arXiv},
       eprint = {2105.02339},
 primaryClass = {astro-ph.GA},
       adsurl = {https://ui.adsabs.harvard.edu/abs/2021AJ....162...45E}
}

@ARTICLE{gilbert2014,
       author = {{Gilbert}, Karoline M. and {Kalirai}, Jason S. and {Guhathakurta}, Puragra and {Beaton}, Rachael L. and {Geha}, Marla C. and {Kirby}, Evan N. and {Majewski}, Steven R. and {Patterson}, Richard J. and {Tollerud}, Erik J. and {Bullock}, James S. and {Tanaka}, Mikito and {Chiba}, Masashi},
        title = "{Global Properties of M31's Stellar Halo from the SPLASH Survey. II. Metallicity Profile}",
      journal = {\apj},
         year = 2014,
        month = dec,
       volume = {796},
       number = {2},
          eid = {76},
        pages = {76},
          doi = {10.1088/0004-637X/796/2/76},
archivePrefix = {arXiv},
       eprint = {1409.3843},
 primaryClass = {astro-ph.GA},
       adsurl = {https://ui.adsabs.harvard.edu/abs/2014ApJ...796...76G}
}

@ARTICLE{fardal2012,
       author = {{Fardal}, Mark A. and {Guhathakurta}, Puragra and {Gilbert}, Karoline M. and {Tollerud}, Erik J. and {Kalirai}, Jason S. and {Tanaka}, Mikito and {Beaton}, Rachael and {Chiba}, Masashi and {Komiyama}, Yutaka and {Iye}, Masanori},
        title = "{A spectroscopic survey of Andromeda's Western Shelf}",
      journal = {\mnras},
         year = 2012,
        month = jul,
       volume = {423},
       number = {4},
        pages = {3134-3147},
          doi = {10.1111/j.1365-2966.2012.21094.x},
archivePrefix = {arXiv},
       eprint = {1206.2619},
 primaryClass = {astro-ph.CO},
       adsurl = {https://ui.adsabs.harvard.edu/abs/2012MNRAS.423.3134F}
}

@ARTICLE{gilbert2009,
       author = {{Gilbert}, Karoline M. and {Guhathakurta}, Puragra and {Kollipara}, Priya and {Beaton}, Rachael L. and {Geha}, Marla C. and {Kalirai}, Jason S. and {Kirby}, Evan N. and {Majewski}, Steven R. and {Patterson}, Richard J.},
        title = "{The Splash Survey: A Spectroscopic Portrait of Andromeda's Giant Southern Stream}",
      journal = {\apj},
         year = 2009,
        month = nov,
       volume = {705},
       number = {2},
        pages = {1275-1297},
          doi = {10.1088/0004-637X/705/2/1275},
archivePrefix = {arXiv},
       eprint = {0909.4540},
 primaryClass = {astro-ph.CO},
       adsurl = {https://ui.adsabs.harvard.edu/abs/2009ApJ...705.1275G}
}

@ARTICLE{gilbert2007,
       author = {{Gilbert}, Karoline M. and {Fardal}, Mark and {Kalirai}, Jasonjot S. and {Guhathakurta}, Puragra and {Geha}, Marla C. and {Isler}, Jedidah and {Majewski}, Steven R. and {Ostheimer}, James C. and {Patterson}, Richard J. and {Reitzel}, David B. and {Kirby}, Evan and {Cooper}, Michael C.},
        title = "{Stellar Kinematics in the Complicated Inner Spheroid of M31: Discovery of Substructure along the Southeastern Minor Axis and Its Relationship to the Giant Southern Stream}",
      journal = {\apj},
         year = 2007,
        month = oct,
       volume = {668},
       number = {1},
        pages = {245-267},
          doi = {10.1086/521094},
archivePrefix = {arXiv},
       eprint = {astro-ph/0703029},
 primaryClass = {astro-ph},
       adsurl = {https://ui.adsabs.harvard.edu/abs/2007ApJ...668..245G}
}

@ARTICLE{gilbert2019,
       author = {{Gilbert}, Karoline M. and {Kirby}, Evan N. and {Escala}, Ivanna and {Wojno}, Jennifer and {Kalirai}, Jason S. and {Guhathakurta}, Puragra},
        title = "{Elemental Abundances in M31: First Alpha and Iron Abundance Measurements in M31's Giant Stellar Stream}",
      journal = {\apj},
         year = 2019,
        month = oct,
       volume = {883},
       number = {2},
          eid = {128},
        pages = {128},
          doi = {10.3847/1538-4357/ab3807},
archivePrefix = {arXiv},
       eprint = {1908.04429},
 primaryClass = {astro-ph.GA},
       adsurl = {https://ui.adsabs.harvard.edu/abs/2019ApJ...883..128G}
}

@ARTICLE{gregersen2015,
       author = {{Gregersen}, Dylan and {Seth}, Anil C. and {Williams}, Benjamin F. and {Lang}, Dustin and {Dalcanton}, Julianne J. and {Girardi}, Le{\'o} and {Skillman}, Evan D. and {Bell}, Eric and {Dolphin}, Andrew E. and {Fouesneau}, Morgan and {Guhathakurta}, Puragra and {Hamren}, Katherine M. and {Johnson}, L.~C. and {Kalirai}, Jason and {Lewis}, Alexia R. and {Monachesi}, Antonela and {Olsen}, Knut},
        title = "{Panchromatic Hubble Andromeda Treasury. XII. Mapping Stellar Metallicity Distributions in M31}",
      journal = {\aj},
         year = 2015,
        month = dec,
       volume = {150},
       number = {6},
          eid = {189},
        pages = {189},
          doi = {10.1088/0004-6256/150/6/189},
archivePrefix = {arXiv},
       eprint = {1511.00006},
 primaryClass = {astro-ph.GA},
       adsurl = {https://ui.adsabs.harvard.edu/abs/2015AJ....150..189G}
}

@ARTICLE{dorazi2024,
       author = {{D'Orazi}, Valentina and {Storm}, Nicholas and {Casey}, Andrew R. and {Braga}, Vittorio F. and {Zocchi}, Alice and {Bono}, Giuseppe and {Fabrizio}, Michele and {Sneden}, Christopher and {Massari}, Davide and {Giribaldi}, Riano E. and {Bergemann}, Maria and {Campbell}, Simon W. and {Casagrande}, Luca and {de Grijs}, Richard and {De Silva}, Gayandhi and {Lugaro}, Maria and {Zucker}, Daniel B. and {Bragaglia}, Angela and {Feuillet}, Diane and {Fiorentino}, Giuliana and {Chaboyer}, Brian and {Dall'Ora}, Massimo and {Marengo}, Massimo and {Mart{\'\i}nez-V{\'a}zquez}, Clara E. and {Matsunaga}, Noriyuki and {Monelli}, Matteo and {Mullen}, Joseph P. and {Nataf}, David and {Tantalo}, Maria and {Thevenin}, Frederic and {Vitello}, Fabio R. and {Kudritzki}, Rolf-Peter and {Bland-Hawthorn}, Joss and {Buder}, Sven and {Freeman}, Ken and {Kos}, Janez and {Lewis}, Geraint F. and {Lind}, Karin and {Martell}, Sarah and {Sharma}, Sanjib and {Stello}, Dennis and {Zwitter}, Toma{\v{z}}},
        title = "{The GALAH survey: tracing the Milky Way's formation and evolution through RR Lyrae stars}",
      journal = {\mnras},
         year = 2024,
        month = jun,
       volume = {531},
       number = {1},
        pages = {137-162},
          doi = {10.1093/mnras/stae1149},
archivePrefix = {arXiv},
       eprint = {2405.04580},
 primaryClass = {astro-ph.GA},
       adsurl = {https://ui.adsabs.harvard.edu/abs/2024MNRAS.531..137D}
}

@ARTICLE{clementini19,
       author = {{Clementini}, G. and {Ripepi}, V. and {Molinaro}, R. and {Garofalo}, A. and {Muraveva}, T. and {Rimoldini}, L. and {Guy}, L.~P. and {Jevardat de Fombelle}, G. and {Nienartowicz}, K. and {Marchal}, O. and {Audard}, M. and {Holl}, B. and {Leccia}, S. and {Marconi}, M. and {Musella}, I. and {Mowlavi}, N. and {Lecoeur-Taibi}, I. and {Eyer}, L. and {De Ridder}, J. and {Regibo}, S. and {Sarro}, L.~M. and {Szabados}, L. and {Evans}, D.~W. and {Riello}, M.},
        title = "{Gaia Data Release 2. Specific characterisation and validation of all-sky Cepheids and RR Lyrae stars}",
      journal = {\aap},
         year = 2019,
        month = feb,
       volume = {622},
          eid = {A60},
        pages = {A60},
          doi = {10.1051/0004-6361/201833374},
archivePrefix = {arXiv},
       eprint = {1805.02079},
 primaryClass = {astro-ph.SR},
       adsurl = {https://ui.adsabs.harvard.edu/abs/2019A&A...622A..60C}
}

@ARTICLE{bensby2014,
       author = {{Bensby}, T. and {Feltzing}, S. and {Oey}, M.~S.},
        title = "{Exploring the Milky Way stellar disk. A detailed elemental abundance study of 714 F and G dwarf stars in the solar neighbourhood}",
      journal = {A\&A},
         year = 2014,
        month = feb,
       volume = {562},
          eid = {A71},
        pages = {A71},
          doi = {10.1051/0004-6361/201322631},
archivePrefix = {arXiv},
       eprint = {1309.2631},
 primaryClass = {astro-ph.GA},
       adsurl = {https://ui.adsabs.harvard.edu/abs/2014A&A...562A..71B}
}

@ARTICLE{sakari2015,
       author = {{Sakari}, Charli M. and {Venn}, Kim A. and {Mackey}, Dougal and {Shetrone}, Matthew D. and {Dotter}, Aaron and {Ferguson}, Annette M.~N. and {Huxor}, Avon},
        title = "{Integrated light chemical tagging analyses of seven M31 outer halo globular clusters from the Pan-Andromeda Archaeological Survey}",
      journal = {\mnras},
         year = 2015,
        month = apr,
       volume = {448},
       number = {2},
        pages = {1314-1334},
          doi = {10.1093/mnras/stv020},
archivePrefix = {arXiv},
       eprint = {1501.04626},
 primaryClass = {astro-ph.GA},
       adsurl = {https://ui.adsabs.harvard.edu/abs/2015MNRAS.448.1314S}
}

@ARTICLE{sakari2021,
       author = {{Sakari}, Charli M. and {Shetrone}, Matthew D. and {McWilliam}, Andrew and {Wallerstein}, George},
        title = "{The massive M31 cluster G1: detailed chemical abundances from integrated light spectroscopy}",
      journal = {\mnras},
         year = 2021,
        month = apr,
       volume = {502},
       number = {4},
        pages = {5745-5761},
          doi = {10.1093/mnras/stab141},
archivePrefix = {arXiv},
       eprint = {2012.03971},
 primaryClass = {astro-ph.GA},
       adsurl = {https://ui.adsabs.harvard.edu/abs/2021MNRAS.502.5745S}
}

@ARTICLE{sakari2022,
       author = {{Sakari}, Charli M. and {Wallerstein}, George},
        title = "{Metallicities of outer halo M31 globular clusters from integrated light calcium-II triplet spectroscopy}",
      journal = {\mnras},
         year = 2022,
        month = jun,
       volume = {512},
       number = {4},
        pages = {4819-4834},
          doi = {10.1093/mnras/stac752},
archivePrefix = {arXiv},
       eprint = {2203.08840},
 primaryClass = {astro-ph.GA},
       adsurl = {https://ui.adsabs.harvard.edu/abs/2022MNRAS.512.4819S}
}

@ARTICLE{mackey2019a,
       author = {{Mackey}, A.~D. and {Ferguson}, A.~M.~N. and {Huxor}, A.~P. and {Veljanoski}, J. and {Lewis}, G.~F. and {McConnachie}, A.~W. and {Martin}, N.~F. and {Ibata}, R.~A. and {Irwin}, M.~J. and {C{\^o}t{\'e}}, P. and {Collins}, M.~L.~M. and {Tanvir}, N.~R. and {Bate}, N.~F.},
        title = "{The outer halo globular cluster system of M31 - III. Relationship to the stellar halo}",
      journal = {\mnras},
         year = 2019,
        month = apr,
       volume = {484},
       number = {2},
        pages = {1756-1789},
          doi = {10.1093/mnras/stz072},
archivePrefix = {arXiv},
       eprint = {1810.10719},
 primaryClass = {astro-ph.GA},
       adsurl = {https://ui.adsabs.harvard.edu/abs/2019MNRAS.484.1756M}
}

@ARTICLE{bobrick2024,
       author = {{Bobrick}, Alexey and {Iorio}, Giuliano and {Belokurov}, Vasily and {Vos}, Joris and {Vu{\v{c}}kovi{\'c}}, Maja and {Giacobbo}, Nicola},
        title = "{RR Lyrae from binary evolution: abundant, young, and metal-rich}",
      journal = {\mnras},
         year = 2024,
        month = feb,
       volume = {527},
       number = {4},
        pages = {12196-12218},
          doi = {10.1093/mnras/stad3996},
archivePrefix = {arXiv},
       eprint = {2208.04332},
 primaryClass = {astro-ph.SR},
       adsurl = {https://ui.adsabs.harvard.edu/abs/2024MNRAS.52712196B}
}

@ARTICLE{ztf2019a,
       author = {{Bellm}, Eric C. and {Kulkarni}, Shrinivas R. and {Graham}, Matthew J. and {Dekany}, Richard and {Smith}, Roger M. and {Riddle}, Reed and {Masci}, Frank J. and {Helou}, George and {Prince}, Thomas A. and {Adams}, Scott M. and {Barbarino}, C. and {Barlow}, Tom and {Bauer}, James and {Beck}, Ron and {Belicki}, Justin and {Biswas}, Rahul and {Blagorodnova}, Nadejda and {Bodewits}, Dennis and {Bolin}, Bryce and {Brinnel}, Valery and {Brooke}, Tim and {Bue}, Brian and {Bulla}, Mattia and {Burruss}, Rick and {Cenko}, S. Bradley and {Chang}, Chan-Kao and {Connolly}, Andrew and {Coughlin}, Michael and {Cromer}, John and {Cunningham}, Virginia and {De}, Kishalay and {Delacroix}, Alex and {Desai}, Vandana and {Duev}, Dmitry A. and {Eadie}, Gwendolyn and {Farnham}, Tony L. and {Feeney}, Michael and {Feindt}, Ulrich and {Flynn}, David and {Franckowiak}, Anna and {Frederick}, S. and {Fremling}, C. and {Gal-Yam}, Avishay and {Gezari}, Suvi and {Giomi}, Matteo and {Goldstein}, Daniel A. and {Golkhou}, V. Zach and {Goobar}, Ariel and {Groom}, Steven and {Hacopians}, Eugean and {Hale}, David and {Henning}, John and {Ho}, Anna Y.~Q. and {Hover}, David and {Howell}, Justin and {Hung}, Tiara and {Huppenkothen}, Daniela and {Imel}, David and {Ip}, Wing-Huen and {Ivezi{\'c}}, {\v{Z}}eljko and {Jackson}, Edward and {Jones}, Lynne and {Juric}, Mario and {Kasliwal}, Mansi M. and {Kaspi}, S. and {Kaye}, Stephen and {Kelley}, Michael S.~P. and {Kowalski}, Marek and {Kramer}, Emily and {Kupfer}, Thomas and {Landry}, Walter and {Laher}, Russ R. and {Lee}, Chien-De and {Lin}, Hsing Wen and {Lin}, Zhong-Yi and {Lunnan}, Ragnhild and {Giomi}, Matteo and {Mahabal}, Ashish and {Mao}, Peter and {Miller}, Adam A. and {Monkewitz}, Serge and {Murphy}, Patrick and {Ngeow}, Chow-Choong and {Nordin}, Jakob and {Nugent}, Peter and {Ofek}, Eran and {Patterson}, Maria T. and {Penprase}, Bryan and {Porter}, Michael and {Rauch}, Ludwig and {Rebbapragada}, Umaa and {Reiley}, Dan and {Rigault}, Mickael and {Rodriguez}, Hector and {van Roestel}, Jan and {Rusholme}, Ben and {van Santen}, Jakob and {Schulze}, S. and {Shupe}, David L. and {Singer}, Leo P. and {Soumagnac}, Maayane T. and {Stein}, Robert and {Surace}, Jason and {Sollerman}, Jesper and {Szkody}, Paula and {Taddia}, F. and {Terek}, Scott and {Van Sistine}, Angela and {van Velzen}, Sjoert and {Vestrand}, W. Thomas and {Walters}, Richard and {Ward}, Charlotte and {Ye}, Quan-Zhi and {Yu}, Po-Chieh and {Yan}, Lin and {Zolkower}, Jeffry},
        title = "{The Zwicky Transient Facility: System Overview, Performance, and First Results}",
      journal = {\pasp},
         year = 2019,
        month = jan,
       volume = {131},
       number = {995},
        pages = {018002},
          doi = {10.1088/1538-3873/aaecbe},
archivePrefix = {arXiv},
       eprint = {1902.01932},
 primaryClass = {astro-ph.IM},
       adsurl = {https://ui.adsabs.harvard.edu/abs/2019PASP..131a8002B}
}

@ARTICLE{ztf2019b,
       author = {{Bellm}, Eric C. and {Kulkarni}, Shrinivas R. and {Barlow}, Tom and {Feindt}, Ulrich and {Graham}, Matthew J. and {Goobar}, Ariel and {Kupfer}, Thomas and {Ngeow}, Chow-Choong and {Nugent}, Peter and {Ofek}, Eran and {Prince}, Thomas A. and {Riddle}, Reed and {Walters}, Richard and {Ye}, Quan-Zhi},
        title = "{The Zwicky Transient Facility: Surveys and Scheduler}",
      journal = {\pasp},
         year = 2019,
        month = jun,
       volume = {131},
       number = {1000},
        pages = {068003},
          doi = {10.1088/1538-3873/ab0c2a},
archivePrefix = {arXiv},
       eprint = {1905.02209},
 primaryClass = {astro-ph.IM},
       adsurl = {https://ui.adsabs.harvard.edu/abs/2019PASP..131f8003B}
}

@ARTICLE{Horta2023,
       author = {{Horta}, Danny and {Schiavon}, Ricardo P. and {Mackereth}, J. Ted and {Weinberg}, David H. and {Hasselquist}, Sten and {Feuillet}, Diane and {O'Connell}, Robert W. and {Anguiano}, Borja and {Allende-Prieto}, Carlos and {Beaton}, Rachael L. and {Bizyaev}, Dmitry and {Cunha}, Katia and {Geisler}, Doug and {Garc{\'\i}a-Hern{\'a}ndez}, D.~A. and {Holtzman}, Jon and {J{\"o}nsson}, Henrik and {Lane}, Richard R. and {Majewski}, Steve R. and {M{\'e}sz{\'a}ros}, Szabolcs and {Minniti}, Dante and {Nitschelm}, Christian and {Shetrone}, Matthew and {Smith}, Verne V. and {Zasowski}, Gail},
        title = "{The chemical characterization of halo substructure in the Milky Way based on APOGEE}",
      journal = {\mnras},
         year = 2023,
        month = apr,
       volume = {520},
       number = {4},
        pages = {5671-5711},
          doi = {10.1093/mnras/stac3179},
archivePrefix = {arXiv},
       eprint = {2204.04233},
 primaryClass = {astro-ph.GA},
       adsurl = {https://ui.adsabs.harvard.edu/abs/2023MNRAS.520.5671H}
}

@ARTICLE{callingham2022,
       author = {{Callingham}, Thomas M. and {Cautun}, Marius and {Deason}, Alis J. and {Frenk}, Carlos S. and {Grand}, Robert J.~J. and {Marinacci}, Federico},
        title = "{The chemo-dynamical groups of Galactic globular clusters}",
      journal = {\mnras},
         year = 2022,
        month = jul,
       volume = {513},
       number = {3},
        pages = {4107-4129},
          doi = {10.1093/mnras/stac1145},
archivePrefix = {arXiv},
       eprint = {2202.00591},
 primaryClass = {astro-ph.GA},
       adsurl = {https://ui.adsabs.harvard.edu/abs/2022MNRAS.513.4107C}
}

@ARTICLE{taam1976,
       author = {{Taam}, R.~E. and {Kraft}, R.~P. and {Suntzeff}, N.},
        title = "{The origin and evolution of RR Lyrae stars of high metal abundance.}",
      journal = {\apj},
         year = 1976,
        month = jul,
       volume = {207},
        pages = {201-208},
          doi = {10.1086/154485},
       adsurl = {https://ui.adsabs.harvard.edu/abs/1976ApJ...207..201T}
}

@ARTICLE{micali2013,
       author = {{Micali}, A. and {Matteucci}, F. and {Romano}, D.},
        title = "{The chemical evolution of the Milky Way: the Three Infall Model}",
      journal = {\mnras},
         year = 2013,
        month = dec,
       volume = {436},
       number = {2},
        pages = {1648-1658},
          doi = {10.1093/mnras/stt1681},
archivePrefix = {arXiv},
       eprint = {1309.1283},
 primaryClass = {astro-ph.GA},
       adsurl = {https://ui.adsabs.harvard.edu/abs/2013MNRAS.436.1648M}
}

@ARTICLE{spitoni2019,
       author = {{Spitoni}, E. and {Cescutti}, G. and {Minchev}, I. and {Matteucci}, F. and {Silva Aguirre}, V. and {Martig}, M. and {Bono}, G. and {Chiappini}, C.},
        title = "{2D chemical evolution model: The impact of Galactic disc asymmetries on azimuthal chemical abundance variations}",
      journal = {\aap},
         year = 2019,
        month = aug,
       volume = {628},
          eid = {A38},
        pages = {A38},
          doi = {10.1051/0004-6361/201834665},
archivePrefix = {arXiv},
       eprint = {1811.11196},
 primaryClass = {astro-ph.GA},
       adsurl = {https://ui.adsabs.harvard.edu/abs/2019A&A...628A..38S}
}

@ARTICLE{borbolato2025,
       author = {{Borbolato}, Lais and {Rossi}, Silvia and {Perottoni}, H{\'e}lio D. and {Limberg}, Guilherme and {Amarante}, Jo{\~a}o A.~S. and {Queiroz}, Anna B.~A. and {Chiappini}, Cristina and {Anders}, Friedrich and {Santucci}, Rafael M. and {Barbosa}, Fabr{\'\i}cia O. and {Nogueira-Santos}, Jo{\~a}o V.},
        title = "{Early Co-formation of the Milky Way's Thin and Thick Disks at Redshift z > 2}",
      journal = {arXiv e-prints},
         year = 2025,
        month = mar,
          eid = {arXiv:2504.00135},
        pages = {arXiv:2504.00135},
          doi = {10.48550/arXiv.2504.00135},
archivePrefix = {arXiv},
       eprint = {2504.00135},
 primaryClass = {astro-ph.GA},
       adsurl = {https://ui.adsabs.harvard.edu/abs/2025arXiv250400135B}
}

@ARTICLE{kinman14,
       author = {{Kinman}, T.~D. and {Brown}, Warren R.},
        title = "{The Identification of RR Lyrae and {\ensuremath{\delta}} Scutti Stars from Variable Galaxy Evolution Explorer Ultraviolet Sources}",
      journal = {\aj},
         year = 2014,
        month = dec,
       volume = {148},
       number = {6},
          eid = {121},
        pages = {121},
          doi = {10.1088/0004-6256/148/6/121},
archivePrefix = {arXiv},
       eprint = {1408.0808},
 primaryClass = {astro-ph.SR},
       adsurl = {https://ui.adsabs.harvard.edu/abs/2014AJ....148..121K}
}

@INPROCEEDINGS{vivaszinn06,
       author = {{Vivas}, A.~K. and {Zinn}, R. and {Gallart}, C.},
        title = "{Structure of the Halo of the Milky Way}",
    booktitle = {Revista Mexicana de Astronomia y Astrofisica Conference Series},
         year = 2006,
       editor = {{Abad}, Carlos and {Bongiovanni}, Angel and {Guillen}, Yaneth},
       series = {Revista Mexicana de Astronomia y Astrofisica Conference Series},
       volume = {25},
        month = jan,
        pages = {37-40},
       adsurl = {https://ui.adsabs.harvard.edu/abs/2006RMxAC..25...37V}
}

@ARTICLE{bhardwaj2022,
       author = {{Bhardwaj}, Anupam},
        title = "{RR Lyrae and Type II Cepheid Variables in Globular Clusters: Optical and Infrared Properties}",
      journal = {Universe},
         year = 2022,
        month = feb,
       volume = {8},
       number = {2},
        pages = {122},
          doi = {10.3390/universe8020122},
archivePrefix = {arXiv},
       eprint = {2202.06982},
 primaryClass = {astro-ph.SR},
       adsurl = {https://ui.adsabs.harvard.edu/abs/2022Univ....8..122B}
}

@ARTICLE{alfaro93,
       author = {{Alfaro}, Emilio J. and {Cabrera-Cano}, Jesus and {Delgado}, Antonio J.},
        title = "{Metallicity Maps of the Globular Cluster System in the Galaxy}",
      journal = {\apjl},
         year = 1993,
        month = jan,
       volume = {402},
        pages = {L53},
          doi = {10.1086/186698},
       adsurl = {https://ui.adsabs.harvard.edu/abs/1993ApJ...402L..53A}
}

@ARTICLE{iorio2019,
       author = {{Iorio}, Giuliano and {Belokurov}, Vasily},
        title = "{The shape of the Galactic halo with Gaia DR2 RR Lyrae. Anatomy of an ancient major merger}",
      journal = {\mnras},
         year = 2019,
        month = jan,
       volume = {482},
       number = {3},
        pages = {3868-3879},
          doi = {10.1093/mnras/sty2806},
archivePrefix = {arXiv},
       eprint = {1808.04370},
 primaryClass = {astro-ph.GA},
       adsurl = {https://ui.adsabs.harvard.edu/abs/2019MNRAS.482.3868I}
}

@ARTICLE{zhu2018,
       author = {{Zhu}, Ling and {van den Bosch}, Remco and {van de Ven}, Glenn and {Lyubenova}, Mariya and {Falc{\'o}n-Barroso}, Jes{\'u}s and {Meidt}, Sharon E. and {Martig}, Marie and {Shen}, Juntai and {Li}, Zhao-Yu and {Yildirim}, Akin and {Walcher}, C. Jakob and {Sanchez}, Sebastian F.},
        title = "{Orbital decomposition of CALIFA spiral galaxies}",
      journal = {\mnras},
         year = 2018,
        month = jan,
       volume = {473},
       number = {3},
        pages = {3000-3018},
          doi = {10.1093/mnras/stx2409},
archivePrefix = {arXiv},
       eprint = {1709.06649},
 primaryClass = {astro-ph.GA},
       adsurl = {https://ui.adsabs.harvard.edu/abs/2018MNRAS.473.3000Z}
}

@ARTICLE{santucci2023,
       author = {{Santucci}, Giulia and {Brough}, Sarah and {van de Sande}, Jesse and {McDermid}, Richard and {Barsanti}, Stefania and {Bland-Hawthorn}, Joss and {Bryant}, Julia J. and {Croom}, Scott M. and {Lagos}, Claudia and {Lawrence}, Jon S. and {Owers}, Matt S. and {van de Ven}, Glenn and {Vaughan}, Sam P. and {Yi}, Sukyoung K.},
        title = "{The SAMI Galaxy Survey: Environmental analysis of the orbital structures of passive galaxies}",
      journal = {\mnras},
         year = 2023,
        month = may,
       volume = {521},
       number = {2},
        pages = {2671-2691},
          doi = {10.1093/mnras/stad713},
archivePrefix = {arXiv},
       eprint = {2303.04161},
 primaryClass = {astro-ph.GA},
       adsurl = {https://ui.adsabs.harvard.edu/abs/2023MNRAS.521.2671S}
}

@ARTICLE{eggen1962,
       author = {{Eggen}, O.~J. and {Lynden-Bell}, D. and {Sandage}, A.~R.},
        title = "{Evidence from the motions of old stars that the Galaxy collapsed.}",
      journal = {\apj},
         year = 1962,
        month = nov,
       volume = {136},
        pages = {748},
          doi = {10.1086/147433},
       adsurl = {https://ui.adsabs.harvard.edu/abs/1962ApJ...136..748E}
}

@ARTICLE{monachesi2019,
       author = {{Monachesi}, Antonela and {G{\'o}mez}, Facundo A. and {Grand}, Robert J.~J. and {Simpson}, Christine M. and {Kauffmann}, Guinevere and {Bustamante}, Sebasti{\'a}n and {Marinacci}, Federico and {Pakmor}, R{\"u}diger and {Springel}, Volker and {Frenk}, Carlos S. and {White}, Simon D.~M. and {Tissera}, Patricia B.},
        title = "{The Auriga stellar haloes: connecting stellar population properties with accretion and merging history}",
      journal = {\mnras},
         year = 2019,
        month = may,
       volume = {485},
       number = {2},
        pages = {2589-2616},
          doi = {10.1093/mnras/stz538},
archivePrefix = {arXiv},
       eprint = {1804.07798},
 primaryClass = {astro-ph.GA},
       adsurl = {https://ui.adsabs.harvard.edu/abs/2019MNRAS.485.2589M}
}

@ARTICLE{kervella2019,
       author = {{Kervella}, Pierre and {Gallenne}, Alexandre and {Evans}, Nancy Remage and {Szabados}, Laszlo and {Arenou}, Fr{\'e}d{\'e}ric and {M{\'e}rand}, Antoine and {Nardetto}, Nicolas and {Gieren}, Wolfgang and {Pietrzynski}, Grzegorz},
        title = "{Multiplicity of Galactic Cepheids and RR Lyrae stars from Gaia DR2. II. Resolved common proper motion pairs}",
      journal = {\aap},
         year = 2019,
        month = mar,
       volume = {623},
          eid = {A117},
        pages = {A117},
          doi = {10.1051/0004-6361/201834211},
archivePrefix = {arXiv},
       eprint = {1908.00545},
 primaryClass = {astro-ph.SR},
       adsurl = {https://ui.adsabs.harvard.edu/abs/2019A&A...623A.117K}
}

@ARTICLE{kervella2019a,
       author = {{Kervella}, Pierre and {Gallenne}, Alexandre and {Remage Evans}, Nancy and {Szabados}, Laszlo and {Arenou}, Fr{\'e}d{\'e}ric and {M{\'e}rand}, Antoine and {Proto}, Yann and {Karczmarek}, Paulina and {Nardetto}, Nicolas and {Gieren}, Wolfgang and {Pietrzynski}, Grzegorz},
        title = "{Multiplicity of Galactic Cepheids and RR Lyrae stars from Gaia DR2. I. Binarity from proper motion anomaly}",
      journal = {\aap},
         year = 2019,
        month = mar,
       volume = {623},
          eid = {A116},
        pages = {A116},
          doi = {10.1051/0004-6361/201834210},
archivePrefix = {arXiv},
       eprint = {1903.03632},
 primaryClass = {astro-ph.SR},
       adsurl = {https://ui.adsabs.harvard.edu/abs/2019A&A...623A.116K}
}

@ARTICLE{stokes2000,
       author = {{Stokes}, Grant H. and {Evans}, Jenifer B. and {Viggh}, Herbert E.~M. and {Shelly}, Frank C. and {Pearce}, Eric C.},
        title = "{Lincoln Near-Earth Asteroid Program (LINEAR)}",
      journal = {\icarus},
         year = 2000,
        month = nov,
       volume = {148},
       number = {1},
        pages = {21-28},
          doi = {10.1006/icar.2000.6493},
       adsurl = {https://ui.adsabs.harvard.edu/abs/2000Icar..148...21S}
}

@ARTICLE{Welsh2005,
       author = {{Welsh}, Barry Y. and {Wheatley}, Jonathan M. and {Heafield}, Kenneth and {Seibert}, Mark and {Browne}, Stanley E. and {Salim}, Samir and {Rich}, R. Michael and {Barlow}, Tom A. and {Bianchi}, Luciana and {Byun}, Yong-Ik and {Donas}, Jose and {Forster}, Karl and {Friedman}, Peter G. and {Heckman}, Timothy M. and {Jelinsky}, Patrick N. and {Lee}, Young-Wook and {Madore}, Barry F. and {Malina}, Roger F. and {Martin}, D. Christopher and {Milliard}, Bruno and {Morrissey}, Patrick and {Neff}, Susan G. and {Schiminovich}, David and {Siegmund}, Oswald H.~W. and {Small}, Todd and {Szalay}, Alex S. and {Wyder}, Ted K.},
        title = "{The GALEX Ultraviolet Variability Catalog}",
      journal = {\aj},
         year = 2005,
        month = aug,
       volume = {130},
       number = {2},
        pages = {825-831},
          doi = {10.1086/431222},
archivePrefix = {arXiv},
       eprint = {astro-ph/0504489},
 primaryClass = {astro-ph},
       adsurl = {https://ui.adsabs.harvard.edu/abs/2005AJ....130..825W}
}

@ARTICLE{sesar2011a,
       author = {{Sesar}, Branimir and {Juri{\'c}}, Mario and {Ivezi{\'c}}, {\v{Z}}eljko},
        title = "{The Shape and Profile of the Milky Way Halo as Seen by the Canada-France-Hawaii Telescope Legacy Survey}",
      journal = {\apj},
         year = 2011,
        month = apr,
       volume = {731},
       number = {1},
          eid = {4},
        pages = {4},
          doi = {10.1088/0004-637X/731/1/4},
archivePrefix = {arXiv},
       eprint = {1011.4487},
 primaryClass = {astro-ph.GA},
       adsurl = {https://ui.adsabs.harvard.edu/abs/2011ApJ...731....4S}
}

@ARTICLE{Sesar2011b,
       author = {{Sesar}, Branimir and {Stuart}, J. Scott and {Ivezi{\'c}}, {\v{Z}}eljko and {Morgan}, Dylan P. and {Becker}, Andrew C. and {Wo{\'z}niak}, Przemys{\l}aw},
        title = "{Exploring the Variable Sky with LINEAR. I. Photometric Recalibration with the Sloan Digital Sky Survey}",
      journal = {\aj},
         year = 2011,
        month = dec,
       volume = {142},
       number = {6},
          eid = {190},
        pages = {190},
          doi = {10.1088/0004-6256/142/6/190},
archivePrefix = {arXiv},
       eprint = {1109.5227},
 primaryClass = {astro-ph.GA},
       adsurl = {https://ui.adsabs.harvard.edu/abs/2011AJ....142..190S}
}

@ARTICLE{koppelman2019a,
       author = {{Koppelman}, Helmer H. and {Helmi}, Amina and {Massari}, Davide and {Roelenga}, Sebastian and {Bastian}, Ulrich},
        title = "{Characterization and history of the Helmi streams with Gaia DR2}",
      journal = {\aap},
         year = 2019,
        month = may,
       volume = {625},
          eid = {A5},
        pages = {A5},
          doi = {10.1051/0004-6361/201834769},
archivePrefix = {arXiv},
       eprint = {1812.00846},
 primaryClass = {astro-ph.GA},
       adsurl = {https://ui.adsabs.harvard.edu/abs/2019A&A...625A...5K}
}

@ARTICLE{apogee_dr14,
       author = {{Abolfathi}, Bela and {Aguado}, D.~S. and {Aguilar}, Gabriela and
         {Allende Prieto}, Carlos and {Almeida}, Andres and
         {Ananna}, Tonima Tasnim and {Anders}, Friedrich and
         {Anderson}, Scott F. and {Andrews}, Brett H. and {Anguiano}, Borja},
        title = "{The Fourteenth Data Release of the Sloan Digital Sky Survey: First Spectroscopic Data from the Extended Baryon Oscillation Spectroscopic Survey and from the Second Phase of the Apache Point Observatory Galactic Evolution Experiment}",
      journal = {\apjs},
         year = "2018",
        month = "Apr",
       volume = {235},
       number = {2},
          eid = {42},
        pages = {42},
          doi = {10.3847/1538-4365/aa9e8a},
archivePrefix = {arXiv},
       eprint = {1707.09322},
 primaryClass = {astro-ph.GA},
       adsurl = {https://ui.adsabs.harvard.edu/abs/2018ApJS..235...42A}
}

@ARTICLE{ibata2014,
       author = {{Ibata}, Rodrigo A. and {Lewis}, Geraint F. and {McConnachie}, Alan W. and {Martin}, Nicolas F. and {Irwin}, Michael J. and {Ferguson}, Annette M.~N. and {Babul}, Arif and {Bernard}, Edouard J. and {Chapman}, Scott C. and {Collins}, Michelle and {Fardal}, Mark and {Mackey}, A.~D. and {Navarro}, Julio and {Pe{\~n}arrubia}, Jorge and {Rich}, R. Michael and {Tanvir}, Nial and {Widrow}, Lawrence},
        title = "{The Large-scale Structure of the Halo of the Andromeda Galaxy. I. Global Stellar Density, Morphology and Metallicity Properties}",
      journal = {\apj},
         year = 2014,
        month = jan,
       volume = {780},
       number = {2},
          eid = {128},
        pages = {128},
          doi = {10.1088/0004-637X/780/2/128},
archivePrefix = {arXiv},
       eprint = {1311.5888},
 primaryClass = {astro-ph.GA},
       adsurl = {https://ui.adsabs.harvard.edu/abs/2014ApJ...780..128I}
}

@ARTICLE{baumgardt2019,
       author = {{Baumgardt}, H. and {Hilker}, M. and {Sollima}, A. and {Bellini}, A.},
        title = "{Mean proper motions, space orbits, and velocity dispersion profiles of Galactic globular clusters derived from Gaia DR2 data}",
      journal = {\mnras},
         year = "2019",
        month = "Feb",
       volume = {482},
        pages = {5138-5155},
          doi = {10.1093/mnras/sty2997},
archivePrefix = {arXiv},
       eprint = {1811.01507},
 primaryClass = {astro-ph.GA},
       adsurl = {https://ui.adsabs.harvard.edu/\#abs/2019MNRAS.482.5138B}
}

@ARTICLE{fernley1996,
       author = {{Fernley}, J. and {Barnes}, T.~G.},
        title = "{Metal abundances of field RR Lyraes.}",
      journal = {\aap},
         year = 1996,
        month = aug,
       volume = {312},
        pages = {957-965},
       adsurl = {https://ui.adsabs.harvard.edu/abs/1996A&A...312..957F}
}

@ARTICLE{nemec2013,
       author = {{Nemec}, James M. and {Cohen}, Judith G. and {Ripepi}, Vincenzo and
         {Derekas}, Aliz and {Moskalik}, Pawel and {Sesar}, Branimir and
         {Chadid}, Merieme and {Bruntt}, Hans},
        title = "{Metal Abundances, Radial Velocities, and Other Physical Characteristics for the RR Lyrae Stars in The Kepler Field}",
      journal = {\apj},
         year = 2013,
        month = aug,
       volume = {773},
       number = {2},
          eid = {181},
        pages = {181},
          doi = {10.1088/0004-637X/773/2/181},
archivePrefix = {arXiv},
       eprint = {1307.5820},
 primaryClass = {astro-ph.SR},
       adsurl = {https://ui.adsabs.harvard.edu/abs/2013ApJ...773..181N}
}

@ARTICLE{andrievsky2018,
       author = {{Andrievsky}, S. and {Wallerstein}, G. and {Korotin}, S. and
         {Lyashko}, D. and {Kovtyukh}, V. and {Tsymbal}, V. and {Davis}, C.~E. and
         {Gomez}, T. and {Huang}, W. and {Farrell}, E.~M.},
        title = "{The Relationship of Sodium and Oxygen in Galactic Field RR Lyrae Stars}",
      journal = {\pasp},
         year = 2018,
        month = feb,
       volume = {130},
       number = {984},
        pages = {024201},
          doi = {10.1088/1538-3873/aa9783},
       adsurl = {https://ui.adsabs.harvard.edu/abs/2018PASP..130b4201A}
}

@ARTICLE{kunder2017,
       author = {{Kunder}, Andrea and {Kordopatis}, Georges and {Steinmetz}, Matthias and {Zwitter}, Toma{\v{z}} and {McMillan}, Paul J. and {Casagrande}, Luca and {Enke}, Harry and {Wojno}, Jennifer and {Valentini}, Marica and {Chiappini}, Cristina and {Matijevi{\v{c}}}, Gal and {Siviero}, Alessandro and {de Laverny}, Patrick and {Recio-Blanco}, Alejandra and {Bijaoui}, Albert and {Wyse}, Rosemary F.~G. and {Binney}, James and {Grebel}, E.~K. and {Helmi}, Amina and {Jofre}, Paula and {Antoja}, Teresa and {Gilmore}, Gerard and {Siebert}, Arnaud and {Famaey}, Benoit and {Bienaym{\'e}}, Olivier and {Gibson}, Brad K. and {Freeman}, Kenneth C. and {Navarro}, Julio F. and {Munari}, Ulisse and {Seabroke}, George and {Anguiano}, Borja and {{\v{Z}}erjal}, Maru{\v{s}}a and {Minchev}, Ivan and {Reid}, Warren and {Bland-Hawthorn}, Joss and {Kos}, Janez and {Sharma}, Sanjib and {Watson}, Fred and {Parker}, Quentin A. and {Scholz}, Ralf-Dieter and {Burton}, Donna and {Cass}, Paul and {Hartley}, Malcolm and {Fiegert}, Kristin and {Stupar}, Milorad and {Ritter}, Andreas and {Hawkins}, Keith and {Gerhard}, Ortwin and {Chaplin}, W.~J. and {Davies}, G.~R. and {Elsworth}, Y.~P. and {Lund}, M.~N. and {Miglio}, A. and {Mosser}, B.},
        title = "{The Radial Velocity Experiment (RAVE): Fifth Data Release}",
      journal = {\aj},
         year = 2017,
        month = feb,
       volume = {153},
       number = {2},
          eid = {75},
        pages = {75},
          doi = {10.3847/1538-3881/153/2/75},
archivePrefix = {arXiv},
       eprint = {1609.03210},
 primaryClass = {astro-ph.SR},
       adsurl = {https://ui.adsabs.harvard.edu/abs/2017AJ....153...75K}
}

@ARTICLE{braga2019,
       author = {{Braga}, V.~F. and {Stetson}, P.~B. and {Bono}, G. and {Dall'Ora}, M. and
         {Ferraro}, I. and {Fiorentino}, G. and {Iannicola}, G. and {Inno}, L. and
         {Marengo}, M. and {Neeley}, J. and {Beaton}, R.~L. and {Buonanno}, R. and
         {Calamida}, A. and {Contreras Ramos}, R. and {Chaboyer}, B. and
         {Fabrizio}, M. and {Freedman}, W.~L. and {Gilligan}, C.~K. and
         {Johnston}, K.~V. and {Lub}, J. and {Madore}, B.~F. and {Magurno}, D. and
         {Marconi}, M. and {Marinoni}, S. and {Marrese}, P.~M. and {Mateo}, M. and
         {Matsunaga}, N. and {Minniti}, D. and {Monson}, A.~J. and
         {Monelli}, M. and {Nonino}, M. and {Persson}, S.~E. and
         {Pietrinferni}, A. and {Sneden}, C. and {Storm}, J. and
         {Walker}, A.~R. and {Valenti}, E. and {Zoccali}, M.},
        title = "{New near-infrared JHK$_{s}$ light-curve templates for RR Lyrae variables}",
      journal = {\aap},
         year = "2019",
        month = "May",
       volume = {625},
          eid = {A1},
        pages = {A1},
          doi = {10.1051/0004-6361/201834893},
archivePrefix = {arXiv},
       eprint = {1812.06372},
 primaryClass = {astro-ph.SR},
       adsurl = {https://ui.adsabs.harvard.edu/abs/2019A&A...625A...1B}
}

@ARTICLE{minniti2016,
   author = {{Minniti}, D. and {Contreras Ramos}, R. and {Zoccali}, M. and 
	{Rejkuba}, M. and {Gonzalez}, O.~A. and {Valenti}, E. and {Gran}, F.
	},
    title = "{Discovery of RR Lyrae Stars in the Nuclear Bulge of the Milky Way}",
  journal = {\apjl},
archivePrefix = "arXiv",
   eprint = {1610.04689},
     year = 2016,
    month = oct,
   volume = 830,
      eid = {L14},
    pages = {L14},
      doi = {10.3847/2041-8205/830/1/L14},
   adsurl = {http://adsabs.harvard.edu/abs/2016ApJ...830L..14M}
}

@ARTICLE{gaia_dr3,
       author = {{Gaia Collaboration} and {Vallenari}, A. and {Brown}, A.~G.~A. and {Prusti}, T. and {de Bruijne}, J.~H.~J. and {Arenou}, F. and {Babusiaux}, C. and {Biermann}, M. and {Creevey}, O.~L. and {Ducourant}, C. and {Evans}, D.~W. and {Eyer}, L. and {Guerra}, R. and {Hutton}, A. and {Jordi}, C. and {Klioner}, S.~A. and {Lammers}, U.~L. and {Lindegren}, L. and {Luri}, X. and {Mignard}, F. and {Panem}, C. and {Pourbaix}, D. and {Randich}, S. and {Sartoretti}, P. and {Soubiran}, C. and {Tanga}, P. and {Walton}, N.~A. and {Bailer-Jones}, C.~A.~L. and {Bastian}, U. and {Drimmel}, R. and {Jansen}, F. and {Katz}, D. and {Lattanzi}, M.~G. and {van Leeuwen}, F. and {Bakker}, J. and {Cacciari}, C. and {Casta{\~n}eda}, J. and {De Angeli}, F. and {Fabricius}, C. and {Fouesneau}, M. and {Fr{\'e}mat}, Y. and {Galluccio}, L. and {Guerrier}, A. and {Heiter}, U. and {Masana}, E. and {Messineo}, R. and {Mowlavi}, N. and {Nicolas}, C. and {Nienartowicz}, K. and {Pailler}, F. and {Panuzzo}, P. and {Riclet}, F. and {Roux}, W. and {Seabroke}, G.~M. and {Sordo}, R. and {Th{\'e}venin}, F. and {Gracia-Abril}, G. and {Portell}, J. and {Teyssier}, D. and {Altmann}, M. and {Andrae}, R. and {Audard}, M. and {Bellas-Velidis}, I. and {Benson}, K. and {Berthier}, J. and {Blomme}, R. and {Burgess}, P.~W. and {Busonero}, D. and {Busso}, G. and {C{\'a}novas}, H. and {Carry}, B. and {Cellino}, A. and {Cheek}, N. and {Clementini}, G. and {Damerdji}, Y. and {Davidson}, M. and {de Teodoro}, P. and {Nu{\~n}ez Campos}, M. and {Delchambre}, L. and {Dell'Oro}, A. and {Esquej}, P. and {Fern{\'a}ndez-Hern{\'a}ndez}, J. and {Fraile}, E. and {Garabato}, D. and {Garc{\'\i}a-Lario}, P. and {Gosset}, E. and {Haigron}, R. and {Halbwachs}, J. -L. and {Hambly}, N.~C. and {Harrison}, D.~L. and {Hern{\'a}ndez}, J. and {Hestroffer}, D. and {Hodgkin}, S.~T. and {Holl}, B. and {Jan{\ss}en}, K. and {Jevardat de Fombelle}, G. and {Jordan}, S. and {Krone-Martins}, A. and {Lanzafame}, A.~C. and {L{\"o}ffler}, W. and {Marchal}, O. and {Marrese}, P.~M. and {Moitinho}, A. and {Muinonen}, K. and {Osborne}, P. and {Pancino}, E. and {Pauwels}, T. and {Recio-Blanco}, A. and {Reyl{\'e}}, C. and {Riello}, M. and {Rimoldini}, L. and {Roegiers}, T. and {Rybizki}, J. and {Sarro}, L.~M. and {Siopis}, C. and {Smith}, M. and {Sozzetti}, A. and {Utrilla}, E. and {van Leeuwen}, M. and {Abbas}, U. and {{\'A}brah{\'a}m}, P. and {Abreu Aramburu}, A. and {Aerts}, C. and {Aguado}, J.~J. and {Ajaj}, M. and {Aldea-Montero}, F. and {Altavilla}, G. and {{\'A}lvarez}, M.~A. and {Alves}, J. and {Anders}, F. and {Anderson}, R.~I. and {Anglada Varela}, E. and {Antoja}, T. and {Baines}, D. and {Baker}, S.~G. and {Balaguer-N{\'u}{\~n}ez}, L. and {Balbinot}, E. and {Balog}, Z. and {Barache}, C. and {Barbato}, D. and {Barros}, M. and {Barstow}, M.~A. and {Bartolom{\'e}}, S. and {Bassilana}, J. -L. and {Bauchet}, N. and {Becciani}, U. and {Bellazzini}, M. and {Berihuete}, A. and {Bernet}, M. and {Bertone}, S. and {Bianchi}, L. and {Binnenfeld}, A. and {Blanco-Cuaresma}, S. and {Blazere}, A. and {Boch}, T. and {Bombrun}, A. and {Bossini}, D. and {Bouquillon}, S. and {Bragaglia}, A. and {Bramante}, L. and {Breedt}, E. and {Bressan}, A. and {Brouillet}, N. and {Brugaletta}, E. and {Bucciarelli}, B. and {Burlacu}, A. and {Butkevich}, A.~G. and {Buzzi}, R. and {Caffau}, E. and {Cancelliere}, R. and {Cantat-Gaudin}, T. and {Carballo}, R. and {Carlucci}, T. and {Carnerero}, M.~I. and {Carrasco}, J.~M. and {Casamiquela}, L. and {Castellani}, M. and {Castro-Ginard}, A. and {Chaoul}, L. and {Charlot}, P. and {Chemin}, L. and {Chiaramida}, V. and {Chiavassa}, A. and {Chornay}, N. and {Comoretto}, G. and {Contursi}, G. and {Cooper}, W.~J. and {Cornez}, T. and {Cowell}, S. and {Crifo}, F. and {Cropper}, M. and {Crosta}, M. and {Crowley}, C. and {Dafonte}, C. and {Dapergolas}, A. and {David}, M. and {David}, P. and {de Laverny}, P. and {De Luise}, F. and {De March}, R.},
        title = "{Gaia Data Release 3. Summary of the content and survey properties}",
      journal = {\aap},
         year = 2023,
        month = jun,
       volume = {674},
          eid = {A1},
        pages = {A1},
          doi = {10.1051/0004-6361/202243940},
archivePrefix = {arXiv},
       eprint = {2208.00211},
 primaryClass = {astro-ph.GA},
       adsurl = {https://ui.adsabs.harvard.edu/abs/2023A&A...674A...1G}
}

@ARTICLE{bono2020,
       author = {{Bono}, G. and {Braga}, V.~F. and {Crestani}, J. and {Fabrizio}, M. and
         {Sneden}, C. and {Marconi}, M. and {Preston}, G.~W. and
         {Mullen}, J.~P. and {Gilligan}, C.~K. and {Fiorentino}, G. and
         {Pietrinferni}, A. and {Altavilla}, G. and {Buonanno}, R. and
         {Chaboyer}, B. and {Silva}, R. da and {Dall'Ora}, M. and
         {Degl'Innocenti}, S. and {Carlo}, E. Di and {Ferraro}, I. and
         {Grebel}, E.~K. and {Iannicola}, G. and {Inno}, L. and {Kovtyukh}, V. and
         {Kunder}, A. and {Lemasle}, B. and {Marengo}, M. and {Marinoni}, S. and
         {Marrese}, P.~M. and {Mart{\'\i}nez-V{\'a}zquez}, C.~E. and
         {Matsunaga}, N. and {Monelli}, M. and {Neeley}, J. and {Nonino}, M. and
         {Moroni}, P.~G. Prada and {Prudil}, Z. and {Stetson}, P.~B. and
         {Th{\'e}venin}, F. and {Tognelli}, E. and {Valenti}, E. and
         {Walker}, A.~R.},
        title = "{On the Metamorphosis of the Bailey Diagram for RR Lyrae Stars}",
      journal = {\apjl},
         year = 2020,
        month = jun,
       volume = {896},
       number = {1},
          eid = {L15},
        pages = {L15},
          doi = {10.3847/2041-8213/ab9538},
archivePrefix = {arXiv},
       eprint = {2005.11566},
 primaryClass = {astro-ph.SR},
       adsurl = {https://ui.adsabs.harvard.edu/abs/2020ApJ...896L..15B}
}

@ARTICLE{bono2019,
   author = {{Bono}, G. and {Iannicola}, G. and {Braga}, V.~F. and {Ferraro}, I. and 
	{Stetson}, P.~B. and {Magurno}, D. and {Matsunaga}, N. and {Beaton}, R.~L. and 
	{Buonanno}, R. and {Chaboyer}, B. and {Dall'Ora}, M. and 
	{Fabrizio}, M. and {Fiorentino}, G. and {Freedman}, W.~L. and 
	{Gilligan}, C.~K. and {Madore}, B.~F. and {Marconi}, M. and 
	{Marengo}, M. and {Marinoni}, S. and {Marrese}, P.~M. and {Martinez-Vazquez}, C.~E. and 
	{Mateo}, M. and {Monelli}, M. and {Neeley}, J.~R. and {Nonino}, M. and 
	{Sneden}, C. and {Thevenin}, F. and {Valenti}, E. and {Walker}, A.~R.
	},
    title = "{On a New Method to Estimate the Distance, Reddening, and Metallicity of RR Lyrae Stars Using Optical/Near-infrared (B, V, I, J, H, K) Mean Magnitudes: {$\omega$} Centauri as a First Test Case}",
  journal = {\apj},
archivePrefix = "arXiv",
   eprint = {1811.07069},
 primaryClass = "astro-ph.SR",
     year = 2019,
    month = jan,
   volume = 870,
      eid = {115},
    pages = {115},
      doi = {10.3847/1538-4357/aaf23f},
   adsurl = {http://adsabs.harvard.edu/abs/2019ApJ...870..115B}
}

@ARTICLE{mullen2023,
       author = {{Mullen}, Joseph P. and {Marengo}, Massimo and {Mart{\'\i}nez-V{\'a}zquez}, Clara E. and {Chaboyer}, Brian and {Bono}, Giuseppe and {Braga}, Vittorio F. and {Dall'Ora}, Massimo and {D'Orazi}, Valentina and {Fabrizio}, Michele and {Monelli}, Matteo and {Th{\'e}venin}, Fr{\'e}d{\'e}ric},
        title = "{RR Lyrae Mid-infrared Period-Luminosity-Metallicity and Period-Wesenheit-Metallicity Relations Based on Gaia DR3 Parallaxes}",
      journal = {\apj},
         year = 2023,
        month = mar,
       volume = {945},
       number = {1},
          eid = {83},
        pages = {83},
          doi = {10.3847/1538-4357/acb20a},
archivePrefix = {arXiv},
       eprint = {2301.03777},
 primaryClass = {astro-ph.GA},
       adsurl = {https://ui.adsabs.harvard.edu/abs/2023ApJ...945...83M}
}

@ARTICLE{sneden2017,
   author = {{Sneden}, C. and {Preston}, G.~W. and {Chadid}, M. and {Adam{\'o}w}, M.
	},
    title = "{The RRc Stars: Chemical Abundances and Envelope Kinematics}",
  journal = {\apj},
archivePrefix = "arXiv",
   eprint = {1709.00494},
 primaryClass = "astro-ph.SR",
     year = 2017,
    month = oct,
   volume = 848,
      eid = {68},
    pages = {68},
      doi = {10.3847/1538-4357/aa8b10},
   adsurl = {http://adsabs.harvard.edu/abs/2017ApJ...848...68S}
}

@ARTICLE{lindegren2021,
       author = {{Lindegren}, L. and {Klioner}, S.~A. and {Hern{\'a}ndez}, J. and {Bombrun}, A. and {Ramos-Lerate}, M. and {Steidelm{\"u}ller}, H. and {Bastian}, U. and {Biermann}, M. and {de Torres}, A. and {Gerlach}, E. and {Geyer}, R. and {Hilger}, T. and {Hobbs}, D. and {Lammers}, U. and {McMillan}, P.~J. and {Stephenson}, C.~A. and {Casta{\~n}eda}, J. and {Davidson}, M. and {Fabricius}, C. and {Gracia-Abril}, G. and {Portell}, J. and {Rowell}, N. and {Teyssier}, D. and {Torra}, F. and {Bartolom{\'e}}, S. and {Clotet}, M. and {Garralda}, N. and {Gonz{\'a}lez-Vidal}, J.~J. and {Torra}, J. and {Abbas}, U. and {Altmann}, M. and {Anglada Varela}, E. and {Balaguer-N{\'u}{\~n}ez}, L. and {Balog}, Z. and {Barache}, C. and {Becciani}, U. and {Bernet}, M. and {Bertone}, S. and {Bianchi}, L. and {Bouquillon}, S. and {Brown}, A.~G.~A. and {Bucciarelli}, B. and {Busonero}, D. and {Butkevich}, A.~G. and {Buzzi}, R. and {Cancelliere}, R. and {Carlucci}, T. and {Charlot}, P. and {Cioni}, M. -R.~L. and {Crosta}, M. and {Crowley}, C. and {del Peloso}, E.~F. and {del Pozo}, E. and {Drimmel}, R. and {Esquej}, P. and {Fienga}, A. and {Fraile}, E. and {Gai}, M. and {Garcia-Reinaldos}, M. and {Guerra}, R. and {Hambly}, N.~C. and {Hauser}, M. and {Jan{\ss}en}, K. and {Jordan}, S. and {Kostrzewa-Rutkowska}, Z. and {Lattanzi}, M.~G. and {Liao}, S. and {Licata}, E. and {Lister}, T.~A. and {L{\"o}ffler}, W. and {Marchant}, J.~M. and {Masip}, A. and {Mignard}, F. and {Mints}, A. and {Molina}, D. and {Mora}, A. and {Morbidelli}, R. and {Murphy}, C.~P. and {Pagani}, C. and {Panuzzo}, P. and {Pe{\~n}alosa Esteller}, X. and {Poggio}, E. and {Re Fiorentin}, P. and {Riva}, A. and {Sagrist{\`a} Sell{\'e}s}, A. and {Sanchez Gimenez}, V. and {Sarasso}, M. and {Sciacca}, E. and {Siddiqui}, H.~I. and {Smart}, R.~L. and {Souami}, D. and {Spagna}, A. and {Steele}, I.~A. and {Taris}, F. and {Utrilla}, E. and {van Reeven}, W. and {Vecchiato}, A.},
        title = "{Gaia Early Data Release 3. The astrometric solution}",
      journal = {\aap},
         year = 2021,
        month = may,
       volume = {649},
          eid = {A2},
        pages = {A2},
          doi = {10.1051/0004-6361/202039709},
archivePrefix = {arXiv},
       eprint = {2012.03380},
 primaryClass = {astro-ph.IM},
       adsurl = {https://ui.adsabs.harvard.edu/abs/2021A&A...649A...2L}
}

@ARTICLE{pagnini2023,
       author = {{Pagnini}, G. and {Di Matteo}, P. and {Khoperskov}, S. and {Mastrobuono-Battisti}, A. and {Haywood}, M. and {Renaud}, F. and {Combes}, F.},
        title = "{The distribution of globular clusters in kinematic spaces does not trace the accretion history of the host galaxy}",
      journal = {\aap},
         year = 2023,
        month = may,
       volume = {673},
          eid = {A86},
        pages = {A86},
          doi = {10.1051/0004-6361/202245128},
archivePrefix = {arXiv},
       eprint = {2210.04245},
 primaryClass = {astro-ph.GA},
       adsurl = {https://ui.adsabs.harvard.edu/abs/2023A&A...673A..86P}
}

@ARTICLE{hansen2011,
       author = {{Hansen}, C.~J. and {Nordstr{\"o}m}, B. and {Bonifacio}, P. and
         {Spite}, M. and {Andersen}, J. and {Beers}, T.~C. and {Cayrel}, R. and
         {Spite}, F. and {Molaro}, P. and {Barbuy}, B. and {Depagne}, E. and
         {Fran{\c{c}}ois}, P. and {Hill}, V. and {Plez}, B. and {Sivarani}, T.},
        title = "{First stars. XIII. Two extremely metal-poor RR Lyrae stars}",
      journal = {\aap},
         year = 2011,
        month = mar,
       volume = {527},
          eid = {A65},
        pages = {A65},
          doi = {10.1051/0004-6361/201015076},
archivePrefix = {arXiv},
       eprint = {1101.2207},
 primaryClass = {astro-ph.SR},
       adsurl = {https://ui.adsabs.harvard.edu/abs/2011A&A...527A..65H}
}

@ARTICLE{liu2018,
       author = {{Liu}, Shuai and {Du}, Cuihua and {Newberg}, Heidi Jo and {Chen}, Yuqin and {Wu}, Zhenyu and {Ma}, Jun and {Zhou}, Xu and {Cao}, Zihuang and {Hou}, Yonghui and {Wang}, Yuefei and {Zhang}, Yong},
        title = "{Metallicity and Kinematics of the Galactic Halo from the LAMOST Sample Stars}",
      journal = {\apj},
         year = 2018,
        month = aug,
       volume = {862},
       number = {2},
          eid = {163},
        pages = {163},
          doi = {10.3847/1538-4357/aacf91},
archivePrefix = {arXiv},
       eprint = {1806.10315},
 primaryClass = {astro-ph.GA},
       adsurl = {https://ui.adsabs.harvard.edu/abs/2018ApJ...862..163L}
}

@ARTICLE{liu2020,
       author = {{Liu}, G. -C. and {Huang}, Y. and {Zhang}, H. -W. and {Xiang}, M. -S. and {Ren}, J. -J. and {Chen}, B. -Q. and {Yuan}, H. -B. and {Wang}, C. and {Yang}, Y. and {Tian}, Z. -J. and {Wang}, F. and {Liu}, X. -W.},
        title = "{Probing the Galactic Halo with RR Lyrae Stars. I. The Catalog}",
      journal = {\apjs},
         year = 2020,
        month = apr,
       volume = {247},
       number = {2},
          eid = {68},
        pages = {68},
          doi = {10.3847/1538-4365/ab72f8},
archivePrefix = {arXiv},
       eprint = {2002.01188},
 primaryClass = {astro-ph.SR},
       adsurl = {https://ui.adsabs.harvard.edu/abs/2020ApJS..247...68L}
}

@ARTICLE{mateu2012,
       author = {{Mateu}, C. and {Vivas}, A.~K. and {Downes}, J.~J. and {Brice{\~n}o}, C. and {Zinn}, R. and {Cruz-Diaz}, G.},
        title = "{The QUEST RR Lyrae Survey - III. The low Galactic latitude catalogue}",
      journal = {\mnras},
         year = 2012,
        month = dec,
       volume = {427},
       number = {4},
        pages = {3374-3395},
          doi = {10.1111/j.1365-2966.2012.21968.x},
archivePrefix = {arXiv},
       eprint = {1208.4599},
 primaryClass = {astro-ph.GA},
       adsurl = {https://ui.adsabs.harvard.edu/abs/2012MNRAS.427.3374M}
}

@ARTICLE{pancino15,
   author = {{Pancino}, E. and {Britavskiy}, N. and {Romano}, D. and {Cacciari}, C. and 
	{Mucciarelli}, A. and {Clementini}, G.},
    title = "{Chemical abundances of solar neighbourhood RR Lyrae stars}",
  journal = {\mnras},
archivePrefix = "arXiv",
   eprint = {1412.4580},
 primaryClass = "astro-ph.SR",
     year = 2015,
    month = mar,
   volume = 447,
    pages = {2404-2419},
      doi = {10.1093/mnras/stu2616},
   adsurl = {http://adsabs.harvard.edu/abs/2015MNRAS.447.2404P}
}

@ARTICLE{for11,
   author = {{For}, B.-Q. and {Sneden}, C. and {Preston}, G.~W.},
    title = "{The Chemical Compositions of Variable Field Horizontal-branch Stars: RR Lyrae Stars}",
  journal = {\apjs},
archivePrefix = "arXiv",
   eprint = {1110.0548},
 primaryClass = "astro-ph.SR",
     year = 2011,
    month = dec,
   volume = 197,
      eid = {29},
    pages = {29},
      doi = {10.1088/0067-0049/197/2/29},
   adsurl = {http://adsabs.harvard.edu/abs/2011ApJS..197...29F}
}

@ARTICLE{ivezic2000,
       author = {{Ivezi{\'c}}, {\v{Z}}eljko and {Goldston}, Josh and {Finlator}, Kristian and {Knapp}, Gillian R. and {Yanny}, Brian and {McKay}, Timothy A. and {Amrose}, Susan and {Krisciunas}, Kevin and {Willman}, Beth and {Anderson}, Scott and {Schaber}, Chris and {Erb}, Dawn and {Logan}, Chelsea and {Stubbs}, Chris and {Chen}, Bing and {Neilsen}, Eric and {Uomoto}, Alan and {Pier}, Jeffrey R. and {Fan}, Xiaohui and {Gunn}, James E. and {Lupton}, Robert H. and {Rockosi}, Constance M. and {Schlegel}, David and {Strauss}, Michael A. and {Annis}, James and {Brinkmann}, Jon and {Csabai}, Istv{\'a}n and {Doi}, Mamoru and {Fukugita}, Masataka and {Hennessy}, Gregory S. and {Hindsley}, Robert B. and {Margon}, Bruce and {Munn}, Jeffrey A. and {Newberg}, Heidi Jo and {Schneider}, Donald P. and {Smith}, J. Allyn and {Szokoly}, Gyula P. and {Thakar}, Aniruddha R. and {Vogeley}, Michael S. and {Waddell}, Patrick and {Yasuda}, Naoki and {York}, Donald G. and {SDSS Collaboration}},
        title = "{Candidate RR Lyrae Stars Found in Sloan Digital Sky Survey Commissioning Data}",
      journal = {\aj},
         year = 2000,
        month = aug,
       volume = {120},
       number = {2},
        pages = {963-977},
          doi = {10.1086/301455},
archivePrefix = {arXiv},
       eprint = {astro-ph/0004130},
 primaryClass = {astro-ph},
       adsurl = {https://ui.adsabs.harvard.edu/abs/2000AJ....120..963I}
}

@ARTICLE{DaflonCunha2004,
       author = {{Daflon}, Simone and {Cunha}, Katia},
        title = "{Galactic Metallicity Gradients Derived from a Sample of OB Stars}",
      journal = {\apj},
         year = 2004,
        month = dec,
       volume = {617},
       number = {2},
        pages = {1115-1126},
          doi = {10.1086/425607},
archivePrefix = {arXiv},
       eprint = {astro-ph/0409084},
 primaryClass = {astro-ph},
       adsurl = {https://ui.adsabs.harvard.edu/abs/2004ApJ...617.1115D}
}

@ARTICLE{roskar2008,
       author = {{Ro{\v{s}}kar}, Rok and {Debattista}, Victor P. and {Quinn}, Thomas R. and {Stinson}, Gregory S. and {Wadsley}, James},
        title = "{Riding the Spiral Waves: Implications of Stellar Migration for the Properties of Galactic Disks}",
      journal = {\apjl},
         year = 2008,
        month = sep,
       volume = {684},
       number = {2},
        pages = {L79},
          doi = {10.1086/592231},
archivePrefix = {arXiv},
       eprint = {0808.0206},
 primaryClass = {astro-ph},
       adsurl = {https://ui.adsabs.harvard.edu/abs/2008ApJ...684L..79R}
}

@ARTICLE{anders2017,
       author = {{Anders}, F. and {Chiappini}, C. and {Minchev}, I. and {Miglio}, A. and {Montalb{\'a}n}, J. and {Mosser}, B. and {Rodrigues}, T.~S. and {Santiago}, B.~X. and {Baudin}, F. and {Beers}, T.~C. and {da Costa}, L.~N. and {Garc{\'\i}a}, R.~A. and {Garc{\'\i}a-Hern{\'a}ndez}, D.~A. and {Holtzman}, J. and {Maia}, M.~A.~G. and {Majewski}, S. and {Mathur}, S. and {Noels-Grotsch}, A. and {Pan}, K. and {Schneider}, D.~P. and {Schultheis}, M. and {Steinmetz}, M. and {Valentini}, M. and {Zamora}, O.},
        title = "{Red giants observed by CoRoT and APOGEE: The evolution of the Milky Way's radial metallicity gradient}",
      journal = {\aap},
         year = 2017,
        month = apr,
       volume = {600},
          eid = {A70},
        pages = {A70},
          doi = {10.1051/0004-6361/201629363},
archivePrefix = {arXiv},
       eprint = {1608.04951},
 primaryClass = {astro-ph.GA},
       adsurl = {https://ui.adsabs.harvard.edu/abs/2017A&A...600A..70A}
}

@ARTICLE{willett2023,
       author = {{Willett}, Emma and {Miglio}, Andrea and {Mackereth}, J. Ted and {Chiappini}, Cristina and {Lyttle}, Alexander J. and {Elsworth}, Yvonne and {Mosser}, Beno{\^\i}t and {Khan}, Saniya and {Anders}, Friedrich and {Casali}, Giada and {Grisoni}, Valeria},
        title = "{The evolution of the Milky Way's thin disc radial metallicity gradient with K2 asteroseismic ages}",
      journal = {\mnras},
         year = 2023,
        month = dec,
       volume = {526},
       number = {2},
        pages = {2141-2155},
          doi = {10.1093/mnras/stad2374},
archivePrefix = {arXiv},
       eprint = {2307.14422},
 primaryClass = {astro-ph.GA},
       adsurl = {https://ui.adsabs.harvard.edu/abs/2023MNRAS.526.2141W}
}

@ARTICLE{Ratcliffe2023,
       author = {{Ratcliffe}, Bridget and {Minchev}, Ivan and {Anders}, Friedrich and {Khoperskov}, Sergey and {Guiglion}, Guillaume and {Buck}, Tobias and {Cunha}, Katia and {Queiroz}, Anna and {Nitschelm}, Christian and {Meszaros}, Szabolcs and {Steinmetz}, Matthias and {de Jong}, Roelof S. and {Nepal}, Samir and {Lane}, Richard R. and {Sobeck}, Jennifer},
        title = "{Unveiling the time evolution of chemical abundances across the Milky Way disc with APOGEE}",
      journal = {\mnras},
         year = 2023,
        month = oct,
       volume = {525},
       number = {2},
        pages = {2208-2228},
          doi = {10.1093/mnras/stad1573},
archivePrefix = {arXiv},
       eprint = {2305.13378},
 primaryClass = {astro-ph.GA},
       adsurl = {https://ui.adsabs.harvard.edu/abs/2023MNRAS.525.2208R}
}

@ARTICLE{pojmanski1997,
   author = {{Pojma{\'n}ski}, G.},
    title = "{The All Sky Automated Survey}",
  journal = {\actaa},
   eprint = {astro-ph/9712146},
     year = 1997,
    month = oct,
   volume = 47,
    pages = {467-481},
   adsurl = {http://adsabs.harvard.edu/abs/1997AcA....47..467P}
}

@ARTICLE{magurno2019,
       author = {{Magurno}, D. and {Sneden}, C. and {Bono}, G. and {Braga}, V.~F. and
         {Mateo}, M. and {Persson}, S.~E. and {Preston}, G. and
         {Th{\'e}venin}, F. and {da Silva}, R. and {Dall'Ora}, M. and
         {Fabrizio}, M. and {Ferraro}, I. and {Fiorentino}, G. and
         {Iannicola}, G. and {Inno}, L. and {Marengo}, M. and {Marinoni}, S. and
         {Marrese}, P.~M. and {Mart{\'\i}nez-V{\'a}zquez}, C.~E. and
         {Matsunaga}, N. and {Monelli}, M. and {Neeley}, J.~R. and {Nonino}, M. and
         {Walker}, A.~R.},
        title = "{Chemical Compositions of Field and Globular Cluster RR Lyrae Stars. II. {\ensuremath{\omega}} Centauri}",
      journal = {\apj},
         year = "2019",
        month = "Aug",
       volume = {881},
       number = {2},
          eid = {104},
        pages = {104},
          doi = {10.3847/1538-4357/ab2e76},
archivePrefix = {arXiv},
       eprint = {1906.08550},
 primaryClass = {astro-ph.SR},
       adsurl = {https://ui.adsabs.harvard.edu/abs/2019ApJ...881..104M}
}

@ARTICLE{magurno2018,
   author = {{Magurno}, D. and {Sneden}, C. and {Braga}, V.~F. and {Bono}, G. and {Mateo}, M. and {Persson}, S.~E. and {Dall'Ora}, M. and {Marengo}, M. and {Monelli}, M. and {Neeley}, J.~R.},
        title = "{Chemical Compositions of Field and Globular Cluster RR Lyrae Stars. I. NGC 3201}",
      journal = {\apj},
         year = 2018,
        month = sep,
       volume = {864},
       number = {1},
          eid = {57},
        pages = {57},
          doi = {10.3847/1538-4357/aad4a3},
archivePrefix = {arXiv},
       eprint = {1807.06681},
 primaryClass = {astro-ph.SR},
       adsurl = {https://ui.adsabs.harvard.edu/abs/2018ApJ...864...57M}
}

@ARTICLE{gomez13,
       author = {{G{\'o}mez}, Facundo A. and {Helmi}, Amina and {Cooper}, Andrew P. and {Frenk}, Carlos S. and {Navarro}, Julio F. and {White}, Simon D.~M.},
        title = "{Streams in the Aquarius stellar haloescarollo07
}",
      journal = {\mnras},
         year = 2013,
        month = dec,
       volume = {436},
       number = {4},
        pages = {3602-3613},
          doi = {10.1093/mnras/stt1838},
archivePrefix = {arXiv},
       eprint = {1307.0008},
 primaryClass = {astro-ph.GA},
       adsurl = {https://ui.adsabs.harvard.edu/abs/2013MNRAS.436.3602G}
}

@ARTICLE{bullock05,
       author = {{Bullock}, James S. and {Johnston}, Kathryn V.},
        title = "{Tracing Galaxy Formation with Stellar Halos. I. Methods}",
      journal = {\apj},
         year = 2005,
        month = dec,
       volume = {635},
       number = {2},
        pages = {931-949},
          doi = {10.1086/497422},
archivePrefix = {arXiv},
       eprint = {astro-ph/0506467},
 primaryClass = {astro-ph},
       adsurl = {https://ui.adsabs.harvard.edu/abs/2005ApJ...635..931B}
}

@ARTICLE{cooper15,
       author = {{Cooper}, Andrew P. and {Parry}, Owen H. and {Lowing}, Ben and {Cole}, Shaun and {Frenk}, Carlos},
        title = "{Formation of in situ stellar haloes in Milky Way-mass galaxies}",
      journal = {\mnras},
         year = 2015,
        month = dec,
       volume = {454},
       number = {3},
        pages = {3185-3199},
          doi = {10.1093/mnras/stv2057},
archivePrefix = {arXiv},
       eprint = {1501.04630},
 primaryClass = {astro-ph.GA},
       adsurl = {https://ui.adsabs.harvard.edu/abs/2015MNRAS.454.3185C}
}

@ARTICLE{font11,
       author = {{Font}, A.~S. and {McCarthy}, I.~G. and {Crain}, R.~A. and {Theuns}, T. and {Schaye}, J. and {Wiersma}, R.~P.~C. and {Dalla Vecchia}, C.},
        title = "{Cosmological simulations of the formation of the stellar haloes around disc galaxies}",
      journal = {\mnras},
         year = 2011,
        month = oct,
       volume = {416},
       number = {4},
        pages = {2802-2820},
          doi = {10.1111/j.1365-2966.2011.19227.x},
archivePrefix = {arXiv},
       eprint = {1102.2526},
 primaryClass = {astro-ph.CO},
       adsurl = {https://ui.adsabs.harvard.edu/abs/2011MNRAS.416.2802F}
}

@ARTICLE{mc12,
       author = {{McCarthy}, I.~G. and {Font}, A.~S. and {Crain}, R.~A. and {Deason}, A.~J. and {Schaye}, J. and {Theuns}, T.},
        title = "{Global structure and kinematics of stellar haloes in cosmological hydrodynamic simulations}",
      journal = {\mnras},
         year = 2012,
        month = mar,
       volume = {420},
       number = {3},
        pages = {2245-2262},
          doi = {10.1111/j.1365-2966.2011.20189.x},
archivePrefix = {arXiv},
       eprint = {1111.1747},
 primaryClass = {astro-ph.GA},
       adsurl = {https://ui.adsabs.harvard.edu/abs/2012MNRAS.420.2245M}
}

@ARTICLE{hattori11,
       author = {{Hattori}, K. and {Yoshii}, Y.},
        title = "{The orbital eccentricity distribution of solar-neighbourhood halo stars}",
      journal = {\mnras},
         year = 2011,
        month = dec,
       volume = {418},
       number = {4},
        pages = {2481-2492},
          doi = {10.1111/j.1365-2966.2011.19639.x},
archivePrefix = {arXiv},
       eprint = {1108.4103},
 primaryClass = {astro-ph.GA},
       adsurl = {https://ui.adsabs.harvard.edu/abs/2011MNRAS.418.2481H}
}

@ARTICLE{kinman12a,
       author = {{Kinman}, T.~D. and {Cacciari}, C. and {Bragaglia}, A. and {Smart}, R. and {Spagna}, A.},
        title = "{The kinematic properties of BHB and RR Lyrae stars towards the Anticentre and the North Galactic Pole: the transition between the inner and the outer halo}",
      journal = {\mnras},
         year = 2012,
        month = may,
       volume = {422},
       number = {3},
        pages = {2116-2144},
          doi = {10.1111/j.1365-2966.2012.20747.x},
archivePrefix = {arXiv},
       eprint = {1203.2146},
 primaryClass = {astro-ph.GA},
       adsurl = {https://ui.adsabs.harvard.edu/abs/2012MNRAS.422.2116K}
}

@ARTICLE{dietz2020,
       author = {{Dietz}, Sarah E. and {Yoon}, Jinmi and {Beers}, Timothy C. and {Placco}, Vinicius M.},
        title = "{The Metallicity Gradient and Complex Formation History of the Outermost Halo of the Milky Way}",
      journal = {\apj},
         year = 2020,
        month = may,
       volume = {894},
       number = {1},
          eid = {34},
        pages = {34},
          doi = {10.3847/1538-4357/ab7fa4},
archivePrefix = {arXiv},
       eprint = {1911.11140},
 primaryClass = {astro-ph.GA},
       adsurl = {https://ui.adsabs.harvard.edu/abs/2020ApJ...894...34D}
}

@article{kim2019,
	author = {Kim, Young Kwang and Lee, Young Sun and Beers, Timothy C.},
	doi = {10.3847/1538-4357/ab3660},
	journal = {The Astrophysical Journal},
	month = {sep},
	number = {2},
	pages = {176},
	publisher = {The American Astronomical Society},
	title = {Dependence of Galactic Halo Kinematics on the Adopted Galactic Potential},
	url = {https://dx.doi.org/10.3847/1538-4357/ab3660},
	volume = {882},
	year = {2019}}

@ARTICLE{xue2015,
       author = {{Xue}, Xiang-Xiang and {Rix}, Hans-Walter and {Ma}, Zhibo and {Morrison}, Heather and {Bovy}, Jo and {Sesar}, Branimir and {Janesh}, William},
        title = "{The Radial Profile and Flattening of the Milky Way{\textquoteright}s Stellar Halo to 80 kpc from the SEGUE K-giant Survey}",
      journal = {\apj},
         year = 2015,
        month = aug,
       volume = {809},
       number = {2},
          eid = {144},
        pages = {144},
          doi = {10.1088/0004-637X/809/2/144},
archivePrefix = {arXiv},
       eprint = {1506.06144},
 primaryClass = {astro-ph.GA},
       adsurl = {https://ui.adsabs.harvard.edu/abs/2015ApJ...809..144X}
}

@ARTICLE{helmi1999,
       author = {{Helmi}, Amina and {White}, Simon D.~M. and {de Zeeuw}, P. Tim and {Zhao}, Hongsheng},
        title = "{Debris streams in the solar neighbourhood as relicts from the formation of the Milky Way}",
      journal = {\nat},
         year = 1999,
        month = nov,
       volume = {402},
       number = {6757},
        pages = {53-55},
          doi = {10.1038/46980},
archivePrefix = {arXiv},
       eprint = {astro-ph/9911041},
 primaryClass = {astro-ph},
       adsurl = {https://ui.adsabs.harvard.edu/abs/1999Natur.402...53H}
}

@ARTICLE{layden1995,
       author = {{Layden}, Andrew C.},
        title = "{The Metallicities and Kinematics of RR Lyrae Variables.II. Galactic Structure and Formation from Local Stars}",
      journal = {\aj},
         year = 1995,
        month = nov,
       volume = {110},
        pages = {2288},
          doi = {10.1086/117690},
       adsurl = {https://ui.adsabs.harvard.edu/abs/1995AJ....110.2288L}
}

@ARTICLE{prudil2024,
       author = {{Prudil}, Z. and {Kunder}, A. and {D{\'e}k{\'a}ny}, I. and {Koch-Hansen}, A.~J.},
        title = "{The Galactic bulge exploration. I. The period-absolute magnitude-metallicity relations for RR Lyrae stars for G$_{BP}$, V, G, G$_{RP}$, I, J, H, and K$_{s}$ passbands using Gaia DR3 parallaxes}",
      journal = {\aap},
         year = 2024,
        month = apr,
       volume = {684},
          eid = {A176},
        pages = {A176},
          doi = {10.1051/0004-6361/202347338},
archivePrefix = {arXiv},
       eprint = {2310.19438},
 primaryClass = {astro-ph.SR},
       adsurl = {https://ui.adsabs.harvard.edu/abs/2024A&A...684A.176P}
}

@ARTICLE{soszynski2019c,
       author = {{Soszy{\'n}ski}, I. and {Udalski}, A. and {Wrona}, M. and
         {Szyma{\'n}ski}, M.~K. and {Pietrukowicz}, P. and {Skowron}, J. and
         {Skowron}, D. and {Poleski}, R. and {Koz{\l}owski}, S. and
         {Mr{\'o}z}, P. and {Ulaczyk}, K. and {Rybicki}, K. and {Iwanek}, P. and
         {Gromadzki}, M.},
        title = "{Over 78 000 RR Lyrae Stars in the Galactic Bulge and Disk from the OGLE Survey}",
      journal = {\actaa},
         year = "2019",
        month = "Dec",
       volume = {69},
       number = {4},
        pages = {321-337},
          doi = {10.32023/0001-5237/69.4.2},
archivePrefix = {arXiv},
       eprint = {2001.00025},
 primaryClass = {astro-ph.SR},
       adsurl = {https://ui.adsabs.harvard.edu/abs/2019AcA....69..321S}
}

@ARTICLE{soszynski2019b,
       author = {{Soszy{\'n}ski}, I. and {Udalski}, A. and {Szyma{\'n}ski}, M.~K. and {Pietrukowicz}, P. and {Skowron}, J. and {Skowron}, D.~M. and {Poleski}, R. and {Koz{\l}owski}, S. and {Mr{\'o}z}, P. and {Ulaczyk}, K. and {Rybicki}, K. and {Iwanek}, P. and {Wrona}, M.},
        title = "{Final Release of the OGLE Collection of Cepheids and RR Lyrae Stars in the Magellanic System. The Outer Regions}",
      journal = {\actaa},
         year = 2019,
        month = jun,
       volume = {69},
       number = {2},
        pages = {87-99},
          doi = {10.32023/0001-5237/69.2.1},
archivePrefix = {arXiv},
       eprint = {1907.02073},
 primaryClass = {astro-ph.SR},
       adsurl = {https://ui.adsabs.harvard.edu/abs/2019AcA....69...87S}
}

@ARTICLE{dekany2018,
   author = {{D{\'e}k{\'a}ny}, I. and {Hajdu}, G. and {Grebel}, E.~K. and 
	{Catelan}, M. and {Elorrieta}, F. and {Eyheramendy}, S. and 
	{Majaess}, D. and {Jord{\'a}n}, A.},
    title = "{A Near-infrared RR Lyrae Census along the Southern Galactic Plane: The Milky Way{\rsquo}s Stellar Fossil Brought to Light}",
  journal = {\apj},
archivePrefix = "arXiv",
   eprint = {1804.01457},
 primaryClass = "astro-ph.SR",
     year = 2018,
    month = apr,
   volume = 857,
      eid = {54},
    pages = {54},
      doi = {10.3847/1538-4357/aab4fa},
   adsurl = {http://adsabs.harvard.edu/abs/2018ApJ...857...54D}
}

@ARTICLE{chadid2017,
   author = {{Chadid}, M. and {Sneden}, C. and {Preston}, G.~W.},
    title = "{Spectroscopic Comparison of Metal-rich RRab Stars of the Galactic Field with their Metal-poor Counterparts}",
  journal = {\apj},
archivePrefix = "arXiv",
   eprint = {1611.02368},
 primaryClass = "astro-ph.SR",
     year = 2017,
    month = feb,
   volume = 835,
      eid = {187},
    pages = {187},
      doi = {10.3847/1538-4357/835/2/187},
   adsurl = {http://adsabs.harvard.edu/abs/2017ApJ...835..187C}
}

@ARTICLE{sesar2012a,
       author = {{Sesar}, Branimir and {Cohen}, Judith G. and {Levitan}, David and {Grillmair}, Carl J. and {Juri{\'c}}, Mario and {Kirby}, Evan N. and {Laher}, Russ R. and {Ofek}, Eran O. and {Surace}, Jason A. and {Kulkarni}, Shrinivas R. and {Prince}, Thomas A.},
        title = "{Two Distant Halo Velocity Groups Discovered by the Palomar Transient Factory}",
      journal = {\apj},
         year = 2012,
        month = aug,
       volume = {755},
       number = {2},
          eid = {134},
        pages = {134},
          doi = {10.1088/0004-637X/755/2/134},
archivePrefix = {arXiv},
       eprint = {1206.0269},
 primaryClass = {astro-ph.GA},
       adsurl = {https://ui.adsabs.harvard.edu/abs/2012ApJ...755..134S}
}

@ARTICLE{gonzalez2012,
   author = {{Gonzalez}, O.~A. and {Rejkuba}, M. and {Zoccali}, M. and {Valenti}, E. and 
	{Minniti}, D. and {Schultheis}, M. and {Tobar}, R. and {Chen}, B.
	},
    title = "{Reddening and metallicity maps of the Milky Way bulge from VVV and 2MASS. II. The complete high resolution extinction map and implications for Galactic bulge studies}",
  journal = {\aap},
archivePrefix = "arXiv",
   eprint = {1204.4004},
     year = 2012,
    month = jul,
   volume = 543,
      eid = {A13},
    pages = {A13},
      doi = {10.1051/0004-6361/201219222},
   adsurl = {http://adsabs.harvard.edu/abs/2012A%26A...543A..13G}
}

@ARTICLE{mcmillan2017,
   author = {{McMillan}, P.~J.},
    title = "{The mass distribution and gravitational potential of the Milky Way}",
  journal = {\mnras},
archivePrefix = "arXiv",
   eprint = {1608.00971},
     year = 2017,
    month = feb,
   volume = 465,
    pages = {76-94},
      doi = {10.1093/mnras/stw2759},
   adsurl = {http://adsabs.harvard.edu/abs/2017MNRAS.465...76M}
}

@ARTICLE{walker1991a,
   author = {{Walker}, Alistair R. and {Terndrup}, Donald M.},
        title = "{The Metallicity of RR Lyrae Stars in Baade's Window}",
      journal = {\apj},
         year = 1991,
        month = sep,
       volume = {378},
        pages = {119},
          doi = {10.1086/170411},
       adsurl = {https://ui.adsabs.harvard.edu/abs/1991ApJ...378..119W}
}

@ARTICLE{butler1976,
   author = {{Butler}, D. and {Carbon}, D. and {Kraft}, R.~P.},
    title = "{Metal abundances of RR Lyrae variables in selected galactic star fields. I - Baade's window}",
  journal = {\apj},
     year = 1976,
    month = nov,
   volume = 210,
    pages = {120-123},
      doi = {10.1086/154809},
   adsurl = {http://adsabs.harvard.edu/abs/1976ApJ...210..120B}
}

@ARTICLE{butler1982,
       author = {{Butler}, D. and {Kemper}, E. and {Kraft}, R.~P. and {Suntzeff}, N.~B.},
        title = "{Metal abundances of RR Lyrae variables in selected galactic star fields. III. The Lick astrographic fields near the galactic anticenter.}",
      journal = {\aj},
         year = 1982,
        month = feb,
       volume = {87},
        pages = {353-359},
          doi = {10.1086/113106},
       adsurl = {https://ui.adsabs.harvard.edu/abs/1982AJ.....87..353B}
}

@ARTICLE{caceres2008,
       author = {{C{\'a}ceres}, C. and {Catelan}, M.},
        title = "{The Period-Luminosity Relation of RR Lyrae Stars in the SDSS Photometric System}",
      journal = {\apjs},
         year = 2008,
        month = nov,
       volume = {179},
       number = {1},
        pages = {242-248},
          doi = {10.1086/591231},
archivePrefix = {arXiv},
       eprint = {0805.3704},
 primaryClass = {astro-ph},
       adsurl = {https://ui.adsabs.harvard.edu/abs/2008ApJS..179..242C}
}

@ARTICLE{bono1997f,
       author = {{Bono}, Giuseppe and {Caputo}, Filippina and {Cassisi}, Santi and
         {Castellani}, Vittorio and {Marconi}, Marcella},
        title = "{Evolutionary and Pulsational Constraints for Super-Metal-rich Stars with Z = 0.04}",
      journal = {\apj},
         year = 1997,
        month = nov,
       volume = {489},
       number = {2},
        pages = {822-847},
          doi = {10.1086/304807},
       adsurl = {https://ui.adsabs.harvard.edu/abs/1997ApJ...489..822B}
}

@ARTICLE{baade58c,
   author = {{Baade}, W.},
    title = "{The Population of the Galactic Nucleus and the Evidence for the Presence of an Old Population Pervading the Whole Disk of our Galaxy}",
  journal = {Ricerche Astronomiche, Specola Vaticana, Proceedings of the conference sponsored by the Pontifical Academy of Science and the Vatican Observatory},
     year = 1958,
   volume = 5,
    pages = {303},
   adsurl = {http://adsabs.harvard.edu/abs/1958RA......5..303B}
}

@ARTICLE{sarajedini2024,
       author = {{Sarajedini}, Ata},
        title = "{The properties of RR Lyrae variable stars in the Local Group dwarf galaxy LGS-3}",
      journal = {\mnras},
         year = 2024,
        month = feb,
       volume = {527},
       number = {4},
        pages = {11751-11755},
          doi = {10.1093/mnras/stad3832},
       adsurl = {https://ui.adsabs.harvard.edu/abs/2024MNRAS.52711751S}
}

@ARTICLE{sarajedini2025,
       author = {{Sarajedini}, Ata},
        title = "{Color{\textendash}Magnitude Diagram of NGC 205 and Its RR Lyrae Variables}",
      journal = {\aj},
         year = 2025,
        month = mar,
       volume = {169},
       number = {3},
          eid = {154},
        pages = {154},
          doi = {10.3847/1538-3881/ada2eb},
       adsurl = {https://ui.adsabs.harvard.edu/abs/2025AJ....169..154S}
}

@ARTICLE{muraveva2025,
       author = {{Muraveva}, Tatiana and {Giannetti}, Andrea and {Clementini}, Gisella and {Garofalo}, Alessia and {Monti}, Lorenzo},
        title = "{Metallicity of RR Lyrae stars from the Gaia Data Release 3 catalogue computed with Machine Learning algorithms}",
      journal = {\mnras},
         year = 2025,
        month = jan,
       volume = {536},
       number = {3},
        pages = {2749-2769},
          doi = {10.1093/mnras/stae2679},
archivePrefix = {arXiv},
       eprint = {2407.05815},
 primaryClass = {astro-ph.SR},
       adsurl = {https://ui.adsabs.harvard.edu/abs/2025MNRAS.536.2749M}
}

@ARTICLE{pietrukowicz2015,
   author = {{Pietrukowicz}, P. and {Koz{\l}owski}, S. and {Skowron}, J. and 
	{Soszy{\'n}ski}, I. and {Udalski}, A. and {Poleski}, R. and 
	{Wyrzykowski}, {\L}. and {Szyma{\'n}ski}, M.~K. and {Pietrzy{\'n}ski}, G. and 
	{Ulaczyk}, K. and {Mr{\'o}z}, P. and {Skowron}, D.~M. and {Kubiak}, M.
	},
    title = "{Deciphering the 3D Structure of the Old Galactic Bulge from the OGLE RR Lyrae Stars}",
  journal = {\apj},
archivePrefix = "arXiv",
   eprint = {1412.4121},
     year = 2015,
    month = oct,
   volume = 811,
      eid = {113},
    pages = {113},
      doi = {10.1088/0004-637X/811/2/113},
   adsurl = {http://adsabs.harvard.edu/abs/2015ApJ...811..113P}
}

@ARTICLE{drake13a,
   author = {{Drake}, A.~J. and {Catelan}, M. and {Djorgovski}, S.~G. and 
	{Torrealba}, G. and {Graham}, M.~J. and {Belokurov}, V. and 
	{Koposov}, S.~E. and {Mahabal}, A. and {Prieto}, J.~L. and {Donalek}, C. and 
	{Williams}, R. and {Larson}, S. and {Christensen}, E. and {Beshore}, E.
	},
    title = "{Probing the Outer Galactic Halo with RR Lyrae from the Catalina Surveys}",
  journal = {\apj},
archivePrefix = "arXiv",
   eprint = {1211.2866},
     year = 2013,
    month = jan,
   volume = {763},
      eid = {32},
    pages = {32},
      doi = {10.1088/0004-637X/763/1/32},
   adsurl = {http://adsabs.harvard.edu/abs/2013ApJ...763...32D}
}

@ARTICLE{torrealba2015,
       author = {{Torrealba}, G. and {Catelan}, M. and {Drake}, A.~J. and {Djorgovski}, S.~G. and {McNaught}, R.~H. and {Belokurov}, V. and {Koposov}, S. and {Graham}, M.~J. and {Mahabal}, A. and {Larson}, S. and {Christensen}, E.},
        title = "{Discovery of {\ensuremath{\sim}}9000 new RR Lyrae in the southern Catalina surveys}",
      journal = {\mnras},
         year = 2015,
        month = jan,
       volume = {446},
       number = {3},
        pages = {2251-2266},
          doi = {10.1093/mnras/stu2274},
archivePrefix = {arXiv},
       eprint = {1410.7653},
 primaryClass = {astro-ph.GA},
       adsurl = {https://ui.adsabs.harvard.edu/abs/2015MNRAS.446.2251T}
}

@ARTICLE{fabrizio2019,
       author = {{Fabrizio}, M. and {Bono}, G. and {Braga}, V.~F. and {Magurno}, D. and
         {Marinoni}, S. and {Marrese}, P.~M. and {Ferraro}, I. and
         {Fiorentino}, G. and {Giuffrida}, G. and {Iannicola}, G. and
         {Monelli}, M. and {Altavilla}, G. and {Chaboyer}, B. and
         {Dall'Ora}, M. and {Gilligan}, C.~K. and {Layden}, A. and
         {Marengo}, M. and {Nonino}, M. and {Preston}, G.~W. and {Sesar}, B. and
         {Sneden}, C. and {Valenti}, E. and {Th{\'e}venin}, F. and {Zoccali}, E.},
        title = "{On the Use of Field RR Lyrae as Galactic Probes. I. The Oosterhoff Dichotomy Based on Fundamental Variables}",
      journal = {\apj},
         year = "2019",
        month = "Sep",
       volume = {882},
       number = {2},
          eid = {169},
        pages = {169},
          doi = {10.3847/1538-4357/ab3977},
archivePrefix = {arXiv},
       eprint = {1908.02064},
 primaryClass = {astro-ph.SR},
       adsurl = {https://ui.adsabs.harvard.edu/abs/2019ApJ...882..169F}
}

@ARTICLE{fiorentino2017b,
       author = {{Fiorentino}, G. and {Monelli}, M. and {Stetson}, P.~B. and {Bono}, G. and {Gallart}, C. and {Mart{\'\i}nez-V{\'a}zquez}, C.~E. and {Bernard}, E.~J. and {Massari}, D. and {Braga}, V.~F. and {Dall'Ora}, M.},
        title = "{Weak Galactic halo-Fornax dSph connection from RR Lyrae stars}",
      journal = {\aap},
         year = 2017,
        month = mar,
       volume = {599},
          eid = {A125},
        pages = {A125},
          doi = {10.1051/0004-6361/201629501},
archivePrefix = {arXiv},
       eprint = {1612.02991},
 primaryClass = {astro-ph.GA},
       adsurl = {https://ui.adsabs.harvard.edu/abs/2017A&A...599A.125F}
}

@article{mullen2021,
	doi = {10.3847/1538-4357/abefd4},
	url = {https://doi.org/10.3847/1538-4357/abefd4},
	year = 2021,
	month = {may},
	publisher = {American Astronomical Society},
	volume = {912},
	number = {2},
	pages = {144},
	author = {Joseph P. Mullen and Massimo Marengo and Clara E. Mart{\'{\i}}nez-V{\'{a}}zquez and Jillian R. Neeley and Giuseppe Bono and Massimo Dall'Ora and Brian Chaboyer and Fr{\'{e}}d{\'{e}}ric Th{\'{e}}venin and Vittorio F. Braga and Juliana Crestani and Michele Fabrizio and Giuliana Fiorentino and Christina K. Gilligan and Matteo Monelli and Peter B. Stetson},
	title = {Metallicity of Galactic {RR} Lyrae from Optical and Infrared Light Curves. I. Period{\textendash}Fourier{\textendash}Metallicity Relations for Fundamental-mode {RR} Lyrae},
	journal = {The Astrophysical Journal},
}

@ARTICLE{crestani2021a,
       author = {{Crestani}, J. and {Fabrizio}, M. and {Braga}, V.~F. and {Sneden}, C. and {Preston}, G. and {Ferraro}, I. and {Iannicola}, G. and {Bono}, G. and {Alves-Brito}, A. and {Nonino}, M. and {D'Orazi}, V. and {Inno}, L. and {Monelli}, M. and {Storm}, J. and {Altavilla}, G. and {Chaboyer}, B. and {Dall'Ora}, M. and {Fiorentino}, G. and {Gilligan}, C. and {Grebel}, E.~K. and {Lala}, H. and {Lemasle}, B. and {Marengo}, M. and {Marinoni}, S. and {Marrese}, P.~M. and {Mart{\'\i}nez-V{\'a}zquez}, C.~E. and {Matsunaga}, N. and {Mullen}, J.~P. and {Neeley}, J. and {Prudil}, Z. and {da Silva}, R. and {Stetson}, P.~B. and {Th{\'e}venin}, F. and {Valenti}, E. and {Walker}, A. and {Zoccali}, M.},
        title = "{On the Use of Field RR Lyrae as Galactic Probes. II. A New {\ensuremath{\Delta}}S Calibration to Estimate Their Metallicity}",
      journal = {\apj},
         year = 2021,
        month = feb,
       volume = {908},
       number = {1},
          eid = {20},
        pages = {20},
          doi = {10.3847/1538-4357/abd183},
archivePrefix = {arXiv},
       eprint = {2012.02284},
 primaryClass = {astro-ph.SR},
       adsurl = {https://ui.adsabs.harvard.edu/abs/2021ApJ...908...20C}
}

@ARTICLE{crestani2021b,
       author = {{Crestani}, J. and {Braga}, V.~F. and {Fabrizio}, M. and {Bono}, G. and {Sneden}, C. and {Preston}, G. and {Ferraro}, I. and {Iannicola}, G. and {Nonino}, M. and {Fiorentino}, G. and {Th{\'e}venin}, F. and {Lemasle}, B. and {Prudil}, Z. and {Alves-Brito}, A. and {Altavilla}, G. and {Chaboyer}, B. and {Dall'Ora}, M. and {D'Orazi}, V. and {Gilligan}, C. and {Grebel}, E.~K. and {Koch-Hansen}, A.~J. and {Lala}, H. and {Marengo}, M. and {Marinoni}, S. and {Marrese}, P.~M. and {Mart{\'\i}nez-V{\'a}zquez}, C. and {Matsunaga}, N. and {Monelli}, M. and {Mullen}, J.~P. and {Neeley}, J. and {da Silva}, R. and {Stetson}, P.~B. and {Salaris}, M. and {Storm}, J. and {Valenti}, E. and {Zoccali}, M.},
        title = "{On the Use of Field RR Lyrae as Galactic Probes. III. The {\ensuremath{\alpha}}-element Abundances}",
      journal = {\apj},
         year = 2021,
        month = jun,
       volume = {914},
       number = {1},
          eid = {10},
        pages = {10},
          doi = {10.3847/1538-4357/abfa23},
archivePrefix = {arXiv},
       eprint = {2104.08113},
 primaryClass = {astro-ph.GA},
       adsurl = {https://ui.adsabs.harvard.edu/abs/2021ApJ...914...10C}
}

@ARTICLE{vasiliev2019,
       author = {{Vasiliev}, Eugene},
        title = "{Proper motions and dynamics of the Milky Way globular cluster system from Gaia DR2}",
      journal = {\mnras},
         year = 2019,
        month = apr,
       volume = {484},
       number = {2},
        pages = {2832-2850},
          doi = {10.1093/mnras/stz171},
archivePrefix = {arXiv},
       eprint = {1807.09775},
 primaryClass = {astro-ph.GA},
       adsurl = {https://ui.adsabs.harvard.edu/abs/2019MNRAS.484.2832V}
}

@ARTICLE{clementini2023,
       author = {{Clementini}, G. and {Ripepi}, V. and {Garofalo}, A. and {Molinaro}, R. and {Muraveva}, T. and {Leccia}, S. and {Rimoldini}, L. and {Holl}, B. and {Jevardat de Fombelle}, G. and {Sartoretti}, P. and {Marchal}, O. and {Audard}, M. and {Nienartowicz}, K. and {Andrae}, R. and {Marconi}, M. and {Szabados}, L. and {Evans}, D.~W. and {Lecoeur-Taibi}, I. and {Mowlavi}, N. and {Musella}, I. and {Eyer}, L.},
        title = "{Gaia Data Release 3. Specific processing and validation of all-sky RR Lyrae and Cepheid stars: The RR Lyrae sample}",
      journal = {\aap},
         year = 2023,
        month = jun,
       volume = {674},
          eid = {A18},
        pages = {A18},
          doi = {10.1051/0004-6361/202243964},
archivePrefix = {arXiv},
       eprint = {2206.06278},
 primaryClass = {astro-ph.SR},
       adsurl = {https://ui.adsabs.harvard.edu/abs/2023A&A...674A..18C}
}

@ARTICLE{samus2017,
   author = {{Samus'}, N.~N. and {Kazarovets}, E.~V. and {Durlevich}, O.~V. and 
	{Kireeva}, N.~N. and {Pastukhova}, E.~N.},
    title = "{General catalogue of variable stars: Version GCVS 5.1}",
  journal = {Astronomy Reports},
     year = 2017,
    month = jan,
   volume = {61},
    pages = {80-88},
      doi = {10.1134/S1063772917010085},
   adsurl = {http://adsabs.harvard.edu/abs/2017ARep...61...80S}
}

@ARTICLE{nemec2011,
       author = {{Nemec}, J.~M. and {Smolec}, R. and {Benk{\H{o}}}, J.~M. and {Moskalik}, P. and {Kolenberg}, K. and {Szab{\'o}}, R. and {Kurtz}, D.~W. and {Bryson}, S. and {Guggenberger}, E. and {Chadid}, M. and {Jeon}, Y. -B. and {Kunder}, A. and {Layden}, A.~C. and {Kinemuchi}, K. and {Kiss}, L.~L. and {Poretti}, E. and {Christensen-Dalsgaard}, J. and {Kjeldsen}, H. and {Caldwell}, D. and {Ripepi}, V. and {Derekas}, A. and {Nuspl}, J. and {Mullally}, F. and {Thompson}, S.~E. and {Borucki}, W.~J.},
        title = "{Fourier analysis of non-Blazhko ab-type RR Lyrae stars observed with the Kepler space telescope}",
      journal = {\mnras},
         year = 2011,
        month = oct,
       volume = {417},
       number = {2},
        pages = {1022-1053},
          doi = {10.1111/j.1365-2966.2011.19317.x},
archivePrefix = {arXiv},
       eprint = {1106.6120},
 primaryClass = {astro-ph.SR},
       adsurl = {https://ui.adsabs.harvard.edu/abs/2011MNRAS.417.1022N}
}

@ARTICLE{lub1977,
       author = {{Lub}, J.},
        title = "{An atlas of light and colour curves of field RR Lyrae stars.}",
      journal = {\aaps},
         year = 1977,
        month = sep,
       volume = {29},
        pages = {345-378},
       adsurl = {https://ui.adsabs.harvard.edu/abs/1977A&AS...29..345L}
}

@ARTICLE{drake2017,
       author = {{Drake}, A.~J. and {Djorgovski}, S.~G. and {Catelan}, M. and
         {Graham}, M.~J. and {Mahabal}, A.~A. and {Larson}, S. and
         {Christensen}, E. and {Torrealba}, G. and {Beshore}, E. and
         {McNaught}, R.~H. and {Garradd}, G. and {Belokurov}, V. and
         {Koposov}, S.~E.},
        title = "{The Catalina Surveys Southern periodic variable star catalogue}",
      journal = {\mnras},
         year = 2017,
        month = aug,
       volume = {469},
       number = {3},
        pages = {3688-3712},
          doi = {10.1093/mnras/stx1085},
       adsurl = {https://ui.adsabs.harvard.edu/abs/2017MNRAS.469.3688D}
}

@ARTICLE{jayasinghe2019,
       author = {{Jayasinghe}, T. and {Stanek}, K.~Z. and {Kochanek}, C.~S. and
         {Shappee}, B.~J. and {Holoien}, T.~W. -S. and {Thompson}, Todd A. and
         {Prieto}, J.~L. and {Dong}, Subo and {Pawlak}, M. and {Pejcha}, O. and
         {Shields}, J.~V. and {Pojma{\'n}ski}, G. and {Otero}, S. and {Hurst}, N. and
         {Britt}, C.~A. and {Will}, D.},
        title = "{The ASAS-SN catalogue of variable stars III: variables in the southern TESS continuous viewing zone}",
      journal = {\mnras},
         year = 2019,
        month = may,
       volume = {485},
       number = {1},
        pages = {961-971},
          doi = {10.1093/mnras/stz444},
archivePrefix = {arXiv},
       eprint = {1901.00009},
 primaryClass = {astro-ph.SR},
       adsurl = {https://ui.adsabs.harvard.edu/abs/2019MNRAS.485..961J}
}

@ARTICLE{zinn2020,
       author = {{Zinn}, R. and {Chen}, X. and {Layden}, A.~C. and
         {Casetti-Dinescu}, D.~I.},
        title = "{Local RR Lyrae stars: native and alien}",
      journal = {\mnras},
         year = 2020,
        month = feb,
       volume = {492},
       number = {2},
        pages = {2161-2176},
          doi = {10.1093/mnras/stz3580},
archivePrefix = {arXiv},
       eprint = {1912.07686},
 primaryClass = {astro-ph.GA},
       adsurl = {https://ui.adsabs.harvard.edu/abs/2020MNRAS.492.2161Z}
}

@ARTICLE{wright2010,
   author = {{Wright}, E.~L. and {Eisenhardt}, P.~R.~M. and {Mainzer}, A.~K. and 
	{Ressler}, M.~E. and {Cutri}, R.~M. and {Jarrett}, T. and {Kirkpatrick}, J.~D. and 
	{Padgett}, D. and {McMillan}, R.~S. and {Skrutskie}, M. and 
	{Stanford}, S.~A. and {Cohen}, M. and {Walker}, R.~G. and {Mather}, J.~C. and 
	{Leisawitz}, D. and {Gautier}, III, T.~N. and {McLean}, I. and 
	{Benford}, D. and {Lonsdale}, C.~J. and {Blain}, A. and {Mendez}, B. and 
	{Irace}, W.~R. and {Duval}, V. and {Liu}, F. and {Royer}, D. and 
	{Heinrichsen}, I. and {Howard}, J. and {Shannon}, M. and {Kendall}, M. and 
	{Walsh}, A.~L. and {Larsen}, M. and {Cardon}, J.~G. and {Schick}, S. and 
	{Schwalm}, M. and {Abid}, M. and {Fabinsky}, B. and {Naes}, L. and 
	{Tsai}, C.-W.},
    title = "{The Wide-field Infrared Survey Explorer (WISE): Mission Description and Initial On-orbit Performance}",
  journal = {\aj},
archivePrefix = "arXiv",
   eprint = {1008.0031},
 primaryClass = "astro-ph.IM",
     year = 2010,
    month = dec,
   volume = 140,
      eid = {1868-1881},
    pages = {1868-1881},
      doi = {10.1088/0004-6256/140/6/1868},
   adsurl = {http://adsabs.harvard.edu/abs/2010AJ....140.1868W}
}

@ARTICLE{drake2009,
       author = {{Drake}, A.~J. and {Djorgovski}, S.~G. and {Mahabal}, A. and {Beshore}, E. and {Larson}, S. and {Graham}, M.~J. and {Williams}, R. and {Christensen}, E. and {Catelan}, M. and {Boattini}, A. and {Gibbs}, A. and {Hill}, R. and {Kowalski}, R.},
        title = "{First Results from the Catalina Real-Time Transient Survey}",
      journal = {\apj},
         year = 2009,
        month = may,
       volume = {696},
       number = {1},
        pages = {870-884},
          doi = {10.1088/0004-637X/696/1/870},
archivePrefix = {arXiv},
       eprint = {0809.1394},
 primaryClass = {astro-ph},
       adsurl = {https://ui.adsabs.harvard.edu/abs/2009ApJ...696..870D}
}

@ARTICLE{neeley2019,
       author = {{Neeley}, Jillian R. and {Marengo}, Massimo and {Freedman}, Wendy L. and {Madore}, Barry F. and {Beaton}, Rachael L. and {Hatt}, Dylan and {Hoyt}, Taylor and {Monson}, Andrew J. and {Rich}, Jeffrey A. and {Sarajedini}, Ata and {Seibert}, Mark and {Scowcroft}, Victoria},
        title = "{Standard Galactic field RR Lyrae II: a Gaia DR2 calibration of the period-Wesenheit-metallicity relation}",
      journal = {\mnras},
         year = 2019,
        month = dec,
       volume = {490},
       number = {3},
        pages = {4254-4270},
          doi = {10.1093/mnras/stz2814},
archivePrefix = {arXiv},
       eprint = {1910.01773},
 primaryClass = {astro-ph.SR},
       adsurl = {https://ui.adsabs.harvard.edu/abs/2019MNRAS.490.4254N}
}

@ARTICLE{mcmahon13,
   author = {{McMahon}, R.~G. and {Banerji}, M. and {Gonzalez}, E. and {Koposov}, S.~E. and 
	{Bejar}, V.~J. and {Lodieu}, N. and {Rebolo}, R. and {VHS Collaboration}},
    title = "{First Scientific Results from the VISTA Hemisphere Survey (VHS)}",
  journal = {The Messenger},
     year = 2013,
    month = dec,
   volume = {154},
    pages = {35-37},
   adsurl = {http://adsabs.harvard.edu/abs/2013Msngr.154...35M}
}

@dataset{mcmahon2021,
       author = {{McMahon}, R.~G. and {Banerji}, M. and {Gonzalez}, E. and {Koposov}, S.~E. and {Bejar}, V.~J. and {Lodieu}, N. and {Rebolo}, R. and {VHS Collaboration}},
        title = "{VizieR Online Data Catalog: The VISTA Hemisphere Survey (VHS) catalog DR5 (McMahon+, 2020)}",
 howpublished = {VizieR On-line Data Catalog: II/367.  Originally published in: 2013Msngr.154...35M},
         year = 2021,
        month = jan,
          eid = {II/367},
       adsurl = {https://ui.adsabs.harvard.edu/abs/2021yCat.2367....0M}
}

@ARTICLE{skrutskie2006,
   author = {{Skrutskie}, M.~F. and {Cutri}, R.~M. and {Stiening}, R. and 
	{Weinberg}, M.~D. and {Schneider}, S. and {Carpenter}, J.~M. and 
	{Beichman}, C. and {Capps}, R. and {Chester}, T. and {Elias}, J. and 
	{Huchra}, J. and {Liebert}, J. and {Lonsdale}, C. and {Monet}, D.~G. and 
	{Price}, S. and {Seitzer}, P. and {Jarrett}, T. and {Kirkpatrick}, J.~D. and 
	{Gizis}, J.~E. and {Howard}, E. and {Evans}, T. and {Fowler}, J. and 
	{Fullmer}, L. and {Hurt}, R. and {Light}, R. and {Kopan}, E.~L. and 
	{Marsh}, K.~A. and {McCallon}, H.~L. and {Tam}, R. and {Van Dyk}, S. and 
	{Wheelock}, S.},
    title = "{The Two Micron All Sky Survey (2MASS)}",
  journal = {\aj},
     year = 2006,
    month = feb,
   volume = 131,
    pages = {1163-1183},
      doi = {10.1086/498708},
   adsurl = {http://adsabs.harvard.edu/abs/2006AJ....131.1163S}
}

@ARTICLE{warren2007,
       author = {{Warren}, S.~J. and {Cross}, N.~J.~G. and {Dye}, S. and {Hambly}, N.~C. and {Almaini}, O. and {Edge}, A.~C. and {Hewett}, P.~C. and {Hodgkin}, S.~T. and {Irwin}, M.~J. and {Jameson}, R.~F. and {Lawrence}, A. and {Lucas}, P.~W. and {Mortlock}, D.~J. and {Adamson}, A.~J. and {Bryant}, J. and {Collins}, R.~S. and {Davis}, C.~J. and {Emerson}, J.~P. and {Evans}, D.~W. and {Gonzales-Solares}, E.~A. and {Hirst}, P. and {Kerr}, T.~H. and {Lewis}, J.~R. and {Mann}, R.~G. and {Rawlings}, M.~G. and {Read}, M.~A. and {Riello}, M. and {Sutorius}, E.~T.~W. and {Varricatt}, W.~P.},
        title = "{The UKIRT Infrared Deep Sky Survey Second Data Release}",
      journal = {arXiv e-prints},
         year = 2007,
        month = mar,
          eid = {astro-ph/0703037},
        pages = {astro-ph/0703037},
          doi = {10.48550/arXiv.astro-ph/0703037},
archivePrefix = {arXiv},
       eprint = {astro-ph/0703037},
 primaryClass = {astro-ph},
       adsurl = {https://ui.adsabs.harvard.edu/abs/2007astro.ph..3037W}
}

@ARTICLE{lucas2008,
       author = {{Lucas}, P.~W. and {Hoare}, M.~G. and {Longmore}, A. and {Schr{\"o}der}, A.~C. and {Davis}, C.~J. and {Adamson}, A. and {Bandyopadhyay}, R.~M. and {de Grijs}, R. and {Smith}, M. and {Gosling}, A. and {Mitchison}, S. and {G{\'a}sp{\'a}r}, A. and {Coe}, M. and {Tamura}, M. and {Parker}, Q. and {Irwin}, M. and {Hambly}, N. and {Bryant}, J. and {Collins}, R.~S. and {Cross}, N. and {Evans}, D.~W. and {Gonzalez-Solares}, E. and {Hodgkin}, S. and {Lewis}, J. and {Read}, M. and {Riello}, M. and {Sutorius}, E.~T.~W. and {Lawrence}, A. and {Drew}, J.~E. and {Dye}, S. and {Thompson}, M.~A.},
        title = "{The UKIDSS Galactic Plane Survey}",
      journal = {\mnras},
         year = 2008,
        month = nov,
       volume = {391},
       number = {1},
        pages = {136-163},
          doi = {10.1111/j.1365-2966.2008.13924.x},
archivePrefix = {arXiv},
       eprint = {0712.0100},
 primaryClass = {astro-ph},
       adsurl = {https://ui.adsabs.harvard.edu/abs/2008MNRAS.391..136L}
}

@ARTICLE{dye2018,
       author = {{Dye}, S. and {Lawrence}, A. and {Read}, M.~A. and {Fan}, X. and {Kerr}, T. and {Varricatt}, W. and {Furnell}, K.~E. and {Edge}, A.~C. and {Irwin}, M. and {Hambly}, N. and {Lucas}, P. and {Almaini}, O. and {Chambers}, K. and {Green}, R. and {Hewett}, P. and {Liu}, M.~C. and {McGreer}, I. and {Best}, W. and {Zhang}, Z. and {Sutorius}, E. and {Froebrich}, D. and {Magnier}, E. and {Hasinger}, G. and {Lederer}, S.~M. and {Bold}, M. and {Tedds}, J.~A.},
        title = "{The UKIRT Hemisphere Survey: definition and J-band data release}",
      journal = {\mnras},
         year = 2018,
        month = feb,
       volume = {473},
       number = {4},
        pages = {5113-5125},
          doi = {10.1093/mnras/stx2622},
archivePrefix = {arXiv},
       eprint = {1707.09975},
 primaryClass = {astro-ph.IM},
       adsurl = {https://ui.adsabs.harvard.edu/abs/2018MNRAS.473.5113D}
}

@ARTICLE{layden2019,
       author = {{Layden}, Andrew C. and {Tiede}, Glenn P. and {Chaboyer}, Brian and {Bunner}, Curtis and {Smitka}, Michael T.},
        title = "{Infrared K-band Photometry of Field RR Lyrae Variable Stars}",
      journal = {\aj},
         year = 2019,
        month = sep,
       volume = {158},
       number = {3},
          eid = {105},
        pages = {105},
          doi = {10.3847/1538-3881/ab2e10},
archivePrefix = {arXiv},
       eprint = {1907.04920},
 primaryClass = {astro-ph.GA},
       adsurl = {https://ui.adsabs.harvard.edu/abs/2019AJ....158..105L}
}

@ARTICLE{preston1959,
   author = {{Preston}, G.~W.},
    title = "{A Spectroscopic Study of the RR Lyrae Stars.}",
  journal = {\apj},
     year = 1959,
    month = sep,
   volume = {130},
    pages = {507},
      doi = {10.1086/146743},
   adsurl = {http://adsabs.harvard.edu/abs/1959ApJ...130..507P}
}

@ARTICLE{preston1961,
       author = {{Preston}, George W.},
        title = "{A Coarse Analysis of Three RR Lyrae Stars.}",
      journal = {\apj},
         year = 1961,
        month = sep,
       volume = {134},
        pages = {633},
          doi = {10.1086/147185},
       adsurl = {https://ui.adsabs.harvard.edu/abs/1961ApJ...134..633P}
}

@ARTICLE{ivezic2008a,
       author = {{Ivezi{\'c}}, {\v{Z}}eljko and {Sesar}, Branimir and {Juri{\'c}}, Mario and {Bond}, Nicholas and {Dalcanton}, Julianne and {Rockosi}, Constance M. and {Yanny}, Brian and {Newberg}, Heidi J. and {Beers}, Timothy C. and {Allende Prieto}, Carlos and {Wilhelm}, Ron and {Lee}, Young Sun and {Sivarani}, Thirupathi and {Norris}, John E. and {Bailer-Jones}, Coryn A.~L. and {Re Fiorentin}, Paola and {Schlegel}, David and {Uomoto}, Alan and {Lupton}, Robert H. and {Knapp}, Gillian R. and {Gunn}, James E. and {Covey}, Kevin R. and {Allyn Smith}, J. and {Miknaitis}, Gajus and {Doi}, Mamoru and {Tanaka}, Masayuki and {Fukugita}, Masataka and {Kent}, Steve and {Finkbeiner}, Douglas and {Munn}, Jeffrey A. and {Pier}, Jeffrey R. and {Quinn}, Tom and {Hawley}, Suzanne and {Anderson}, Scott and {Kiuchi}, Furea and {Chen}, Alex and {Bushong}, James and {Sohi}, Harkirat and {Haggard}, Daryl and {Kimball}, Amy and {Barentine}, John and {Brewington}, Howard and {Harvanek}, Mike and {Kleinman}, Scott and {Krzesinski}, Jurek and {Long}, Dan and {Nitta}, Atsuko and {Snedden}, Stephanie and {Lee}, Brian and {Harris}, Hugh and {Brinkmann}, Jonathan and {Schneider}, Donald P. and {York}, Donald G.},
        title = "{The Milky Way Tomography with SDSS. II. Stellar Metallicity}",
      journal = {\apj},
         year = 2008,
        month = sep,
       volume = {684},
       number = {1},
        pages = {287-325},
          doi = {10.1086/589678},
archivePrefix = {arXiv},
       eprint = {0804.3850},
 primaryClass = {astro-ph},
       adsurl = {https://ui.adsabs.harvard.edu/abs/2008ApJ...684..287I}
}

@ARTICLE{zolotov09,
   author = {{Zolotov}, A. and {Willman}, B. and {Brooks}, A.~M. and {Governato}, F. and 
	{Brook}, C.~B. and {Hogg}, D.~W. and {Quinn}, T. and {Stinson}, G.
	},
    title = "{The Dual Origin of Stellar Halos}",
  journal = {\apj},
archivePrefix = "arXiv",
   eprint = {0904.3333},
 primaryClass = "astro-ph.GA",
     year = 2009,
    month = sep,
   volume = 702,
    pages = {1058-1067},
      doi = {10.1088/0004-637X/702/2/1058},
   adsurl = {http://adsabs.harvard.edu/abs/2009ApJ...702.1058Z}
}

@ARTICLE{zolotov10,
       author = {{Zolotov}, Adi and {Willman}, Beth and {Brooks}, Alyson M. and {Governato}, Fabio and {Hogg}, David W. and {Shen}, Sijing and {Wadsley}, James},
        title = "{The Dual Origin of Stellar Halos. II. Chemical Abundances as Tracers of Formation History}",
      journal = {\apj},
         year = 2010,
        month = sep,
       volume = {721},
       number = {1},
        pages = {738-743},
          doi = {10.1088/0004-637X/721/1/738},
archivePrefix = {arXiv},
       eprint = {1004.3789},
 primaryClass = {astro-ph.GA},
       adsurl = {https://ui.adsabs.harvard.edu/abs/2010ApJ...721..738Z}
}

@ARTICLE{martinezvazquez16a,
   author = {{Mart{\'{\i}}nez-V{\'a}zquez}, C.~E. and {Monelli}, M. and {Gallart}, C. and 
	{Bono}, G. and {Bernard}, E.~J. and {Stetson}, P.~B. and {Ferraro}, I. and 
	{Walker}, A.~R. and {Dall'Ora}, M. and {Fiorentino}, G. and 
	{Iannicola}, G.},
    title = "{Probing the early chemical evolution of the Sculptor dSph with purely old stellar tracers}",
  journal = {\mnras},
archivePrefix = "arXiv",
   eprint = {1605.02768},
     year = 2016,
    month = sep,
   volume = 461,
    pages = {L41-L45},
      doi = {10.1093/mnrasl/slw093},
   adsurl = {http://adsabs.harvard.edu/abs/2016MNRAS.461L..41M}
}

@ARTICLE{fiorentino15a,
   author = {{Fiorentino}, G. and {Bono}, G. and {Monelli}, M. and {Stetson}, P.~B. and 
	{Tolstoy}, E. and {Gallart}, C. and {Salaris}, M. and {Mart{\'{\i}}nez-V{\'a}zquez}, C.~E. and 
	{Bernard}, E.~J.},
    title = "{Weak Galactic Halo-Dwarf Spheroidal Connection from RR Lyrae Stars}",
  journal = {\apjl},
archivePrefix = "arXiv",
   eprint = {1411.7300},
 primaryClass = "astro-ph.SR",
     year = 2015,
    month = jan,
   volume = {798},
      eid = {L12},
    pages = {L12},
      doi = {10.1088/2041-8205/798/1/L12},
   adsurl = {http://adsabs.harvard.edu/abs/2015ApJ...798L..12F}
}

@ARTICLE{fiorentino15b,
   author = {{Fiorentino}, G. and {Marconi}, M. and {Bono}, G. and {Dalessandro}, E. and 
	{Ferraro}, F.~R. and {Lanzoni}, B. and {Lovisi}, L. and {Mucciarelli}, A.
	},
    title = "{Blue Straggler Masses from Pulsation Properties. II. Topology of the Instability Strip}",
  journal = {\apj},
archivePrefix = "arXiv",
   eprint = {1507.06603},
 primaryClass = "astro-ph.SR",
     year = 2015,
    month = sep,
   volume = {810},
      eid = {15},
    pages = {15},
      doi = {10.1088/0004-637X/810/1/15},
   adsurl = {http://adsabs.harvard.edu/abs/2015ApJ...810...15F}
}

@ARTICLE{dambis2013,
       author = {{Dambis}, A.~K. and {Berdnikov}, L.~N. and {Kniazev}, A.~Y. and {Kravtsov}, V.~V. and {Rastorguev}, A.~S. and {Sefako}, R. and {Vozyakova}, O.~V.},
        title = "{RR Lyrae variables: visual and infrared luminosities, intrinsic colours and kinematics}",
      journal = {\mnras},
         year = 2013,
        month = nov,
       volume = {435},
       number = {4},
        pages = {3206-3220},
          doi = {10.1093/mnras/stt1514},
archivePrefix = {arXiv},
       eprint = {1308.4727},
 primaryClass = {astro-ph.GA},
       adsurl = {https://ui.adsabs.harvard.edu/abs/2013MNRAS.435.3206D}
}

@ARTICLE{dambis14,
   author = {{Dambis}, A.~K. and {Rastorguev}, A.~S. and {Zabolotskikh}, M.~V.
	},
    title = "{Mid-infrared period-luminosity relations for globular cluster RR Lyrae}",
  journal = {\mnras},
archivePrefix = "arXiv",
   eprint = {1401.5523},
 primaryClass = "astro-ph.GA",
     year = 2014,
    month = apr,
   volume = 439,
    pages = {3765-3774},
      doi = {10.1093/mnras/stu226},
   adsurl = {http://adsabs.harvard.edu/abs/2014MNRAS.439.3765D}
}

@ARTICLE{vargas14,
       author = {{Vargas}, Luis C. and {Gilbert}, Karoline M. and {Geha}, Marla and {Tollerud}, Erik J. and {Kirby}, Evan N. and {Guhathakurta}, Puragra},
        title = "{[{\ensuremath{\alpha}}/Fe] Abundances of Four Outer M31 Halo Stars}",
      journal = {\apjl},
         year = 2014,
        month = dec,
       volume = {797},
       number = {1},
          eid = {L2},
        pages = {L2},
          doi = {10.1088/2041-8205/797/1/L2},
archivePrefix = {arXiv},
       eprint = {1410.3475},
 primaryClass = {astro-ph.GA},
       adsurl = {https://ui.adsabs.harvard.edu/abs/2014ApJ...797L...2V}
}

@ARTICLE{nissen2010,
       author = {{Nissen}, P.~E. and {Schuster}, W.~J.},
        title = "{Two distinct halo populations in the solar neighborhood. Evidence from stellar abundance ratios and kinematics}",
      journal = {\aap},
         year = 2010,
        month = feb,
       volume = {511},
          eid = {L10},
        pages = {L10},
          doi = {10.1051/0004-6361/200913877},
archivePrefix = {arXiv},
       eprint = {1002.4514},
 primaryClass = {astro-ph.GA},
       adsurl = {https://ui.adsabs.harvard.edu/abs/2010A&A...511L..10N}
}

@ARTICLE{zoccali2004,
       author = {{Zoccali}, M. and {Barbuy}, B. and {Hill}, V. and {Ortolani}, S. and {Renzini}, A. and {Bica}, E. and {Momany}, Y. and {Pasquini}, L. and {Minniti}, D. and {Rich}, R.~M.},
        title = "{The metal content of the bulge globular cluster NGC 6528}",
      journal = {\aap},
         year = 2004,
        month = aug,
       volume = {423},
        pages = {507-516},
          doi = {10.1051/0004-6361:20041014},
archivePrefix = {arXiv},
       eprint = {astro-ph/0405475},
 primaryClass = {astro-ph},
       adsurl = {https://ui.adsabs.harvard.edu/abs/2004A&A...423..507Z}
}

@ARTICLE{mainzer2011,
       author = {{Mainzer}, A. and {Bauer}, J. and {Grav}, T. and {Masiero}, J. and
         {Cutri}, R.~M. and {Dailey}, J. and {Eisenhardt}, P. and
         {McMillan}, R.~S. and {Wright}, E. and {Walker}, R. and {Jedicke}, R. and
         {Spahr}, T. and {Tholen}, D. and {Alles}, R. and {Beck}, R. and {Brand
        enburg}, H. and {Conrow}, T. and {Evans}, T. and {Fowler}, J. and
         {Jarrett}, T. and {Marsh}, K. and {Masci}, F. and {McCallon}, H. and
         {Wheelock}, S. and {Wittman}, M. and {Wyatt}, P. and {DeBaun}, E. and
         {Elliott}, G. and {Elsbury}, D. and {Gautier}, T., IV and
         {Gomillion}, S. and {Leisawitz}, D. and {Maleszewski}, C. and
         {Micheli}, M. and {Wilkins}, A.},
        title = "{Preliminary Results from NEOWISE: An Enhancement to the Wide-field Infrared Survey Explorer for Solar System Science}",
      journal = {\apj},
         year = 2011,
        month = apr,
       volume = {731},
       number = {1},
          eid = {53},
        pages = {53},
          doi = {10.1088/0004-637X/731/1/53},
archivePrefix = {arXiv},
       eprint = {1102.1996},
 primaryClass = {astro-ph.EP},
       adsurl = {https://ui.adsabs.harvard.edu/abs/2011ApJ...731...53M}
}

@MISC{cutri2013,
       author = {{Cutri}, R.~M. and {Wright}, E.~L. and {Conrow}, T. and {Fowler}, J.~W. and {Eisenhardt}, P.~R.~M. and {Grillmair}, C. and {Kirkpatrick}, J.~D. and {Masci}, F. and {McCallon}, H.~L. and {Wheelock}, S.~L. and {Fajardo-Acosta}, S. and {Yan}, L. and {Benford}, D. and {Harbut}, M. and {Jarrett}, T. and {Lake}, S. and {Leisawitz}, D. and {Ressler}, M.~E. and {Stanford}, S.~A. and {Tsai}, C.~W. and {Liu}, F. and {Helou}, G. and {Mainzer}, A. and {Gettings}, D. and {Gonzalez}, A. and {Hoffman}, D. and {Marsh}, K.~A. and {Padgett}, D. and {Skrutskie}, M.~F. and {Beck}, R.~P. and {Papin}, M. and {Wittman}, M.},
        title = "{Explanatory Supplement to the AllWISE Data Release Products}",
 howpublished = {Explanatory Supplement to the AllWISE Data Release Products, by R. M. Cutri et al.},
         year = 2013,
        month = nov,
        pages = {1},
       adsurl = {https://ui.adsabs.harvard.edu/abs/2013wise.rept....1C}
}

@ARTICLE{searle78,
   author = {{Searle}, L. and {Zinn}, R.},
    title = "{Compositions of halo clusters and the formation of the galactic halo}",
  journal = {\apj},
     year = 1978,
    month = oct,
   volume = 225,
    pages = {357-379},
      doi = {10.1086/156499},
   adsurl = {http://adsabs.harvard.edu/abs/1978ApJ...225..357S}
}

@ARTICLE{brusadin13,
   author = {{Brusadin}, G. and {Matteucci}, F. and {Romano}, D.},
    title = "{Modeling the chemical evolution of the Galaxy halo}",
  journal = {\aap},
archivePrefix = "arXiv",
   eprint = {1304.4385},
 primaryClass = "astro-ph.GA",
     year = 2013,
    month = jun,
   volume = 554,
      eid = {A135},
    pages = {A135},
      doi = {10.1051/0004-6361/201220884},
   adsurl = {http://adsabs.harvard.edu/abs/2013A%26A...554A.135B}
}

@ARTICLE{tissera14,
       author = {{Tissera}, Patricia B. and {Beers}, Timothy C. and {Carollo}, Daniela and {Scannapieco}, Cecilia},
        title = "{Stellar haloes in Milky Way mass galaxies: from the inner to the outer haloes}",
      journal = {\mnras},
         year = 2014,
        month = apr,
       volume = {439},
       number = {3},
        pages = {3128-3138},
          doi = {10.1093/mnras/stu181},
archivePrefix = {arXiv},
       eprint = {1309.3609},
 primaryClass = {astro-ph.CO},
       adsurl = {https://ui.adsabs.harvard.edu/abs/2014MNRAS.439.3128T}
}

@BOOK{baade1963,
       author = {{Baade}, Walter},
        title = "{Evolution of Stars and Galaxies}",
         year = 1963,
          doi = {10.4159/harvard.9780674280311},
       adsurl = {https://ui.adsabs.harvard.edu/abs/1963esg..book.....B}
}

@ARTICLE{steinmetz2006,
       author = {{Steinmetz}, M. and {Zwitter}, T. and {Siebert}, A. and {Watson}, F.~G. and {Freeman}, K.~C. and {Munari}, U. and {Campbell}, R. and {Williams}, M. and {Seabroke}, G.~M. and {Wyse}, R.~F.~G. and {Parker}, Q.~A. and {Bienaym{\'e}}, O. and {Roeser}, S. and {Gibson}, B.~K. and {Gilmore}, G. and {Grebel}, E.~K. and {Helmi}, A. and {Navarro}, J.~F. and {Burton}, D. and {Cass}, C.~J.~P. and {Dawe}, J.~A. and {Fiegert}, K. and {Hartley}, M. and {Russell}, K.~S. and {Saunders}, W. and {Enke}, H. and {Bailin}, J. and {Binney}, J. and {Bland-Hawthorn}, J. and {Boeche}, C. and {Dehnen}, W. and {Eisenstein}, D.~J. and {Evans}, N.~W. and {Fiorucci}, M. and {Fulbright}, J.~P. and {Gerhard}, O. and {Jauregi}, U. and {Kelz}, A. and {Mijovi{\'c}}, L. and {Minchev}, I. and {Parmentier}, G. and {Pe{\~n}arrubia}, J. and {Quillen}, A.~C. and {Read}, M.~A. and {Ruchti}, G. and {Scholz}, R. -D. and {Siviero}, A. and {Smith}, M.~C. and {Sordo}, R. and {Veltz}, L. and {Vidrih}, S. and {von Berlepsch}, R. and {Boyle}, B.~J. and {Schilbach}, E.},
        title = "{The Radial Velocity Experiment (RAVE): First Data Release}",
      journal = {\aj},
         year = 2006,
        month = oct,
       volume = {132},
       number = {4},
        pages = {1645-1668},
          doi = {10.1086/506564},
archivePrefix = {arXiv},
       eprint = {astro-ph/0606211},
 primaryClass = {astro-ph},
       adsurl = {https://ui.adsabs.harvard.edu/abs/2006AJ....132.1645S}
}

@ARTICLE{steinmetz2020a,
       author = {{Steinmetz}, Matthias and {Matijevi{\v{c}}}, Gal and {Enke}, Harry and {Zwitter}, Toma{\v{z}} and {Guiglion}, Guillaume and {McMillan}, Paul J. and {Kordopatis}, Georges and {Valentini}, Marica and {Chiappini}, Cristina and {Casagrande}, Luca and {Wojno}, Jennifer and {Anguiano}, Borja and {Bienaym{\'e}}, Olivier and {Bijaoui}, Albert and {Binney}, James and {Burton}, Donna and {Cass}, Paul and {de Laverny}, Patrick and {Fiegert}, Kristin and {Freeman}, Kenneth and {Fulbright}, Jon P. and {Gibson}, Brad K. and {Gilmore}, Gerard and {Grebel}, Eva K. and {Helmi}, Amina and {Kunder}, Andrea and {Munari}, Ulisse and {Navarro}, Julio F. and {Parker}, Quentin and {Ruchti}, Gregory R. and {Recio-Blanco}, Alejandra and {Reid}, Warren and {Seabroke}, George M. and {Siviero}, Alessandro and {Siebert}, Arnaud and {Stupar}, Milorad and {Watson}, Fred and {Williams}, Mary E.~K. and {Wyse}, Rosemary F.~G. and {Anders}, Friedrich and {Antoja}, Teresa and {Birko}, Danijela and {Bland-Hawthorn}, Joss and {Bossini}, Diego and {Garc{\'\i}a}, Rafael A. and {Carrillo}, Ismael and {Chaplin}, William J. and {Elsworth}, Yvonne and {Famaey}, Benoit and {Gerhard}, Ortwin and {Jofre}, Paula and {Just}, Andreas and {Mathur}, Savita and {Miglio}, Andrea and {Minchev}, Ivan and {Monari}, Giacomo and {Mosser}, Benoit and {Ritter}, Andreas and {Rodrigues}, Thaise S. and {Scholz}, Ralf-Dieter and {Sharma}, Sanjib and {Sysoliatina}, Kseniia and {RAVE Collaboration}},
        title = "{The Sixth Data Release of the Radial Velocity Experiment (RAVE). I. Survey Description, Spectra, and Radial Velocities}",
      journal = {\aj},
         year = 2020,
        month = aug,
       volume = {160},
       number = {2},
          eid = {82},
        pages = {82},
          doi = {10.3847/1538-3881/ab9ab9},
archivePrefix = {arXiv},
       eprint = {2002.04377},
 primaryClass = {astro-ph.SR},
       adsurl = {https://ui.adsabs.harvard.edu/abs/2020AJ....160...82S}
}

@ARTICLE{steinmetz2020b,
       author = {{Steinmetz}, Matthias and {Guiglion}, Guillaume and {McMillan}, Paul J. and {Matijevi{\v{c}}}, Gal and {Enke}, Harry and {Kordopatis}, Georges and {Zwitter}, Toma{\v{z}} and {Valentini}, Marica and {Chiappini}, Cristina and {Casagrande}, Luca and {Wojno}, Jennifer and {Anguiano}, Borja and {Bienaym{\'e}}, Olivier and {Bijaoui}, Albert and {Binney}, James and {Burton}, Donna and {Cass}, Paul and {de Laverny}, Patrick and {Fiegert}, Kristin and {Freeman}, Kenneth and {Fulbright}, Jon P. and {Gibson}, Brad K. and {Gilmore}, Gerard and {Grebel}, Eva K. and {Helmi}, Amina and {Kunder}, Andrea and {Munari}, Ulisse and {Navarro}, Julio F. and {Parker}, Quentin and {Ruchti}, Gregory R. and {Recio-Blanco}, Alejandra and {Reid}, Warren and {Seabroke}, George M. and {Siviero}, Alessandro and {Siebert}, Arnaud and {Stupar}, Milorad and {Watson}, Fred and {Williams}, Mary E.~K. and {Wyse}, Rosemary F.~G. and {Anders}, Friedrich and {Antoja}, Teresa and {Birko}, Danijela and {Bland-Hawthorn}, Joss and {Bossini}, Diego and {Garc{\'\i}a}, Rafael A. and {Carrillo}, Ismael and {Chaplin}, William J. and {Elsworth}, Yvonne and {Famaey}, Benoit and {Gerhard}, Ortwin and {Jofre}, Paula and {Just}, Andreas and {Mathur}, Savita and {Miglio}, Andrea and {Minchev}, Ivan and {Monari}, Giacomo and {Mosser}, Benoit and {Ritter}, Andreas and {Rodrigues}, Thaise S. and {Scholz}, Ralf-Dieter and {Sharma}, Sanjib and {Sysoliatina}, Kseniia and {RAVE Collaboration}},
        title = "{The Sixth Data Release of the Radial Velocity Experiment (RAVE). II. Stellar Atmospheric Parameters, Chemical Abundances, and Distances}",
      journal = {\aj},
         year = 2020,
        month = aug,
       volume = {160},
       number = {2},
          eid = {83},
        pages = {83},
          doi = {10.3847/1538-3881/ab9ab8},
archivePrefix = {arXiv},
       eprint = {2002.04512},
 primaryClass = {astro-ph.SR},
       adsurl = {https://ui.adsabs.harvard.edu/abs/2020AJ....160...83S}
}

@ARTICLE{sesar13b,
       author = {{Sesar}, Branimir and {Grillmair}, Carl J. and {Cohen}, Judith G. and
         {Bellm}, Eric C. and {Bhalerao}, Varun B. and {Levitan}, David and
         {Laher}, Russ R. and {Ofek}, Eran O. and {Surace}, Jason A. and
         {Tang}, Sumin and {Waszczak}, Adam and {Kulkarni}, Shrinivas R. and
         {Prince}, Thomas A.},
        title = "{Tracing the Orphan Stream to 55 kpc with RR Lyrae Stars}",
      journal = {\apj},
         year = 2013,
        month = oct,
       volume = {776},
       number = {1},
          eid = {26},
        pages = {26},
          doi = {10.1088/0004-637X/776/1/26},
archivePrefix = {arXiv},
       eprint = {1308.0857},
 primaryClass = {astro-ph.GA},
       adsurl = {https://ui.adsabs.harvard.edu/abs/2013ApJ...776...26S}
}

@ARTICLE{akerlof2000,
       author = {{Akerlof}, C. and {Amrose}, S. and {Balsano}, R. and {Bloch}, J. and {Casperson}, D. and {Fletcher}, S. and {Gisler}, G. and {Hills}, J. and {Kehoe}, R. and {Lee}, B. and {Marshall}, S. and {McKay}, T. and {Pawl}, A. and {Schaefer}, J. and {Szymanski}, J. and {Wren}, J.},
        title = "{ROTSE All-Sky Surveys for Variable Stars. I. Test Fields}",
      journal = {\aj},
         year = 2000,
        month = apr,
       volume = {119},
       number = {4},
        pages = {1901-1913},
          doi = {10.1086/301321},
archivePrefix = {arXiv},
       eprint = {astro-ph/0001388},
 primaryClass = {astro-ph},
       adsurl = {https://ui.adsabs.harvard.edu/abs/2000AJ....119.1901A}
}

@ARTICLE{pojmanski2002,
       author = {{Pojma{\'n}ski}, G.},
        title = "{The All Sky Automated Survey. Catalog of Variable Stars.  I. 0 h - 6 hQuarter of the Southern Hemisphere}",
      journal = {\actaa},
         year = 2002,
        month = dec,
       volume = {52},
        pages = {397-427},
archivePrefix = {arXiv},
       eprint = {astro-ph/0210283},
 primaryClass = {astro-ph},
       adsurl = {https://ui.adsabs.harvard.edu/abs/2002AcA....52..397P}
}

@ARTICLE{pojmanski2014,
       author = {{Pojma{\'n}ski}, G.},
        title = "{The all sky automated survey}",
      journal = {Contributions of the Astronomical Observatory Skalnate Pleso},
         year = 2014,
        month = mar,
       volume = {43},
       number = {3},
        pages = {523-523},
       adsurl = {https://ui.adsabs.harvard.edu/abs/2014CoSka..43..523P}
}

@article{carollo10,
	author = {Carollo, Daniela and Beers, Timothy C. and Chiba, Masashi and Norris, John E. and Freeman, Ken C. and Lee, Young Sun and Ivezi{\'c}, {\v Z}eljko and Rockosi, Constance M. and Yanny, Brian},
	doi = {10.1088/0004-637X/712/1/692},
	journal = {The Astrophysical Journal},
	month = {mar},
	number = {1},
	pages = {692},
	publisher = {The American Astronomical Society},
	title = {STRUCTURE AND KINEMATICS OF THE STELLAR HALOS AND THICK DISKS OF THE MILKY WAY BASED ON CALIBRATION STARS FROM SLOAN DIGITAL SKY SURVEY DR7},
	url = {https://dx.doi.org/10.1088/0004-637X/712/1/692},
	volume = {712},
	year = {2010}}

@ARTICLE{carollo07,
   author = {{Carollo}, D. and {Beers}, T.~C. and {Lee}, Y.~S. and {Chiba}, M. and 
	{Norris}, J.~E. and {Wilhelm}, R. and {Sivarani}, T. and {Marsteller}, B. and 
	{Munn}, J.~A. and {Bailer-Jones}, C.~A.~L. and {Fiorentin}, P.~R. and 
	{York}, D.~G.},
    title = "{Two stellar components in the halo of the Milky Waydietz2020}",
  journal = {\nat},
archivePrefix = "arXiv",
   eprint = {0706.3005},
     year = 2007,
    month = dec,
   volume = 450,
    pages = {1020-1025},
      doi = {10.1038/nature06460},
   adsurl = {http://adsabs.harvard.edu/abs/2007Natur.450.1020C}
}

@ARTICLE{garofalo13,
   author = {{Garofalo}, A. and {Cusano}, F. and {Clementini}, G. and {Ripepi}, V. and 
	{Dall'Ora}, M. and {Moretti}, M.~I. and {Coppola}, G. and {Musella}, I. and 
	{Marconi}, M.},
    title = "{Variable Stars in the Ultra-faint Dwarf Spheroidal Galaxy Ursa Major I}",
  journal = {\apj},
archivePrefix = "arXiv",
   eprint = {1302.3230},
 primaryClass = "astro-ph.GA",
     year = 2013,
    month = apr,
   volume = 767,
      eid = {62},
    pages = {62},
      doi = {10.1088/0004-637X/767/1/62},
   adsurl = {http://adsabs.harvard.edu/abs/2013ApJ...767...62G}
}

@ARTICLE{bonifacio2024,
       author = {{Bonifacio}, P. and {Caffau}, E. and {Monaco}, L. and {Sbordone}, L. and {Spite}, M. and {Mucciarelli}, A. and {Fran{\c{c}}ois}, P. and {Lombardo}, L. and {Matas Pinto}, A. d. M.},
        title = "{High-speed stars. II. An unbound star, young stars, bulge metal-poor stars, and Aurora candidates}",
      journal = {\aap},
         year = 2024,
        month = apr,
       volume = {684},
          eid = {A91},
        pages = {A91},
          doi = {10.1051/0004-6361/202347865},
archivePrefix = {arXiv},
       eprint = {2402.02876},
 primaryClass = {astro-ph.GA},
       adsurl = {https://ui.adsabs.harvard.edu/abs/2024A&A...684A..91B}
}

@ARTICLE{lane2022,
       author = {{Lane}, James M.~M. and {Bovy}, Jo and {Mackereth}, J. Ted},
        title = "{The kinematic properties of Milky Way stellar halo populations}",
      journal = {\mnras},
         year = 2022,
        month = mar,
       volume = {510},
       number = {4},
        pages = {5119-5141},
          doi = {10.1093/mnras/stab3755},
archivePrefix = {arXiv},
       eprint = {2106.09699},
 primaryClass = {astro-ph.GA},
       adsurl = {https://ui.adsabs.harvard.edu/abs/2022MNRAS.510.5119L}
}

@ARTICLE{kinemuchi2006,
       author = {{Kinemuchi}, K. and {Smith}, H.~A. and {Wo{\'z}niak}, P.~R. and {McKay}, T.~A. and {ROTSE Collaboration}},
        title = "{Analysis of RR Lyrae Stars in the Northern Sky Variability Survey}",
      journal = {\aj},
         year = 2006,
        month = sep,
       volume = {132},
       number = {3},
        pages = {1202-1220},
          doi = {10.1086/506198},
archivePrefix = {arXiv},
       eprint = {astro-ph/0606092},
 primaryClass = {astro-ph},
       adsurl = {https://ui.adsabs.harvard.edu/abs/2006AJ....132.1202K}
}

@ARTICLE{Abbas14,
       author = {{Abbas}, Mohamad and {Grebel}, Eva K. and {Martin}, N.~F. and {Kaiser}, N. and {Burgett}, W.~S. and {Huber}, M.~E. and {Waters}, C.},
        title = "{An Optimized Method to Identify RR Lyrae Stars in the SDSS{\texttimes}Pan-STARRS1 Overlapping Area Using a Bayesian Generative Technique}",
      journal = {\aj},
         year = 2014,
        month = jul,
       volume = {148},
       number = {1},
          eid = {8},
        pages = {8},
          doi = {10.1088/0004-6256/148/1/8},
archivePrefix = {arXiv},
       eprint = {1405.3537},
 primaryClass = {astro-ph.GA},
       adsurl = {https://ui.adsabs.harvard.edu/abs/2014AJ....148....8A}
}

@ARTICLE{fernandezalvar2017,
       author = {{Fern{\'a}ndez-Alvar}, E. and {Carigi}, L. and {Allende Prieto}, C. and {Hayden}, M.~R. and {Beers}, T.~C. and {Fern{\'a}ndez-Trincado}, J.~G. and {Meza}, A. and {Schultheis}, M. and {Santiago}, B.~X. and {Queiroz}, A.~B. and {Anders}, F. and {da Costa}, L.~N. and {Chiappini}, C.},
        title = "{Chemical trends in the Galactic halo from APOGEE data}",
      journal = {\mnras},
         year = 2017,
        month = feb,
       volume = {465},
       number = {2},
        pages = {1586-1600},
          doi = {10.1093/mnras/stw2861},
archivePrefix = {arXiv},
       eprint = {1611.01249},
 primaryClass = {astro-ph.SR},
       adsurl = {https://ui.adsabs.harvard.edu/abs/2017MNRAS.465.1586F}
}

@ARTICLE{conroy2019a,
       author = {{Conroy}, Charlie and {Bonaca}, Ana and {Cargile}, Phillip and {Johnson}, Benjamin D. and {Caldwell}, Nelson and {Naidu}, Rohan P. and {Zaritsky}, Dennis and {Fabricant}, Daniel and {Moran}, Sean and {Rhee}, Jaehyon and {Szentgyorgyi}, Andrew and {Berlind}, Perry and {Calkins}, Michael L. and {Kattner}, ShiAnne and {Ly}, Chun},
        title = "{Mapping the Stellar Halo with the H3 Spectroscopic Survey}",
      journal = {\apj},
         year = 2019,
        month = sep,
       volume = {883},
       number = {1},
          eid = {107},
        pages = {107},
          doi = {10.3847/1538-4357/ab38b8},
archivePrefix = {arXiv},
       eprint = {1907.07684},
 primaryClass = {astro-ph.GA},
       adsurl = {https://ui.adsabs.harvard.edu/abs/2019ApJ...883..107C}
}

@ARTICLE{conroy2019b,
       author = {{Conroy}, Charlie and {Naidu}, Rohan P. and {Zaritsky}, Dennis and {Bonaca}, Ana and {Cargile}, Phillip and {Johnson}, Benjamin D. and {Caldwell}, Nelson},
        title = "{Resolving the Metallicity Distribution of the Stellar Halo with the H3 Survey}",
      journal = {\apj},
         year = 2019,
        month = dec,
       volume = {887},
       number = {2},
          eid = {237},
        pages = {237},
          doi = {10.3847/1538-4357/ab5710},
archivePrefix = {arXiv},
       eprint = {1909.02007},
 primaryClass = {astro-ph.GA},
       adsurl = {https://ui.adsabs.harvard.edu/abs/2019ApJ...887..237C}
}

@ARTICLE{layden94,
   author = {{Layden}, Andrew C.},
        title = "{The Metallicities and Kinematics of RR Lyrae Variables. I. New Observations of Local Stars}",
      journal = {\aj},
         year = 1994,
        month = sep,
       volume = {108},
        pages = {1016},
          doi = {10.1086/117132},
       adsurl = {https://ui.adsabs.harvard.edu/abs/1994AJ....108.1016L}
}

@ARTICLE{caldwell2011,
       author = {{Caldwell}, Nelson and {Schiavon}, Ricardo and {Morrison}, Heather and {Rose}, James A. and {Harding}, Paul},
        title = "{Star Clusters in M31. II. Old Cluster Metallicities and Ages from Hectospec Data}",
      journal = {\aj},
         year = 2011,
        month = feb,
       volume = {141},
       number = {2},
          eid = {61},
        pages = {61},
          doi = {10.1088/0004-6256/141/2/61},
archivePrefix = {arXiv},
       eprint = {1101.3779},
 primaryClass = {astro-ph.GA},
       adsurl = {https://ui.adsabs.harvard.edu/abs/2011AJ....141...61C}
}

@ARTICLE{caldwell2016,
       author = {{Caldwell}, Nelson and {Romanowsky}, Aaron J.},
        title = "{Star Clusters in M31. VII. Global Kinematics and Metallicity Subpopulations of the Globular Clusters}",
      journal = {\apj},
         year = 2016,
        month = jun,
       volume = {824},
       number = {1},
          eid = {42},
        pages = {42},
          doi = {10.3847/0004-637X/824/1/42},
archivePrefix = {arXiv},
       eprint = {1603.06947},
 primaryClass = {astro-ph.GA},
       adsurl = {https://ui.adsabs.harvard.edu/abs/2016ApJ...824...42C}
}

@ARTICLE{miceli08,
   author = {{Miceli}, Antonino and {Rest}, Armin and {Stubbs}, Christopher W. and {Hawley}, Suzanne L. and {Cook}, Kem H. and {Magnier}, Eugene A. and {Krisciunas}, Kevin and {Bowell}, Edward and {Koehn}, Bruce},
        title = "{Evidence for Distinct Components of the Galactic Stellar Halo from 838 RR Lyrae Stars Discovered in the LONEOS-I Survey}",
      journal = {\apj},
         year = 2008,
        month = may,
       volume = {678},
       number = {2},
        pages = {865-887},
          doi = {10.1086/533484},
archivePrefix = {arXiv},
       eprint = {0706.1583},
 primaryClass = {astro-ph},
       adsurl = {https://ui.adsabs.harvard.edu/abs/2008ApJ...678..865M}
}

@ARTICLE{helmi2020,
       author = {{Helmi}, Amina},
        title = "{Streams, Substructures, and the Early History of the Milky Way}",
      journal = {\araa},
         year = 2020,
        month = aug,
       volume = {58},
        pages = {205-256},
          doi = {10.1146/annurev-astro-032620-021917},
archivePrefix = {arXiv},
       eprint = {2002.04340},
 primaryClass = {astro-ph.GA},
       adsurl = {https://ui.adsabs.harvard.edu/abs/2020ARA&A..58..205H}
}

@ARTICLE{belokurov06,
   author = {{Belokurov}, V. and {Zucker}, D.~B. and {Evans}, N.~W. and {Wilkinson}, M.~I. and 
	{Irwin}, M.~J. and {Hodgkin}, S. and {Bramich}, D.~M. and {Irwin}, J.~M. and 
	{Gilmore}, G. and {Willman}, B. and {Vidrih}, S. and {Newberg}, H.~J. and 
	{Wyse}, R.~F.~G. and {Fellhauer}, M. and {Hewett}, P.~C. and 
	{Cole}, N. and {Bell}, E.~F. and {Beers}, T.~C. and {Rockosi}, C.~M. and 
	{Yanny}, B. and {Grebel}, E.~K. and {Schneider}, D.~P. and {Lupton}, R. and 
	{Barentine}, J.~C. and {Brewington}, H. and {Brinkmann}, J. and 
	{Harvanek}, M. and {Kleinman}, S.~J. and {Krzesinski}, J. and 
	{Long}, D. and {Nitta}, A. and {Smith}, J.~A. and {Snedden}, S.~A.
	},
    title = "{A Faint New Milky Way Satellite in Bootes}",
  journal = {\apjl},
   eprint = {astro-ph/0604355},
     year = 2006,
    month = aug,
   volume = 647,
    pages = {L111-L114},
      doi = {10.1086/507324},
   adsurl = {http://adsabs.harvard.edu/abs/2006ApJ...647L.111B}
}

@ARTICLE{salvadori07,
       author = {{Salvadori}, Stefania and {Schneider}, Raffaella and {Ferrara}, Andrea},
        title = "{Cosmic stellar relics in the Galactic halo}",
      journal = {\mnras},
         year = 2007,
        month = oct,
       volume = {381},
       number = {2},
        pages = {647-662},
          doi = {10.1111/j.1365-2966.2007.12133.x},
archivePrefix = {arXiv},
       eprint = {astro-ph/0611130},
 primaryClass = {astro-ph},
       adsurl = {https://ui.adsabs.harvard.edu/abs/2007MNRAS.381..647S}
}

@ARTICLE{lee2008_sspp,
       author = {{Lee}, Young Sun and {Beers}, Timothy C. and {Sivarani}, Thirupathi and
         {Allende Prieto}, Carlos and {Koesterke}, Lars and {Wilhelm}, Ronald and
         {Re Fiorentin}, Paola and {Bailer-Jones}, Coryn A.~L. and
         {Norris}, John E. and {Rockosi}, Constance M. and {Yanny}, Brian and
         {Newberg}, Heidi J. and {Covey}, Kevin R. and {Zhang}, Hao-Tong and
         {Luo}, A. -Li},
        title = "{The SEGUE Stellar Parameter Pipeline. I. Description and Comparison of Individual Methods}",
      journal = {\aj},
         year = 2008,
        month = nov,
       volume = {136},
       number = {5},
        pages = {2022-2049},
          doi = {10.1088/0004-6256/136/5/2022},
archivePrefix = {arXiv},
       eprint = {0710.5645},
 primaryClass = {astro-ph},
       adsurl = {https://ui.adsabs.harvard.edu/abs/2008AJ....136.2022L}
}

@ARTICLE{tolstoy09,
   author = {{Tolstoy}, E. and {Hill}, V. and {Tosi}, M.},
    title = "{Star-Formation Histories, Abundances, and Kinematics of Dwarf Galaxies in the Local Group}",
  journal = {\araa},
archivePrefix = "arXiv",
   eprint = {0904.4505},
 primaryClass = "astro-ph.CO",
     year = 2009,
    month = sep,
   volume = 47,
    pages = {371-425},
      doi = {10.1146/annurev-astro-082708-101650},
   adsurl = {http://adsabs.harvard.edu/abs/2009ARA%26A..47..371T}
}

@ARTICLE{bono94c,
       author = {{Bono}, G. and {Caputo}, F. and {Stellingwerf}, R.~F.},
        title = "{On the Application of the Baade-Wesselink Method to RR Lyrae Stars}",
      journal = {\apjl},
         year = 1994,
        month = sep,
       volume = {432},
        pages = {L51},
          doi = {10.1086/187509},
       adsurl = {https://ui.adsabs.harvard.edu/abs/1994ApJ...432L..51B}
}

@ARTICLE{schlafly11,
   author = {{Schlafly}, E.~F. and {Finkbeiner}, D.~P.},
    title = "{Measuring Reddening with Sloan Digital Sky Survey Stellar Spectra and Recalibrating SFD}",
  journal = {\apj},
archivePrefix = "arXiv",
   eprint = {1012.4804},
 primaryClass = "astro-ph.GA",
     year = 2011,
    month = aug,
   volume = 737,
      eid = {103},
    pages = {103},
      doi = {10.1088/0004-637X/737/2/103},
   adsurl = {http://adsabs.harvard.edu/abs/2011ApJ...737..103S}
}

@ARTICLE{marconi15,
   author = {{Marconi}, M. and {Coppola}, G. and {Bono}, G. and {Braga}, V. and 
	{Pietrinferni}, A. and {Buonanno}, R. and {Castellani}, M. and 
	{Musella}, I. and {Ripepi}, V. and {Stellingwerf}, R.~F.},
    title = "{On a New Theoretical Framework for RR Lyrae Stars. I. The Metallicity Dependence}",
  journal = {\apj},
archivePrefix = "arXiv",
   eprint = {1505.02531},
 primaryClass = "astro-ph.SR",
     year = 2015,
    month = jul,
   volume = 808,
      eid = {50},
    pages = {50},
      doi = {10.1088/0004-637X/808/1/50},
   adsurl = {http://adsabs.harvard.edu/abs/2015ApJ...808...50M}
}

@ARTICLE{braga2021,
       author = {{Braga}, V.~F. and {Crestani}, J. and {Fabrizio}, M. and {Bono}, G. and {Sneden}, C. and {Preston}, G.~W. and {Storm}, J. and {Kamann}, S. and {Latour}, M. and {Lala}, H. and {Lemasle}, B. and {Prudil}, Z. and {Altavilla}, G. and {Chaboyer}, B. and {Dall'Ora}, M. and {Ferraro}, I. and {Gilligan}, C.~K. and {Fiorentino}, G. and {Iannicola}, G. and {Inno}, L. and {Kwak}, S. and {Marengo}, M. and {Marinoni}, S. and {Marrese}, P.~M. and {Mart{\'\i}nez-V{\'a}zquez}, C.~E. and {Monelli}, M. and {Mullen}, J.~P. and {Matsunaga}, N. and {Neeley}, J. and {Stetson}, P.~B. and {Valenti}, E. and {Zoccali}, M.},
        title = "{On the Use of Field RR Lyrae as Galactic Probes. V. Optical and Radial Velocity Curve Templates}",
      journal = {\apj},
         year = 2021,
        month = oct,
       volume = {919},
       number = {2},
          eid = {85},
        pages = {85},
          doi = {10.3847/1538-4357/ac1074},
archivePrefix = {arXiv},
       eprint = {2107.00923},
 primaryClass = {astro-ph.SR},
       adsurl = {https://ui.adsabs.harvard.edu/abs/2021ApJ...919...85B}
}

@ARTICLE{gilligan2021,
       author = {{Gilligan}, Christina K. and {Chaboyer}, Brian and {Marengo}, Massimo and {Mullen}, Joseph P. and {Bono}, Giuseppe and {Braga}, Vittorio F. and {Crestani}, Juliana and {Dall'Ora}, Massimo and {Fiorentino}, Giuliana and {Monelli}, Matteo and {Neeley}, Jill R. and {Fabrizio}, Michele and {Mart{\'\i}nez-V{\'a}zquez}, Clara E. and {Th{\'e}venin}, Fr{\'e}d{\'e}ric and {Sneden}, Christopher},
        title = "{Metallicities from high-resolution spectra of 49 RR Lyrae variables}",
      journal = {\mnras},
         year = 2021,
        month = jun,
       volume = {503},
       number = {4},
        pages = {4719-4733},
          doi = {10.1093/mnras/stab857},
archivePrefix = {arXiv},
       eprint = {2103.11012},
 primaryClass = {astro-ph.SR},
       adsurl = {https://ui.adsabs.harvard.edu/abs/2021MNRAS.503.4719G}
}

@ARTICLE{fabrizio2021,
       author = {{Fabrizio}, M. and {Braga}, V.~F. and {Crestani}, J. and {Bono}, G. and {Ferraro}, I. and {Fiorentino}, G. and {Iannicola}, G. and {Preston}, G.~W. and {Sneden}, C. and {Th{\'e}venin}, F. and {Altavilla}, G. and {Chaboyer}, B. and {Dall'Ora}, M. and {da Silva}, R. and {Grebel}, E.~K. and {Gilligan}, C.~K. and {Lala}, H. and {Lemasle}, B. and {Magurno}, D. and {Marengo}, M. and {Marinoni}, S. and {Marrese}, P.~M. and {Mart{\'\i}nez-V{\'a}zquez}, C.~E. and {Matsunaga}, N. and {Monelli}, M. and {Mullen}, J.~P. and {Neeley}, J. and {Nonino}, M. and {Prudil}, Z. and {Salaris}, M. and {Stetson}, P.~B. and {Valenti}, E. and {Zoccali}, M.},
        title = "{On the Use of Field RR Lyrae As Galactic Probes: IV. New Insights Into and Around the Oosterhoff Dichotomy}",
      journal = {\apj},
         year = 2021,
        month = oct,
       volume = {919},
       number = {2},
          eid = {118},
        pages = {118},
          doi = {10.3847/1538-4357/ac1115},
archivePrefix = {arXiv},
       eprint = {2107.00919},
 primaryClass = {astro-ph.SR},
       adsurl = {https://ui.adsabs.harvard.edu/abs/2021ApJ...919..118F}
}

@ARTICLE{colucci2014,
       author = {{Colucci}, Janet E. and {Bernstein}, Rebecca A. and {Cohen}, Judith G.},
        title = "{The Detailed Chemical Properties of M31 Star Clusters. I. Fe, Alpha and Light Elements}",
      journal = {\apj},
         year = 2014,
        month = dec,
       volume = {797},
       number = {2},
          eid = {116},
        pages = {116},
          doi = {10.1088/0004-637X/797/2/116},
archivePrefix = {arXiv},
       eprint = {1411.0696},
 primaryClass = {astro-ph.GA},
       adsurl = {https://ui.adsabs.harvard.edu/abs/2014ApJ...797..116C}
}

@ARTICLE{medina2021,
       author = {{Medina}, Gustavo E. and {Lemasle}, Bertrand and {Grebel}, Eva K.},
        title = "{A revisited study of Cepheids in open clusters in the Gaia era}",
      journal = {\mnras},
         year = 2021,
        month = jul,
       volume = {505},
       number = {1},
        pages = {1342-1366},
          doi = {10.1093/mnras/stab1267},
archivePrefix = {arXiv},
       eprint = {2104.14565},
 primaryClass = {astro-ph.GA},
       adsurl = {https://ui.adsabs.harvard.edu/abs/2021MNRAS.505.1342M}
}

@ARTICLE{medina2023,
       author = {{Medina}, Gustavo E. and {Hansen}, Camilla J. and {Mu{\~n}oz}, Ricardo R. and {Grebel}, Eva K. and {Vivas}, A. Katherina and {Carlin}, Jeffrey L. and {Mart{\'\i}nez-V{\'a}zquez}, Clara E.},
        title = "{RR Lyrae stars as probes of the outer Galactic halo: chemical and kinematic analysis of a pilot sample}",
      journal = {\mnras},
         year = 2023,
        month = mar,
       volume = {519},
       number = {4},
        pages = {5689-5722},
          doi = {10.1093/mnras/stac3800},
archivePrefix = {arXiv},
       eprint = {2309.03271},
 primaryClass = {astro-ph.GA},
       adsurl = {https://ui.adsabs.harvard.edu/abs/2023MNRAS.519.5689M}
}

@ARTICLE{medina2025a,
       author = {{Medina}, Gustavo E. and {Li}, Ting S. and {Koposov}, Sergey E. and {Riley}, A.~H. and {Beraldo e Silva}, L. and {Valluri}, M. and {Wang}, W. and {Bystr{\"o}m}, A. and {Gnedin}, O.~Y. and {Carlberg}, R.~G. and {Kizhuprakkat}, N. and {Weaver}, B.~A. and {Aguilar}, J. and {Ahlen}, S. and {Bianchi}, D. and {Brooks}, D. and {Claybaugh}, T. and {Cooper}, A.~P. and {de la Macorra}, A. and {Dey}, A. and {Doel}, P. and {Font-Ribera}, A. and {Forero-Romero}, J.~E. and {Gazta{\~n}aga}, E. and {Gontcho}, S. Gontcho A and {Gutierrez}, G. and {Guy}, J. and {Honscheid}, K. and {Ishak}, M. and {Kisner}, T. and {Landriau}, M. and {Le Guillou}, L. and {Meisner}, A. and {Miquel}, R. and {Myers}, A.~D. and {Nadathur}, S. and {Poppett}, C. and {Prada}, F. and {P{\'e}rez-R{\`a}fols}, I. and {Rossi}, G. and {Sanchez}, E. and {Seo}, H. and {Sprayberry}, D. and {Tarl{\'e}}, G. and {Wechsler}, R.~H. and {Zhou}, R. and {Zou}, H.},
        title = "{The DESI Y1 RR Lyrae catalog I: Empirical modeling of the cyclic variation of spectroscopic properties and a chemodynamical analysis of the outer halo}",
      journal = {arXiv e-prints},
         year = 2025,
        month = apr,
          eid = {arXiv:2504.02924},
        pages = {arXiv:2504.02924},
          doi = {10.48550/arXiv.2504.02924},
archivePrefix = {arXiv},
       eprint = {2504.02924},
 primaryClass = {astro-ph.GA},
       adsurl = {https://ui.adsabs.harvard.edu/abs/2025arXiv250402924M}
}

@ARTICLE{medina2025b,
       author = {{Medina}, Gustavo E. and {Li}, Ting S. and {Allende Prieto}, C. and {Beraldo e Silva}, L. and {Bystrom}, A. and {Carlberg}, R.~G. and {Koposov}, S.~E. and {Lambert}, M. and {Najita}, J.~R. and {Rockosi}, C.~M. and {Kizhuprakkat}, N. and {Riley}, A. and {Aguilar}, J. and {Ahlen}, S. and {Bianchi}, D. and {Brooks}, D. and {Claybaugh}, T. and {Cooper}, A.~P. and {de la Macorra}, A. and {Dey}, A. and {Doel}, P. and {Forero-Romero}, J. and {Gazta{\~n}aga}, E. and {Gontcho}, S. Gontcho A and {Gutierrez}, G. and {Ishak}, M. and {Kehoe}, R. and {Kisner}, T. and {Landriau}, M. and {Le Guillou}, L. and {Meisner}, A. and {Miquel}, R. and {Prada}, F. and {Perez-Rafols}, I. and {Rossi}, G. and {Sanchez}, E. and {Schlegel}, D.~J. and {Silber}, J.~H. and {Sprayberry}, D. and {Tarle}, G. and {Weaver}, B.~A. and {Zhou}, R.},
        title = "{The DESI Y1 RR Lyrae catalog II: The metallicity dependency of pulsational properties, the shape of the RR Lyrae instability strip, and metal rich RR Lyrae}",
      journal = {arXiv e-prints},
         year = 2025,
        month = may,
          eid = {arXiv:2505.10614},
        pages = {arXiv:2505.10614},
          doi = {10.48550/arXiv.2505.10614},
archivePrefix = {arXiv},
       eprint = {2505.10614},
 primaryClass = {astro-ph.GA},
       adsurl = {https://ui.adsabs.harvard.edu/abs/2025arXiv250510614M}
}

@ARTICLE{bono97a,
   author = {{Bono}, G. and {Caputo}, F. and {Cassisi}, S. and {Castellani}, V. and 
	{Marconi}, M.},
    title = "{Metal-rich RR Lyrae Variables. I. The Evolutionary Scenario}",
  journal = {\apj},
   eprint = {arXiv:astro-ph/9609153},
     year = 1997,
    month = apr,
   volume = 479,
    pages = {279-+},
      doi = {10.1086/303872},
   adsurl = {http://adsabs.harvard.edu/abs/1997ApJ...479..279B}
}

@ARTICLE{bono97b,
   author = {{Bono}, G. and {Caputo}, F. and {Cassisi}, S. and {Incerpi}, R. and 
	{Marconi}, M.},
    title = "{Metal-rich RR Lyrae Variables. II. The Pulsational Scenario}",
  journal = {\apj},
   eprint = {arXiv:astro-ph/9702083},
     year = 1997,
    month = jul,
   volume = 483,
    pages = {811-+},
      doi = {10.1086/304284},
   adsurl = {http://adsabs.harvard.edu/abs/1997ApJ...483..811B}
}

@ARTICLE{bono07,
   author = {{Bono}, G. and {Caputo}, F. and {Di Criscienzo}, M.},
    title = "{RR Lyrae stars in Galactic globular clusters. VI. The period-amplitude relation}",
  journal = {\aap},
archivePrefix = "arXiv",
   eprint = {0709.3177},
     year = 2007,
    month = dec,
   volume = 476,
    pages = {779-790},
      doi = {10.1051/0004-6361:20078206},
   adsurl = {http://adsabs.harvard.edu/abs/2007A%26A...476..779B}
}

@ARTICLE{saha1984,
       author = {{Saha}, A. and {Oke}, J.~B.},
        title = "{Spectroscopy and spectrophotometry of distant halo RR Lyrae stars.}",
      journal = {\apj},
         year = 1984,
        month = oct,
       volume = {285},
        pages = {688-697},
          doi = {10.1086/162546},
       adsurl = {https://ui.adsabs.harvard.edu/abs/1984ApJ...285..688S}
}

@ARTICLE{saha1985,
       author = {{Saha}, A.},
        title = "{RR Lyrae stars and the distant galactic halo : distribution, chemical composition, kinematics and dynamics.}",
      journal = {\apj},
         year = 1985,
        month = feb,
       volume = {289},
        pages = {310-319},
          doi = {10.1086/162890},
       adsurl = {https://ui.adsabs.harvard.edu/abs/1985ApJ...289..310S}
}

@ARTICLE{Brown06,
   author = {{Brown}, T.~M. and {Smith}, E. and {Ferguson}, H.~C. and {Rich}, R.~M. and 
	{Guhathakurta}, P. and {Renzini}, A. and {Sweigart}, A.~V. and 
	{Kimble}, R.~A.},
    title = "{The Detailed Star Formation History in the Spheroid, Outer Disk, and Tidal Stream of the Andromeda Galaxy}",
  journal = {\apj},
   eprint = {arXiv:astro-ph/0607637},
     year = 2006,
    month = nov,
   volume = 652,
    pages = {323-353},
      doi = {10.1086/508015},
   adsurl = {http://adsabs.harvard.edu/abs/2006ApJ...652..323B}
}

@ARTICLE{Brown09,
   author = {{Brown}, T.~M. and {Smith}, E. and {Ferguson}, H.~C. and {Guhathakurta}, P. and 
	{Kalirai}, J.~S. and {Kimble}, R.~A. and {Renzini}, A. and {Rich}, R.~M. and 
	{Sweigart}, A.~V. and {Vanden Berg}, D.~A.},
    title = "{Deep Optical Photometry of Six Fields in the Andromeda Galaxy}",
  journal = {\apjs},
archivePrefix = "arXiv",
   eprint = {0908.0476},
 primaryClass = "astro-ph.CO",
     year = 2009,
    month = sep,
   volume = 184,
    pages = {152-157},
      doi = {10.1088/0067-0049/184/1/152},
   adsurl = {http://adsabs.harvard.edu/abs/2009ApJS..184..152B}
}

@INPROCEEDINGS{cacciari03a,
   author = {{Cacciari}, C. and {Clementini}, G.},
    title = "{Globular Cluster Distances from RR Lyrae Stars}",
booktitle = {Stellar Candles for the Extragalactic Distance Scale},
     year = 2003,
   series = {Lecture Notes in Physics, Berlin Springer Verlag},
   volume = 635,
   editor = {{Alloin}, D. and {Gieren}, W.},
    pages = {105},
   adsurl = {http://adsabs.harvard.edu/abs/2003LNP...635..105C}
}

@ARTICLE{harris79,
       author = {{Harris}, W.~E. and {Canterna}, R.},
        title = "{On the abundance gradient in the galactic halo.}",
      journal = {\apjl},
         year = 1979,
        month = jul,
       volume = {231},
        pages = {L19-L23},
          doi = {10.1086/182997},
       adsurl = {https://ui.adsabs.harvard.edu/abs/1979ApJ...231L..19H}
}

@ARTICLE{harris96,
   author = {{Harris}, W.~E.},
    title = "{A Catalog of Parameters for Globular Clusters in the Milky Way}",
  journal = {\aj},
     year = 1996,
    month = oct,
   volume = 112,
    pages = {1487-+},
      doi = {10.1086/118116},
   adsurl = {http://adsabs.harvard.edu/abs/1996AJ....112.1487H}
}

@ARTICLE{dejong2010,
       author = {{de Jong}, Jelte T.~A. and {Yanny}, Brian and {Rix}, Hans-Walter and {Dolphin}, Andrew E. and {Martin}, Nicolas F. and {Beers}, Timothy C.},
        title = "{Mapping the Stellar Structure of the Milky Way Thick Disk and Halo Using SEGUE Photometry}",
      journal = {\apj},
         year = 2010,
        month = may,
       volume = {714},
       number = {1},
        pages = {663-674},
          doi = {10.1088/0004-637X/714/1/663},
archivePrefix = {arXiv},
       eprint = {0911.3900},
 primaryClass = {astro-ph.GA},
       adsurl = {https://ui.adsabs.harvard.edu/abs/2010ApJ...714..663D}
}

@ARTICLE{kunder2014,
       author = {{Kunder}, A. and {Bono}, G. and {Piffl}, T. and {Steinmetz}, M. and {Grebel}, E.~K. and {Anguiano}, B. and {Freeman}, K. and {Kordopatis}, G. and {Zwitter}, T. and {Scholz}, R. and {Gibson}, B.~K. and {Bland-Hawthorn}, J. and {Seabroke}, G. and {Boeche}, C. and {Siebert}, A. and {Wyse}, R.~F.~G. and {Bienaym{\'e}}, O. and {Navarro}, J. and {Siviero}, A. and {Minchev}, I. and {Parker}, Q. and {Reid}, W. and {Gilmore}, G. and {Munari}, U. and {Helmi}, A.},
        title = "{Spectroscopic signatures of extratidal stars around the globular clusters NGC 6656 (M 22), NGC 3201, and NGC 1851 from RAVE}",
      journal = {\aap},
         year = 2014,
        month = dec,
       volume = {572},
          eid = {A30},
        pages = {A30},
          doi = {10.1051/0004-6361/201424113},
archivePrefix = {arXiv},
       eprint = {1408.6236},
 primaryClass = {astro-ph.SR},
       adsurl = {https://ui.adsabs.harvard.edu/abs/2014A&A...572A..30K}
}

@ARTICLE{armandroff92,
       author = {{Armandroff}, Taft E. and {Da Costa}, G.~S. and {Zinn}, Robert},
        title = "{Metallicities for the Outer-Halo Globular Cluster PAL 3,4 and 14}",
      journal = {\aj},
         year = 1992,
        month = jul,
       volume = {104},
        pages = {164},
          doi = {10.1086/116228},
       adsurl = {https://ui.adsabs.harvard.edu/abs/1992AJ....104..164A}
}

@ARTICLE{romano2010,
       author = {{Romano}, D. and {Karakas}, A.~I. and {Tosi}, M. and {Matteucci}, F.},
        title = "{Quantifying the uncertainties of chemical evolution studies. II. Stellar yields}",
      journal = {\aap},
         year = 2010,
        month = nov,
       volume = {522},
          eid = {A32},
        pages = {A32},
          doi = {10.1051/0004-6361/201014483},
archivePrefix = {arXiv},
       eprint = {1006.5863},
 primaryClass = {astro-ph.GA},
       adsurl = {https://ui.adsabs.harvard.edu/abs/2010A&A...522A..32R}
}

@ARTICLE{zinn14,
   author = {{Zinn}, R. and {Horowitz}, B. and {Vivas}, A.~K. and {Baltay}, C. and 
	{Ellman}, N. and {Hadjiyska}, E. and {Rabinowitz}, D. and {Miller}, L.
	},
    title = "{La Silla QUEST RR Lyrae Star Survey: Region I}",
  journal = {\apj},
archivePrefix = "arXiv",
   eprint = {1312.1602},
 primaryClass = "astro-ph.GA",
     year = 2014,
    month = jan,
   volume = 781,
      eid = {22},
    pages = {22},
      doi = {10.1088/0004-637X/781/1/22},
   adsurl = {http://adsabs.harvard.edu/abs/2014ApJ...781...22Z}
}

@ARTICLE{Suntzeff91,
   author = {{Suntzeff}, N.~B. and {Kinman}, T.~D. and {Kraft}, R.~P.},
    title = "{Metal abundances of RR Lyrae variables in selected Galactic star fields. V - The Lick Astrographic fields at intermediate Galactic latitudes}",
  journal = {\apj},
     year = 1991,
    month = feb,
   volume = 367,
    pages = {528-546},
      doi = {10.1086/169650},
   adsurl = {http://adsabs.harvard.edu/abs/1991ApJ...367..528S}
}

@ARTICLE{Suntzeff94,
       author = {{Suntzeff}, Nicholas B. and {Kraft}, Robert P. and {Kinman}, T.~D.},
        title = "{Summary of Delta S Metallicity Measurements for Bright RR Lyrae Variables Observed at Lick Observatory and KPNO between 1972 and 1987}",
      journal = {\apjs},
         year = 1994,
        month = jul,
       volume = {93},
        pages = {271},
          doi = {10.1086/192055},
       adsurl = {https://ui.adsabs.harvard.edu/abs/1994ApJS...93..271S}
}

@ARTICLE{werner2004,
       author = {{Werner}, M.~W. and {Roellig}, T.~L. and {Low}, F.~J. and {Rieke}, G.~H. and {Rieke}, M. and {Hoffmann}, W.~F. and {Young}, E. and {Houck}, J.~R. and {Brandl}, B. and {Fazio}, G.~G. and {Hora}, J.~L. and {Gehrz}, R.~D. and {Helou}, G. and {Soifer}, B.~T. and {Stauffer}, J. and {Keene}, J. and {Eisenhardt}, P. and {Gallagher}, D. and {Gautier}, T.~N. and {Irace}, W. and {Lawrence}, C.~R. and {Simmons}, L. and {Van Cleve}, J.~E. and {Jura}, M. and {Wright}, E.~L. and {Cruikshank}, D.~P.},
        title = "{The Spitzer Space Telescope Mission}",
      journal = {\apjs},
         year = 2004,
        month = sep,
       volume = {154},
       number = {1},
        pages = {1-9},
          doi = {10.1086/422992},
archivePrefix = {arXiv},
       eprint = {astro-ph/0406223},
 primaryClass = {astro-ph},
       adsurl = {https://ui.adsabs.harvard.edu/abs/2004ApJS..154....1W}
}

@ARTICLE{deLuciaHelmi08,
       author = {{De Lucia}, Gabriella and {Helmi}, Amina},
        title = "{The Galaxy and its stellar halo: insights on their formation from a hybrid cosmological approach}",
      journal = {\mnras},
         year = 2008,
        month = nov,
       volume = {391},
       number = {1},
        pages = {14-31},
          doi = {10.1111/j.1365-2966.2008.13862.x},
archivePrefix = {arXiv},
       eprint = {0804.2465},
 primaryClass = {astro-ph},
       adsurl = {https://ui.adsabs.harvard.edu/abs/2008MNRAS.391...14D}
}

@ARTICLE{fabrizio15,
   author = {{Fabrizio}, M. and {Nonino}, M. and {Bono}, G. and {Primas}, F. and 
	{Th{\'e}venin}, F. and {Stetson}, P.~B. and {Cassisi}, S. and 
	{Buonanno}, R. and {Coppola}, G. and {da Silva}, R.~O. and {Dall'Ora}, M. and 
	{Ferraro}, I. and {Genovali}, K. and {Gilmozzi}, R. and {Iannicola}, G. and 
	{Marconi}, M. and {Monelli}, M. and {Romaniello}, M. and {Walker}, A.~R.
	},
    title = "{The Carina Project. VIII. The {$\alpha$}-element abundances}",
  journal = {\aap},
archivePrefix = "arXiv",
   eprint = {1505.06597},
     year = 2015,
    month = aug,
   volume = 580,
      eid = {A18},
    pages = {A18},
      doi = {10.1051/0004-6361/201525753},
   adsurl = {http://adsabs.harvard.edu/abs/2015A%26A...580A..18F}
}

@ARTICLE{galah_dr3,
       author = {{Buder}, Sven and {Sharma}, Sanjib and {Kos}, Janez and {Amarsi}, Anish M. and {Nordlander}, Thomas and {Lind}, Karin and {Martell}, Sarah L. and {Asplund}, Martin and {Bland-Hawthorn}, Joss and {Casey}, Andrew R. and {de Silva}, Gayandhi M. and {D'Orazi}, Valentina and {Freeman}, Ken C. and {Hayden}, Michael R. and {Lewis}, Geraint F. and {Lin}, Jane and {Schlesinger}, Katharine J. and {Simpson}, Jeffrey D. and {Stello}, Dennis and {Zucker}, Daniel B. and {Zwitter}, Toma{\v{z}} and {Beeson}, Kevin L. and {Buck}, Tobias and {Casagrande}, Luca and {Clark}, Jake T. and {{\v{C}}otar}, Klemen and {da Costa}, Gary S. and {de Grijs}, Richard and {Feuillet}, Diane and {Horner}, Jonathan and {Kafle}, Prajwal R. and {Khanna}, Shourya and {Kobayashi}, Chiaki and {Liu}, Fan and {Montet}, Benjamin T. and {Nandakumar}, Govind and {Nataf}, David M. and {Ness}, Melissa K. and {Spina}, Lorenzo and {Tepper-Garc{\'\i}a}, Thor and {Ting}, Yuan-Sen and {Traven}, Gregor and {Vogrin{\v{c}}i{\v{c}}}, Rok and {Wittenmyer}, Robert A. and {Wyse}, Rosemary F.~G. and {{\v{Z}}erjal}, Maru{\v{s}}a and {Galah Collaboration}},
        title = "{The GALAH+ survey: Third data release}",
      journal = {\mnras},
         year = 2021,
        month = sep,
       volume = {506},
       number = {1},
        pages = {150-201},
          doi = {10.1093/mnras/stab1242},
archivePrefix = {arXiv},
       eprint = {2011.02505},
 primaryClass = {astro-ph.GA},
       adsurl = {https://ui.adsabs.harvard.edu/abs/2021MNRAS.506..150B}
}

@ARTICLE{cardelli89,
   author = {{Cardelli}, J.~A. and {Clayton}, G.~C. and {Mathis}, J.~S.},
    title = "{The relationship between infrared, optical, and ultraviolet extinction}",
  journal = {\apj},
     year = 1989,
    month = oct,
   volume = 345,
    pages = {245-256},
      doi = {10.1086/167900},
   adsurl = {http://adsabs.harvard.edu/abs/1989ApJ...345..245C}
}

@ARTICLE{carretta09,
   author = {{Carretta}, E. and {Bragaglia}, A. and {Gratton}, R. and {D'Orazi}, V. and 
	{Lucatello}, S.},
    title = "{Intrinsic iron spread and a new metallicity scale for globular clusters}",
  journal = {\aap},
archivePrefix = "arXiv",
   eprint = {0910.0675},
 primaryClass = "astro-ph.GA",
     year = 2009,
    month = dec,
   volume = 508,
    pages = {695-706},
      doi = {10.1051/0004-6361/200913003},
   adsurl = {http://adsabs.harvard.edu/abs/2009A%26A...508..695C}
}

@ARTICLE{Bragaetal2016,
       author = {{Braga}, V.~F. and {Stetson}, P.~B. and {Bono}, G. and {Dall'Ora}, M. and {Ferraro}, I. and {Fiorentino}, G. and {Freyhammer}, L.~M. and {Iannicola}, G. and {Marengo}, M. and {Neeley}, J. and {Valenti}, E. and {Buonanno}, R. and {Calamida}, A. and {Castellani}, M. and {da Silva}, R. and {Degl'Innocenti}, S. and {Di Cecco}, A. and {Fabrizio}, M. and {Freedman}, W.~L. and {Giuffrida}, G. and {Lub}, J. and {Madore}, B.~F. and {Marconi}, M. and {Marinoni}, S. and {Matsunaga}, N. and {Monelli}, M. and {Persson}, S.~E. and {Piersimoni}, A.~M. and {Pietrinferni}, A. and {Prada-Moroni}, P. and {Pulone}, L. and {Stellingwerf}, R. and {Tognelli}, E. and {Walker}, A.~R.},
        title = "{On the RR Lyrae Stars in Globulars. IV. {\ensuremath{\omega}} Centauri Optical UBVRI Photometry}",
      journal = {\aj},
         year = 2016,
        month = dec,
       volume = {152},
       number = {6},
          eid = {170},
        pages = {170},
          doi = {10.3847/0004-6256/152/6/170},
archivePrefix = {arXiv},
       eprint = {1609.04916},
 primaryClass = {astro-ph.SR},
       adsurl = {https://ui.adsabs.harvard.edu/abs/2016AJ....152..170B}
}

@ARTICLE{Bragaetal2021,
       author = {{Braga}, V.~F. and {Crestani}, J. and {Fabrizio}, M. and {Bono}, G. and {Sneden}, C. and {Preston}, G.~W. and {Storm}, J. and {Kamann}, S. and {Latour}, M. and {Lala}, H. and {Lemasle}, B. and {Prudil}, Z. and {Altavilla}, G. and {Chaboyer}, B. and {Dall'Ora}, M. and {Ferraro}, I. and {Gilligan}, C.~K. and {Fiorentino}, G. and {Iannicola}, G. and {Inno}, L. and {Kwak}, S. and {Marengo}, M. and {Marinoni}, S. and {Marrese}, P.~M. and {Mart{\'\i}nez-V{\'a}zquez}, C.~E. and {Monelli}, M. and {Mullen}, J.~P. and {Matsunaga}, N. and {Neeley}, J. and {Stetson}, P.~B. and {Valenti}, E. and {Zoccali}, M.},
        title = "{On the Use of Field RR Lyrae as Galactic Probes. V. Optical and Radial Velocity Curve Templates}",
      journal = {\apj},
         year = 2021,
        month = oct,
       volume = {919},
       number = {2},
          eid = {85},
        pages = {85},
          doi = {10.3847/1538-4357/ac1074},
archivePrefix = {arXiv},
       eprint = {2107.00923},
 primaryClass = {astro-ph.SR},
       adsurl = {https://ui.adsabs.harvard.edu/abs/2021ApJ...919...85B}
}

@ARTICLE{aumer2017a,
       author = {{Aumer}, Michael and {Binney}, James},
        title = "{The structural evolution of galaxies with both thin and thick discs}",
      journal = {\mnras},
         year = 2017,
        month = sep,
       volume = {470},
       number = {2},
        pages = {2113-2132},
          doi = {10.1093/mnras/stx1300},
archivePrefix = {arXiv},
       eprint = {1705.09240},
 primaryClass = {astro-ph.GA},
       adsurl = {https://ui.adsabs.harvard.edu/abs/2017MNRAS.470.2113A}
}

@ARTICLE{aumer2017b,
       author = {{Aumer}, Michael and {Binney}, James and {Sch{\"o}nrich}, Ralph},
        title = "{Migration and kinematics in growing disc galaxies with thin and thick discs}",
      journal = {\mnras},
         year = 2017,
        month = sep,
       volume = {470},
       number = {3},
        pages = {3685-3706},
          doi = {10.1093/mnras/stx1483},
archivePrefix = {arXiv},
       eprint = {1706.04042},
 primaryClass = {astro-ph.GA},
       adsurl = {https://ui.adsabs.harvard.edu/abs/2017MNRAS.470.3685A}
}

@ARTICLE{Parikh2021,
       author = {{Parikh}, Taniya and {Thomas}, Daniel and {Maraston}, Claudia and {Westfall}, Kyle B. and {Andrews}, Brett H. and {Boardman}, Nicholas Fraser and {Drory}, Niv and {Oyarzun}, Grecco},
        title = "{SDSS-IV MaNGA: radial gradients in stellar population properties of early-type and late-type galaxies}",
      journal = {\mnras},
         year = 2021,
        month = apr,
       volume = {502},
       number = {4},
        pages = {5508-5527},
          doi = {10.1093/mnras/stab449},
archivePrefix = {arXiv},
       eprint = {2102.06703},
 primaryClass = {astro-ph.GA},
       adsurl = {https://ui.adsabs.harvard.edu/abs/2021MNRAS.502.5508P}
}

@ARTICLE{chiappini1997,
       author = {{Chiappini}, C. and {Matteucci}, F. and {Gratton}, R.},
        title = "{The Chemical Evolution of the Galaxy: The Two-Infall Model}",
      journal = {\apj},
         year = 1997,
        month = mar,
       volume = {477},
       number = {2},
        pages = {765-780},
          doi = {10.1086/303726},
archivePrefix = {arXiv},
       eprint = {astro-ph/9609199},
 primaryClass = {astro-ph},
       adsurl = {https://ui.adsabs.harvard.edu/abs/1997ApJ...477..765C}
}

@ARTICLE{Duffau2014,
       author = {{Duffau}, Sonia and {Vivas}, A. Katherina and {Zinn}, Robert and {M{\'e}ndez}, Ren{\'e} A. and {Ruiz}, Mar{\'\i}a T.},
        title = "{A comprehensive view of the Virgo stellar stream}",
      journal = {\aap},
         year = 2014,
        month = jun,
       volume = {566},
          eid = {A118},
        pages = {A118},
          doi = {10.1051/0004-6361/201219654},
archivePrefix = {arXiv},
       eprint = {1403.2388},
 primaryClass = {astro-ph.GA},
       adsurl = {https://ui.adsabs.harvard.edu/abs/2014A&A...566A.118D}
}

@ARTICLE{ForSneden2010,
       author = {{For}, Bi-Qing and {Sneden}, Christopher},
        title = "{The Chemical Compositions of Non-variable Red and Blue Field Horizontal Branch Stars}",
      journal = {\aj},
         year = 2010,
        month = dec,
       volume = {140},
       number = {6},
        pages = {1694-1718},
          doi = {10.1088/0004-6256/140/6/1694},
archivePrefix = {arXiv},
       eprint = {1009.4461},
 primaryClass = {astro-ph.SR},
       adsurl = {https://ui.adsabs.harvard.edu/abs/2010AJ....140.1694F}
}

@ARTICLE{Innoetal2013,
       author = {{Inno}, L. and {Matsunaga}, N. and {Bono}, G. and {Caputo}, F. and {Buonanno}, R. and {Genovali}, K. and {Laney}, C.~D. and {Marconi}, M. and {Piersimoni}, A.~M. and {Primas}, F. and {Romaniello}, M.},
        title = "{On the Distance of the Magellanic Clouds Using Cepheid NIR and Optical-NIR Period-Wesenheit Relations}",
      journal = {\apj},
         year = 2013,
        month = feb,
       volume = {764},
       number = {1},
          eid = {84},
        pages = {84},
          doi = {10.1088/0004-637X/764/1/84},
archivePrefix = {arXiv},
       eprint = {1212.4376},
 primaryClass = {astro-ph.SR},
       adsurl = {https://ui.adsabs.harvard.edu/abs/2013ApJ...764...84I}
}

@ARTICLE{kinman2007,
       author = {{Kinman}, T.~D. and {Cacciari}, C. and {Bragaglia}, A. and {Buzzoni}, A. and {Spagna}, A.},
        title = "{Kinematic structure in the Galactic halo at the North Galactic Pole: RR Lyrae and blue horizontal branch stars show different kinematics}",
      journal = {\mnras},
         year = 2007,
        month = mar,
       volume = {375},
       number = {4},
        pages = {1381-1398},
          doi = {10.1111/j.1365-2966.2006.11394.x},
archivePrefix = {arXiv},
       eprint = {astro-ph/0612416},
 primaryClass = {astro-ph},
       adsurl = {https://ui.adsabs.harvard.edu/abs/2007MNRAS.375.1381K}
}

@ARTICLE{kinman2000,
       author = {{Kinman}, T. and {Castelli}, F. and {Cacciari}, C. and {Bragaglia}, A. and {Harmer}, D. and {Valdes}, F.},
        title = "{A spectroscopic study of field BHB star candidates}",
      journal = {\aap},
         year = 2000,
        month = dec,
       volume = {364},
        pages = {102-136},
          doi = {10.48550/arXiv.astro-ph/0006179},
archivePrefix = {arXiv},
       eprint = {astro-ph/0006179},
 primaryClass = {astro-ph},
       adsurl = {https://ui.adsabs.harvard.edu/abs/2000A&A...364..102K}
}

@ARTICLE{kinman1994,
       author = {{Kinman}, T.~D. and {Suntzeff}, N.~B. and {Kraft}, R.~P.},
        title = "{The Structure of the Galactic Halo Outside the Solar Circle as Traced by the Blue Horizontal Branch Stars}",
      journal = {\aj},
         year = 1994,
        month = nov,
       volume = {108},
        pages = {1722},
          doi = {10.1086/117191},
       adsurl = {https://ui.adsabs.harvard.edu/abs/1994AJ....108.1722K}
}

\end{document}